\documentclass[twoside,english,3p]{elsarticle}
\usepackage[T1]{fontenc}
\usepackage{geometry}
\usepackage{amssymb}
\usepackage{amsmath}

\usepackage{amsthm}
\usepackage{stmaryrd}
\usepackage{graphicx}
\usepackage{booktabs}
\usepackage{multirow}
\usepackage{makecell}
\usepackage{xurl}
\usepackage{esint}
\usepackage{algorithm}
\usepackage{algpseudocode}
\usepackage{rotating}
\usepackage{adjustbox}
\makeatletter
\theoremstyle{plain}

\theoremstyle{boldremark} 

\ifx\proof\undefined

\providecommand{\proofname}{Proof}
\fi

\renewenvironment{proof}[1][\proofname]{%
	\par\pushQED{\qed}\normalfont%
	\topsep6\p@\@plus6\p@\relax
	\trivlist\item[\hskip\labelsep\bfseries#1\@addpunct{.}]%
	\ignorespaces
}{%
	\popQED\endtrivlist\@endpefalse
}

\journal{Elsevier}

\usepackage{hyperref}
\hypersetup{colorlinks = true, allcolors = blue}

\usepackage[nameinlink]{cleveref}

\crefname{figure}{Fig.}{Figs.}
\crefformat{equation}{Eq.~#2(#1)#3}
\crefformat{section}{Section~#2#1#3}
\AtBeginDocument{%
	\let\citet\cite
}

\usepackage[labelfont=bf]{caption}
\@ifundefined{showcaptionsetup}{}{%
	\PassOptionsToPackage{caption=false}{subfig}}
\usepackage{subfig}
\makeatother

\usepackage{babel}
\providecommand{\remarkname}{Remark}
\providecommand{\theoremname}{Theorem}

\begin{document}
	
	\begin{frontmatter}{}
		
		\title{Artificial intelligence for partial differential equations in computational mechanics: A review }
		
		\author[rvt,rvt3]{Yizheng Wang}
		
		\ead{wang-yz19@tsinghua.org.cn}
		
		\author[rvt,rvt4,rvt5]{Jinshuai Bai}
		
		\author[rvt]{Zhongya Lin}
		
		\author[rvt6]{Qimin Wang}
		
		\author[rvt3]{Cosmin Anitescu}
		
		\author[rvt2]{Jia Sun}

		\author[rvt6]{Mohammad Sadegh Eshaghi}
		\author[rvt4,rvt5]{Yuantong Gu}
		
		\author[rvt]{Xi-Qiao Feng}
		
		\author[rvt6]{Xiaoying Zhuang}
		
		\author[rvt3]{Timon Rabczuk\corref{cor1}}
		
		\ead{timon.rabczuk@uni-weimar.de}
		
		\author[rvt]{Yinghua Liu\corref{cor1}}
		
		\ead{yhliu@mail.tsinghua.edu.cn}
		\cortext[cor1]{Corresponding author}
		\address[rvt]{Department of Engineering Mechanics, Tsinghua University, Beijing 100084, China}

		\address[rvt3]{Institute of Structural Mechanics, Bauhaus-Universit\"{a}t Weimar, Marienstr. 15, D-99423 Weimar, Germany}	
		
		\address[rvt6]{Chair of Computational Science and Simulation Technology, Institute of Photonics, Leibniz University Hannover, Hannover 30167, Germany}
		
		\address[rvt2]{Drilling Mechanical Department, CNPC Engineering Technology RD Company Limited, Beijing 102206, China}
		
		\address[rvt4]{School of Mechanical, Medical and Process Engineering, Queensland University of Technology, Brisbane, QLD 4000, Australia}
		
		\address[rvt5]{ARC Industrial Transformation Training Centre—Joint Biomechanics, Queensland University of Technology, Brisbane, QLD 4000, Australia}
		%
		%
		%
		
		\begin{abstract}
			In recent years, Artificial intelligence (AI) has become ubiquitous, empowering various fields, especially integrating artificial intelligence and traditional science (AI for Science: Artificial intelligence for science), which has attracted
			widespread attention. In AI for Science, using artificial intelligence algorithms to solve partial
			differential equations (AI for PDEs: Artificial intelligence for partial differential equations) has become a focal point in computational mechanics. The core of AI for PDEs
			is the fusion of data and partial differential equations (PDEs), which can solve almost any PDEs. 
			Therefore, this article provides a comprehensive review of the research on AI for PDEs, summarizing the existing algorithms and theories. The article discusses the applications of AI for PDEs in computational mechanics, including solid mechanics, fluid mechanics, and biomechanics.
			The existing AI for PDEs algorithms include
			those based on Physics-Informed Neural Networks (PINNs), Deep Energy Methods (DEM), Operator Learning,
			and Physics-Informed Neural Operator (PINO). AI for PDEs represents a new method of scientific simulation
			that provides approximate solutions to specific problems using large amounts of data, then fine-tuning
			according to specific physical laws, avoiding the need to compute from scratch like traditional algorithms.
			Thus, AI for PDEs is the prototype for future foundation models in computational mechanics, capable of significantly
			accelerating traditional numerical algorithms. 
		\end{abstract}
		
	\end{frontmatter}{}

	\section{Introduction}
\begin{figure}
		\begin{centering}
			\includegraphics[scale=0.55]{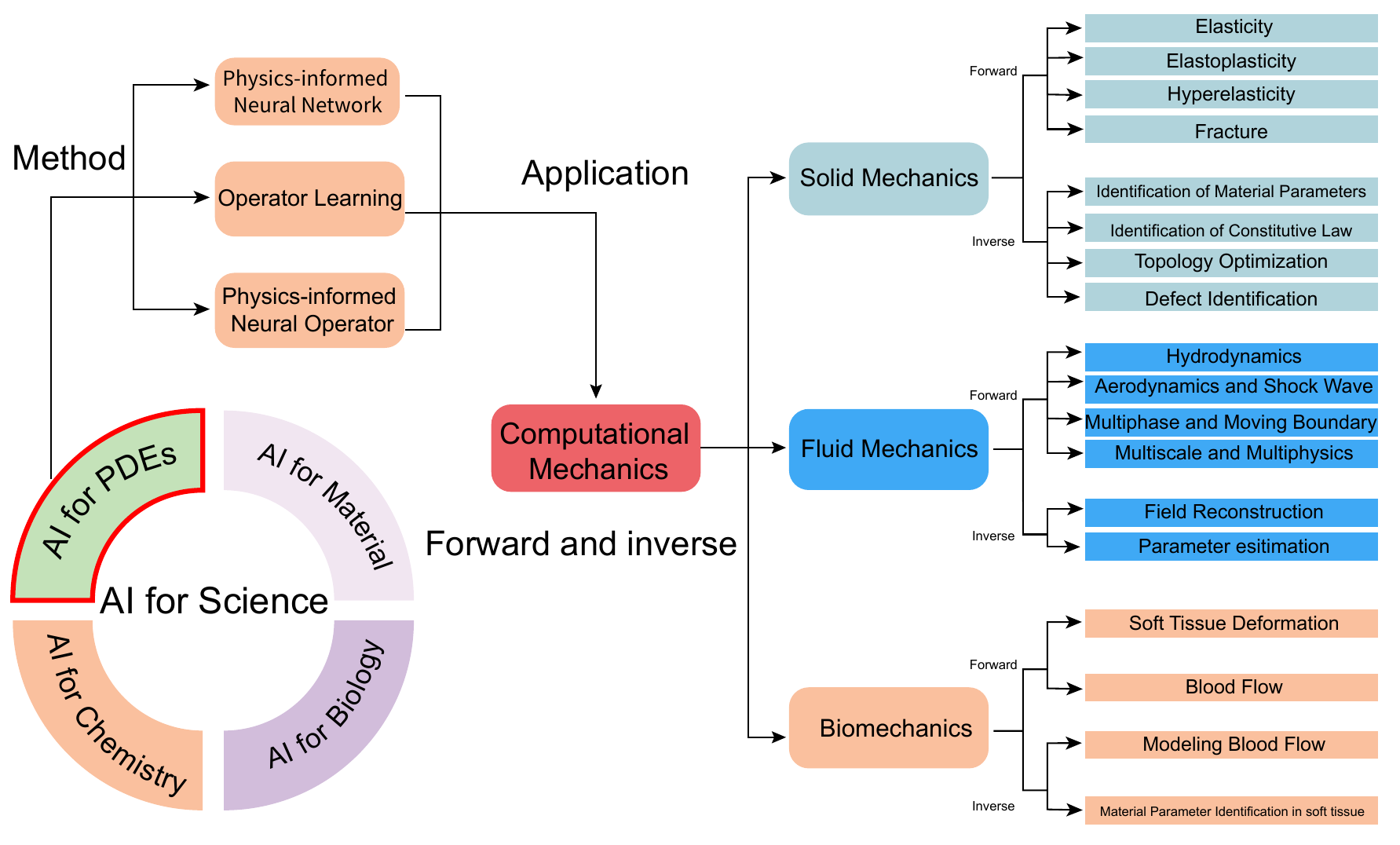}
			\par\end{centering}
		\caption{The role of AI4PDEs in AI4Science, along with an introduction to AI4PDEs in computational mechanics, including solid mechanics, fluid mechanics, and biomechanics.\label{fig:AI4PDEs}}
	\end{figure}

	Recently, the integration of AI and traditional
	science (AI4Science: AI for Science) has drawn broad attention \cite{zhang2023artificial}. Due to many physical computing problems
	being closely related to the solution of partial differential equations (PDEs), using artificial intelligence
	algorithms to solve PDEs \cite{PINN_original_paper}, a key research area known as AI4PDEs, has gained substantial interest among
	researchers in computational physics and mathematics \cite{PINN_review}. This area is recognized as one of the most important
	research directions within the broader field of AI4Science \cite{wang2023scientific}. \Cref{fig:AI4PDEs} displays the relationship
	between AI4PDEs and AI4Science.
	This review specifically focuses on the role of AI4PDEs in computational mechanics. As shown in \Cref{fig:AI4PDEs}, recent progress has rapidly expanded across several major branches of computational mechanics. It is worth emphasizing that computational mechanics is a broad discipline. Accordingly, this review concentrates on representative scenarios where AI4PDEs interacts with solid mechanics, fluid mechanics, and biomechanics, with particular attention to methodological developments, practical challenges, and emerging opportunities in these domains.
	
	AI contributes to PDE-based simulation in two complementary ways. First, it enables the acceleration of existing workflows through surrogate models, neural operators \cite{bi2023accurate} and hybrid approaches \cite{wang2026pretrain} that reduce computational cost while maintaining accuracy. Second, it opens the possibility of scientific discovery \cite{wang2023scientific}, where machine learning models are used to identify unknown parameters, infer constitutive relations or extract governing principles from data. While the former direction is already demonstrating practical impact, the latter remains more challenging due to issues such as data scarcity, noise, and identifiability, particularly in engineering applications \cite{cuomo2022scientific}. In the long term, these developments may contribute to more general and adaptive models for computational mechanics, capable of transferring knowledge across geometries, materials and physical processes. However, such a vision remains far from realization. Current limitations in data availability, generalization and reliability highlight the need for approaches that combine data-driven learning with strong physical structure rather than relying on purely data-driven paradigms.
	
	The field of computational mechanics has also been impacted by AI4Science. The integration of AI4Science
	and computational mechanics mainly occurs in two aspects. One is through deep learning methods, utilizing real experimental data or reliable numerical results \cite{li2019predicting} (such as those obtained from finite element methods)
	and then using neural networks to construct surrogate models. This is a type of implicit programming
	\cite{ill_gradient}, where \textquotedbl implicit programming\textquotedbl{} refers to inputting data
	and utilizing the strong fitting ability of AI algorithms to output \textquotedbl programs,\textquotedbl{}
	which consist of neural network parameters, thus eliminating the need for humans to write
	programs to solve particular problems specifically. This traditional method of explicitly writing programs to solve
	specific problems often requires designing programs to transform abstract principles into computer-readable
	code. Artificial intelligence algorithms represent a form of implicit programming that uses data to build
	surrogate models. The core idea is to use the fitting power of neural networks to model
	the abstract relationships between data \cite{li2019predicting}. If training is successful, and the
	test set and training set are consistent in data distribution, surrogate models often exhibit very high
	computational efficiency and accuracy on the test set. However, this method has some unavoidable problems,
	such as requiring large amounts of data, and the quality and quantity of data determining the effectiveness
	of the model. This integration method relies entirely on data, necessitating the use of existing methods
	to obtain relevant data. Therefore, it faces challenges like the curse of dimensionality \cite{bc-pinn,high_dimension}.
	Additionally, due to the lack of physical constraints, the interpretability is poor, leading to consequences
	that make it difficult to improve the model's effectiveness. This disadvantage causes a significant amount
	of manual, trial and error based parameter tuning, so using neural networks to establish surrogate models is essentially a
	black-box approach. The choice of surrogate models often benefits from computer vision algorithms
	to better integrate with current physical problems. For instance, computer vision algorithms like
	the Swin Transformer were used to predict weather in the PanGu foundation model \cite{bi2023accurate}, or to predict
	the equivalent modulus of non-uniform materials \cite{li2019predicting}. Overall, the first aspect
	is a completely data-driven modeling method.
	The limitation of this type of algorithm is that it only has the potential to be faster, but it is difficult to have the potential to be more accurate, because the data-driven modeling method is based on fitting existing high-precision data, and the accuracy can only be infinitely close to the existing high-precision data.
	
	The other important aspect of the integration between AI4Science and mechanics is the incorporation of
	physical laws into the loss function of neural networks, which is a core part of AI4PDEs. Using physical
	information, by which the need for high-quality data is reduced \cite{hp-VPINN}, often reflected in
	solving PDEs, i.e., using AI4PDEs algorithms to solve PDEs of computational mechanics. The
	introduction of physical equations reduces the data requirement because it incorporates inductive biases,
	which refers to the assumptions on which algorithm models are based, helping the model to make predictions
	and reducing the amount of data needed. PDEs also represent a special form of inductive bias, as physical
	equations are often derived using more basic assumptions (such as assumptions of linearity, small deformations,
	and continuity in elasticity mechanics) and then derived with mathematical tools to describe the laws
	of material motion. Thus, PDEs are a form of advanced inductive bias \cite{lu2022comprehensive}. 
	In this review, AI for PDEs is primarily organized into three major methodological paradigms: Physics-Informed Neural Networks (PINNs), operator learning, and physics-informed neural operators, which currently constitute the dominant research directions in computational mechanics.
	Beyond these three mainstream paradigms, emerging directions such as reinforcement-learning-based PDE control \cite{yu2021reinforcement}, generative surrogate modeling \cite{bastek2024physics}, and foundation-model-assisted scientific computing \cite{jiang2025deepseek} are also attracting increasing attention, although they remain less mature in computational mechanics applications at present.
	Moreover, it is important to note that AI for PDEs do not always replace classical numerical techniques such as finite element methods (FEM). In many applications, AI for PDEs are integrated within traditional PDE solvers, for example, by accelerating multiscale simulations \cite{wang2024homogenius}. These hybrid approaches complement classical methods while leveraging the efficiency and generalization capabilities of AI.

	The use of deep learning to solve PDEs (AI4PDEs) first appeared with the  Physics-Informed Neural Networks (PINNs), a term coined by Raissi et al. in 2019 \citet{PINN_original_paper}.
	However, the concept of using neural networks to solve PDEs dates back to 1990 \cite{use_neural_network_to_solve_PDE}, 1994 \cite{dissanayake1994neural} and 1998 \cite{admissible_earliest_paper}. Because deep neural networks (DNNs), powerful GPU computing, and associated software tools were not available at that time, these ideas did not receive sufficient attention.
	Recently, with the rapid development of computer hardware, advancements in machine learning algorithms, and the convenient implementation of automatic differentiation algorithms within artificial intelligence frameworks like PyTorch and TensorFlow, the PINNs algorithm has garnered considerable attention, leading to an increase in the complexity of the problems that can be solved \cite{cuomo2022scientific}.
	Compared to traditional numerical methods, PINNs has two main advantages: firstly, utilizing the powerful fitting capabilities of neural
	networks for solving challenging PDEs problems, such as those involving significant nonlinearity, convection-dominated,
	or shock problems, provides a new method of scientific computation \cite{cuomo2022scientific}. For
	forward problems, since PINNs are a type of mesh-free method where the approximate function is a neural
	network and the solution is obtained for the entire domain at once, they share similar advantages with
	mesh-free methods, such as handling large deformation problems with significant mesh distortion. For
	inverse problems, the code format can be easily adapted, allowing for straightforward application with
	automatic differentiation (AD: Automatic Differentiation) \cite{mcclenny2023self}. The second advantage
	is its suitability for solving high-dimensional problems, mainly due to the construction of the PINNs
	loss function using the Monte Carlo method, which itself is a powerful tool for high-dimensional integral
	calculations. Therefore, the advantages of PINNs in solving high-dimensional problems do not originate
	from neural networks but from the Monte Carlo method used in the loss function. In theory, if traditional
	methods also adopt the Monte Carlo approach for integral calculations, they would possess the same capabilities
	as PINNs in solving high-dimensional problems.
	PINNs can solve almost all PDE
	problems because the core of PINNs involves setting the approximate function as a neural network, so
	the computational approach is similar to traditional methods. Thus, any PDEs problem that traditional
	methods can solve, PINNs can also address, such as linear PDEs \cite{PINN_energy_form_to_solve_C0_without_subdomains,deep_ritz},
	non-linear PDEs \cite{PINN_solid_mechanics,abueidda2021meshless}, stochastic PDEs \cite{yang2020physics,zhang2019quantifying},
	and integral differential equations (IDEs) \cite{yuan2022pinn}. 
	PINNs also have drawbacks, including a lack of robustness, accuracy, and computational efficiency \cite{wang2022cenn} in many applications. For instance, even when the forward problem is well-posed and linear, the PINN formulation can lead to ill-posed (nonlinear) optimization problems \cite{wang2023dcm}.

\begin{figure}
		\begin{centering}
			\includegraphics[scale=0.65]{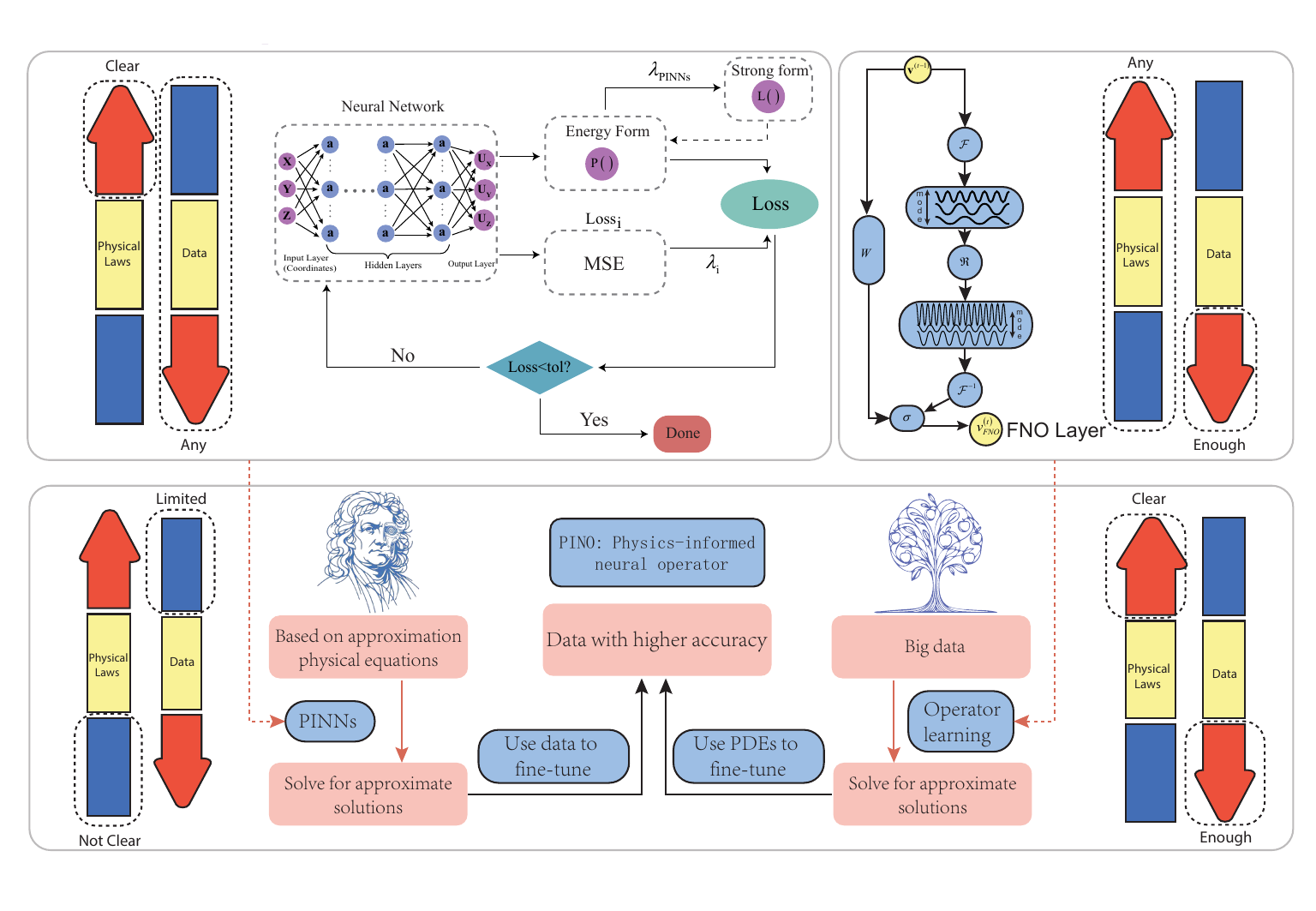}
			\par\end{centering}
		\caption{Main methods of AI4PDEs: Physics-informed neural networks \citet{PINN_original_paper,loss_is_minimum_potential_energy,deep_ritz}, operator learning \cite{DeepOnet,li2020fourier,kovachki2023neural}, and physics-informed
			neural operators \cite{li2021physics,chakraborty2021transfer}.
			PINNs are primarily trained using governing physical equations and do not intrinsically require observational data. Operator learning is a data-driven approach for approximating solution operators from input-output pairs, while PINO integrates physical equations into operator learning, enabling physics-informed training of neural operators.
			\label{fig:AI4PDEs_method}}
	\end{figure}

	Mathematically, PINNs are divided into
	two formats: the original strong form of PINNs \cite{PINN_original_paper} and the energy form of PINNs
	\cite{deep_ritz}. The strong form and energy form of PINNs are explained in detail in \Cref{subsec:Strong-Form}
	and \Cref{subsec:Energy-Form}, respectively. The energy form is the deep Ritz method introduced by Yu
	in 2018 \cite{deep_ritz}, which was later applied to the field of computational mechanics by Samaniego
	et al. (2020) \cite{loss_is_minimum_potential_energy}, proposing the Deep Energy Method (DEM). 
	The core idea of DEM is minimizing the potential energy of the system. In contrast to the strong form of PINNs, which utilizes PDEs, DEM approaches the problem through the system's total energy using the variational principle.
	Notably,
	although DEM mainly follows the energy form, it also incorporates the idea of the strong form of PINNs
	in the original review \cite{loss_is_minimum_potential_energy}, called the deep collocation method
	(DCM), which is the strong form of PINNs. It can be said that DEM is the first major summary and large-scale
	application of PINNs in computational mechanics. The main difference between the strong form and energy
	form of PINNs lies in the composition of the loss function. The loss function of the strong form combines
	PDEs directly through weighted residuals, while the loss function of DEM is based on the principle of
	least action, such as the principle of minimum potential energy in mechanics. Both the strong form and
	energy form of PINNs are shown in the upper left corner of \Cref{fig:AI4PDEs_method}. A significant advantage
	of PINNs is that they can combine data and physical equations, making them a semi-supervised algorithm
	particularly suitable for problems involving both data and equations. Additionally, PINNs can be easily
	adapted for inverse problems, with the programming code being nearly identical to that for forward problems.
	It only requires setting the variables to be solved in the inverse problem as trainable optimization
	variables and incorporating them into the loss function. Thus, compared to traditional inverse problem algorithms, PINNs is very easy to program for solving inverse problems \cite{PINN_review}.
	
	Overall, PINNs leverage existing physical laws, significantly reducing the need for datasets and offering
	better interpretability than purely data-driven approaches.
	However, assigning physical meaning to the hyperparameters in PINNs remains challenging, especially in comparison to the finite element method (FEM).
	The accuracy of PINNs in solving
	forward problems was difficulties in surpassing traditional methods. The main reason is the non-convex
	nature of neural networks. Although the non-convexity of neural networks provides powerful fitting capabilities,
	it also introduces difficulties in optimization. PINNs are suitable for problems with clear physical
	laws and small data volumes. Of course, if the data volume is large enough, PINNs can be used, but in
	such cases, operator learning would be more appropriate (please note that the original PINNs do not require
	a specific data volume). 
	
	In recent years, operator learning has become a hot research topic, as shown
	in the upper right corner of \Cref{fig:AI4PDEs_method}. Representative works include DeepONet \cite{DeepOnet} and FNO (Fourier Neural Operator) \cite{li2020fourier}.
	The initial proposals for operator learning were purely data-driven, making them highly suitable for
	large data volume problems (even if the physical processes are unclear, operator learning can still rely
	on learning the abstract complex relationships within the data). The difference from traditional data-driven
	methods lies in the reliable mathematical theory supporting operator learning and the advantage of discretization-invariance in operator learning. In terms of theoretical support, for example, DeepONet is based on the
	theory proposed by Chen et al. (1995) \cite{chen1995universal} that neural networks can fit any continuous
	operator, which led to the design of the algorithm. FNO is based on Fourier transforms; discretization-invariance
	roughly refers to the ability to train and test on any grid, and after the mesh size changes, retraining
	is not required. This will be explained in detail in \Cref{subsec:Operator-Learning}. 
	It is important to note that some neural operator algorithms actually learn the mapping from a discrete input function to a discrete output function space, rather than the differential operator itself. To address this, Bartolucci et al. \cite{bartolucci2024representation} proposed RENOs (Representation Equivalent Neural Operators) to learn the operator mapping of PDEs.
	
	Existing approaches in computational mechanics differ primarily in how they handle generalization and solution strategies \cite{kahana2023geometry}. Neural operator \cite{lu2019deeponet,li2020fourier} and transformer-based models \cite{hao2023gnot} aim to learn parametric mappings between problem inputs, such as geometry, material parameters and boundary conditions, and the corresponding solution fields. Once trained, these models enable fast evaluation across families of problems without resolving each instance from more detailed simulations. In contrast, classical numerical methods such as the finite element or finite volume methods, solve each problem instance independently through discretization and iterative solution procedures \cite{wang2021learning}. While modern simulation software platforms provide broad applicability across geometries and physics, the underlying solution process remains instance-specific and computationally intensive. Bridging these two paradigms, combining the generalization capabilities of learned operators with the robustness and accuracy of classical solvers, represents a central challenge in AI-assisted scientific computing \cite{wang2026pretrain, eshaghi2025nows}.
	
	A key obstacle in this context is not only the solution of the governing equations, but the representation and processing of the geometry \cite{wu2024transolver}. This challenge extends beyond classical CAD-based workflows and includes a wide range of geometric representations, such as surface triangulations, voxelized data, level-set descriptions, and point clouds obtained from imaging or scanning technologies \cite{hughes2005isogeometric}. Traditional simulation pipelines require these representations to be converted into watertight volumetric meshes, a process that is often brittle, time-consuming and difficult to automate.
		Recent developments in hybrid neural operator-based methods such as PFEM \cite{wang2026pretrain} suggest an alternative perspective, where geometry can be treated in a more flexible and direct manner. In such approaches, geometry is encoded through implicit or discrete representations and processed together with physical parameters to predict solution fields. Hybrid strategies that combine these models with classical solvers provide a promising direction, where learned operators can serve as fast predictors or initial guesses, while numerical solvers ensure robustness and accuracy \cite{eshaghi2025nows,wang2024homogenius}.
		In this setting, the choice of physical constraints plays a critical role. Recent studies based formulations based on variational principles or energy minimization provide a natural and effective way to enforce physical consistency \cite{eshaghi2025variational}. Such approaches can improve stability, reduce the need for large training datasets, and enable closer integration between learning-based models and classical numerical methods.
	
	Moreover, many complex engineering
	problems face the challenge of unclear physical processes. Even if PDEs are used to describe them, they
	are not completely accurate, and the data volume is also limited. The latest development of physics-informed
	neural operators (PINO: Physics-informed neural operator \cite{li2021physics}) can handle these issues
	of vague physics and insufficient data to some extent, as shown in the lower left corner of \Cref{fig:AI4PDEs_method}.
	Initially, an approximate solution is obtained by relying on approximate physical equations and corresponding
	boundary initial conditions, which is then fine-tuned using limited data \cite{chakraborty2021transfer}.
	Additionally, for problems with large data volumes and clear physical processes, physics-informed neural
	operators can also be effectively applied to these issues \cite{goswami2022physics,li2021physics,wang2021learning,wang2023dcm},
	as shown in the lower right corner of \Cref{fig:AI4PDEs_method}. The approach is to first obtain a good
	approximate solution using operator learning and then fine-tune it based on clear physical processes.
	Since solving PDEs relies on operator learning partly, it can significantly accelerate PDEs computations,
	thousands to tens of thousands of times faster than traditional numerical methods \cite{hao2020ai}.
	At the same time, iterating the approximate solution by operator learning according to the physical equation
	does not require much time because the initial solution (approximate solution) is not far from the true
	solution. This simultaneously possesses the speed of operator learning and the accuracy of physical equations.
	Therefore, it can reduce the dependence on supercomputers for solving complex PDEs and solve larger problems
	with fixed computing power. Thus, the potential of PINO is substantial not only in academia but also
	in industry. A detailed explanation will be provided in \Cref{subsec:PINO=00FF1A-Physics-Informed-Neural}.
	It is worth noting that PINO is unlike the initially proposed PINNs which can only solve specific problems,
	i.e., once boundary conditions, geometric shapes, or materials change, a new solution is required in
	PINNs. However, PINO learns a series of mappings of PDEs family, allowing for rapid attainment of target
	solutions even if the aforementioned conditions change. Therefore, this will be one of the most promising
	directions in the field of computational mechanics. The essence of PINO is the combination of operators
	and physical equations, incorporating the ideas of PINNs in the physical equations and operator learning
	using data-driven like DeepONet or FNO. Therefore, PINO simultaneously includes
	PINNs and operator learning. If the data volume is large enough, PINO can be reduced to operator learning.
	If the physical process is clear, it can be reduced to PINNs.
	
	Finally, a fundamental practical limitation in AI-driven computational mechanics is the availability of high-quality data \cite{li2021physics}. Unlike fields such as computer vision, where large standardized datasets are readily available \cite{deng2009imagenet}, data in computational mechanics often requires expensive numerical simulations or experiments. This limitation motivates the development of physics-informed and hybrid approaches that reduce reliance on large datasets while maintaining predictive accuracy \cite{eshaghi2025nows,wang2026pretrain}.    
	
	The outline of this article is shown in \Cref{fig:AI4PDEs}. \Cref{sec:AI4PDEs_method} systematically summarizes AI4PDEs algorithms, including the strong form of PINNs,
	energy form, operator learning, and physics-informed neural operators.  \Cref{sec:AI4PDEs_theory} introduces some theoretical
	work on AI4PDEs.  \Cref{sec:AI4PDEs_forward,sec:AI4PDEs_inverse}  discuss
	some applications of AI4PDEs in computational mechanics, including forward  and inverse problems in solid,
	fluid, and biomechanics\footnote{Please note that most of the images in \Cref{sec:AI4PDEs_forward} and \Cref{sec:AI4PDEs_inverse} are adapted from the most important papers closely related to this review, combined with text and formulas, to clearly explain their important ideas. Due
		to the limited length of this article, for details on the algorithms, please refer to the original papers.}.   \Cref{sec:Conclusion} predicts the likely emergence of the foundation model in
	computational mechanics and the potential opportunities and challenges AI4PDEs may face in the future.
	
	\section{AI for PDEs: Methodology}\label{sec:AI4PDEs_method}
	
	A multitude of physical phenomena rely on PDEs for modeling. Once the boundary and
	initial conditions become complex, the analytical solution of PDEs is often difficult to obtain. At this
	point, various numerical methods are employed to achieve approximate solutions, such as the commonly used
	finite element methods \cite{finite_element_book,hughes2012finite,bathe2006finite,reddy2019introduction},
	mesh-free methods \cite{zhang2016material,liu2003mesh,rabczuk2004cracking,rabczuk2007three,rabczuk2019extended,nguyen2008meshless},
	finite difference methods \cite{leveque2007finitedifferentialmethod}, finite volume methods \cite{darwish2016finitevolumemethod},
	boundary element methods \cite{brebbia2012boundary}, and spectral methods \cite{karniadakis2005spectral}. Despite the great success of traditional numerical
	methods in computational mechanics over the past 50 years, we still struggle to integrate data into traditional
	algorithms, especially in industries where multimodal, multi-resolution, and high-error complex data are
	common \cite{cai2021physics}. Moreover, finding a model that can describe such complex systems is also
	a challenge \cite{kovachki2023neural}, as PDEs describing these systems are also models \cite{chakraborty2021transfer}.
	Therefore, integrating data into computational models is of significant importance to describe complex
	systems. Additionally, when solving inverse problems and large, complex nonlinear problems, traditional
	algorithms often involve large computational loads and complex formats, especially when dealing with
	highly complex nonlinear issues, which require vast amounts of code and are not friendly to updates \cite{PINN_review}, such as OpenFOAM \cite{jasak2007openfoam}, which has more than 100,000 lines of code.
	For low-dimensional and relatively simple geometric problems, traditional numerical methods can offer high computational accuracy and efficiency. 
	
	While these methods, such as the FEM, are quite effective in dealing with many complex problems, challenges can arise when dealing with high-dimensional problems and complex geometric shapes \cite{high_dimension,complex_PINN_a_method_to_construct_admissible_function}.
	However, the latest AlphaGeometry \cite{AlphaGeometry} may assist mesh generators in the future.
	Additionally, traditional FEM often require the selection of specific basis functions (shape functions). 
	
	AI for PDEs, unlike traditional numerical methods, refers to a class of algorithms that use deep learning
	to solve PDEs. From the perspective of deep learning, it can be seen as learning a
	mapping relationship, which maps the input field through a neural network to the interested physical
	field and this mapping is achieved through composite functions, which is different from the current
	the predominance of additive methods for approximating mappings. The reason why deep learning is effective in solving PDEs boils down to the following points:
	\begin{itemize}
		\item Powerful approximation capability of neural networks: Under the condition of monotonically bounded activation
		functions, with a certain distribution of hidden layer neurons (at least one hidden layer), any continuous
		function can be approximated \cite{super_approximation,hornik1989multilayer}. Therefore, for some field
		functions with drastic changes, AI4PDEs can theoretically select the best basis functions automatically
		through the powerful fitting capabilities of neural networks, thereby avoiding the need to design shape
		functions like finite elements, which undoubtedly reduces the intellectual cost \cite{wang2022cenn}.
		Additionally, neural networks can also fit any continuous operator \cite{chen1995universal}, with DeepONet
		proposed by Lu et al. \cite{DeepOnet} utilizing this theory. This theory \cite{chen1995universal}
		provides a good theoretical prospect for the current hot research topic of operator learning.
		\item Direct integration of data and physical laws: AI4PDEs can utilize the powerful fitting capabilities of
		neural networks to learn the abstract rules behind data and physical equations, especially for complex
		physical processes where the physical laws are not very clear, and where the data volume is not large.
		An approximate solution can be obtained based on the physical equation and then adjusted with a small
		amount of data to achieve a solution closer to the real laws compared to traditional numerical methods
		\cite{chakraborty2021transfer}. Additionally, for domains with large data volumes and clear physical
		laws, operator learning can be used to pre-train the data and then fine-tune it according to the physical
		equation, significantly improving the speed of traditional numerical methods because the initial solution
		obtained by operator learning needs fewer iteration steps compared to a random initial solution by traditional
		numerical methods. This direction is currently one of the most cutting-edge in AI4PDEs \cite{wang2021learning,goswami2022physics,li2021physics}.
		\item Precision and convenience of derivatives by automatic differentiation: This is because backpropagation
		uses the method of analytic gradient flow to calculate gradients, rather than using numerical methods,
		thereby obtaining precise gradient values. This feature has been efficiently implemented in deep learning
		frameworks, making backpropagation very easy, thus AI4PDEs are simple in computation format and conducive
		to updates and iterations \cite{cai2021physics}. Although the mathematical format of AI4PDEs
		is not simple, thanks to the current optimization frameworks, such as Pytorch and Tensorflow, these frameworks
		facilitate the programming of AI4PDEs \cite{PINNlibrary,bai2023introduction}, and the core solver usually does not exceed a hundred lines
		of code.  Additionally, due to the wide applicability and adaptability of these frameworks, it reduces the need for people to be deeply familiar with traditional algorithms, especially inverse problem \cite{PINN_review}. 
	\end{itemize}
	Due to the powerful fitting capabilities of neural networks as trial functions, the new paradigm of integrating
	data and equations, coupled with convenient code implementation, these unique advantages make AI4PDEs
	important and highly promising in solving mechanical PDEs, i.e., computational mechanics. AI4PDEs represent
	a class of numerical algorithms independent of finite elements, providing another possibility for computational
	mechanics. AI4PDEs have two important modules, the physical and data-driven modules, which we will introduce
	separately. AI4PDEs' physical module is based on the idea of PINNs, with the two mainstream methods being
	the strong form and energy form of PINNs, both of which essentially use neural networks to replace the
	approximate functions in the weighted residual method. The strong form of PINNs often writes the domain
	equation, boundary conditions, and initial conditions into the loss function through hyperparameter-weighted
	residuals for optimization, different choices of approximate functions (trial functions), and weight functions (test functions) result in
	different numerical methods of PINNs strong form, similar to the research idea of different numerical
	methods arising from different trial and test functions in finite elements. PINNs' strong
	form often does not rely on numerical integration, and since all PDEs have a weighted residual form,
	the strong form of PINNs is general, almost any PDEs can be solved. However, PINNs' strong form has numerous hyperparameters, often requiring tuning, and when facing specific problems, the optimal hyperparameters
	are often unknown. Another important method of PINNs is the energy form, which essentially uses the variational
	principle, converting the strong form into an integral functional. The advantage of PINNs energy form
	is the dramatic reduction in hyperparameters, often higher accuracy and efficiency (the variational principle
	has fewer derivative orders). The drawback of PINNs energy form is also very obvious: it is highly dependent
	on the choice of numerical integration scheme, also, not all PDEs have a well-behaved
	extremum variational principle, so its generality is not as good as PINNs strong form. AI4PDEs' data-driven
	module is based on the recently developed operator learning, which is actually a special kind of
	pure data-driven algorithm. Using operator learning to learn the mapping of PDEs families, this 
	mapping can be understood as the same type of PDEs under different parameters. The reason why it is possible
	to use neural networks to fit PDEs solutions is because we can understand PDEs as a kind of implicit mapping,
	which maps boundary conditions, geometric shapes, material fields to the solution
	fields. However, the mapping is very complex and difficult to write out in an analytic explicit form, therefore, it might be possible to use neural networks
	to learn the mapping. If successful, then inputting different boundary conditions, geometric
	shapes, and material fields can quickly obtain the needed field variables (such as displacement field),
	and the high-precision data obtained using traditional methods is the premise for learning PDEs' implicit
	mapping. Some recent works hope to combine physical equations into the operator \cite{wang2021learning,goswami2022physics,li2021physics,xu2024worth},
	which can use the physical equation to fine-tune the initial solution given by operator learning, thereby
	achieving lower computational cost compared to purely physics-based models. The current direction is still explored
	and has great research prospects, but the key is how to better combine prior knowledge and data together
	to better train the model.
	
	Considering that the physical and data modules are the two most important parts of AI4PDEs, this chapter
	will review the AI4PDEs physical module (PINNs: Physics-informed neural networks), including the strong
	form and energy form. In addition, the AI4PDEs data module will be reviewed, including operator learning
	and the integration of data and physical modules (PINO: Physics-informed neural operators).
	
	\subsection{PINNs: Physics-informed neural networks}
	
	\subsubsection{Strong form\label{subsec:Strong-Form}}

	PINNs consist of two main aspects:
	\begin{itemize}
		\item The first is the use of neural networks to replace the approximate functions. The core of finite element
		in computational mechanics involves the selection of trial functions and test functions, and different
		selections form various types of finite element methods. PINNs mainly change shape function of FEM to neural networks in the trial function. 
		The solution procedures of FEM and PINNs are fundamentally different. PINNs solve a non-convex optimization problem, whereas FEM addresses a linearized system of equations.
		
		\item The second aspect is the adoption of convenient automatic differentiation technology from artificial
		intelligence algorithm frameworks in the optimization algorithm to implement the physical equation loss
		function formed by PDEs. Using PDEs to construct the loss function essentially restricts the optimization
		space of neural networks, thereby obtaining an approximate solution space \cite{cuomo2022scientific}. 
	\end{itemize}
	
	The idea of PINNs is very simple and straightforward, fundamentally introducing pre-known knowledge into
	the neural network, reducing the optimization space, and thereby improving accuracy and efficiency. The
	idea of combining traditional knowledge into algorithms has been very common in traditional machine
	learning, for example, Lauer et al. (2008) \cite{lauer2008incorporating} embedded the prior knowledge
	of functions and their derivatives as constraints into support vector regression (SVR: Support Vector
	Regression) to reduce approximation errors.
	
\begin{figure}
		\begin{centering}
			\includegraphics[scale=1.2]{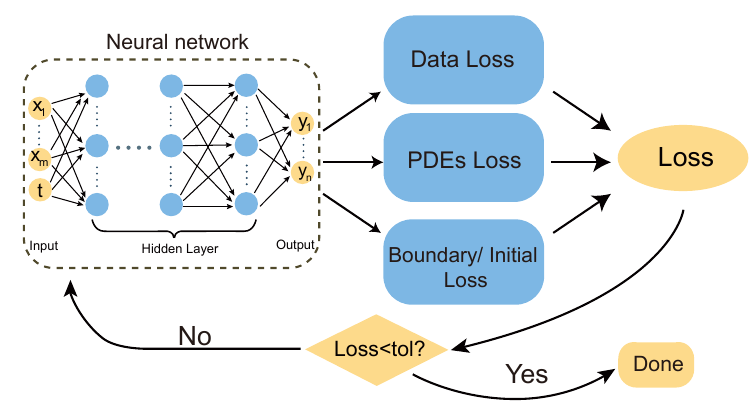}
			\par\end{centering}
		\caption{AI for PDEs method: Schematic of PINNs strong form \citet{PINN_original_paper}.\label{fig:PINN_strong_form}}
	\end{figure}

	PINNs were first introduced in the strong form, as shown in \Cref{fig:PINN_strong_form}. The strong form
	of PINNs is an approximation function using neural networks in the weighted residual method. Considering the specific PDE, the original
	PDEs is:
	\begin{equation}
		\begin{cases}
			\text{Domain PDEs: } \boldsymbol{L}(\boldsymbol{u}(\boldsymbol{x},t))=\boldsymbol{f}(\boldsymbol{x},t) & \forall(\boldsymbol{x},t)\in\Omega\vartimes(0,T]\\
			\text{Boundary condition: } \boldsymbol{u}(\boldsymbol{x},t)=\boldsymbol{h}(\boldsymbol{x},t) & \forall(\boldsymbol{x},t)\in\partial\Omega\vartimes(0,T]\\
			\text{Initial condition: } \boldsymbol{u}(\boldsymbol{x})=\boldsymbol{g}(\boldsymbol{x}) & \forall(\boldsymbol{x},t)\in\Omega\vartimes\{t=0\}
		\end{cases}.
	\end{equation}
	We change the original equation into residual form, divided into three parts:
	
	\begin{equation}
		\begin{cases}
			\text{Domain PDEs residual: } \boldsymbol{r}_{d}(\tilde{\boldsymbol{u}}(\boldsymbol{x},t))=\boldsymbol{L}(\tilde{\boldsymbol{u}}(\boldsymbol{x},t))-\boldsymbol{f}(\boldsymbol{x},t) & \forall(\boldsymbol{x},t)\in\Omega\vartimes(0,T]\\
			\text{Boundary condition residual: } \boldsymbol{r}_{b}(\boldsymbol{x},t)=\tilde{\boldsymbol{u}}(\boldsymbol{x},t)-\boldsymbol{h}(\boldsymbol{x},t) & \forall(\boldsymbol{x},t)\in\partial\Omega\vartimes(0,T]\\
			\text{Initial condition residual: } \boldsymbol{r}_{ini}(\boldsymbol{x})=\tilde{\boldsymbol{u}}(\boldsymbol{x},t=0)-\boldsymbol{g}(\boldsymbol{x}) & \forall(\boldsymbol{x},t)\in\Omega\vartimes\{t=0\}
		\end{cases}, 
	\end{equation}
	where $\boldsymbol{L}$ is the differential operator, $\boldsymbol{f}$ is the non-homogeneous term
	of the PDEs, $\Omega$ is the spatial domain, $\partial\Omega$ is the boundary,
	$\boldsymbol{x}$ is the spatial coordinate, $t$ is time. $\boldsymbol{u}$ is the field of interest approximated by the neural network, and controlled by the parameters
	of the neural network. The neural network is denoted as $\boldsymbol{u}(\boldsymbol{x},t;\boldsymbol{W})$, where $\boldsymbol{W}$
	represents the parameters of the neural network.  $\boldsymbol{h}$ and $\boldsymbol{g}$ are the conditions that must be satisfied by the boundary
	conditions and initial conditions, respectively, which can often be analytically represented as functions
	$\boldsymbol{h}(\boldsymbol{x},t)$ and $\boldsymbol{g}(\boldsymbol{x})$, respectively. $\boldsymbol{r}_{d}$,
	$\boldsymbol{r}_{b}$, and $\boldsymbol{r}_{ini}$ are the domain residual, boundary condition residual,
	and initial condition residual, respectively.
	
	The above residuals $\boldsymbol{r}_{d}$, $\boldsymbol{r}_{b}$, and $\boldsymbol{r}_{ini}$ are numerically
	summed using weight functions:
	\begin{equation}
		\begin{cases}
			\text{Domain PDEs residual weighted integral: } \boldsymbol{R}_{i}^{d}=\sum_{I=1}^{N_{d}}\boldsymbol{v}_{i}^{d}(\boldsymbol{x}_{I},t_{I})\varodot\boldsymbol{r}_{d}(\tilde{\boldsymbol{u}}(\boldsymbol{x}_{I},t_{I})) & \forall(\boldsymbol{x},t)\in\Omega\vartimes(0,T]\\
			\text{Boundary condition residual weighted integral: } \boldsymbol{R}_{i}^{b}=\sum_{I=1}^{N_{b}}\boldsymbol{v}_{i}^{b}(\boldsymbol{x}_{I},t_{I})\varodot\boldsymbol{r}_{b}(\tilde{\boldsymbol{u}}(\boldsymbol{x}_{I},t_{I})) & \forall(\boldsymbol{x},t)\in\partial\Omega\vartimes(0,T]\\
			\text{Initial condition residual weighted integral: } \boldsymbol{R}_{i}^{ini}=\sum_{I=1}^{N_{ini}}\boldsymbol{v}_{i}^{ini}(\boldsymbol{x}_{I})\varodot\boldsymbol{r}_{ini}(\tilde{\boldsymbol{u}}(\boldsymbol{x}_{I})) & \forall(\boldsymbol{x},t)\in\Omega\vartimes\{t=0\}
		\end{cases},  \label{eq:PINNs_strong_form_residual}
	\end{equation}
	where $\boldsymbol{v}_{i}^{d}$, $\boldsymbol{v}_{i}^{b}$, and  $\boldsymbol{v}_{i}^{ini}$ are the domain,
	boundary, and initial weight functions, respectively, whose dimensions correspond one-to-one with $\boldsymbol{r}_{d}$,
	$\boldsymbol{r}_{b}$, and $\boldsymbol{r}_{ini}$. $\boldsymbol{R}_{i}^{d}$, $\boldsymbol{R}_{i}^{b}$, and $\boldsymbol{R}_{i}^{ini}$
	are the domain residual integral, boundary condition residual integral, and initial condition residual
	integral, respectively. $\odot$ is the element-wise operation, meaning corresponding elements are multiplied
	without changing the tensor's shape. $N_{d}$, $N_{b}$, and $N_{ini}$ are the total number of numerical
	summation points in the domain, boundary, and initial areas, respectively. All of the above integral
	forms of residuals are weighted summed to form the loss function of PINNs in the strong form:
	
	\begin{equation}
		\begin{alignedat}{1}\mathcal{L} & =\lambda_{pdes}\mathcal{L}_{pdes}+\lambda_{b}\mathcal{L}_{b}+\lambda_{data}\mathcal{L}_{data}\\
			\mathcal{L}_{pdes} & =\sum_{j=1}^{S^{d}}\boldsymbol{\beta}_{j}^{d}\cdot(\boldsymbol{R}_{j}^{d})^{2}\\
			\mathcal{L}_{b} & =[\sum_{j=1}^{S^{b}}\boldsymbol{\beta}_{j}^{b}\cdot(\boldsymbol{R}_{j}^{b})^{2}+\sum_{j=1}^{S^{ini}}\boldsymbol{\beta}_{j}^{ini}\cdot(\boldsymbol{R}_{j}^{i})^{2}]\\
			\mathcal{L}_{data} & =\sum_{I=1}^{N_{data}}[\tilde{\boldsymbol{u}}(\boldsymbol{x}_{I},t_{I})-\bar{\boldsymbol{u}}(\boldsymbol{x}_{I},t_{I})]^{2},
		\end{alignedat}
		\label{eq:PINNs_strong_form_loss}
	\end{equation}
	where $S^{d}$, $S^{b}$, and $S^{ini}$ are the number of weight functions in the domain, boundary,
	and initial conditions, respectively; $\boldsymbol{\beta}_{j}^{d}$, $\boldsymbol{\beta}_{j}^{b}$, and
	$\boldsymbol{\beta}_{j}^{ini}$ are the weights of the domain, boundary, and initial condition residual
	integrals, respectively. $\lambda_{pdes}$, $\lambda_{b}$, and $\lambda_{data}$ are the weights of the PDEs, boundary (initial), and
	data loss, respectively. $N_{data}$ is the number of existing high-precision data points. The training
	process involves adjusting the trainable weights of the neural network to minimize the loss function.
	
	All important methods of PINNs in the strong form are almost always developed around \Cref{eq:PINNs_strong_form_residual}
	and \Cref{eq:PINNs_strong_form_loss}. This article summarizes the existing important methods of PINNs
	in strong form from the perspectives of whether the trial functions and test functions are partitioned,
	types of trial functions and test functions, and the method of imposing essential boundary conditions,
	as shown in \Cref{tab:PINNs_strong_form_review}, where partitioning refers to dividing different areas
	into different neural networks. 
	
	\begin{table}
		\caption{Current status of PINNs strong form methods\label{tab:PINNs_strong_form_review}}
		
		\centering{}%
		\begin{adjustbox}{max width=\textwidth}
			\begin{tabular}{cccccc}
				\toprule 
				Methods & Trial function: subdomains & Test function: subdomains & Trial function: type & Test function: type & Method of Imposing Essential Boundary\tabularnewline
				\midrule
				PINNs \citet{PINN_original_paper} & No (Global) & No (Global) & Fully Connected & Dirac delta & Penalty Function\tabularnewline
				DGM \citet{sirignano2018dgm} & No (Global) & No (Global) & Fully Connected & Least Squares & Penalty Function\tabularnewline
				VPINNs \citet{kharazmi2019variational}  & No (Global) & No (Global) & Fully Connected & Mixed Weight Functions & Penalty Function\tabularnewline
				PIELM \citet{dwivedi2020physics} & Yes (Subdomains) & No (Global) & ELM\cite{huang2006extreme} & Dirac delta & Penalty Function\tabularnewline
				cPINN \citet{CPINN} & Yes (Subdomains) & No (Global) & Fully Connected & Dirac delta & Penalty Function\tabularnewline
				XPINN \citet{XPINN} & Yes (Subdomains) & No (Global) & Fully Connected & Dirac delta & Penalty Function\tabularnewline
				hp-VPINNs \citet{hp-VPINN}  & No (Global) & Yes (Subdomains) & Fully Connected & Orthogonal Polynomials & Penalty Function\tabularnewline
				PhyGeoNet \citet{gao2021phygeonet} & No (Global) & No (Global) & Convolutional Network (CNN) & Dirac delta & Strictly Enforced\tabularnewline
				PIGCN \citet{gao2022physics} & No (Global) & No (Global) & Graph Convolution (GCN) & Trial Function Space & Strictly Enforced\tabularnewline
				SPINN \citet{ramabathiran2021spinn} & No (Global) & No (Global) & Radial Basis Function & Trial Function Space & Penalty Function\tabularnewline
				BINN \citet{sun2023binn} & No (Global) & No (Global) & Fully Connected & Fundamental Solution of PDEs & Boundary Integral Terms\tabularnewline
				KINN \citet{wang2024kolmogorov} & No (Global) & No (Global) & Kolmogorov Arnold Network \cite{liu2024kan} & Dirac delta & Penalty Function\tabularnewline
				\bottomrule
			\end{tabular}
		\end{adjustbox}
	\end{table}
	
	The strong form of PINNs fundamentally uses a collocation method, where the approximate function is replaced
	with a neural network. PINNs in the strong form can be applied not only to solve forward problems of
	PDEs but also to determine unknown coefficients in inverse problems. In inverse
	problems, some data combined with a data-driven approach is often required, and the loss function includes
	a norm of the error between predictions and true values, as shown in \Cref{eq:PINNs_strong_form_loss}
	for $\mathcal{L}_{data}$. Unknown parameters of the inverse problem, such as elastic modulus and Poisson's
	ratio, are treated as trainable variables to be optimized. By solving for the gradients of the optimization
	variables in inverse problems, the parameters needed for the inverse problems are inferred to solve the
	inverse problem. It is noteworthy that in PINNs, the optimization variables of inverse problems are optimized
	using $\mathcal{L}_{pdes}$. The optimization of inverse problems in PINNs can also be split into two
	parts: initially fitting the data with a neural network without considering the physical equations, and
	after the fit, optimizing the variables needed for the inverse problem using the PDEs loss. However,
	in traditional PINNs, the loss from data and PDEs is optimized together through hyperparameters. It is
	important to note that whether these are split or not, the mathematical optimization goals are the same,
	essentially solving the same problem but with different optimization strategies. Moreover, traditional
	PINNs in strong form solving inverse problems (optimizing PDEs loss and data loss together) generally
	do not manage to make both $\mathcal{L}_{pdes}$ and $\mathcal{L}_{data}$ zero simultaneously; the optimization
	process is a trade-off between $\mathcal{L}_{pdes}$ and $\mathcal{L}_{data}$, often resulting in $\mathcal{L}_{data}$
	reaching zero faster, while $\mathcal{L}_{pdes}$ tends to optimize more slowly.  Lu et al. (2021) \cite{lu2021physics} proposed the HPINN algorithm, which handles constraints using the
	augmented Lagrangian method \cite{hestenes1969multiplier}, thereby enhancing the accuracy of solving
	inverse problems.
	
	The strong form of PINNs does not require labeled data when addressing forward problems, although
	having labeled data can increase the efficiency and accuracy of the algorithm. The strong form
	of PINNs only requires spatial coordinates to use the powerful approximation capabilities of neural networks
	to map spacetime fields to the solution space \cite{PINN_original_paper}. From the perspective of computational
	mechanics, it is a type of trial function represented by a fully connected neural network and a Dirac delta
	mesh-free method for test functions. In their original paper \cite{PINN_original_paper}, the boundary initial conditions were
	added by a soft constraint method using penalty functions, which involves adjusting extra hyperparameters.
	Around the same time, Sirignano et al. (2018) \cite{sirignano2018dgm} introduced the DGM, which is
	very similar to PINNs, except that DGM does not address inverse problems. Additionally, DGM proposed
	using Monte Carlo algorithms instead of AD algorithms for derivative approximation, a method that can
	reduce the computational demands of high-dimensional problems \cite{he2023learning}, and DGM mathematically
	proved the convergence of PINNs in solving quasi-linear parabolic equations. Subsequently, Kharazmi et
	al. (2019) \cite{kharazmi2019variational} drew on traditional finite element and numerical analysis
	ideas to propose VPINNs, a variational form of the weighted residual method that changes the test functions
	in PINNs from Dirac delta to polynomials and trigonometric functions, thus extending the weighted residual
	method of PINNs strong formulation. However, VPINNs has limitations; if the test functions use Legendre
	orthogonal polynomials, higher-order orthogonal polynomials that satisfy the loss function to zero must
	exist. Given the strong fitting capabilities of neural networks, there's no reason why the network would
	not fit higher-order orthogonal polynomials, thereby causing non-uniqueness in the solution \citet{fang2021high}.
	Later, the same authors \cite{hp-VPINN}, improved VPINNs and further introduced
	hp-VPINNs, which are based on VPINNs but involve partitioning, meaning different areas have different
	test functions. This method found that orthogonal polynomials have high accuracy in this form, and
	it also compared different orders of variational weak forms. Jagtap et al. (2020)
	proposed cPINN \citet{CPINN} and XPINNs \citet{XPINN}, which partition the global domain like finite elements into different areas,
	each using a different neural network as the trial function approximation. However, special handling
	is required at the interfaces, where continuity conditions for the interface and gradient continuity
	conditions (related to the highest derivative of the PDEs) must be added to
	the loss function. For second-order mechanical PDEs, continuity conditions
	for displacement and force must be added as interface loss function terms. cPINN is suitable for handling
	problems with drastic changes, using deeper neural networks where changes are large and shallow networks
	where changes are mild. Shukla et al. \cite{XPINN_parallel} integrated parallel algorithms into cPINN
	and XPINNs, enhancing the efficiency of the partitioned PINNs algorithms.
	
	Most of the work discussed revolves around whether trial functions and test functions are partitioned
	and the types of test functions. Additionally, some work focuses on changing the type of network used
	for the trial functions, essentially modifying the trial function's approximation. Dwivedi et. al (2020)
	\cite{dwivedi2020physics} used Extreme Learning Machines (ELM) for PINNs in strong form. ELMs are a
	type of single hidden layer fully connected neural network, where the single hidden layer is obtained
	through traditional nonlinear mapping followed by a linear transformation to produce the final output.
	Notably, unlike traditional FNNs, the nonlinear transformation of the single hidden layer in ELM is randomly
	initialized and requires no training; ELM only trains the final linear transformation. This brings many
	benefits, as the optimization in ELM is not iterative but can be accomplished through direct methods
	to obtain the weights of the linear transformation, significantly reducing the computational load. Therefore,
	Dwivedi et. al (2020) \cite{dwivedi2020physics} proposed using ELM to replace the approximation function
	in the strong form of PINNs, calling this method PIELM. Since ELM sacrifices the expressive power
	of FNNs for efficiency, PIELM often does not perform well in cases of complex exact solutions. In such
	cases, partitioning ideas are needed to enhance the approximation capability of PIELM, thus improving
	its accuracy. Undoubtedly, since PIELM determines the weights of the neural network through direct solution
	methods, it is more efficient than most works in PINNs based on iterative algorithms. Additionally, some researchers have used new
	neural network structures to replace fully connected neural networks, such as Gao et al. (2021) \cite{gao2021phygeonet},
	who initially proposed PhyGeoNet, which essentially replaces fully connected neural networks with convolutional
	neural networks. The backward derivation in PhyGeoNet is achieved using fixed-weight convolutional kernels,
	and PhyGeoNet uses the idea of parametric transformation to solve problems in irregular domains that
	are challenging for physically based convolutional neural networks. The boundary conditions in this algorithm
	are strongly enforced. Later, Gao et al. (2022) \cite{gao2022physics} continued to improve PhyGeoNet
	and introduced PIGCN, which uses graph neural networks as approximation functions, and the loss function
	is constructed in weak form. 
	Graph-based learned simulators have emerged as an important direction in computational mechanics. By operating directly on mesh connectivity, graph neural networks naturally accommodate irregular discretizations and preserve local geometric relations, making them particularly attractive for engineering simulations on unstructured meshes \cite{pfaff2020learning}.
	Ramabathiran et al. (2021) \cite{ramabathiran2021spinn} proposed SPINN,
	which uses radial basis functions (RBF) as approximation functions, aiming to reduce the complexity of the neural network.
	Bai et al. (2023) \cite{bai2023physicsrbf} further validated the use of RBF networks for solving nonlinear equations through systematic verification via NTK (Neural Tangent Kernel)  theory \cite{NTK_PINN}.
	Sun et al. (2023) \cite{sun2023binn} first combined boundary elements with PINNs methods,
	using the residual of the boundary integral equation as the loss function. Overall, the experience of
	changing the network structure of PINNs strong form is that smaller neural networks often lack expressive
	power, but more complex neural networks are typically difficult to train, so choosing the best neural
	network structure requires considering the complexity of the problem \cite{cuomo2022scientific}, similar
	to choosing element types in finite elements, as different element types in finite elements essentially choose
	different approximation functions, and PINNs choosing network structures essentially also choose different
	approximation functions. Wang et al. (2024) proposed KINN \cite{wang2024kolmogorov}, which uses KAN
	proposed by Liu et al. (2024) \cite{liu2024kan} to replace MLP in PINNs, utilizing
	the strong form, energy form, and inverse form for the first time. Compared with traditional PINNs, the convergence speed
	and the accuracy of KINN were greatly improved.

	In time-dependent PINNs strong form algorithms, 
	there are two ways to solve time-dependent problems: expanding by one dimension or using a time-marching scheme (commonly FDM) \cite{bai2022general}. 
	Mattey et al. (2022) \cite{bc-pinn} proposed bc-PINN,
	which enhances the computational efficiency and accuracy of PINNs in solving strong nonlinear and high-order
	dynamic equations. Meng et al. (2020) \cite{meng2020ppinn} proposed PPINN, which uses a small
	network to enhance the efficiency of PINNs algorithms for long-term PDEs problems. Additionally, for data of different
	accuracies, Meng et al. (2020) \cite{meng2020composite} proposed MPINN, which combines high-precision
	data with low-precision data. In the area of stochastic PDEs, Yang et al. (2020) \cite{yang2020physics}
	proposed PIGAN, which combines physical equations with GANs (Generative Adversarial Networks) to solve
	the direct and inverse problems of stochastic PDEs (SPDEs).
	
	There are several PINN libraries available, such as DeepXDE  based on TensorFlow \cite{PINNlibrary}, SciANN based on Keras \cite{haghighat2021sciann}, NeuralPDE  based on Julia \cite{zubov2021neuralpde}, and SimNet \cite{hennigh2021nvidia}, which facilitate the application of PINNs in solving PDEs and are suitable for both academia and industry. These libraries facilitate the application of PINNs in solving PDEs. 
	
	\Cref{tab:AI4PDEcodes} lists several open-source AI for PDEs codes developed for computational mechanics.
	\begin{sidewaystable}
		\centering{}\caption{The codes of AI for PDEs in computational mechanics\label{tab:AI4PDEcodes}}
		\begin{adjustbox}{max height=\textheight, max width=\textwidth}
			\begin{tabular}{ccccc}
				\toprule 
				&  & Reference & The link of code & Brief Description of Method\tabularnewline
				\midrule
				\multirow{23}{*}{PINNs} & \multirow{12}{*}{The strong form of PINNs} & \cite{PINN_original_paper} & \url{https://github .com /maziarraissi /PINNs} & The original code\tabularnewline
				&  & \cite{hp-VPINN} & \url{https://github.com/ehsankharazmi/hp-VPINNs} & Variational PINNs\tabularnewline
				&  & \cite{CPINN} & \url{https://github.com/AmeyaJagtap/Conservative_PINNs} & PINNs with subdomains\tabularnewline
				&  & \cite{lu2021physics} & \url{https://github.com/lululxvi/hpinn} & Inverse design\tabularnewline
				&  & \cite{meng2020ppinn} & \url{https://github.com/XuhuiM/PPINN} & Time-dependent PDEs\tabularnewline
				&  & \cite{PINN_solid_mechanics} & \url{https://github.com/sciann/sciann-applications/tree/master/SciANN-SolidMechanics-BCs} & Solid mechanics\tabularnewline
				&  & \cite{PINNstrong_form_in_elastodynamics} & \url{https://github.com/Raocp/PINN-elastodynamics } & Elastodynamics\tabularnewline
				&  & \cite{sahin2024solving} & \url{https://github.com/imcs-compsim/pinns_for_comp_mech} & Contact mechanics\tabularnewline	
				&  & \cite{sun2020surrogate} & \url{https://github.com/Jianxun-Wang/LabelFree-DNN-Surrogate } & Incompressible NS \tabularnewline
				&  & \cite{wang2021deep} & \url{https://github.com/PredictiveIntelligenceLab/DeepStefan } & Multiphase and moving boundary problems\tabularnewline
				&  & \cite{rao2020physics} & \url{https://github.com/Raocp/PINN-laminar-flow} & Incompressible laminar flows\tabularnewline
				&  & \cite{cai2021artificial} & \url{https://github.com/shengzesnail/AIV_MAOAC} & Blood flow\tabularnewline
				\cmidrule{3-5} \cmidrule{4-5} \cmidrule{5-5} 
				& \multirow{12}{*}{DEM: Deep Energy Method} & \cite{loss_is_minimum_potential_energy} & \url{https://github.com/ISM-Weimar/DeepEnergyMethods} & The original code\tabularnewline
				&  & \cite{wang2023dcm} & \url{https://github.com/yizheng-wang} & DCEM\tabularnewline
				&  & \cite{wang2022cenn} & \url{https://github.com/yizheng-wang} & DEM with subdomains\tabularnewline
				&  & \cite{zhuang2021deep,the_comparision_of_strong_and_energy_form} & \url{https://github.com/weili101/Deep Plates} & Kirchhoff plate\tabularnewline
				&  & \cite{goswami2020adaptive,goswami2020transfer} & \url{https://github.com/somdattagoswami/IGAPack-PhaseField} & Phase field for fracture mechanics\tabularnewline
				&  & \cite{PINN_hyperelasticity} & \url{https://github.com/MinhNguyenIKM/dem_hyperelasticity} & Hyperelasticity\tabularnewline
				&  & \cite{bai2024robust} & \url{https://github.com/JinshuaiBai/RPIM_NNS} & Hyperelasticity\tabularnewline
				&  & \cite{he2023deep} & \url{https://github.com/Jasiuk-Research-Group/DEM_for_J2_plasticity} & J2 elastoplasticity\tabularnewline
				&  & \cite{paradeepenergy} & \url{https://github.com/MinhNguyenIKM/parametric-deep-energy-method} & Parametric DEM\tabularnewline
				&  & \cite{he2023deep_o} & \url{https://github.com/Jasiuk-Research-Group/DeepEnergy-TopOpt} & Topology optimization \tabularnewline
				&  & \cite{buoso2021personalising} & \url{https://github.com/sbuoso/Cardio-PINN/} & Heart\tabularnewline
				&  & \cite{dalton2023physics} & \url{https://github.com/dodaltuin/soft-tissue-pignn} & Soft tissue\tabularnewline
				\midrule 
				\multirow{7}{*}{Operator learning} & \multirow{3}{*}{DeepONet} & \cite{DeepOnet} & \url{https://github.com/lululxvi/deeponet} & The original code\tabularnewline
				&  & \cite{zhu2023reliable} & \url{https://github.com/lu-group/deeponet-extrapolation} & DeepONet with physics or sparse observations\tabularnewline	
				&  & \cite{he2023novel} & \url{https://github.com/Jasiuk-Research-Group/ResUNet-DeepONet-Plasticity} & DeepONet for  elastoplastic\tabularnewline
				\cmidrule{3-5} \cmidrule{4-5} \cmidrule{5-5} 
				& \multirow{4}{*}{FNO} & \cite{li2020fourier} & \url{https://github.com/neuraloperator/neuraloperator} & The original code\tabularnewline
				&  & \cite{es4846935deepnetbeam} & \url{https://github.com/eshaghi-ms/DeepNetBeam} & Functionally Graded Porous Beams\tabularnewline
				&  & \cite{wen2022u} & \url{https://github.com/gegewen/ufno} & Multiphase flow\tabularnewline
				&  & \cite{li2022fourier} & \url{https://github.com/neuraloperator/Geo-FNO} & FNO for general geometries\tabularnewline
				\midrule 
				\multirow{4}{*}{PINO} &  & \cite{li2021physics} & \url{https://github.com/neuraloperator/physics_informed} & PINNs with FNO\tabularnewline
				&  & \cite{wang2021learning} & \url{https://github.com/PredictiveIntelligenceLab/Physics-informed-DeepONets} & PINNs with DeepONet\tabularnewline
				&  & \cite{goswami2022physics} & \url{https://github.com/somdattagoswami/IGAPack-PhaseField} & DEM with DeepONet\tabularnewline
				&  & \cite{wang2023dcm} & \url{https://github.com/yizheng-wang} & DCEM with DeepONet\tabularnewline
				\bottomrule
			\end{tabular}
		\end{adjustbox}
	\end{sidewaystable}
	
	\subsubsection{Energy form\label{subsec:Energy-Form}}
\begin{figure}
		\begin{centering}
			\includegraphics[scale=0.45]{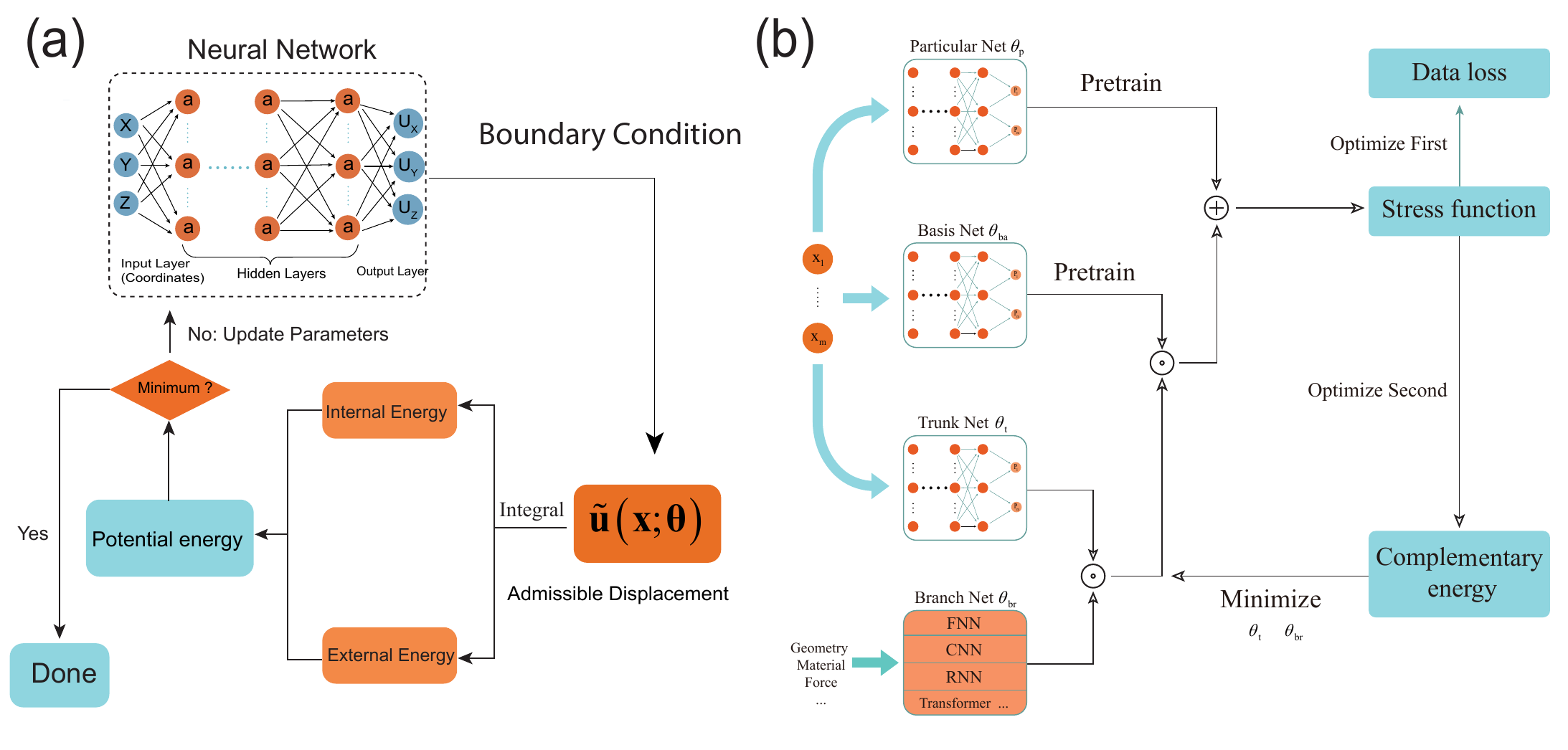}
			\par\end{centering}
		\caption{AI for PDEs method: Schematic of PINNs energy form \cite{loss_is_minimum_potential_energy} (a) DEM based on the principle of minimum potential
			energy \cite{loss_is_minimum_potential_energy}, (b) DCEM based on the principle of minimum complementary
			energy \cite{wang2023dcm}.\label{fig:PINN_energy_form}}
	\end{figure}

	DEM is a deep learning method for solving PDEs, designed based on the principle of least action in physics.
	When applied to a mechanical system, the principle of least action is manifested as Hamilton's principle:
	
	\begin{equation}
		\mathcal{H}=\int_{t_{0}}^{t_{1}}Ldt
	\end{equation}
	where \(\mathcal{H}\) represents the Hamiltonian action, which is the spacetime functional of the entire system. \(t_{0}\) and \(t_{1}\) represent the initial and final times respectively.
	\(L\) is the Lagrangian function:
	\begin{equation}
		L  =T-V
	\end{equation}
	where \(T\) and \(V\) represent the kinetic energy and potential energy of the system, respectively. For an elastic system, the expressions for kinetic energy \(T\) and potential energy \(V\) are given as:
	\begin{equation}
		\begin{aligned}T & =\int_{\Omega}\frac{1}{2}\rho\boldsymbol{v}\cdot\boldsymbol{v}d\Omega\\
			V & =\int_{\Omega}\Psi d\Omega-\int_{\Omega}\boldsymbol{f}\cdot\boldsymbol{u}d\Omega-\int_{\Gamma^{t}}\bar{\boldsymbol{t}}\cdot\boldsymbol{u}d\Gamma
		\end{aligned}
	\end{equation}
	where $\rho$, $\boldsymbol{v}$,  and $\Psi$ is the density, velocity and the strain energy density respectively.  $\mathbf{f}$ and $\bar{\mathbf{t}}$ are the body force and the surface force vector at
	the traction boundary condition $\Gamma^{t}$, respectively. 
	In DEM, the entire functional \(\mathcal{H}\) is optimized as the loss function. If we consider a static system where \(T = 0\),  DEM uses the potential energy $V$ as the loss function as illustrated in \Cref{fig:PINN_energy_form}. The mathematical form of the DEM loss function is given by:
	\begin{equation}
		\mathcal{L} = \mathcal{H} = \int_{t_{0}}^{t_{1}}Ldt=\int_{\Omega}\Psi d\Omega-\int_{\Omega}\boldsymbol{f}\cdot\boldsymbol{u}d\Omega-\int_{\Gamma^{t}}\bar{\boldsymbol{t}}\cdot\boldsymbol{u}d\Gamma
	\end{equation}

	The principle of DEM is to use the minimum
	potential energy principle under the condition that the displacement field satisfies the essential displacement
	conditions, i.e., to select the displacement field that minimizes the total potential energy $\mathcal{L}$
	among all admissible displacements. Note that DEM requires the defintion of an incremental energy and thus not only limited to static problems but it can also deal with transient and history dependent problems including plasticity \cite{he2023deep}, viscoelasticity \cite{lin2026physics}, damage and fracture \cite{wang2025towards,goswami2020transfer}. In particular, DEM has been applied to phase field models of fracture \cite{goswami2020transfer}. DEM directly optimizes the energy without needing to express  virtual work in the corresponding weak form like
	finite elements. The admissible displacement field is constructed through a neural network, but this
	admissible displacement field often needs to be pre-built to naturally satisfy the enforced displacement
	boundary conditions. The construction method for the admissible displacement field $\tilde{\boldsymbol{u}}(\boldsymbol{x})$ is commonly adopted
	as: 
	\begin{equation}
		\tilde{\boldsymbol{u}}(\boldsymbol{x})=\boldsymbol{P}(\boldsymbol{x})+\boldsymbol{D}(\boldsymbol{x})*\boldsymbol{u}(\boldsymbol{x};\boldsymbol{\theta}),
	\end{equation}
	where $\boldsymbol{P}(\mathbf{x})$ is a boundary network that satisfies the PDEs displacement boundary
	conditions, i.e., if the coordinate point falls exactly on the essential boundary condition, it outputs
	the value of the essential boundary condition; $\boldsymbol{D}(\mathbf{x})$ is a distance network, indicating
	the shortest distance from the current coordinate point to the nearest essential boundary condition location;
	$\boldsymbol{u}(\mathbf{x};\mathbf{\boldsymbol{\theta}})$ is a generalized network that needs to be
	trained by substituting it into the minimum potential energy loss function. 
	The idea of the admissible displacement field to satisfy the essential boundary conditions also exists in meshless methods \cite{li2002meshfree}.
	Note that boundary and distance
	networks do not necessarily need to be neural networks; choosing any other fitting function, such as
	Radial Basis Function (RBF) \cite{wang2022cenn} is also possible. We discuss the construction method
	of the admissible displacement field in detail in \Cref{subsec:Distance-Networks}.
	
	In the study of PINNs energy form, we provide a review of the status of PINNs energy form from four perspectives:
	whether subdomains are partitioned (whether trial functions are partitioned), types of trial functions,
	how essential boundary conditions are applied, and what kind of energy principle is used, as shown in
	\Cref{tab:PINN_energy_form} on the current state of research on PINNs energy form. Sheng et al. (2021)
	\cite{admissible_in_PINN_energy_form} extended Deep Ritz and proposed PFNN, using distance, boundary,
	and generalized networks to construct admissible displacement fields. Fuhg et al. (2022) \cite{fuhg2022mixed} proposed a classical mixed formulation mDEM,
	which integrates strong form, energy form, and constitutive equations into the loss function for optimization. Nguyen-Thanh et al. (2021) \cite{paradeepenergy} expanded DEM with P-DEM, borrowing
	the idea of parametric elements from finite elements to address complex geometric boundary problems.
	Wang et al. (2022) \cite{wang2022cenn} expanded DEM and proposed CENN, proposing a subdomain form of
	the deep energy method. In terms of neural network structures, He et al. (2023) \cite{he2023use} introduced
	GCN-DEM, replacing the original fully connected neural network with a graph convolutional neural network.
	As current deep energy forms are based on the minimum potential energy principle, Wang et al. (2023)
	\cite{wang2023dcm} first used the minimum complementary energy principle to propose DCEM, which parallels
	the deep energy method DEM, and combined operator learning with physical equations, making it an important
	complementary form to DEM. Wang et al. (2024) proposed KINN \cite{wang2024kolmogorov}, which utilizes
	KAN, as proposed by Liu et al. (2024) \cite{liu2024kan}, to replace MLP in PINNs,
	and employed strong form, energy form, and inverse form for the first time. 
	
	\begin{table}
		\caption{Current status of PINNs energy form methods\label{tab:PINN_energy_form}}
		\centering{}
		\begin{adjustbox}{max width=\textwidth}
			\begin{tabular}{ccccc}
				\toprule 
				Methods & Trial function: subdomains & Trial function: type & Method of Imposing Essential Boundary & Energy Principle\tabularnewline
				\midrule
				Deep Ritz \citet{deep_ritz} & No (Global) & Fully Connected & Penalty Function & Variational Principle\tabularnewline
				DEM \citet{loss_is_minimum_potential_energy} & No (Global) & Fully Connected & distance, boundary, and generalized networks & Minimum Potential Energy\tabularnewline
				PFNN \citet{admissible_in_PINN_energy_form} & No (Global) & Fully Connected & distance, boundary, and generalized networks & Variational Principle\tabularnewline
				mDEM \citet{fuhg2022mixed} & No (Global) & Fully Connected & distance, boundary, and generalized networks & Mixed Form\tabularnewline
				P-DEM \citet{paradeepenergy} & No (Global) & Fully Connected & Penalty Function & Minimum Potential Energy\tabularnewline
				CENN \citet{wang2022cenn} & Yes (Subdomain) & Fully Connected & distance, boundary, and generalized networks & Minimum Potential Energy\tabularnewline
				GCN-DEM \citet{he2023use} & No (Global) & Graph Convolutional (GCN) & distance, boundary, and generalized networks & Minimum Potential Energy\tabularnewline
				PIRBN \citet{bai2023physicsrbf} & No (Global) & Radial Basis Function & distance, boundary, and generalized networks & Minimum Potential Energy\tabularnewline
				DCEM \citet{wang2023dcm} & No (Global) & Fully Connected & distance, boundary, and generalized networks & Minimum Complementary Energy\tabularnewline
				KINN \citet{wang2024kolmogorov} & No (Global) & Kolmogorov Arnold Network \cite{liu2024kan} & distance, boundary, and generalized networks & Minimum Potential Energy\tabularnewline
				\bottomrule
			\end{tabular}
		\end{adjustbox}
	\end{table}
	
	\subsection{Operator learning\label{subsec:Operator-Learning}}
	
	Neural operators learn the mapping between different functions, which has extensive applications in science
	and engineering \cite{kovachki2023neural}. For instance, the input function could be the boundary conditions
	of a PDEs, and the output is the function field of interest. Although functions are infinite-dimensional
	in mathematical terms, in practice, we often input discretized data on a finite mesh to approximate continuous
	functions.  
	
	Unlike previous data-driven methods based on neural networks, operator learning is discretization-invariant, characterized by: 
	\begin{itemize}
		\item The ability to input at any resolution. 
		\item The capability to output at any location. 
		\item Convergence of results as the mesh is refined.
	\end{itemize}
	
	Traditional neural network-based data-driven algorithms are often related to the degree of
	discretization; once the resolution of the input-output functions is changed, the model usually needs
	to be retrained. However, operator learning does not need to restart training for data of different resolutions due to the characteristic of the discretization-invariance.  Additionally,
	not all fitting algorithms can be termed as neural operators. Neural operators not only need to satisfy
	discretization-invariance but also meet the requirements of the generalized approximation \cite{chen1995universal}.
	It can be said that neural operators are a superior class of algorithms within operator learning, such
	as the Fourier Neural Operator (FNO) \cite{li2020fourier}. Operator learning can be applied not only
	in deterministic models but also in probabilistic models, like the diffusion model \cite{yang2022diffusion},
	to learn the probability density in function spaces, such as solving stochastic PDEs.
	Physics-informed diffusion models combine score-based generation with physical constraints and have recently shown strong potential in scientific computing tasks, especially for structural topology optimization \cite{bastek2024physics}.
	
	\begin{figure}
		\begin{centering}
			\includegraphics[scale=0.70]{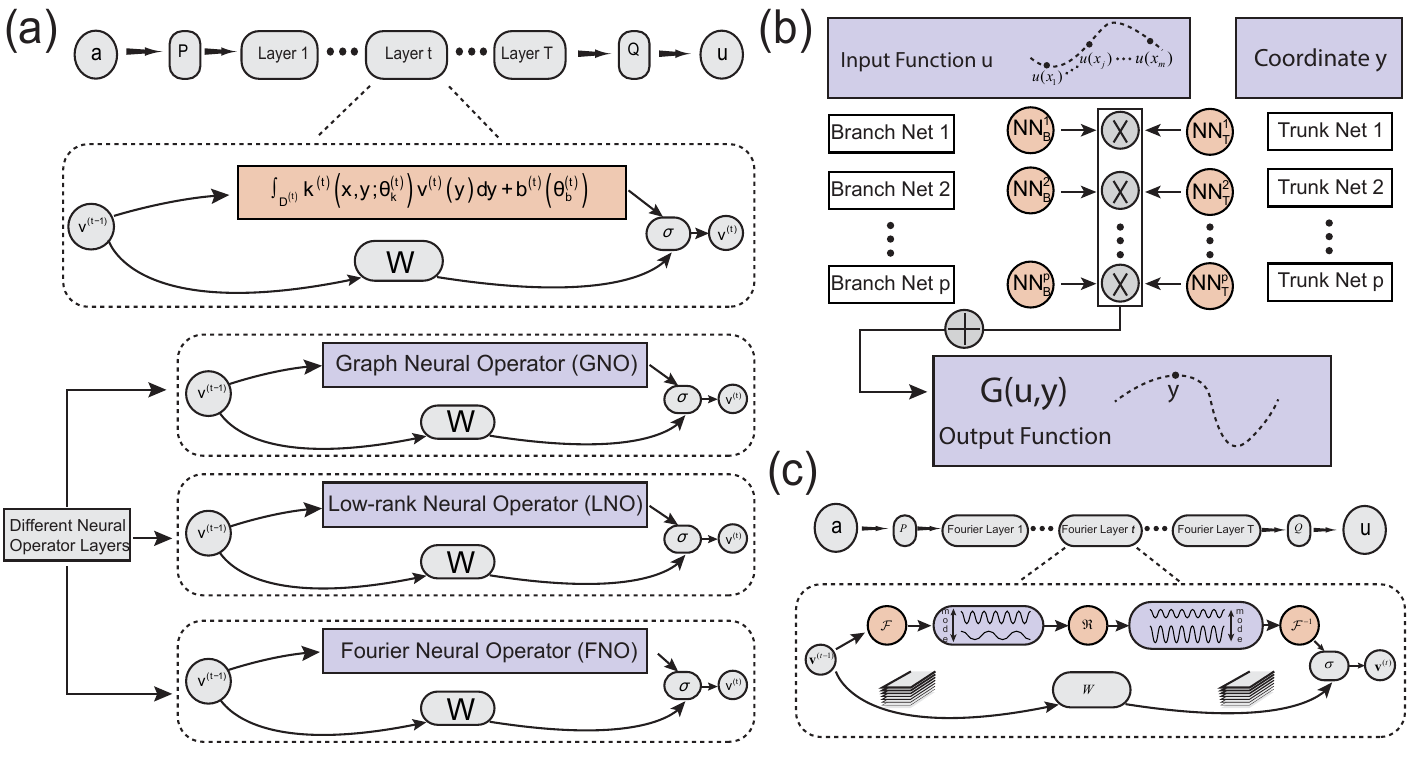}
			\par\end{centering}
		\caption{AI for PDEs Method: Schematic of Operator Learning.
			(a) Neural Operator Layer Structure: Graph Neural Operator (GNO) \cite{li2020neural}, Local Neural Operator (LNO), and Fourier Neural Operator (FNO) \cite{li2020fourier} can serve as the core of the neural operator architecture \cite{kovachki2023neural}.
			(b) Details of DeepONet \cite{DeepOnet}.
			(c) Details of FNO \cite{li2020fourier}.\label{fig:neural_operator_schematic}}
	\end{figure}
	
	The design philosophy of operator learning algorithms is shown in \Cref{fig:neural_operator_schematic}a,
	with the core being the kernel integral method in the neural operator layers: 
	\begin{equation}
		\mathcal{\boldsymbol{H}}^{(t)}(\boldsymbol{v}^{(t)};\boldsymbol{\theta}_{k}^{(t)})(\boldsymbol{x})=\int_{D^{(t)}}\boldsymbol{k}^{(t)}(\boldsymbol{x},\boldsymbol{y};\boldsymbol{\theta}_{k}^{(t)})\boldsymbol{v}^{(t)}(\boldsymbol{y})d\boldsymbol{y},\label{eq:kernel function}
	\end{equation}
	where $\boldsymbol{v}^{(t)}(\boldsymbol{y})$ is the input to the neural operator layer; $\boldsymbol{k}^{(t)}(\boldsymbol{x},\boldsymbol{y})$
	is the kernel function, similar to the Green's function in linear PDEs, and $\boldsymbol{k}^{(t)}(\boldsymbol{x},\boldsymbol{y})$
	is approximated using neural networks, where $\boldsymbol{\theta}_{k}^{(t)}$ are the parameters for
	approximating the kernel function. The reason it's a kernel integral method is that the kernel operator
	$\mathcal{\boldsymbol{H}}^{(t)}$ acts on the function $\boldsymbol{v}^{(t)}$, and through integration,
	$\mathcal{\boldsymbol{H}}^{(t)}$ transforms $\boldsymbol{v}^{(t)}$ into another class of functions $\mathcal{\boldsymbol{H}}^{(t)}(\boldsymbol{v}^{(t)})(\boldsymbol{x})$.
	$\mathcal{\boldsymbol{H}}^{(t)}$ is similar to integral transformations in mathematical equations. In
	terms of methods of mathematical physics, $\mathcal{\boldsymbol{H}}^{(t)}(\boldsymbol{v}^{(t)})(\boldsymbol{x})$
	is referred to the image function, $\boldsymbol{v}^{(t)}(\boldsymbol{y})$ is the original function,
	and $\boldsymbol{k}^{(t)}(\boldsymbol{x},\boldsymbol{y})$ is the kernel function. Note that the $t$
	index indicates the $t$-th neural operator layer. Hence, the overall computation process of the operator
	learning is:
	\begin{equation}
		\begin{alignedat}{1}\boldsymbol{G}_{\theta}(\boldsymbol{a}(\boldsymbol{x});\boldsymbol{\theta})= & \boldsymbol{Q}(\boldsymbol{v}^{(T+1)};\boldsymbol{\theta}_{Q})\circ\sigma[\mathcal{\boldsymbol{H}}^{(T)}(\boldsymbol{v}^{(T)};\boldsymbol{\theta}_{k}^{(T)})+\boldsymbol{W}^{(T)}(\boldsymbol{v}^{(T)};\boldsymbol{\theta}_{w}^{(T)})+\boldsymbol{b}^{(T)}(\boldsymbol{\theta}_{b}^{(T)})]\circ\cdots\\
			& \circ\sigma[\mathcal{\boldsymbol{H}}^{(1)}(\boldsymbol{v}^{(1)};\boldsymbol{\theta}_{k}^{(1)})+\boldsymbol{W}^{(1)}(\boldsymbol{v}^{(1)};\boldsymbol{\theta}_{w}^{(1)})+\boldsymbol{b}^{(1)}(\boldsymbol{\theta}_{b}^{(1)})]\circ\boldsymbol{P}(\boldsymbol{a}(\boldsymbol{x});\boldsymbol{\theta}_{P})
		\end{alignedat}
		,\label{eq:neural operator algorithm}
	\end{equation}
	where $\boldsymbol{P}$ and $\boldsymbol{Q}$ are respectively for elevating and reducing dimensions
	of the original data, with corresponding learnable parameters $\boldsymbol{\theta}_{P}$ and $\boldsymbol{\theta}_{Q}$.
	$\mathcal{\boldsymbol{H}}^{(t)}$, $\boldsymbol{W}^{(t)}$, and $\boldsymbol{b}^{(t)}$ are respectively
	the kernel integral, linear transformation, and bias, with the corresponding learnable parameters being
	$\boldsymbol{\theta}_{k}^{(t)}$, $\boldsymbol{\theta}_{w}^{(t)}$, and $\boldsymbol{\theta}_{b}^{(t)}$.
	Note that these learnable parameters are generally approximated using neural networks. 
	
	Different choices of kernel functions constitute various operator learning algorithms, hence designing the kernel function
	is central to operator learning. When specifically forming a Fourier transform, it results in the Fourier
	Neural Operator (FNO). There is theoretical evidence that \Cref{eq:neural operator algorithm} meets the
	general approximation requirements \cite{de2022generic,kovachki2021universal}, hence operator learning
	algorithms designed according to \Cref{eq:neural operator algorithm} possess excellent convergence properties.
	DeepONet \cite{DeepOnet} also meets the general approximation requirements \cite{chen1995universal},
	essentially designed through the theory proposed by \cite{chen1995universal}. Although the FNO and DeepONet are designed differently, they both exhibit excellent convergence properties. The two most typical
	algorithms in operator learning are FNO and DeepONet, thus this paper will focus on these two algorithms
	below.

	\subsubsection{FNO: Fourier neural operator\label{subsec:FNO:-Fourier-Neural}}
	
	The Fourier Neural Operator (FNO) is an operator learning algorithm introduced by Li et al. (2020) \cite{li2020fourier},
	which has received considerable attention. The initially proposed FNO is a completely data-driven model.
	Even earlier than FNO, the same first author, Li et al. \cite{li2020neural}, used graph neural
	networks to learn relationships between field variables (GNO), but FNO demonstrated better performance
	than GNO. The essence of FNO involves integrating Fourier transforms into operator learning, representing
	\Cref{eq:kernel function} with a Fourier transform: 
	\begin{equation}
		\mathcal{\boldsymbol{H}}^{(t)}(\boldsymbol{v}^{(t)};\boldsymbol{\theta}_{k}^{(t)})(\boldsymbol{x})=\int_{D^{(t)}}\boldsymbol{k}^{(t)}(\boldsymbol{x},\boldsymbol{y};\boldsymbol{\theta}_{k}^{(t)})\boldsymbol{v}^{(t)}(\boldsymbol{y})d\boldsymbol{y}=\mathcal{F}^{-1}\circ\Re\circ\mathcal{F}(\boldsymbol{v}^{(t)}(\boldsymbol{x})),
	\end{equation}
	where $\mathcal{F}$ and $\mathcal{F}^{-1}$ are the Fourier transform and its inverse, respectively,
	$\mathfrak{R}$ is a linear transformation. To clarify the process of the FNO algorithm, we combine the
	framework and examples of the FNO algorithm. The framework of the FNO algorithm is shown in \Cref{fig:neural_operator_schematic}c.
	
	As a simple illustrative example of the capability of FNO, we consider the one-dimensional Burgers equation, which is commonly used as a benchmark problem because of its nonlinear and space-time-dependent characteristics.  $u$ denotes the field variable to be solved, $\nu$ is the viscosity coefficient, and the boundary conditions are fixed. For a given initial condition $u_{0}$, the corresponding solution $u$ is uniquely determined. Thus, in this problem, the input is the initial condition $u_{0}\in\mathbb{R}^{d_{i}}$ (i.e.,
	$a(x)$ in \cite{li2020fourier}), and the output is $u(x,t_{1})$ at a future time $t_{1}$. The FNO
	algorithm process first uses a fully connected neural network $\boldsymbol{P}$ to elevate the dimension
	of the input $a(x)$ to $d_{c}$ (i.e., $P:\mathbb{R}\rightarrow\mathbb{R}^{d_{c}}$, $d_{c}$ similar
	to channels in computer vision): 
	\begin{equation}
		\boldsymbol{v}^{(0)}(x)=\boldsymbol{P}(a(x))\in\mathbb{R}^{d_{i}*d_{c}}.
	\end{equation}
	Next, $v^{(0)}(x)$ is sent into Fourier layer, where each channel of $v^{(0)}(x)$ undergoes a Fourier
	transform, transforming into: 
	\begin{equation}
		\boldsymbol{f}^{(1)}(x)=\mathcal{F}(\boldsymbol{v}^{(0)}(x))\in\mathbb{\mathbb{C}}^{d_{i}*d_{c}}.
	\end{equation}
	High frequencies are filtered out, retaining low frequencies, and cutting off the frequency at $d_{m}(1\leq d_{m}\leq d_{i})$,
	thus the data structure transforms into $f^{(1)}(x)\in\mathbb{C}^{d_{m}\times d_{c}}$. A linear transformation
	$\mathfrak{R}$ is applied to each frequency of all channels: 
	\begin{equation}
		\boldsymbol{f}_{L}^{(1)}(x)=\Re(\boldsymbol{f}^{(1)}(x))\in\mathbb{\mathbb{C}}^{d_{m}*d_{c}}.
	\end{equation}
	Since different frequencies use different linear transformations, there are $d_{m}$ linear transformation
	matrices, and the linear transformation is from $d_{c}$ to $d_{c}$, so the linear transformation matrix
	is $\mathbb{R}^{d_{c}\times d_{c}\times d_{m}}$. After the linear transformation $\mathfrak{R}$, the
	data's dimensional structure is not changed. Since the next step involves Fourier inverse transformation,
	high frequencies removed by cutting are replaced by zeros. Fourier inverse transformation is performed
	on $\boldsymbol{f}_{L}^{(1)}(x)$: 
	\begin{equation}
		\boldsymbol{v}_{f}^{(1)}=\mathcal{F}^{-1}(\boldsymbol{f}_{L}^{(1)}(x))\in\mathbb{\mathbb{\mathbb{R}}}^{d_{i}*d_{c}}.
	\end{equation}
	The advantage of the linear transformation $\mathfrak{R}$ is that it learns the mapping relationships
	between data in the Fourier frequency space while filtering out high frequencies. $\mathfrak{R}$ is a special form of regularization
	that reduces overfitting and enhances the model's generalization capability. 
	
	Furthermore, using the idea of residual networks from \cite{he2016deep}, another linear transformation is
	directly applied to the channels of $v^{(0)}(x)$, the same linear transformation for all different frequencies,
	only one linear transformation, i.e., $\boldsymbol{W}\in\mathbb{R}^{d_{c}\times d_{c}}$: 
	\begin{equation}
		\boldsymbol{v}_{L}^{(1)}=\boldsymbol{W}(\boldsymbol{v}^{(0)}(x))\in\mathbb{\mathbb{\mathbb{R}}}^{d_{i}*d_{c}}.
	\end{equation}
	$\boldsymbol{W}$ uses the idea of residual networks to reintroduce high-frequency components filtered
	out by the Fourier transform back into the network for learning, preventing a decrease in network prediction
	performance due to the high frequencies filtered out by the Fourier transform. The results of the Fourier
	transform and the linear transformation are added together to obtain: 
	\begin{equation}
		\boldsymbol{v}^{(1)}(x)=\sigma(\boldsymbol{v}_{f}^{(1)}+\boldsymbol{v}_{L}^{(1)}+\boldsymbol{b}^{(1)})\in\mathbb{\mathbb{\mathbb{R}}}^{d_{i}*d_{c}},
	\end{equation}
	where $\boldsymbol{b}^{(1)}$ is the bias. Adding a nonlinear transformation $\sigma$ not only increases the
	network's expressive capability but also further enhances the high-frequency fitting capability. The
	Fourier layer is the computational core of FNO, and the above operations are repeated $T$ times to obtain
	$v^{(T)}(x)$. Finally, the dimension $d_{c}$ is reduced to the target dimension by $\boldsymbol{Q}:\mathbb{R}^{d_{c}}\rightarrow\mathbb{R}$.
	
	The two major advantages of FNO are, first, since the Fourier transform filters out high-frequency modes,
	it can enhance the model's speed and generalization ability, reducing overfitting; second, discretization-invariance, since all operations in FNO are unrelated to the data's mesh and are based on point-wise
	operations, so it has the advantage of discretization-invariance. The initially proposed FNO has two
	major flaws. First, the Fourier Neural Operator (FNO) employs the fast Fourier transform, which requires the input data to be defined over a regular domain. Although it is feasible to enclose an irregular domain within a larger regular one, this approach often yields suboptimal results. To address this, Li et al. \cite{li2022fourier} introduced Geo-FNO, specifically designed to handle irregular domains effectively. Second, while FNO requires data points to be positioned on a uniform mesh, points on an uneven mesh must be converted into a uniform mesh using interpolation. Thus, the error of interpolation decreases the accuracy of FNO. An alternative to Geo-FNO is $\phi$-FEM-FNO \cite{duprez2025varphi}, where the geometry is described by signed distance functions (SDFs), usually level sets as in CUTFEM \cite{burman2015cutfem}. This approach avoids interpolation entirely because the training data is generated on the same background grid and the SDF implicitly encodes the geometry without remeshing. It inherits the same difficulties as CUTFEM. Most importantly, for problems requiring geometry evolution, the level set must be updated, and the error from the transport equation solves accumulates. This occurs frequently in fluid mechanics and problems in solid mechanics involving history dependent constitutive models with large deformations and fracture.

	\subsubsection{DeepONet\label{subsec:DeepONet}}
	
	The general operator learning framework introduced in \Cref{eq:neural operator algorithm} constructs neural operators through iterative kernel integral layers, with the key distinction among methods being the parameterization of the kernel function (for example, FNO employs Fourier-based integral kernels). As an alternative and equally important implementation of operator learning, DeepONet follows a different design principle: it builds upon the universal operator approximation theorem \cite{chen1995universal} and decomposes the operator into an inner product of two subnetworks, the Branch and Trunk networks. 
	The mathematical structure of DeepONet
	is analogous to a Taylor series:
	\begin{equation}
		f(\boldsymbol{u})(\boldsymbol{x})=\sum_{i=1}^{n}\alpha_{i}(\boldsymbol{u})\phi_{i}(\boldsymbol{x}).\label{eq:taylor}
	\end{equation}
	
	As illustrated in \Cref{fig:neural_operator_schematic}b, the Trunk net employs neural networks to
	approximate the basis functions $\phi_{i}(\boldsymbol{x})$. Trunk net shares a common idea with PINNs,
	where neural networks fit from the coordinate space to the target function. However, in DeepONet, it is the basis functions $\phi_{i}(\boldsymbol{x})$ that are fitted by the Trunk net, rather than the target functions.
	The coefficients before the basis functions $\alpha_{i}(\boldsymbol{u})$ are fitted by the Branch net. It
	is noteworthy that inputting the same function into the Branch net produces fixed weights $\alpha_{i}(\boldsymbol{u})$,
	aligning with the concept of function approximation in numerical analysis. Unlike traditional function
	approximation algorithms that pre-select basis functions, DeepONet uses neural networks to adaptively
	select suitable basis functions and weights based on data.
	
	Since the Trunk network is similar to PINNs, we can apply automatic differentiation algorithms in PINNs
	to construct partial differential operators and obtain new loss functions (strong or energy form).
	If the Branch network is fixed, DeepONet can be reduced to PINNs.
	
	DeepONet's network structure is constructed according to the theory of universal operator approximation
	from \cite{chen1995universal}. The mathematical form of this approximation theory in \cite{chen1995universal}
	is: 
	\begin{equation}
		\mid G(\boldsymbol{u},\boldsymbol{y})-\sum_{k=1}^{p}\sum_{i=1}^{n}c_{i}^{k}\sigma\left(\sum_{j=1}^{m}w_{ij}^{k}\boldsymbol{u}(\boldsymbol{x}_{j})+b_{i}^{k}\right)\sigma(\boldsymbol{W}^{k}\cdot\boldsymbol{y}+\boldsymbol{B}^{k})\mid<\epsilon.
	\end{equation}
	DeepONet replaces the terms in the above equation with neural networks: 
	\begin{equation}
		\begin{alignedat}{1}NN_{B}^{k}(\boldsymbol{u}(\boldsymbol{x});\boldsymbol{\theta}_{NN}^{B}) & =\sum_{i=1}^{n}c_{i}^{k}\sigma\left(\sum_{j=1}^{m}w_{ij}^{k}\boldsymbol{u}(\boldsymbol{x}_{j})+b_{i}^{k}\right)\\
			NN_{T}^{k}(\boldsymbol{y};\boldsymbol{\theta}_{NN}^{T}) & =\sigma(\boldsymbol{W}^{k}\cdot\boldsymbol{y}+\boldsymbol{B}^{k})\\
			NN_{O}(\boldsymbol{y};\boldsymbol{u}) & =\sum_{k=1}^{p}NN_{B}^{k}(\boldsymbol{u}(\boldsymbol{x});\boldsymbol{\theta}_{NN}^{B})\cdot NN_{T}^{k}(\boldsymbol{y};\boldsymbol{\theta}_{NN}^{T})
		\end{alignedat}
		,
	\end{equation}
	where $NN_{B}^{k}$, $NN_{T}^{k}$, and $NN_{O}$ are respectively the Branch network, Trunk network,
	and the output of DeepONet. $\boldsymbol{u}(\boldsymbol{x})$ can be approximated from sensors $\boldsymbol{x}$
	(discrete points approximating a continuous function). For instance, if the input $\boldsymbol{u}(\boldsymbol{x})$
	of $NN_{B}$ is the gravitational potential energy of a three-dimensional cubic material over $[0,1]^{3}$,
	then the gravitational potential energy field within $[0,1]^{3}$ might be sampled at 0.01 intervals
	taking $101^{3}$ points. Data structure of $\boldsymbol{u}(\boldsymbol{x})$ can vary, such as data
	structures similar to images. Therefore, we can use network structures for the Branch network that are
	appropriate for the type of data, such as CNNs for local features similar to image data, and RNNs for
	time-related features. It is important that the network structure of the Branch network must consider
	the particularities of the problem to choose the most appropriate network structure. Thus, $\boldsymbol{u}(\boldsymbol{x})$ determines the output of the Branch network in DeepONet,
	which is related to the weights $\alpha_{i}(\boldsymbol{u})$ in \Cref{eq:taylor} and is independent
	of the coordinates of interest. Similarly, the coordinates fully determine the output of the Trunk network,
	precisely defining the basis functions $\phi_{i}(\boldsymbol{x})$ in \Cref{eq:taylor}. Therefore, DeepONet
	shares many similarities with traditional function approximation. It can be said that DeepONet can adaptively
	learn basis functions and weights based on data. For more details and extensions on DeepONet, see \cite{DeepOnet}.
	
	For the comparison between FNO and DeepONet, Lu et al. \cite{lu2022comprehensive} systematically analyzed both methods. They introduced enhancements to FNO to handle mappings of different dimensionalities and complex geometries, while also improving DeepONet to accelerate training and enhance accuracy, supplemented by theoretical comparisons.

	\subsection{PINO: Physics-informed neural operator\label{subsec:PINO=00FF1A-Physics-Informed-Neural}}
	
\begin{figure}
		\begin{centering}
			\includegraphics[scale=0.60]{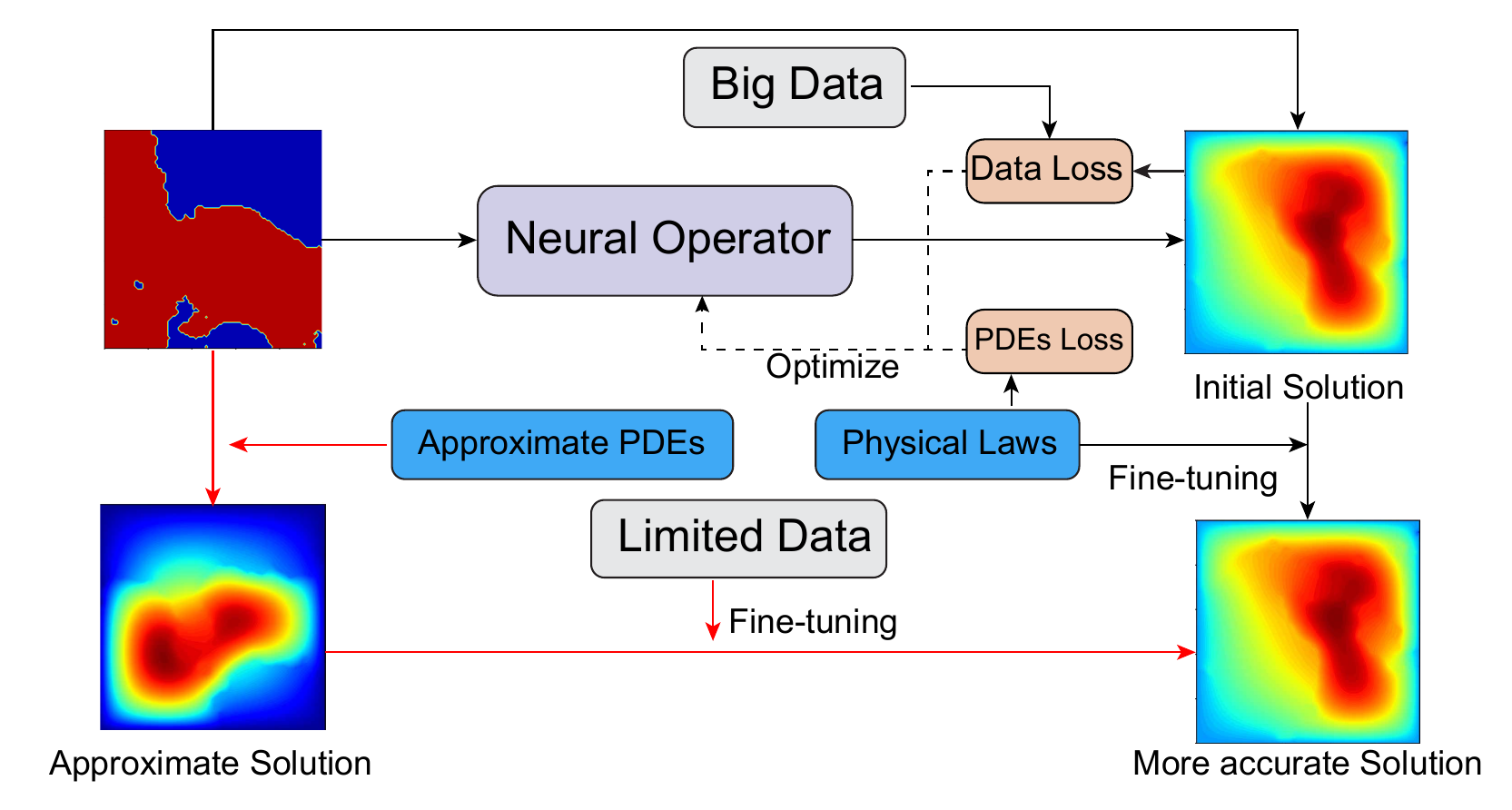}
			\par\end{centering}
		\caption{AI for PDEs method: Schematic of Physics-Informed Neural Operator (PINO).
			There are two implementation approaches. The first, depicted by the black lines, is suitable for scenarios with precise physical equations. Initially, the neural operator is trained using extensive data and physical equations. The method involves using the neural operator to provide an initial solution and establishing a data loss with the data labels, followed by constructing a PDEs loss based on the physical equations. The combination of data loss and PDEs loss allows for more effective training of the neural operator compared to purely data-driven methods \cite{li2021physics}. After training, the neural operator is fine-tuned with the physical equations of specific problems to provide highly accurate solutions.
			The second approach, depicted by the red lines, is suited for cases with only approximate physical equations, such as complex phenomena not well understood by humans. This method starts with an approximate solution derived from the approximate PDEs, which is then fine-tuned using limited data \cite{chakraborty2021transfer}. \label{fig:PINO_schematic}}
	\end{figure}

	The concept of the Physics-Informed Neural Operator (PINO) was proposed by Z. Li et al. (2021) \cite{li2021physics}.
	As shown by the black lines in \Cref{fig:PINO_schematic}, the idea is very concise: first, operator learning is trained
	using big data, then a good initial solution is provided on the test set using operator learning, which
	is then fine-tuned based on physical equations. Since the initial solution is not arbitrarily given but
	inferred from historical data, it only needs some adjustment to be fine-tuned to the true solution.
	Therefore, theoretically, the PINO algorithm not only possesses very high precision (since it is controlled
	by physical equations) but also leverages the efficiency advantage of operator learning. 
	The above process involves using operator learning to provide an initial solution, followed by fine-tuning with PDEs. 
	For example, \cite{wang2023dcm} isolates the training of physical equations and data. The approach first uses purely data-driven training for operator learning and then fine-tunes with PDEs for specific problems.
	Additionally, PDEs and data can be combined together to train an operator constrained by physical equations, similar to the training mode of PINNs.
	For example, Wang et al. (2021) \cite{wang2021learning} proposed combining DeepONet with physical equations, and combined the data loss with PDEs loss together.  Later, Goswami et al. (2022) \cite{goswami2022physics} built on the idea proposed by \citet{wang2021learning}
	to predict crack propagation (by combining DeepONet with the phase field method of predicting crack path).
	Currently, there is no conclusion as to which method is better; both are still being
	explored. 
	The advantage of isolating the training of PDEs and data is that it can avoid catastrophic forgetting and reduce hyperparameters, specifically those used to balance data and PDE losses. The downside is that it requires additional fine-tuning on new PDEs, thereby increasing the testing time. 
	On the other hand, integrating PDEs and data during training has the advantage of high efficiency during the testing period, as PDEs are already incorporated into the operator during training. However, this undoubtedly increases the training time. Additionally, it introduces extra hyperparameters to balance the data and PDEs, and faces the risk of catastrophic forgetting \cite{goodfellow2013empirical}. This occurs because the PDEs under the current parameters can significantly lose the knowledge and performance of the PDEs under previous parameters, thereby increasing the training cost.
	
	Although incorporating PDEs to train neural operators increases computational cost, it remains an effective approach in scenarios where data is scarce but PDEs are available. Eshaghi et al. \cite{eshaghi2025variational} proposed the Variational Physics-Informed Neural Operator (VINO). Unlike PINO, VINO leverages the variational form of PDEs to train a fully Fourier Neural Operator, and further employs finite element shape functions to replace automatic differentiation (AD) in computing differential operators, thereby enhancing the stability and accuracy of derivative computations through more precise integration.
	Similarly, Wang et al. \cite{wang2026pretrain} proposed the Pretrained Finite Element Method (PFEM), a physics-driven framework that bridges the efficiency of neural operator learning with the accuracy and robustness of classical finite element methods (FEM). Like VINO, PFEM fully exploits the governing PDEs to train the neural operator. PFEM employs Transolver \cite{wu2024transolver} as the backbone operator. Unlike FNO, which requires a structured rectangular grid, Transolver operates directly on point clouds, providing a flexible and efficient approach for future large-scale computational mechanics models. In addition, PFEM uses the Transolver-predicted solution as an initial guess for the iterative FEM solver, significantly reducing the number of iterations required while maintaining high accuracy.
	
	Another way of combining data and physical equations is for some complex physical processes \cite{chakraborty2021transfer},
	which refers to those that cannot be accurately described based on current understanding and can only
	be approximated by some physical equations. We can use PINNs to solve for an approximate solution (low-fidelity)
	based on approximate physical equations and boundary conditions, and then adjust the deep parameters
	of the neural network according to the available data, thereby obtaining a data-based
	solution (high-fidelity) that does not require exact knowledge of the physical process. This allows us
	to fuse experimental data with approximate physical equations, as shown by red lines in \Cref{fig:PINO_schematic}.
	Although this method of combination currently does not use operator learning, it can be easily modified
	using operator learning algorithms when obtaining high-fidelity results in the future, which is a promising
	direction, specifically targeting those complex physical processes and combining operators
	with physical equations.
	
	Given that the concept of Physical Neural Operator (PINO) was only recently introduced, numerous research challenges remain to be explored. Key questions include how to more effectively integrate physical equations into operator learning, how to handle noisy data, and how to perform learning with limited samples. In this paper, we focus on the first issue, as we have encountered significant difficulties in integrating physical equations into operator learning. Specifically, after training the Fourier Neural Operator (FNO) with large datasets, we observed a sudden increase in the loss function when incorporating physical equations via finite difference methods. This increase in loss reduces efficiency. In some extreme cases, the combination of data and PDEs may even be worse than using physical equations directly (without data). We hypothesize that this issue arises due to a non-overlapping optimization space between data loss and physical equation loss, which leads to difficulties in their integration. Despite these challenges, we believe that PINO holds substantial promise for academic research. Its theoretical foundation is both robust and promising, providing ample opportunities for exploration and innovation. 
	
	Since the PINO inherently combines operator learning with physical equations, it can be seen as encompassing both Physics-Informed Neural Networks (PINNs) and operator learning. If only physical equations are utilized, PINO reduces to PINNs; if solely data-driven methods are applied, PINO reduces to operator learning.

	\subsection{Summary}
	
	This chapter introduced the methodologies of AI for PDEs, including PINNs, operator learning, and physics-informed neural operators. Specific algorithms have been reviewed in their respective sections. Here, we discuss notable points and potential directions for future research:
	
	The advantage of the strong form of PINNs in solving high-dimensional problems arises not from the neural network itself but from the way the loss function is computed, specifically through Monte Carlo integration. Monte Carlo integration is a well-established area in applied statistics, and many concepts from this field can be borrowed to enhance PINNs, such as variance reduction techniques. These techniques involve altering the probability density of sampling points to reduce the variance of the Monte Carlo integral, which is crucial for the convergence speed of PINNs.
	The smaller the variance of the loss function, the faster the convergence of PINNs.
	The selection of collocation points has always been a core issue for PINNs. Therefore, combining Monte Carlo integration techniques from applied statistics with PINNs is essential for optimizing point selection and reducing variance, thereby accelerating convergence.
	
	Although energy-based PINNs often achieve higher accuracy and computational efficiency than strong-form PINNs in many mechanics problems, and usually involve fewer manually balanced hyperparameters \cite{the_comparision_of_strong_and_energy_form}, their applicability is restricted by the existence and mathematical properties of the underlying variational functional. Not all PDEs admit a corresponding energy formulation, and energy-based approaches generally impose higher requirements on numerical integration accuracy.
	A fundamental mathematical limitation arises when the variational formulation leads to a stationary saddle-point problem rather than a true minimization problem. In such cases, the loss function seeks a saddle point instead of a minimum, which substantially increases optimization difficulty. Although dedicated saddle-point optimization methods such as the Alternating Direction Method of Multipliers (ADMM) may provide partial improvement, reliable convergence remains challenging in practice.
	In addition, neural-network-based optimization introduces a second level of difficulty: the loss landscape itself is highly non-convex and contains numerous spurious stationary points. As a result, even when a stationary point is reached, it is often difficult to determine whether it corresponds to the physically correct saddle-point solution of the original variational problem.
	This issue becomes particularly severe when generalized variational principles are adopted, such as the Hu--Washizu three-field variational principle or the Hellinger--Reissner mixed variational principle, whose stationary solutions are not naturally associated with easily identifiable optimization targets in neural parameter space. Therefore, energy-based PINNs are most naturally applicable when the governing functional is bounded below and locally coercive, for example when $\delta^{2}\mathcal{L}>0$, where $\mathcal{L}$ denotes the variational functional.
	By contrast, strong-form PINNs do not rely on variational structure and therefore remain more broadly applicable across general PDE classes, although they introduce their own challenges such as residual balancing and boundary-condition enforcement. Future progress in PINNs will depend not only on developments in computational mechanics formulations, but also on advances in optimization algorithms for highly non-convex scientific machine learning problems.

	Operator learning algorithms essentially learn the implicit mapping of PDE families through big data. The main operator learning algorithms currently are FNO and DeepONet, both supported by corresponding theories. The core of these algorithms is the kernel integral method, and different kernel integral schemes can give rise to various operator learning algorithms. While Fourier transformation is currently in use, other transformations such as Laplace transform and Z-transform can be considered special cases of kernel integration, leading to diverse neural operator algorithms. 
	Additionally, DeepONet is based on the mathematical proof of approximating continuous operators
	by neural networks from 1995 \cite{chen1995universal}, substituting this mathematical proof's convergence criteria with neural
	networks. We believe that other theories have proven the same thing, that neural networks can approximate any continuous operator. As a result, other theories can be exploited to create various algorithms similar to DeepONet in the future.

	The algorithm based on the physics-informed neural operator is particularly interesting and has significant research and industrial application prospects. This is because this method can continually self-learn.
	The algorithm can become faster over time, and with gradually increasing data, it can train increasingly accurate operators. Essentially, the connection is universal; although the problems we are currently solving have not precisely appeared in the training set, they share similarities, and
	we can leverage existing data for unknown tests, and then correct them based on physical equations.  The larger the dataset, the more accurate
	the initial solution predicted by operator learning will be, and thus the physical equation correction
	time will be shorter. Another combination method is for unknown phenomena, where an approximate equation can provide an approximate solution, which is then fine-tuned based on data, making it very suitable for complex phenomena, such as biomechanics.
	
	In this chapter, we primarily introduced some methodologies of AI for PDEs. In the next chapter, we will introduce some theoretical research on AI for PDEs, focusing on PINNs.
	In addition, recent studies have begun exploring the use of foundation models for assisting AI4PDE development through automatic scientific programming \cite{jiang2025deepseek}. Beyond code generation, AI agents are attracting increasing attention because they offer the possibility of end-to-end scientific workflow automation, including problem formulation, solver construction, execution, and result interpretation \cite{lu2024ai}. 
	In computational mechanics, geometry modeling is widely recognized as one of the most time-consuming components in simulation pipelines, often requiring substantial user interaction during preprocessing. Recently, Wang et al. \cite{wang2026deep} integrated large language models into the deep energy method framework, demonstrating that foundation models can assist geometry generation and preprocessing under natural-language instructions.
	Although the integration of AI agents with AI4PDEs is still at an early exploratory stage, it represents a highly promising future direction. In particular, foundation models may gradually support automatic hyperparameter adjustment in PINNs, adaptive balancing of multiple loss terms, architecture selection, and solver strategy optimization, thereby improving both usability and efficiency of AI4PDE solvers.

	\section{AI for PDEs: Theoretical Research}\label{sec:AI4PDEs_theory}
	
	\subsection{Activation functions}
	
	The choice of activation function affects the function representation of PINNs. Essentially, the activation function determines the degree of nonlinearity, thereby altering the neural network's fitting ability.
	The $tanh$ function is the most commonly
	selected activation function for PINNs, though other non-linear activation functions are also viable.
	PINNS must avoid activation functions with discontinuous higher-order derivatives or those
	whose higher-order derivatives are identically equal to zero, because PDEs often contain higher-order
	derivatives. The chain rule results in an ineffective optimization when the derivative of the activation
	function is not satisfied with the above condition. Therefore, activation functions like ReLu, which
	have a second-order derivative of zero, should not be chosen in higher-order derivative PDEs, especially when the highest order of the derivatives in the PDEs is greater than or equal to 2. The above is achieved by specifically changing
	the type of activation function to alter the accuracy and efficiency of PINNs. However, Jagtap et al.
	\cite{Adaptive_activation_functions} proposed a strategy applicable to any activation function, involving
	the multiplication of a trainable parameter $a$ after each neural network linear transformation layer,
	and the idea behind the strategy is similar to KAN \cite{liu2024kan}. This parameter alters the activation
	function shape to achieve faster convergence, expressed mathematically as: 
	\begin{equation}
		u(x)=(L_{k}\circ\sigma\circ naL_{k-1}\circ\sigma\circ naL_{k-2}\cdots\circ\sigma\circ naL_{1})(x),
	\end{equation}
	where $n$ is a non-trainable parameter set to determine the basic shape of the activation function,
	$L_{i}$ (where i ranges from 1 to k) represents the linear mapping layers of the network, and $\sigma$
	is the non-linear activation function. Later, Jagtap et al. \cite{LAAF} further refined the algorithm
	by independently adjusting each linear layer's $a$ and added a loss function term for $a$, aiming to
	increase $a$ to prevent gradient vanishing and thus further speeding up convergence. Essentially, the
	concept behind adjusting the activation function shape resonates with regularization in machine learning.
	
	\subsection{Errors}
	
	The errors in AI for PDEs primarily consist of four components: approximation error, optimization error, generalization error \cite{DeepOnet}, and integral error.
	
	Although neural networks have the capability to approximate any
	continuous function, a specific neural network structure must still be selected. Therefore,
	there will inevitably be some discrepancy between the solution space of the neural network and the actual
	solution space, which constitutes the approximation error. Optimization error refers to the error arising
	from the local optima due to the highly non-convex nature of neural networks. Generalization error is
	generally not specific to PINNs but is raised concerning operator learning because PINNs solve specific
	problems described by physical equations, whereas operator learning studies families of PDEs. Thus, operator
	learning deals with physical problems across different geometries, constitutive models, and boundary
	conditions, making generalization error a measure of how well operator learning approximates a family
	of PDEs.
	PINNs involve an integral error, which is determined by the number and distribution of collocation
	points. In the strong form of PINNs, integral error refers to the discrepancy between the integral of
	the least squares loss function and the exact integral. In the energy form of PINNs, integral error refers
	to the discrepancy between the integral over the energy functional and the exact energy functional. Most loss functions in PINNs typically use Monte Carlo integration for approximation. The expectation
	of Monte Carlo integration generally matches the exact value, and the variance measures the convergence
	efficiency of the integration. The expectation and variance of Monte Carlo integration are determined
	by the way collocation points are arranged. Therefore, researching how to reduce errors and enhance accuracy
	in PINNs by studying collocation methods is crucial, as these methods can reduce integral error.
	
\begin{figure}
		\begin{centering}
			\includegraphics[scale=0.60]{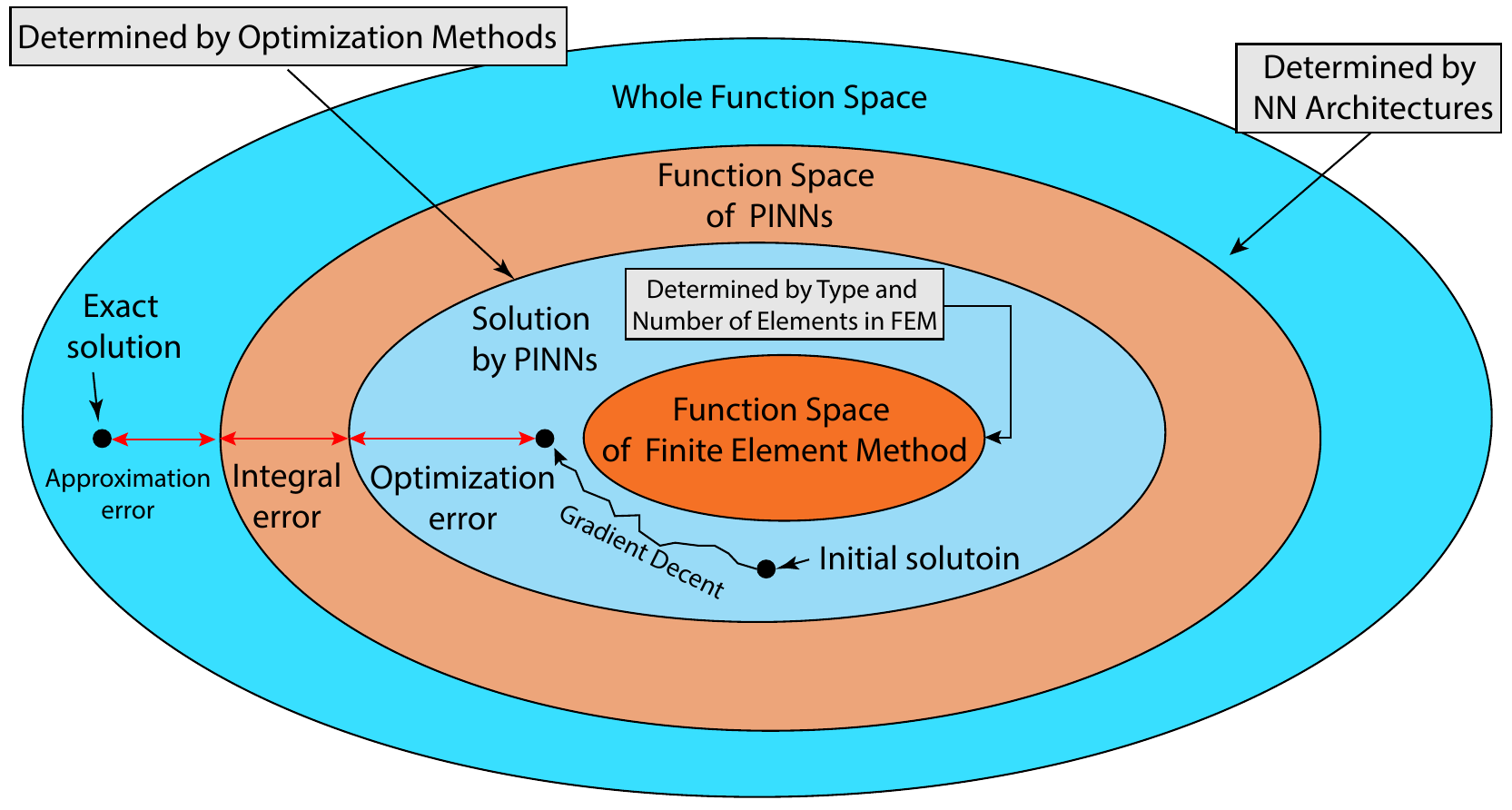}
			\par\end{centering}
		\caption{Error comparison between PINNs and FEM.\label{fig:PINO_FEM}}
	\end{figure}

	\Cref{fig:PINO_FEM} shows a visual representation for an intuitive understanding of errors in PINNs. For
	better comprehension, we compare PINNs with traditional computational mechanics methods. In traditional
	computational mechanics, such as finite element methods, the size of the function space for approximation
	is determined by the type and number of elements. In contrast, PINNs use neural networks as approximation
	functions, so the function space is significantly larger due to the universal approximation \cite{pinkus1999approximation}. However, due to the optimization methods and
	the highly non-convex nature of neural networks, the full potential of PINNs\textquoteright{} powerful
	fitting capabilities is not realized, leading to optimization error. Additionally, the integral error
	introduced by the collocation method determines the extent to which the PINNs' loss function approximates
	the true loss function, thus also limiting the approximation capability of PINNs. Finally, because the
	structure of the neural network is specifically given, the mathematical operations behind PINNs are very
	clear, meaning the approximation function space is already determined and the gap between the function
	space of NN and the exact solution introduces approximation error. From this perspective, the structure
	of the neural network plays a role similar to the element type in finite elements, as both determine
	the approximation function space. Displacement finite elements obtain nodal displacement values which
	is used to acquire the physical field of the entire domain through interpolation with shape functions,
	while PINNs simulate field variables by optimizing neural network parameters. Thus, the parameters of the neural network in PINNs are analogous to the nodal displacements in finite element methods.
	
	To better understand the integral error in PINNs, we provide the following example to illustrate. Consider
	a simple ordinary differential equation example \citet{fang2021high}: 
	\begin{equation}
		\begin{cases}
			u^{''}(x)=0 & x\in(-1,1)\\
			u(-1)=0 & ,u(1)=1
		\end{cases}. \label{eq:loss_of_strange_error}
	\end{equation}
	
	If we select $N_{u}+2$ points between -1 (inclusive) and 1 (inclusive), excluding $x=0$, to construct
	the loss function for the strong form of PINNs, we observe an interesting phenomenon. The loss function
	is: 
	\begin{equation}
		Loss=\frac{1}{N_{u}}\sum_{i=1}^{N_{u}}|u^{''}(x_{i})|^{2}+u(-1)^{2}+[u(1)-1]^{2}. \label{eq:loss_error}
	\end{equation}
	As $N_{u}$ approaches infinity, it is easy to verify that the piece function in \Cref{eq:piece_function} can make the loss function in \Cref{eq:loss_error} zero.
	\begin{equation}
		u(x)=\begin{cases}
			0 & x<0\\
			x & x\geq0
		\end{cases} \label{eq:piece_function}
	\end{equation}
	However, the exact solution to \Cref{eq:loss_of_strange_error}
	is $u(x)=x+1/2$. Due to the powerful fitting capabilities of neural networks, theoretically, exact solution and \Cref{eq:piece_function} both
	could potentially be optimized (in fact, many other solutions that satisfy the zero loss function also
	exist). The phenomenon of non-unique solutions is not only due to the powerful fitting capabilities of neural networks but also
	due to the integral error in PINNs. The neural network brings theoretical
	advantages of a large approximation function space but also introduces the drawback of non-uniqueness in numerical solutions (traditional methods are very robust and stable due to the limitations of approximation
	function space). Because the approximation function space of the neural network is very large, regularization techniques are
	crucial for AI for PDEs \cite{cuomo2022scientific}. 
	
	We summarizes articles related to mathematical proofs of convergence for AI for PDEs. As shown
	in \Cref{tab:AI for PDEs Convergence Proofs}, the proofs are divided into three main parts: the first
	part concerns the convergence proofs for both direct and inverse problems in PINNs, along with proofs
	that neural networks can approximate any continuous function and articles on uncertainty estimates in
	PINNs; 
	the second part relates to operator learning, which includes proofs that neural operators can
	approximate any continuous operator, specific proofs of convergence for the Fourier Neural Operator (FNO)
	within neural operators, and proofs that DeepONet can approximate any continuous operator, meaning it can approximate any family of continuous PDEs; the third part discusses how Physics-Informed Neural Operators (PINO) can approximate
	any continuous function or any continuous operator.

	\begin{figure}
		\begin{centering}
			\includegraphics[scale=0.55]{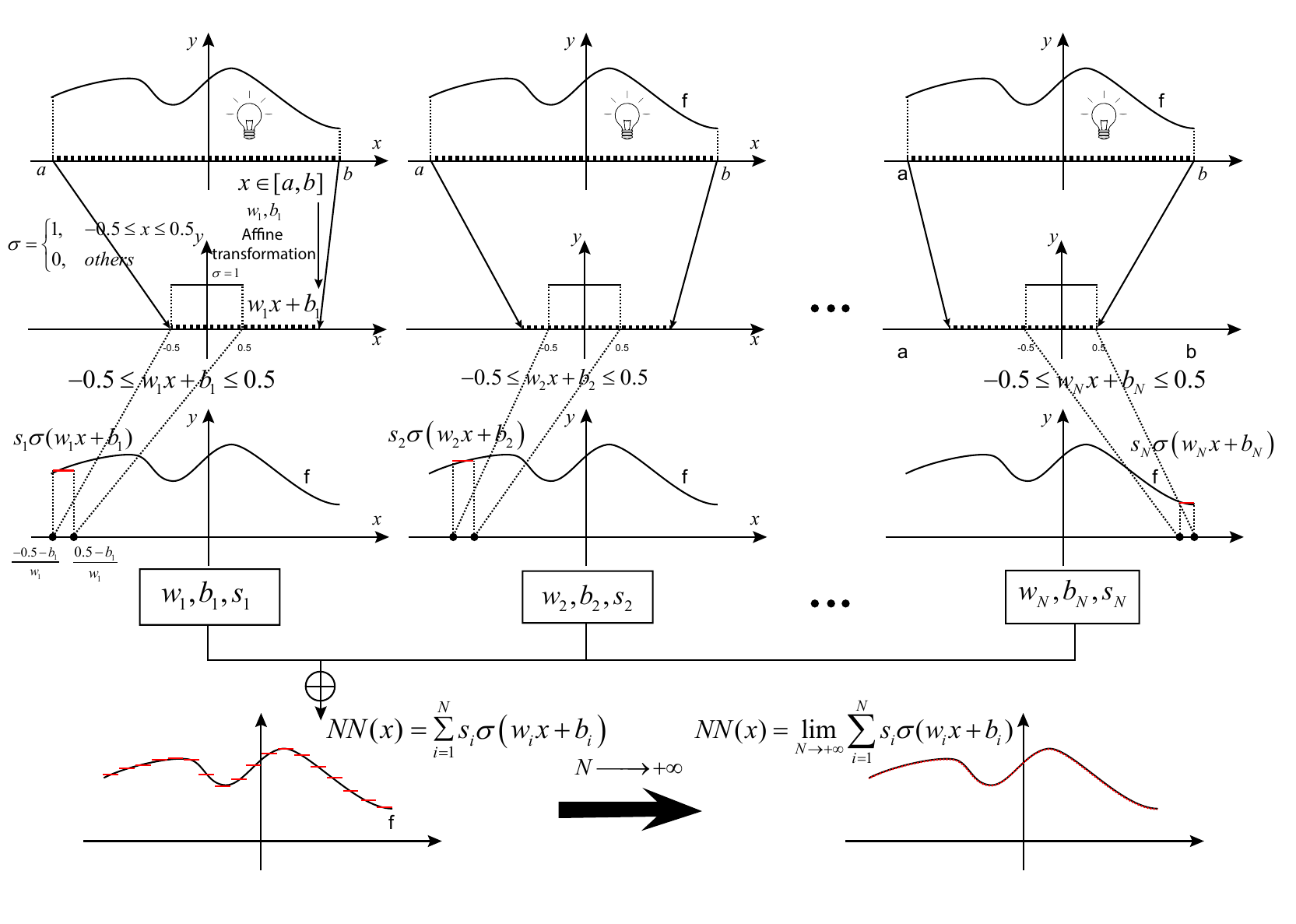}
			\par\end{centering}
		\caption{A brief explanation of the universal approximation theorem for neural networks.\label{fig:NN_approximation}}
	\end{figure}
	
	In summary, a single-layer feedforward network can represent any continuous function (universal approximation theory) \citet{hornik1991approximation}. \Cref{fig:NN_approximation} briefly illustrates the core concept of the universal approximation theory. To explain it briefly, we consider an arbitrary one-dimensional continuous function \( f(x) \), where \( x \in [a,b] \). We approximate it using a fully connected neural network \( NN(x) = \sum_{i=1}^{N}s_{i}\sigma(w_{i}x + b_{i}) \), where the activation function is chosen as:
	
	\begin{equation}
		\sigma(x) = 
		\begin{cases}
			1 & \text{if } -0.5 \leq x \leq 0.5\\
			0 & \text{otherwise}
		\end{cases}
	\end{equation}
	
	It is easy to observe that \( w_{i}x + b_{i} \) is non-zero only when it lies within the range \([-0.5, 0.5]\); outside this range, the result is zero. Therefore, after applying the activation function \( \sigma \), the range of \( x \) where \( \sigma(w_{i}x + b_{i}) \) is non-zero can be expressed as:
	
	\begin{equation}
		\frac{-0.5 - b_{i}}{w_{i}} \leq x \leq \frac{0.5 - b_{i}}{w_{i}}
	\end{equation}
	
	Next, we only need to adjust \( s_{i} \) such that:
	
	\begin{equation}
		||f(x) - s_{i}\sigma(w_{i}x + b_{i})||_{L^{\infty}} \leq \varepsilon_{i}, \; \text{where} \; \frac{-0.5 - b_{i}}{w_{i}} \leq x \leq \frac{0.5 - b_{i}}{w_{i}}
	\end{equation}
	where \( ||\cdot||_{L^{\infty}} \) represents the maximum value of the function. Clearly, by controlling the affine transformations \( w_{i} \) and \( b_{i} \), we can achieve a piecewise approximation, as shown in \Cref{fig:NN_approximation}. Therefore, with a sufficient number of parameters \( \{w_{i}, b_{i}, s_{i}\}_{i=1}^{N} \), we can approximate \( f \) within any desired error \( \varepsilon \). Mathematically, this can be expressed as follows: for any given error \( \varepsilon \), there exists an \( N \) such that:
	
	\begin{equation}
		||f(x) - \sum_{i=1}^{N} s_{i}\sigma(w_{i}x + b_{i})||_{L^{\infty}} \leq \varepsilon = \max\{\varepsilon_{i}\}_{i=1}^{N}, \; \text{where} \; a \leq x \leq b
	\end{equation}

	The above explanation is an informal but intuitive interpretation of the universal approximation theorem. This proof only applies to single-layer networks with a one-dimensional target function, not to higher dimensions. A formal proof can be found in \Cref{tab:AI for PDEs Convergence Proofs} under "Proofs of the strong fitting capabilities of NNs." Empirically, using deeper networks can achieve the same function with fewer units and improve generalization error \cite{goodfellow2016deep}. However, neural networks may require an impractically large number of units and may struggle with optimization problems. Most of the convergence proofs demonstrate that a neural network exists which can approximate a function or operator within \( \varepsilon \), but they do not provide a method for determining its parameters and hyperparameters. Therefore, future research could focus on developing theoretical frameworks that offer guidance on neural network architecture design to achieve better convergence.

	\begin{table}
		\centering{}\caption{AI for PDEs convergence proofs\label{tab:AI for PDEs Convergence Proofs}}
		\begin{tabular}{cc}
			\toprule 
			Literature & Brief Description of Proof\tabularnewline
			\midrule
			\citet{hornik1989multilayer}\citet{super_approximation}\citet{pinkus1999approximation}\citet{cohen2016expressive} & Proofs of the strong fitting capabilities of NNs\tabularnewline
			\citet{shin2020convergence} & Convergence proofs for elliptic second-order linear and parabolic PDEs using PINNs\tabularnewline
			\citet{mishra2022estimates} & Proofs of approximation properties of inverse problems using PINNs\tabularnewline
			\citet{psaros2023uncertainty} & Estimations of uncertainty in PINNs\tabularnewline
			\citet{kovachki2023neural} & Proofs that neural operators can approximate any continuous operator\tabularnewline
			\citet{kovachki2021universal} & Proofs that the FNO in neural operators can approximate any continuous operator\tabularnewline
			\citet{chen1995universal} & Theoretical support for DeepONet: NNs can approximate any continuous operator\tabularnewline
			\citet{lanthaler2022error} & Proofs that DeepONet can approximate any continuous operator\tabularnewline
			\citet{de2022generic} & PINO can approximate any continuous function or any continuous operator\tabularnewline
			\midrule
			& \tabularnewline
		\end{tabular}
	\end{table}
	
	\subsection{Weight Selection}
	
	The optimization process of PINNs is essentially a multi-task learning problem, where balancing different loss terms is crucial for algorithm efficiency and accuracy. From an optimization perspective, gradient-based algorithms tend to prioritize components with larger gradient magnitudes, potentially neglecting others \cite{mcclenny2023self}. The multi-task loss function can be expressed as:
	\begin{equation}
		\mathcal{L}=\lambda_{r}\mathcal{L}_{r}+\lambda_{i}\mathcal{L}_{i}+\lambda_{b}\mathcal{L}_{b}+\lambda_{d}\mathcal{L}_{d},\label{eq:loss_function_PINNs}
	\end{equation}
	where $\{\lambda_{r},\mathcal{L}_{r}\}$, $\{\lambda_{i},\mathcal{L}_{i}\}$, $\{\lambda_{b},\mathcal{L}_{b}\}$, and $\{\lambda_{d},\mathcal{L}_{d}\}$ represent the weights and loss functions for PDE residuals, initial conditions, boundary conditions, and data, respectively. Research on weight selection primarily focuses on how to adaptively determine these $\lambda$ values.
	\Cref{tab:weight_selection} summarizes representative studies on PINNs weight selection. We categorize them into the following four types.
	
	\begin{table}
		\centering{}\caption{Current research on PINNs weight selection \label{tab:weight_selection}}
		\begin{tabular}{cc}
			\toprule 
			Reference & Brief Description of Method\tabularnewline
			\midrule
			\citet{ill_gradient} & Selecting weights by comparing the gradients of different components of the loss function\tabularnewline
			\citet{NTK_to_get_hyperparameter_of_PINN} & Using NTK theory to select weights\tabularnewline
			\citet{NTK_PINN} & Identifying the frequency tendencies of neural networks using NTK theory\tabularnewline
			\citet{liu2021dual} & Modifying the PINNs loss function into a saddle point problem\tabularnewline
			\citet{xu2023transfer} & Modifying the loss function using maximum likelihood estimation\tabularnewline
			\bottomrule
		\end{tabular}
	\end{table}
	
	\paragraph{Gradient-based methods}
	Wang et al. \cite{ill_gradient} proposed adaptively adjusting weights by comparing gradients. The weight for each loss component is computed as the ratio between the maximum gradient of the PDE residual loss and the mean gradient of that component. A moving average scheme, similar to RMSprop and Adam optimizers, is then applied to stabilize the weight updates during training.
	
	\paragraph{NTK-based methods}
	The Neural Tangent Kernel (NTK) theory \cite{jacot2018neural} provides a framework for analyzing neural network training dynamics. Wang et al. \cite{NTK_to_get_hyperparameter_of_PINN} leveraged NTK to adjust loss weights based on the convergence rate of different components. The NTK matrix captures how changes at one training point affect predictions at others, with larger eigenvalues indicating faster convergence. By balancing the trace of NTK matrices corresponding to different loss terms, optimal weights can be determined. Subsequently, Wang et al. \cite{NTK_PINN} used NTK to analyze the spectral bias of neural networks---the tendency to learn low frequencies first---and proposed input coordinate transformations to address this issue in multi-scale PDE problems.
	
	\paragraph{Saddle point formulation}
	Liu et al. \cite{liu2021dual} transformed the loss function into a saddle point problem using constrained Lagrange multiplier methods:
	\begin{equation}
		\min_{\boldsymbol{\theta}}\max_{\boldsymbol{\alpha}}\mathcal{L}(\boldsymbol{\theta},\boldsymbol{\alpha})=\sum_{j}\lambda_{j}(\boldsymbol{\alpha})\mathcal{L}_{j}(\boldsymbol{\theta}),\label{eq:saddle_pinns}
	\end{equation}
	where $\boldsymbol{\theta}$ are network parameters, $\boldsymbol{\alpha}$ are trainable parameters, and $\lambda_{j}$ are obtained by applying softmax to $\boldsymbol{\alpha}$. This formulation generalizes the traditional weighted loss and allows the model to automatically learn optimal weights during training.
	
	\paragraph{Maximum likelihood estimation}
	Xu et al. \cite{xu2023transfer} adapted an uncertainty weighting method \cite{kendall2018multi} for PINNs. By assuming that model predictions follow Gaussian distributions with task-specific uncertainties $\sigma_{i}$, the maximum likelihood estimation leads to the loss function:
	\begin{equation}
		\mathcal{L}=\sum_{i=1}^{k}\left[\frac{\mathcal{L}_{i}}{2\sigma_{i}^{2}}+\frac{1}{2}\ln\sigma_{i}^{2}\right]\label{eq:maximum_likelihood_PINNs},
	\end{equation}
	where $\mathcal{L}_{i}$ are the individual loss components. Tasks with higher uncertainty (larger $\sigma_{i}^{2}$) automatically receive lower weights, enabling dynamic balancing during training without pre-specified hyperparameters. The uncertainty parameters $\sigma_{i}$ are learned jointly with the network parameters.
	
	\subsection{Distance Networks\label{subsec:Distance-Networks}}
	In addition to the constraints of the PDEs, there are also the constraints of boundary conditions.
	The simplest implementation is through penalty factors to satisfy the constraints of boundary conditions:
	\begin{equation}
		Loss_{b} = \beta \sum_{i=1}^{N} \left| NN(\boldsymbol{x}_{i}; \boldsymbol{\theta}) - \bar{u}(\boldsymbol{x}_{i}) \right|^2 
	\end{equation}
	where $Loss_{b}$ is the boundary loss, \( NN(\boldsymbol{x}_{i}; \boldsymbol{\theta}) \) is the neural network of the variable of interest, \( \boldsymbol{\theta} \) is the trainable parameter of the neural network, and \( \bar{u}(\boldsymbol{x}_{i}) \) is the boundary condition at \( \boldsymbol{x}_{i} \). \( \beta \) is the penalty factor.
	
	However, the penalty factors introduce additional hyperparameters and
	can compromise the uniqueness of the solution. 
	The penalty factor is a method of satisfying boundary conditions as a soft constraint, which is an approximate way of meeting boundary conditions. Therefore, using a penalty factor to approximately satisfy boundary conditions can theoretically lead to non-unique solutions in PINNs.
	
	Therefore, strictly enforcing
	boundary conditions is a very important direction for PINNs.
	One method of strict enforcement is through the use of distance networks which is very similar to the idea of imposing boundary conditions in immersed and meshfree methods proposed in the early 1990s. Consider the following mechanical
	equation:
	
	\begin{equation}
		\begin{cases}
			\text{Domain Control Equation: }\boldsymbol{L}(\boldsymbol{u}(\boldsymbol{x}))=\boldsymbol{f}(\boldsymbol{x}) & \forall\boldsymbol{x}\in\Omega\\
			\text{Displacement Boundary Condition: }\boldsymbol{u}(\boldsymbol{x})=\boldsymbol{h}(\boldsymbol{x}) & \forall\boldsymbol{x}\in\Gamma^{u}\\
			\text{Force Boundary Condition: }\boldsymbol{\sigma}(\boldsymbol{x})\cdot\boldsymbol{n}(\boldsymbol{x})=\boldsymbol{t}(\boldsymbol{x}) & \forall\boldsymbol{x}\in\Gamma^{t}
		\end{cases},
	\end{equation}
	where $\boldsymbol{L}$ is the differential operator, $\boldsymbol{u}(\boldsymbol{x})$ is the displacement at $\boldsymbol{x}$, $\boldsymbol{f}(\boldsymbol{x})$ is the body force, $\boldsymbol{h}(\boldsymbol{x})$ is the Dirichlet boundary ($\Gamma^{u}$) condition, $\boldsymbol{t}(\boldsymbol{x})$ is the Neumann boundary ($\Gamma^{t}$) condition, and $\boldsymbol{\sigma}(\boldsymbol{x})$ is the stress tensor.
	To solve this PDEs, we construct the displacement as follows: 
	\begin{equation}
		\boldsymbol{u}(\boldsymbol{x})=\boldsymbol{h}(\boldsymbol{x})+\boldsymbol{D}(\boldsymbol{x})*\boldsymbol{u}_{g}(\boldsymbol{x};\boldsymbol{\theta}),\label{eq:admissible_distance}
	\end{equation}
	where an approximation function fits $\boldsymbol{h}(\boldsymbol{x})$ as particular network, $\boldsymbol{D}(\boldsymbol{x})$
	is the distance network, which outputs the shortest distance to the displacement boundary based
	on the coordinate $\boldsymbol{x}$:
	\begin{equation}
		\boldsymbol{D}(\boldsymbol{x})=\min_{\boldsymbol{y}\in\Gamma^{u}}\sqrt{(\boldsymbol{x}-\boldsymbol{y})\cdot(\boldsymbol{x}-\boldsymbol{y})}.
	\end{equation}
	This distance can be pre-calculated for a finite set of points using nearest neighbor algorithms
	and then approximated by a fitter $D(\boldsymbol{x})$. Finally, $\boldsymbol{u}_{g}(\boldsymbol{x};\boldsymbol{\theta})$
	is a neural network that learns based on the PDEs and the force boundary condition.
	The advantage of the distance network is that if the coordinate point lies on the displacement boundary $\Gamma^{u}$, where the displacement network $D(\boldsymbol{x})$ equals zero, the output of the displacement field from \Cref{eq:admissible_distance} is exactly $\boldsymbol{h}(\boldsymbol{x})$.
	This means that during
	the training of PINNs in the strong form, there is no need to consider the displacement boundary condition.

	Thus, distance networks are important for PINNs, both in the strong and energy forms. Research on distance
	networks primarily focuses on exploring admissible displacement field compositions to better train PINNs.
	The following is a review of distance networks: Lagaris et al. (1998) \cite{admissible_earliest_paper}
	initially proposed the concept of PINNs as well as the construction of admissible displacement fields
	that meet essential boundary conditions, but only provided a general mathematical form and specific construction
	methods in regular domains, i.e., converting the distance network into a multiplication by coordinates.
	Later, Lagaris et al. (2000) \cite{lagaris2000neural} proposed the construction of admissible displacement
	fields for irregular domains. McFall et al. (2009) \cite{mcfall2006artificial,mcfall2009artificial}
	discussed in detail the procedure for solving boundary value problems of arbitrary boundary
	shapes using neural networks. With PINNs receiving significant attention in recent years, this research
	area has been revitalized, and Berg et al. (2018) \cite{complex_PINN_a_method_to_construct_admissible_function}
	proposed the form of admissible displacement fields in \Cref{eq:admissible_distance}, where the particular
	solution network $\boldsymbol{h}(\boldsymbol{x})$, distance network $\boldsymbol{D}(\boldsymbol{x})$,
	and general network $\boldsymbol{u}_{g}(\boldsymbol{x};\boldsymbol{\theta})$ are all approximated through
	fully connected neural networks, essentially similar to the method proposed by Lagaris et al. (2000)
	\cite{lagaris2000neural}. In DEM, Samaniego et al. \cite{loss_is_minimum_potential_energy} were the first to employ distance function in order to impose Dirichlet boundary conditions. Later, Rao et al. (2021) \cite{PINNstrong_form_in_elastodynamics} extended
	the form of admissible displacement fields to linear elastic dynamics; Sheng et al. (2021) \cite{admissible_in_PINN_energy_form}
	proposed using RBF (Radial Basis Function) instead of the fully connected form of the distance network
	to solve the energy form of PINNs. It is worth mentioning that Sukumar et al. (2022) \cite{boundary_conditions_distance_functions}
	discussed PINNs' distance networks in great detail, even satisfying force boundary conditions in addition
	to displacement boundary conditions.
	
	In summary, research on distance networks seems to have reached its developmental limit at the methodological
	level. In the future, the concept of distance networks is expected to be more widely integrated into
	various PINNs and operator learning algorithms, aiming to enhance the accuracy of these algorithms and
	reduce their reliance on hyperparameters.
	
	\subsection{Transfer Learning}
	
	Transfer learning is a crucial concept and application in machine learning, and also applied in PINNs.
	The main advantage of using transfer learning in PINNs is due to iterative algorithms, which can inherit
	previously trained parameters, thus requiring fewer iterations for similar new tasks. 
	The application approach of transfer learning in PINNs is generally the same, which involves directly inheriting the parameters from the previous task and then iterating in new tasks.
	There are two main methods of parameter iteration:
	
	The first method is very simple; it involves completely inheriting the parameters and then re-iterating \cite{PINN_solid_mechanics}. The second method involves freezing the first few layers of the neural network and only training the later layers to achieve the effect of transfer learning. For example, Goswami et al. (2020) \cite{goswami2020transfer} proposed using transfer learning to reduce the computational cost of phase-field modeling in fracture mechanics. In phase-field modeling, incremental step loading is required, so recalculating each step would be computationally expensive. By transferring the parameters learned by the PINNs energy method from the previous step to the next incremental step and training only the later layers, the iteration time for subsequent steps is reduced. Similarly, Chakraborty (2021) \cite{chakraborty2021transfer} proposed a transfer learning approach that incorporates approximate PDEs, and freezed the parameters of the initial layers, and uses high-precision experimental data to train the parameters of the later layers of the neural network that have been trained by PDEs. Xu et al. (2023) \cite{xu2023transfer} suggested first training PINNs on a simple geometric model, and freezing the parameters of the shallow layers of the fully connected neural network while only training the deep layers in real tests with complex geometries. The rationale is that most geometric problems in structural mechanics are similar, and the basic features extracted by the shallow layers can be combined with the deep layers to form similar geometries, allowing the parameters to be directly transferred and reducing the number of iterations in PINNs.
	Recently, Wang et al. incorporated LoRA (Low-Rank Adaptation) \citet{hu2021lora} into both the strong and energy forms of PINNs across different boundary conditions, materials, and geometries \cite{wang2025transfer}. In theory, LoRA can be regarded as a generalized form of the two aforementioned transfer learning strategies, since the rank in LoRA provides a flexible mechanism to control the extent of parameter adaptation during transfer.
	
	In summary, these two methods in transfer learning are quite common in machine learning. This approach allows us to leverage models that have been pre-trained on large datasets and use transfer learning to address new tasks with smaller datasets or tasks that are slightly different from the original ones.

	\subsection{Gradient Approximation}
	
	The core of AI for PDEs lies in the computation of differential operators for PDEs. Therefore, the calculation
	of gradients is particularly crucial for solving PDEs using neural networks. There are generally three
	methods to compute gradients in this field: numerical difference approximation, symbolic computation,
	and automatic differentiation \cite{baydin2018automatic}. Each of these methods is reviewed below.
	
	\paragraph{Numerical Difference Approximation} This method essentially approximates the limit form of derivatives,
	ignoring higher-order terms to achieve an approximate differential. The approach to approximate gradients
	using the numerical difference in AI for PDEs aligns with traditional numerical analysis and doesn't fundamentally
	differ. Numerical differences include forward, backward, and central differences, which differ in terms
	of error order. The advantage of numerical difference lies in its simplicity and wide applicability;
	it does not require knowledge of the derivative's analytic form. However, its drawbacks include significant
	truncation and rounding errors. Therefore, despite its broad applicability, its error characteristics
	must be carefully considered. Recently, some methods have used CNNs to approximate gradients, essentially
	using fixed convolution kernel weights to simulate finite difference \cite{fang2021high}.
	
	\paragraph{Symbolic Computation} This method obtains analytical solutions of gradients through derivative operations
	(e.g., product rule, chain rule), and then inserts specific values to achieve the final gradients. Typically
	implemented recursively, this method offers high precision but can become inefficient with complex functions
	due to expression swelling \cite{noor1979computerized}, which is challenging to simplify.
	
	\paragraph{Automatic Differentiation} Automatic differentiation is similar to symbolic computation but avoids
	the problem of expression swelling by breaking down complex functions into combinations of simple functions
	and using known derivatives of basic functions through the chain rule. During derivative computation,
	node values from forward operations are used in the chain rule formulas, thus eliminating the need for
	a complete analytical expression of gradients, thereby addressing the issue of expression swelling seen
	in symbolic computation. There are two modes of implementation for automatic differentiation: forward
	mode and reverse mode. Forward mode computes function values and gradients simultaneously in one forward
	operation, which requires less memory and is suitable for problems with fewer parameters. However, reverse
	mode, which computes derivatives layer by layer from the output back inward, is more common in deep learning
	due to its suitability for problems with smaller output dimensions and larger parameter sets. Overall,
	automatic differentiation not only provides precise derivatives, similar to symbolic computation, but
	also solves the weaknesses of symbolic methods, such as expression swelling. Furthermore, the reverse
	mode is particularly well-suited to the small-output-dimensional context of deep learning. In AI for PDEs, it is common to compute the derivatives of a single scalar (such as the loss function) concerning a large number of weights, making the reverse mode very suitable for this scenario.
	Thus, the reverse
	mode is the predominant choice for derivative computation in the AI4PDEs domain.
	
	In recent years, apart from automatic differentiation, another method of gradient approximation has emerged
	through statistical approaches \cite{he2023learning,sirignano2018dgm}. This method does not require
	backward gradient propagation for differentiation.

	\subsection{All Roads Lead to Rome}
	
\begin{figure}
		\begin{centering}
			\includegraphics[scale=0.7]{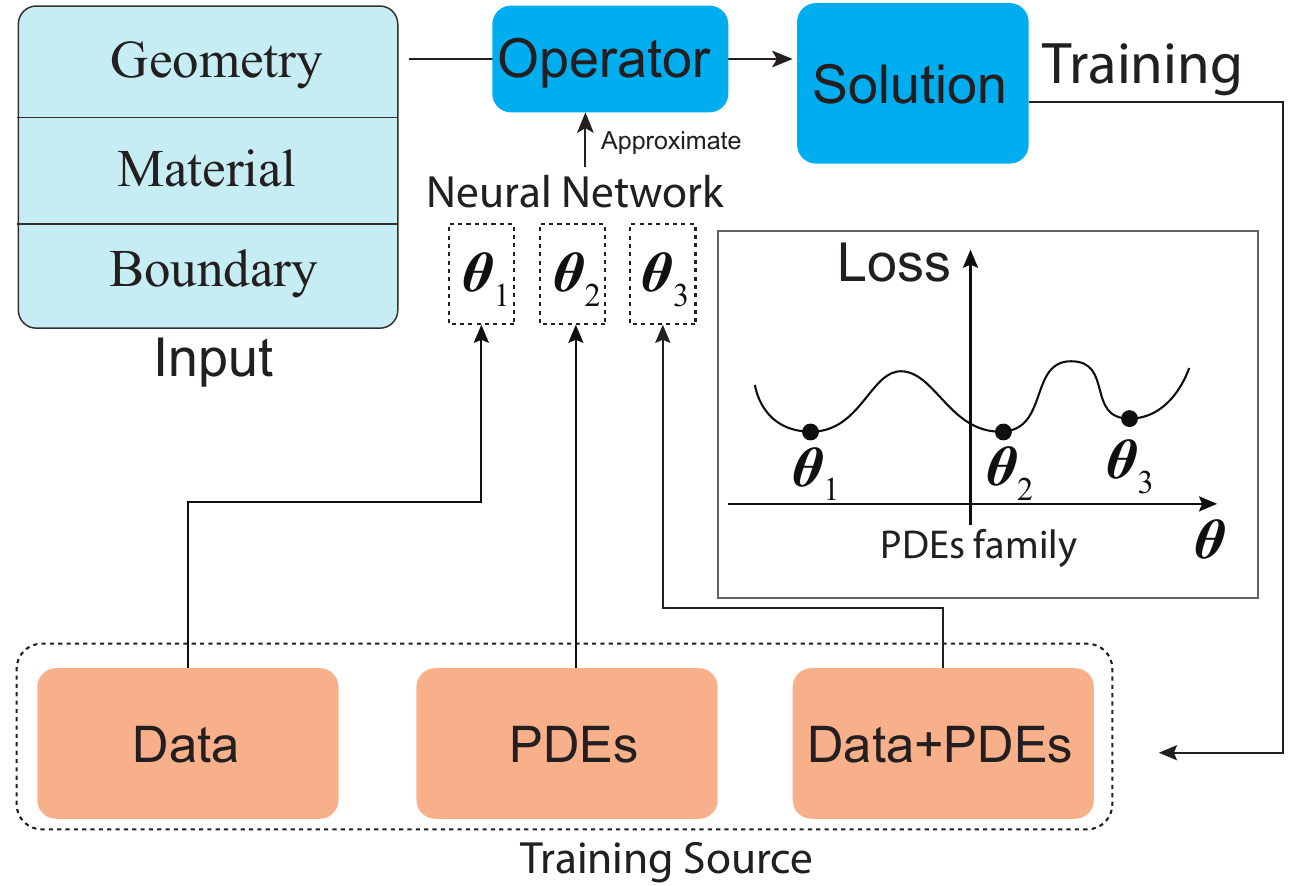}
			\par\end{centering}
		\caption{Illustration of training PDEs family using neural networks: $\boldsymbol{\theta}_{1}$ denotes the parameters of the neural network trained purely with data; $\boldsymbol{\theta}_{2}$ denotes the parameters trained purely with PDEs; $\boldsymbol{\theta}_{3}$ denotes the parameters obtained by combining data and PDEs for training.\label{fig:PDEs-family}}
	\end{figure}

	In theory, there are multiple ways to train an operator mapping from the input function space to the solution space, as illustrated in \Cref{fig:PDEs-family}. For example, FNO can successfully learn a PDEs family operator purely from data \cite{li2020fourier}; VINO can also successfully learn the PDEs family operator solely from the governing PDEs \cite{eshaghi2025variational}; PINO, by combining data and the physical equations, can also successfully learn the PDEs family operator \cite{li2021physics}. This implies that there are multiple ways to achieve the training of a PDEs family operator, i.e., ``All Roads Lead to Rome.''
	
	The approach combining data and physical equations has more advantages compared to using data-only or physics-only methods. Training purely with data does not require additional effort to construct a PDE loss function, while the physics-based approach can compensate for insufficient training data. Therefore, theoretically, the data-physics hybrid approach is the optimal way to train a PDEs family operator. However, this approach may face a trade-off between the data loss and the PDE loss, which could necessitate additional hyperparameter tuning during training.
	
	The idea of ``All Roads Lead to Rome'' is similar to the observation that there is not a single way to realize intelligence. For example, it has been found that modifying just one pixel in an image can completely fool a computer vision algorithm for classification tasks, whereas the human brain remains unaffected. This indicates that although both human brains and artificial intelligence can perform well on computer vision tasks, the underlying mechanisms by which intelligence is realized are different.
	
	\subsection{Benefits of Operator Learning as Good Initial Guesses}
	
	The predictions provided by operator learning can serve as initial guesses for iterative solvers, significantly reducing the required number of iterations compared to traditional numerical algorithms. We explain this by analyzing the convergence of iterative methods for linear systems.
	
	Consider the iterative scheme
	\begin{equation}
		\boldsymbol{X}^{(k)} = \boldsymbol{B}\boldsymbol{X}^{(k-1)} + \boldsymbol{f} \label{eq:iterative_way}
	\end{equation}
	Clearly, the exact solution $\boldsymbol{X}^{*}$ satisfies
	\begin{equation}
		\boldsymbol{X}^{*} = \boldsymbol{B}\boldsymbol{X}^{*} + \boldsymbol{f} \label{eq:exact_solution}
	\end{equation}
	Define the error as
	\begin{equation}
		\boldsymbol{e}^{(k)} = \boldsymbol{X}^{(k)} - \boldsymbol{X}^{*} \label{eq:iter_error}
	\end{equation}
	Substituting \Cref{eq:iterative_way} and \Cref{eq:exact_solution} into \Cref{eq:iter_error} yields
	\begin{equation}
		\begin{aligned}
			\boldsymbol{e}^{(k)} &= (\boldsymbol{B}\boldsymbol{X}^{(k-1)} + \boldsymbol{f}) - (\boldsymbol{B}\boldsymbol{X}^{*} + \boldsymbol{f}) \\
			&= \boldsymbol{B} (\boldsymbol{X}^{(k-1)} - \boldsymbol{X}^{*}) = \boldsymbol{B} \boldsymbol{e}^{(k-1)}
		\end{aligned}
	\end{equation}
	Recursively, we have
	\begin{equation}
		\boldsymbol{e}^{(k)} = \boldsymbol{B}^{k} \boldsymbol{e}^{(0)}
	\end{equation}
	Note that the quality of the initial guess is mathematically captured by $\boldsymbol{e}^{(0)} = \boldsymbol{X}^{(0)} - \boldsymbol{X}^{*}$. That is, a better initial guess results in a smaller norm $||\boldsymbol{e}^{(0)}||$. Using the subordinate matrix norm, we have
	\begin{equation}
		||\boldsymbol{B}^{k}|| = \max_{\boldsymbol{e}^{(0)}} \frac{||\boldsymbol{B}^{k} \boldsymbol{e}^{(0)}||}{||\boldsymbol{e}^{(0)}||} \ge \frac{||\boldsymbol{B}^{k} \boldsymbol{e}^{(0)}||}{||\boldsymbol{e}^{(0)}||}
	\end{equation}
	\begin{equation}
		||\boldsymbol{e}^{(k)}|| = ||\boldsymbol{B}^{k} \boldsymbol{e}^{(0)}|| \le ||\boldsymbol{B}^{k}|| \, ||\boldsymbol{e}^{(0)}|| \label{eq:iter_error_evolution}
	\end{equation}
	This clearly implies that, to achieve the same accuracy $||\boldsymbol{e}^{(k)}||$, a better initial guess reduces the upper bound on the number of iterations $k$. This theoretically explains why operator learning can provide better initial guesses, thereby reducing the iteration count in the fine-tuning phase, especially when traditional iterative algorithms cannot provide good initial values.
	
	The benefit diminishes as the tolerance (tol) becomes smaller. Rewriting \Cref{eq:iter_error_evolution}, we obtain
	\begin{equation}
		\frac{||\boldsymbol{e}^{(k)}||}{||\boldsymbol{e}^{(0)}||} \le ||\boldsymbol{B}^{k}|| \label{eq:ratio_error}
	\end{equation}
	Let $\varepsilon = ||\boldsymbol{B}^{k}||$, then \Cref{eq:ratio_error} becomes
	\begin{equation}
		\frac{||\boldsymbol{e}^{(k)}||}{||\boldsymbol{e}^{(0)}||} \le \varepsilon \label{eq:elison}
	\end{equation}
	Taking logarithms:
	\begin{align}
		||\boldsymbol{B}^{k}||^{1/k} &= \varepsilon^{1/k} \nonumber \\
		\ln ||\boldsymbol{B}^{k}||^{1/k} &= \ln \varepsilon^{1/k} \nonumber \\
		k &= \frac{-\ln \varepsilon}{-\ln ||\boldsymbol{B}^{k}||^{1/k}} \label{eq:iter_num_discussion}
	\end{align}
	The asymptotic convergence rate is defined as
	\begin{equation}
		R(\boldsymbol{B}) = \lim_{k\to +\infty} -\ln ||\boldsymbol{B}^{k}||^{1/k} \label{eq:liter_ratio}
	\end{equation}
	which depends only on the iteration matrix $\boldsymbol{B}$, not on $k$ or the initial guess. For large $k$, substituting \Cref{eq:liter_ratio} into \Cref{eq:iter_num_discussion} gives
	\begin{equation}
		k = \frac{-\ln \varepsilon}{R(\boldsymbol{B})} \label{eq:K_elison}
	\end{equation}
	
	The difference in iteration tolerance (tol) essentially corresponds to a difference in $||\boldsymbol{e}^{(k)}||$. Let $tol = 10^{-m}$, and approximate $||\boldsymbol{e}^{(k)}|| \approx tol$. Then \Cref{eq:elison} becomes
	\begin{align}
		\frac{10^{-m}}{||\boldsymbol{e}^{(0)}||} &\le \varepsilon \nonumber \\
		\ln \frac{10^{-m}}{||\boldsymbol{e}^{(0)}||} &\le \ln \varepsilon \nonumber \\
		-m \ln 10 - \ln ||\boldsymbol{e}^{(0)}|| &\le \ln \varepsilon \label{eq:elison_tol}
	\end{align}
	Substituting \Cref{eq:elison_tol} into \Cref{eq:K_elison} yields
	\begin{align}
		k &= \frac{-\ln \varepsilon}{R(\boldsymbol{B})} \nonumber \\
		&\le \frac{m \ln 10 + \ln ||\boldsymbol{e}^{(0)}||}{R(\boldsymbol{B})} = \frac{m \ln 10}{R(\boldsymbol{B})} + \frac{\ln ||\boldsymbol{e}^{(0)}||}{R(\boldsymbol{B})} \label{eq:k_R}
	\end{align}
	
	From \Cref{eq:k_R}, we see that $m \ln 10 / R(\boldsymbol{B})$ quantifies the impact of the iteration tolerance (tol) on the number of iterations, while $\ln ||\boldsymbol{e}^{(0)}|| / R(\boldsymbol{B})$ captures the influence of the initial guess. Clearly, the smaller the tolerance (larger $m$), the smaller the effect of the initial guess on the iteration count.

	\section{AI for PDEs in the application of forward problems of computational mechanics} \label{sec:AI4PDEs_forward}
	
	In this chapter, we review the applications of AI for PDEs in 
	computational mechanics. Specifically, we focus on their application in addressing forward problems across
	various domains such as solid mechanics, fluid dynamics, and biomechanics.

	\subsection{Solid mechanics}
	
	We focus on discussing the latest research developments in the field of AI for PDEs within solid mechanics.
	Therefore, we detail various examples of applications of AI for PDEs in solid mechanics. For the strong
	form of PINNs, it can solve almost any forward problem in solid mechanics, but for the energy form, it
	is only applicable to solid mechanics problems that adhere to energy principles, such as linear elastic
	statics, hyperelastic problems, and phase field method simulations of fractures.
	
	\subsubsection{Linear elasticity mechanics}
	
	The domain PDEs in linear elasticity mechanics are composed of three sets of equations: the equilibrium
	equations, constitutive equations, and kinematic equations:
	
	\begin{equation}
		\begin{cases}
			\text{Equilibrium equations: }\sigma_{ij,j}+f_{i}=\rho\ddot{u}_{i} & \forall(\boldsymbol{x},t)\in\Omega\vartimes[0,T]\\
			\text{Constitutive equations: }\sigma_{ij}=2G\varepsilon_{ij}+\lambda\varepsilon_{kk}\delta_{ij} & \forall(\boldsymbol{x},t)\in\Omega\vartimes[0,T]\\
			\text{Kinematic equations: }\varepsilon_{ij}=\frac{1}{2}(u_{i,j}+u_{j,i}) & \forall(\boldsymbol{x},t)\in\Omega\vartimes[0,T]
		\end{cases},\label{eq:elastic_pdes}
	\end{equation}
	where $\lambda$ and $G$ are the Lame parameters. $\boldsymbol{u}$, $\boldsymbol{\epsilon}$, and $\sigma$ are the displacement, strain, and stress respectively.
	Boundary and initial conditions include initial displacement conditions, initial velocity conditions,
	displacement boundary conditions, and force boundary conditions: 
	\begin{equation}
		\begin{cases}
			\text{Initial displacement condition: }\boldsymbol{u}(\boldsymbol{x},0)=\boldsymbol{u}_{0}(\boldsymbol{x}) & \forall\boldsymbol{x}\in\Omega\\
			\text{Initial velocity condition: }\dot{\boldsymbol{u}}(\boldsymbol{x},0)=\boldsymbol{v}_{0}(\boldsymbol{x}) & \forall\boldsymbol{x}\in\Omega\\
			\text{Displacement boundary condition: }\boldsymbol{u}(\boldsymbol{x},t)=\bar{\boldsymbol{u}}(\boldsymbol{x},t) & \forall(\boldsymbol{x},t)\in\Gamma^{u}\vartimes[0,T]\\
			\text{Force boundary condition: }\boldsymbol{t}(\boldsymbol{x},t)=\bar{\boldsymbol{t}}(\boldsymbol{x},t) & \forall(\boldsymbol{x},t)\in\Gamma^{t}\vartimes[0,T]
		\end{cases}.\label{eq:elastic_bound}
	\end{equation}
	
	The above set of equations represents all the control equations and boundary conditions in linear elasticity
	mechanics. If we consider the structural characteristics of the problem, these equations can be simplified,
	such as in plate and shell mechanics. However, the essence of the core solution remains solving PDEs in \Cref{eq:elastic_pdes} with boundary conditions in \Cref{eq:elastic_bound}. In the strong form of PINNs, \Cref{eq:elastic_pdes} and \Cref{eq:elastic_bound} are
	directly coupled through different hyperparameters into one loss function for solving. In the energy form
	of PINNs, these equations are transformed into an energy functional, such as the principle of minimum
	potential energy or minimum complementary energy. However, only static problems can be transformed. Although dynamic problems also involve the Hamiltonian functional, the functional of dynamic problems is stationary problems. The requirement of minimum potential and complementary energies is an extremum problem. Due to the highly non-convex
	nature of neural networks, solving the Hamiltonian functional in dynamics faces immense optimization
	difficulties due to infinite saddle points, making it extremely challenging to obtain stationary points.
	Additionally, a significant difficulty with the Hamiltonian functional's admissible displacement field
	is that it requires the displacement field to satisfy the final moment displacement solution (a necessity
	in Hamiltonian theory). Unfortunately, we do not know what the final moment displacement field is.
	Thus, the high degree of non-convexity and the unknown nature of the final moment displacement field
	pose significant challenges for the energy method of PINNs in dynamic problems in linear elasticity.
	The above analysis is just an application of PINNs in linear elasticity problems, while AI for PDEs encompasses
	not only PINNs but also operator learning and PINO. Therefore, we will next provide an
	overview of AI for PDEs in the forward problems of linear elasticity in solid mechanics.
	
\begin{figure}
		\begin{centering}
			\includegraphics[scale=0.55]{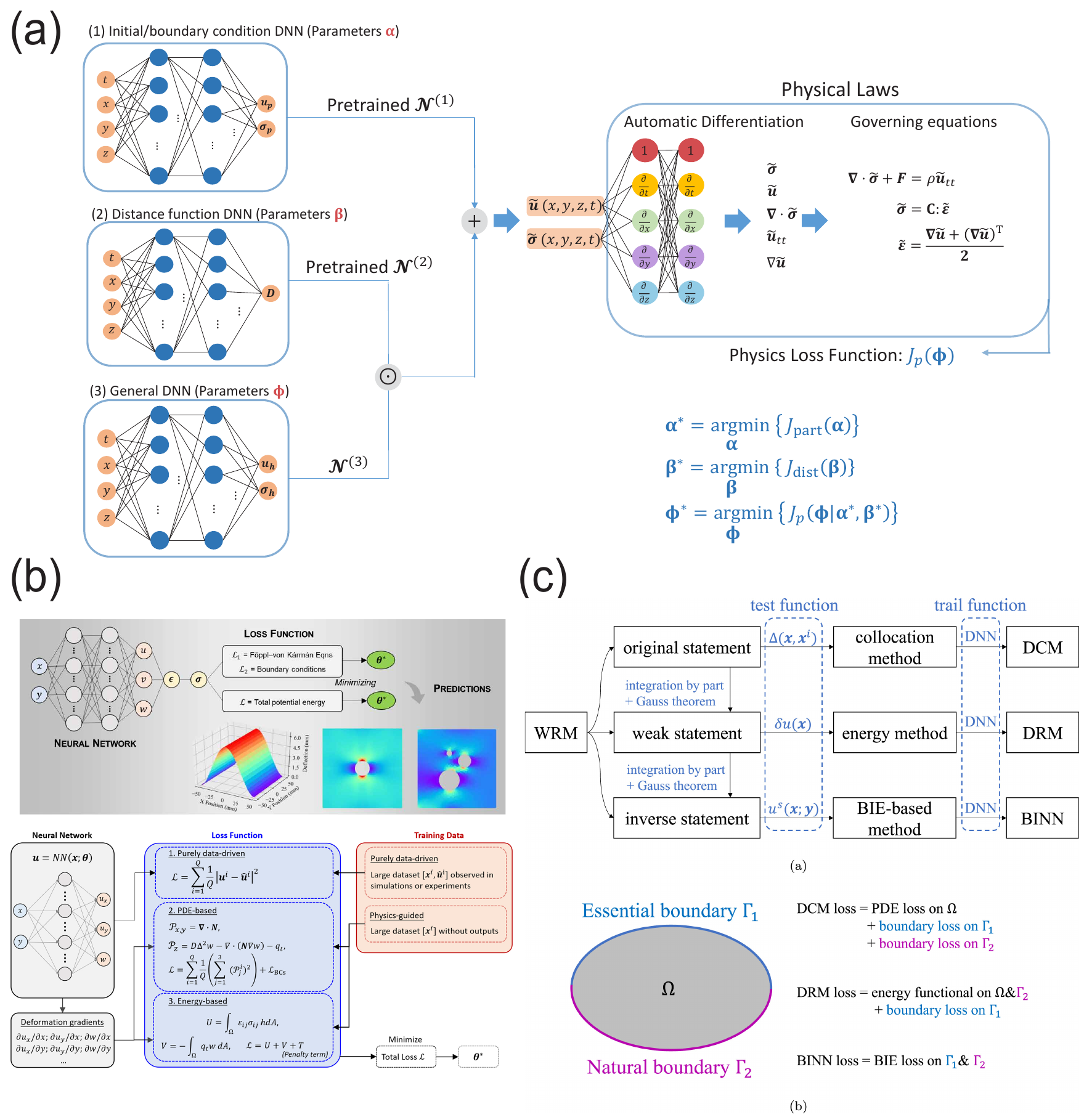}
			\par\end{centering}
		\caption{Applications of PINNs in elastic mechanics. (a) Strong form: linear elastic dynamics \cite{PINNstrong_form_in_elastodynamics}
			, (b) Strong and energy form: Kirchhoff plates \cite{the_comparision_of_strong_and_energy_form}, (c) Inverse form: elasticity problems \cite{sun2023binn} \label{fig:application_elastic}}
	\end{figure}

	Rao et al. (2021) \citet{PINNstrong_form_in_elastodynamics} used the strong form of PINNs algorithm
	and the construction of admissible displacement fields to solve linear elasticity dynamics problems,
	as shown in \Cref{fig:application_elastic}a. Guo et al. (2021) \cite{guo2021deep} used the strong
	form of PINNs algorithm for the first time in Kirchhoff plates. Subsequently, Zhuang et al. (2021) \cite{zhuang2021deep}
	applied the energy form of PINNs using the principle of minimum potential energy in Kirchhoff plates.
	Li et al. (2021) \citet{the_comparision_of_strong_and_energy_form} compared the strong and energy forms
	of PINNs in Kirchhoff plates, as illustrated in \Cref{fig:application_elastic}b. 
	Bai et al. \cite{bai2023physics} studied modified loss functions in strong form, solving elastic problems with geometric nonlinearity and constitutive linearity.
	Later, Sun et al.
	(2023) \cite{sun2023binn} combined PINNs with boundary element methods. Mathematically, the strong
	form of PDEs can be converted into the weak form through Gauss's divergence theorem. If
	the differential equation involves a linear self-adjoint operator, the weak form can be further transformed
	into an energy form. If the weak form is subjected to Gauss's divergence theorem again until the highest
	derivatives in the test functions are achieved, thereby reducing the differential equation to its lowest
	order, it then transforms into the inverse form. Boundary elements leverage the inverse form
	to formulate boundary integral equations. Thus, the strong form, energy form (or weak
	form), and inverse form are all different representations of the same PDEs. However,
	due to limitations in the fundamental solutions of Green's functions in boundary elements, currently,
	only problems in linear elasticity can be solved, as shown in \Cref{fig:application_elastic}c.

	The above literature review discusses using PINNs to replace traditional numerical methods, such as the finite element method, for solving linear elasticity problems.
	Additionally, operator learning combined with finite element methods
	has also been applied to linear elasticity problems, especially for multiscale problem. For instance, Wang et al. (2026) \cite{wang2024homogenius}
	proposed using the Fourier Neural Operator (FNO) to replace the calculation process of elastic tensors
	in mechanical homogenization. In the original problem, traditional methods required finite element calculations
	of displacement fields under six loading conditions for different geometries, followed by numerical homogenization
	formulas to obtain the fourth-order equivalent elastic tensor. Since the boundary conditions in the numerical
	homogenization process are fixed and only the geometry changes, a large amount of data can be calculated
	using finite elements for different geometries, then operator learning can be used to learn the mapping
	from geometry to displacement fields, and finally, it can be input into traditional numerical homogenization
	programs, significantly improving computational efficiency (about 1000 times faster compared to traditional
	finite element methods). 
	The solution predicted by the FNO is then used as the initial guess for the finite element iterative solver. Since this initial guess is much closer to the reference solution, the number of required numerical iterations is greatly reduced compared to traditional finite element methods. This strategy thus achieves both the computational efficiency of operator learning and the accuracy of the classical finite element method.

	\subsubsection{Elastoplastic mechanics}
	
	Elastoplastic theories encompass various models that describe the relationship between strain increments
	and stress during the plastic deformation phase. Abueidda et al. (2021) \cite{abueidda2021meshless}
	utilized PINNs with J2 flow theory, kinematic hardening, and isotropic hardening models to address elastoplastic
	problems. Here, we illustrate how PINNs solve elastoplastic issues, starting by specifying parameters
	for elastoplastic problems: the Lame constants $\lambda$ and $G$, yield stress $\sigma_{s}$, parameters
	$K$ describing isotropic hardening of the yield surface, and parameter $H$ for back stress evolution.
	Using traditional PINNs, sufficient points are sampled on the domain displacement boundaries, and force boundaries.  We initialize variables that will evolve iteratively. The variables to be initialized include initial plastic strain $\boldsymbol{e}_{o}^{p}$,
	initial back stress $\boldsymbol{q}_{o}$, and initial internal variable $\alpha_{0}$ for isotropic
	hardening. According to J2 flow theory,  the yield condition with the kinematic hardening and the isotropic hardening models is:
	\begin{equation}
		\mathcal{F}(\boldsymbol{\sigma},\boldsymbol{q},\alpha)=\sqrt{\boldsymbol{\eta}:\boldsymbol{\eta}}-\sqrt{\frac{2}{3}}(\sigma_{y}+K\alpha),
	\end{equation}
	where $\boldsymbol{\eta}=\boldsymbol{s}-\boldsymbol{q}$, with $\boldsymbol{s}$ representing the deviatoric
	stress tensor reduced by the hydrostatic stress, and $\boldsymbol{q}$ is the back stress. Given the
	path-dependent nature of plastic mechanics, we typically employ incremental loading for numerical simulation.
	Setting boundary conditions incrementally, we consider a pseudo-time $t=1$ scenario, using the PINNs
	network (inputting coordinates, outputting displacement fields) to output displacement fields at points
	within the domain and on displacement and force boundaries. Using automatic differentiation, we derive
	the gradient of displacement. From the relationship between strain and displacement gradients, we compute
	the strain at $t=1$, denoted $\boldsymbol{\boldsymbol{\varepsilon}}_{1}$, and from this strain, compute
	the deviatoric strain $\boldsymbol{e}_{1}$.
	
\begin{figure}
		\begin{centering}
			\includegraphics[scale=0.6]{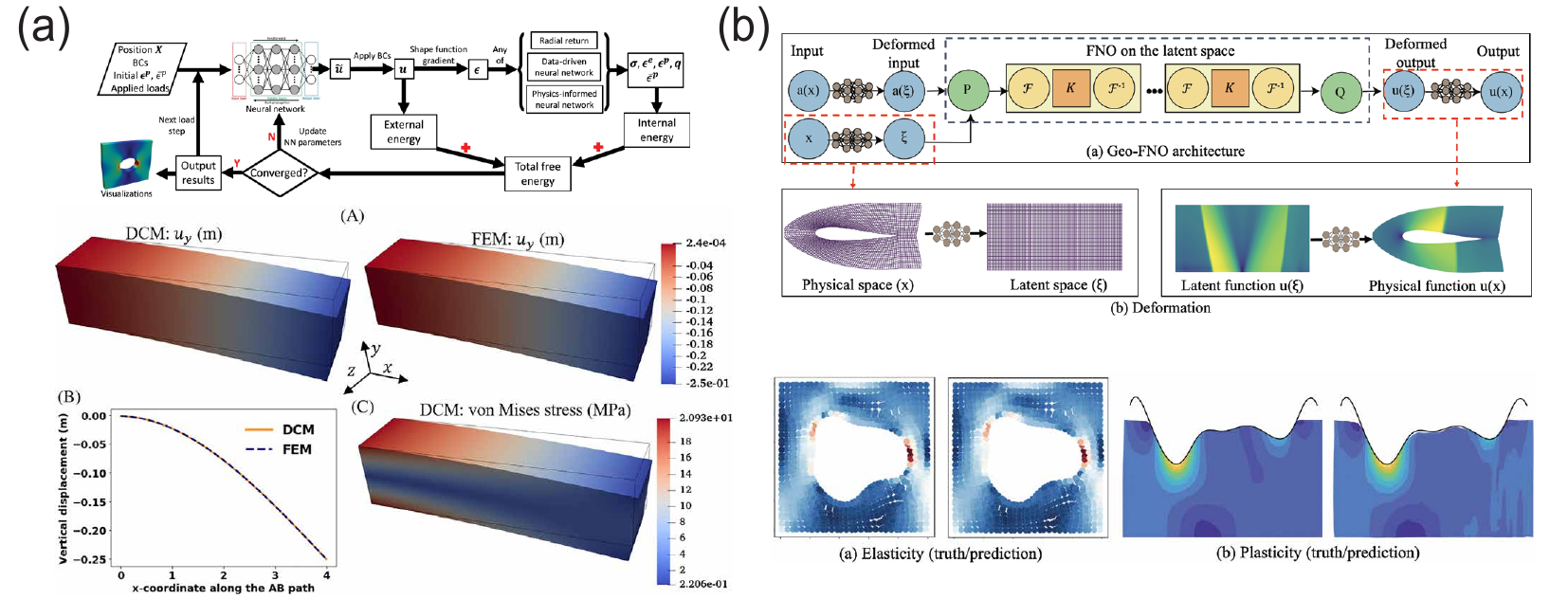}
			\par\end{centering}
		\caption{Applications of PINNs and operator learning in elastoplastic mechanics: (a) PINNs in strong form \cite{abueidda2021meshless}
			and energy form \cite{he2023deep}, (b) Operator learning with Geo-FNO \cite{li2022fourier}.\label{fig:application_plasticity}}
	\end{figure}

	According to the predefined variables, including initial plastic strain $\boldsymbol{e}_{o}^{p}$, initial back stress $\boldsymbol{q}_{o}$,
	and initial isotropic hardening internal variable $\alpha_{0}$, we calculate the trial deviatoric stress
	$\boldsymbol{s}_{1}^{trial}$, trial relative stress $\boldsymbol{\eta}_{1}^{trial}$, and the trial state
	of the yield condition $\mathcal{F}_{1}^{trial}$:
	\begin{equation}
		\begin{aligned}\boldsymbol{s}_{1}^{trial} & =2G(\boldsymbol{e}_{1}-\boldsymbol{e}_{0}^{p})\\
			\boldsymbol{\eta}_{1}^{trial} & =\boldsymbol{s}_{1}^{trial}-\boldsymbol{q}_{o}\\
			\mathcal{F}_{1}^{trial} & =\sqrt{\boldsymbol{\eta}_{1}^{trial}:\boldsymbol{\eta}_{1}^{trial}}-\sqrt{\frac{2}{3}}(\sigma_{y}+K\alpha_{0})
		\end{aligned}.
	\end{equation}
	We determine whether the material enters the plastic phase by checking if $\mathcal{F}_{1}^{trial}$ is less than
	or equal to zero. If  $\mathcal{F}_{1}^{trial}$ is less than or equal to zero, the material is considered to be in
	the elastic phase, and stress is calculated directly using $\boldsymbol{s}_{1}^{trial}$: 
	\begin{equation}
		\boldsymbol{\sigma}_{1}=\lambda trace(\boldsymbol{\varepsilon}_{1})\boldsymbol{I}+\boldsymbol{s}_{1}^{trial}.
	\end{equation}
	Using this stress and displacement field, we compute the loss function for $t=1$ using the balance
	equations, displacement boundary conditions, and force boundary conditions, and optimize the parameters
	of the PINNs network, verifying in each optimization cycle whether the material remains elastic. If $\mathcal{F}_{1}^{trial}$ exceeds zero, we must adjust the previous steps, notably changing how stress is computed.
	Using $\boldsymbol{\eta}_{1}^{trial}$ and $\mathcal{F}_{1}^{trial}$, we compute the flow direction $\boldsymbol{n}_{1}$,
	plastic flow increment $\bigtriangleup\gamma_{1}$, and updated isotropic hardening internal variable
	$\alpha_{1}$ at $t=1$: 
	\begin{equation}
		\begin{aligned}\boldsymbol{n}_{1} & =\frac{\boldsymbol{\eta}_{1}^{trial}}{\sqrt{\boldsymbol{\eta}_{1}^{trial}:\boldsymbol{\eta}_{1}^{trial}}}\\
			\bigtriangleup\gamma_{1} & =\frac{\mathcal{F}_{1}^{trial}}{2(G+\frac{H}{3}+\frac{K}{3})}\\
			\alpha_{1} & =\alpha_{0}+\sqrt{\frac{2}{3}}\bigtriangleup\gamma_{1}
		\end{aligned}.
	\end{equation}
	The back stress $\boldsymbol{q}$, plastic strain $\boldsymbol{e}^{p}$, and stress $\boldsymbol{\sigma}$
	are updated accordingly: 
	\begin{equation}
		\begin{aligned}\boldsymbol{q}_{1} & =\boldsymbol{q}_{0}+\frac{2}{3}\bigtriangleup\gamma_{1}H\boldsymbol{n}_{1}\\
			\boldsymbol{e}_{1}^{p} & =\boldsymbol{e}_{0}^{p}+\bigtriangleup\gamma_{1}\boldsymbol{n}_{1}\\
			\boldsymbol{\sigma}_{1} & =\lambda trace(\boldsymbol{\varepsilon}_{1})\boldsymbol{I}+\boldsymbol{s}_{1}^{trial}-2G\bigtriangleup\gamma_{1}\boldsymbol{n}_{1} \label{eq:return_mapping}
		\end{aligned}.
	\end{equation}
	The loss function is optimized in the same way as during the elastic phase, using the revised stress
	and displacement fields along with corresponding equations and boundary conditions. This process continues
	iteratively through incremental loading steps until the loss function converges, marking the end of the
	elastoplastic simulation in the strong form of PINNs. The simulation results are compared with FEM solutions,
	focusing on displacement fields as illustrated in \Cref{fig:application_plasticity}a.
	
	The steps involved in solving elastoplastic problems using the strong form of PINNs
	are similar to traditional computational mechanics, except that the approximation function has shifted
	from the shape functions in FEM to neural networks. Due to the abundance of plasticity theories,
	aside from the J2 flow rule, other theories can also be simulated using PINNs, thus facilitating the
	evaluation of PINNs models in handling plastomechanical issues. Furthermore, He et al. (2023) \cite{he2023deep} proposed using the energy method with the plastic variational formulation for calculating elastoplastic problems, as shown in the process diagram in \Cref{fig:application_plasticity}a. The plastic variational formulation is based on the theory proposed by Simo et al. (2006) \citet{simo2006computational}.
	Since elastoplastic problems are related to the historical loading path, both the strong form and energy form of PINNs iteratively address each loading step.
	
	Additionally, data-driven simulations of elastoplastic issues have been conducted, with Li et al. (2023)
	\cite{li2022fourier} using Geo-FNO (an enhanced version of FNO capable of operator learning for any
	geometry) to simulate forming and stamping processes. The algorithmic framework and simulation results
	are depicted in \Cref{fig:application_plasticity}b.
	
	The return mapping in \Cref{eq:return_mapping} is a numerical algorithm for the elastoplastic KKT conditions. Mathematically, the KKT can be easily transformed into an optimization problem using e.g. Fischer-Burmeister replacement functions \cite{zhou2022open}, which allows for much larger load steps and can lead to computational savings. Therefore, in the future, a PINNs algorithm based on Fischer-Burmeister replacement functions could be developed to solve elastoplastic problems.

	\subsubsection{Hyperelastic mechanics}
	
	Hyperelastic problems not only involve geometrical nonlinearity but also material nonlinearity, and they
	can be computed using the minimum potential energy principle in the energy form of PINNs. Traditional
	hyperelastic problems are path-independent, making them a simpler type of nonlinear mechanics problem.
	Therefore, many studies in mechanics using PINNs often choose hyperelastic problems as examples. There
	are two main considerations in the PINNs simulation of hyperelastic problems: the first is the selection
	of the potential function according to the constitutive law of hyperelasticity material, and the second is that the kinematic equations must
	consider the nonlinear terms of large deformations. To illustrate, let us consider the potential function
	of the Neo-Hookean constitutive model: 
	\begin{equation}
		\begin{aligned}\Psi & =\frac{1}{2}\lambda(\ln J)^{2}-G\ln J+\frac{1}{2}G(trace(\boldsymbol{C})-3)\end{aligned},
		\label{eq:loss_hyper}
	\end{equation}
	where $J$ is the determinant of the deformation gradient $\boldsymbol{F}$, and $\boldsymbol{C}$ is
	the right Cauchy-Green deformation tensor ($\boldsymbol{C}=\boldsymbol{F}^{T}\boldsymbol{F}$). We first
	establish a neural network mapping from material coordinates $\boldsymbol{X}$ to spatial coordinates
	$\mathbf{x}$. We scatter points within the domain and then use automatic differentiation to calculate
	the deformation gradient $\boldsymbol{F}=\partial\boldsymbol{x}/\partial\boldsymbol{X}$, which is substituted
	into \Cref{eq:loss_hyper} to determine the density of the hyperelastic strain energy. Once $\boldsymbol{F}$
	is known, hyperelastic strain energy can be obtained through some simple algebraic operations. Since
	the potential energy principle must consider the work done by external forces and the displacement field
	must satisfy displacement boundary conditions in advance, we consider
	displacement boundary conditions when hypothesizing admissible displacement fields, employing the construction
	method of \Cref{eq:admissible_distance}. However, if the structure's shape is simple, the distance function
	can be directly constructed analytically (often by multiplying by coordinates, as suggested by \citet{PINN_hyperelasticity}.)
	For the work of external forces, the construction is entirely the same as in linear elasticity problems.
	Finally, we construct the overall loss function: 
	\begin{equation}
		\mathcal{L}=\int_{\Omega}(\Psi-\boldsymbol{f}\cdot\boldsymbol{u})d\Omega-\int_{\Gamma^{t}}\bar{\boldsymbol{t}}\cdot\boldsymbol{u}d\Gamma.
	\end{equation}
	By optimizing the above loss function, we can obtain the neural network parameters for the mapping from
	material coordinates $\mathbf{X}$ to spatial coordinates $\mathbf{x}$, thus determining the corresponding
	mapping relationship. Once the mapping from material coordinates $\mathbf{X}$ to spatial coordinates
	$\mathbf{x}$ is known, all mechanical quantities become clear, including the displacement field, the
	first Piola-Kirchhoff stress $\mathbf{P}$, and the second Piola-Kirchhoff stress $\mathbf{S}$: 
	\begin{equation}
		\begin{aligned}\boldsymbol{u} & =\boldsymbol{x}-\boldsymbol{X}\\
			\boldsymbol{P} & =\frac{\partial\Psi}{\partial\boldsymbol{F}}=G\boldsymbol{F}^{T}+[\lambda\ln J-G]\boldsymbol{F}^{-1}\\
			\boldsymbol{S} & =\boldsymbol{P}\cdot\boldsymbol{F}^{-T}
		\end{aligned}.
		\label{eq:loss_hyper-1}
	\end{equation}
	
	\begin{figure}
		\begin{centering}
			\includegraphics[scale=0.55]{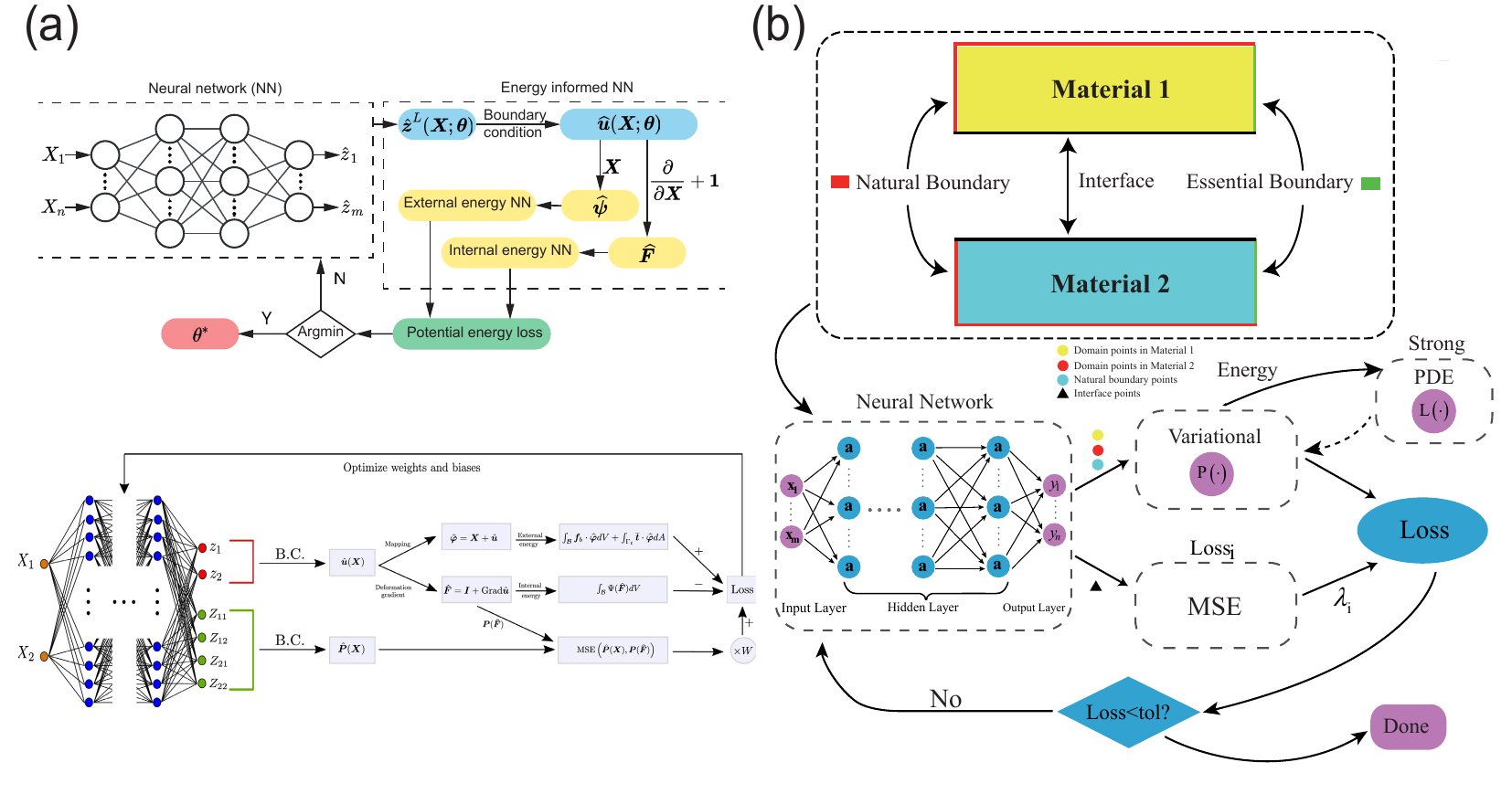}
			\par\end{centering}
		\caption{Applications of PINNs in energy form for hyperelastic mechanics: (a) PINNs in energy form \cite{PINN_hyperelasticity,fuhg2022mixed}
			(b) PINNs in energy form with subdomains \cite{wang2022cenn}.\label{fig:application_hyperelastic}}
	\end{figure}
	
	It is worth emphasizing that numerical integration in the energy form of PINNs is crucial, as it directly
	affects the calculation of functional. Nguyen-Thanh et al. (2020) \citet{PINN_hyperelasticity}
	were the first to apply the PINNs energy principle to hyperelastic problems and introduced traditional
	numerical integration schemes into the PINNs energy form (DEM), as shown in \Cref{fig:application_hyperelastic}a.
	Subsequently, Fuhg et al. (2022) \cite{fuhg2022mixed} extended the PINNs energy form by dividing the
	loss function into two parts: one based on the minimum potential energy principle and the other constructed
	through the equilibrium equation, as shown in \Cref{fig:application_hyperelastic}a below. Wang et al.
	(2022) \cite{wang2022cenn} extended the PINNs energy form algorithm to a subdomain form to solve hyperelastic
	problems, as shown in \Cref{fig:application_hyperelastic}b.
	Bai et al. (2024) \cite{bai2024robust} use Radial Basis Function (RBF) networks to solve hyperelasticity torsional buckling cases in energy form.
	Although Abueidda et al. (2022) \cite{abueidda2022deep}
	were not the first to use the PINNs energy method to solve hyperelastic problems, they were the first
	to propose using the PINNs energy method to solve viscoelastic problems. Of course, we can also use the
	PINNs strong form to simulate hyperelastic problems, but the order of derivatives will increase, thereby
	reducing computational efficiency and accuracy, such as Abueidda et al. (2021) \cite{abueidda2021meshless}
	using the PINNs strong form, directly considering the balance equation and boundary conditions to construct
	the loss function. It can be seen that the core difference between the PINNs strong form and the energy
	form is in the construction of the loss function.

	\subsubsection{Fracture Mechanics}
	
	The Deep Energy Method (DEM), also referred to as the energy-based PINNs approach, is particularly attractive for fracture mechanics, where the governing equations naturally derive from an energy functional \cite{goswami2020transfer}. DEM has been applied to both discrete fracture models and continuous damage models, with several representative studies. For discrete fracture models, Zhao et al. employed DEM to simulate crack propagation \cite{zhao2025denns}, although their approach required additional collocation points along the crack path. Chen et al. \cite{chen2024crack} applied strong-form PINNs combined with asymptotic fracture solutions to simulate fatigue crack growth but did not exploit the energy formulation.
	
	In the context of continuous damage models, Goswami et al. first introduced DEM for phase-field fracture, applying PINNs to both second-order \cite{goswami2020transfer} and fourth-order phase-field models \cite{goswami2020adaptive}. Nevertheless, modeling shear failure (mode-II) remained challenging. Later, Goswami et al. integrated DEM with DeepONet \cite{goswami2022physics,kiyani2025predicting}. Zheng \cite{zheng2022physics} proposed a FEM-inspired approach, where nodal values were predicted by neural networks and interpolated via shape functions to construct displacement and phase fields. However, this approach still required mesh refinement along the crack path, similar to FEM. Building on this idea, Manav et al. \cite{manav2024phase} conducted a more systematic study, applying DEM to crack nucleation, propagation, kinking, branching, and coalescence. Compared with traditional FEM and IGA, a key advantage of DEM is the ability to use larger load increments \cite{goswami2020transfer}, thus enhancing the efficiency of crack propagation simulations.
	
\begin{figure}[H]
		\centering
		\includegraphics[scale=0.6]{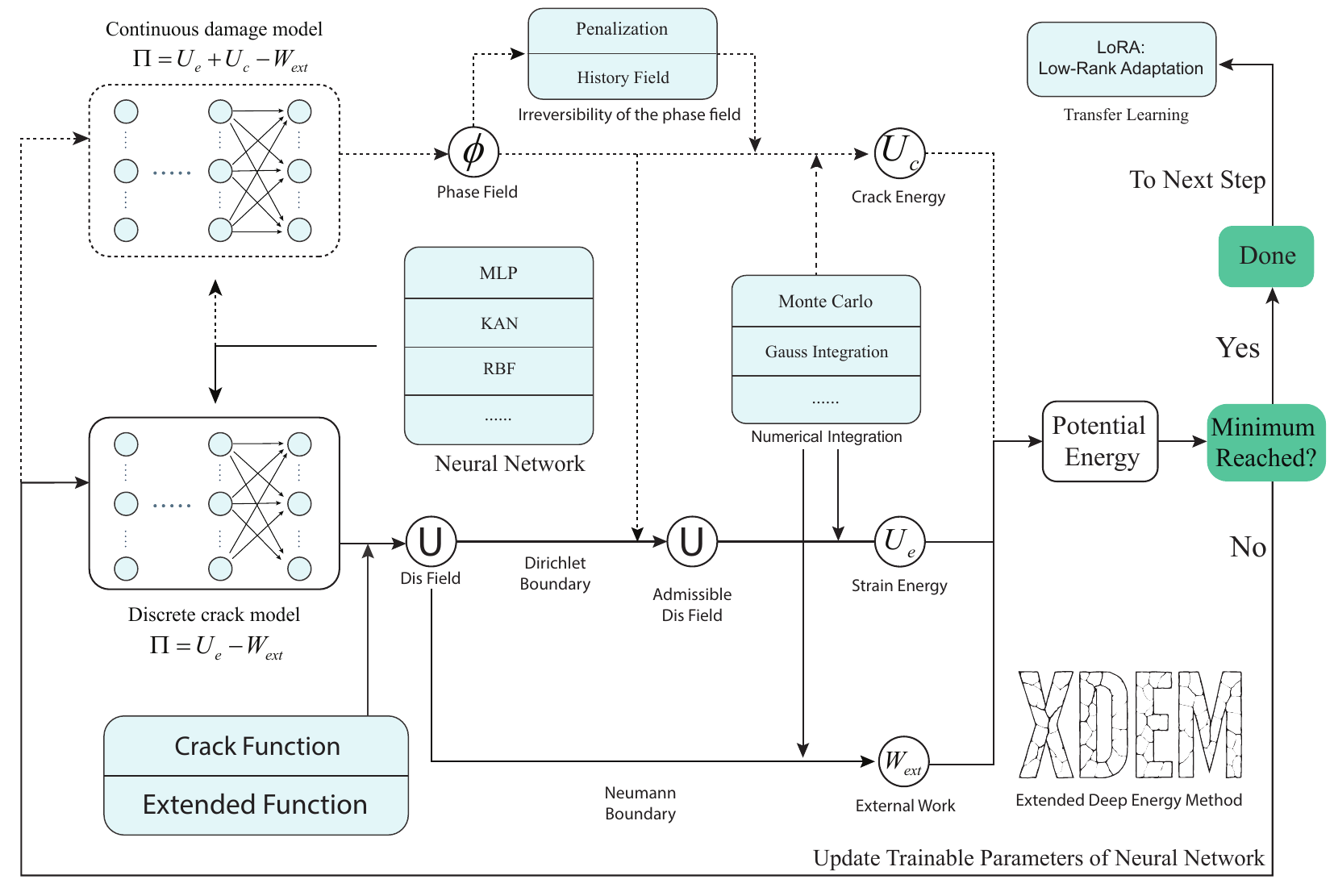}
		\caption{Applications of PINNs in fracture mechanics, which consist of discrete and continuous models. The continuous formulation is indicated by dashed lines \cite{wang2025towards}.\label{fig:application_fracture}}
	\end{figure}

	Recently, Wang et al. \cite{wang2025towards} proposed the Extended Deep Energy Method (XDEM) shown in \Cref{fig:application_fracture}, a unified deep learning framework that incorporates both displacement discontinuities and crack-tip asymptotics in the discrete setting, while flexibly coupling displacement and phase fields in the continuous setting. XDEM leverages Heaviside functions and crack-tip asymptotic enrichment from XFEM, significantly enhancing the robustness and accuracy of DEM in solving fracture mechanics problems.
	
	Below, we provide a introduction to XDEM, which currently represents one of the most promising directions for solving fracture mechanics problems using PINNs. XDEM can be formulated in both discrete and continuous forms.
	
	The discrete crack model of XDEM is governed by the PDEs of linear elasticity:
	\begin{equation}
		\begin{cases}
			\sigma_{ij,j} + f_i = 0, & \boldsymbol{x} \in \Omega, \\
			\sigma_{ij} = C_{ijkl}\varepsilon_{kl}, & \boldsymbol{x} \in \Omega, \\
			\varepsilon_{ij} = \frac{1}{2}(u_{i,j} + u_{j,i}), & \boldsymbol{x} \in \Omega \setminus \Gamma^c, \\
			u_i^+ \not\equiv u_i^-, & \boldsymbol{x} \in \Gamma^c, \\
			\sigma_{ij} n_j = \bar{t}_i, & \boldsymbol{x} \in \Gamma^{\boldsymbol{t}}, \\
			u_i = \bar{u}_i, & \boldsymbol{x} \in \Gamma^{\boldsymbol{u}}.
		\end{cases}
		\label{eq:dis_PDEs}
	\end{equation}
	Here, $\boldsymbol{\sigma}$, $\boldsymbol{C}$, $\boldsymbol{\varepsilon}$, and $\boldsymbol{u}$ denote the stress tensor, stiffness tensor, strain tensor, and displacement vector, respectively. $\Omega$ is the domain, while $\Gamma^{\boldsymbol{u}}$, $\Gamma^{\boldsymbol{t}}$, and $\Gamma^c$ denote Dirichlet boundaries, Neumann boundaries, and the crack surface. The body force, prescribed traction, and prescribed displacement are denoted by $\boldsymbol{f}$, $\bar{\boldsymbol{t}}$, and $\bar{\boldsymbol{u}}$, respectively. The symbol $\not\equiv$ indicates potential displacement discontinuities across $\Gamma^c$. In this study, we consider only Dirichlet and Neumann boundary conditions, with traction-free conditions on the crack surface.
	
	Using the principle of minimum potential energy, the system in \Cref{eq:dis_PDEs} can be reformulated as:
	\begin{equation}
		\begin{aligned}
			\boldsymbol{u} &= \arg\min_{\boldsymbol{u}} \Pi, \\
			\Pi &= U_e - W_{ext}, \\
			U_e &= \int_\Omega \frac{1}{2} \boldsymbol{\varepsilon} : \boldsymbol{C} : \boldsymbol{\varepsilon} \, dV, \\
			W_{ext} &= \int_\Omega \boldsymbol{f} \cdot \boldsymbol{u} \, dV + \int_{\Gamma^{\boldsymbol{t}}} \bar{\boldsymbol{t}} \cdot \boldsymbol{u} \, dS, \\
			\text{s.t. } & u_i = \bar{u}_i, \ \boldsymbol{x} \in \Gamma^{\boldsymbol{u}}; \quad u_i^+ \not\equiv u_i^-, \ \boldsymbol{x} \in \Gamma^c.
		\end{aligned}
		\label{eq:dis_energy}
	\end{equation}
	It is straightforward to show that $\delta\Pi=0$ is equivalent to \Cref{eq:dis_PDEs}.  
	
	By solving \Cref{eq:dis_energy}, the displacement field can be obtained, from which the stress field follows via the constitutive and geometric relations. To further simulate crack propagation, a fracture criterion must be specified. Common criteria include the maximum circumferential stress criterion \citet{erdogan1963crack}, the maximum energy release rate criterion \citet{hussain1974strain}, and the minimum strain energy density criterion \citet{sih1974strain}. XDEM adopt the maximum circumferential stress criterion, whereby the crack propagates in the direction of the maximum hoop tensile stress.

	The discrete XDEM optimization problem is:
	\begin{equation}
		\begin{aligned}
			\boldsymbol{u}^{n+1} &= \arg\min_{\boldsymbol{u}} \Pi, \\
			\Pi &= U_e - W_{ext}, \\
			U_e &= \int_\Omega \frac{1}{2} \boldsymbol{\varepsilon}(\boldsymbol{x}; \boldsymbol{\theta}_{\boldsymbol{u}}) : \boldsymbol{C} : \boldsymbol{\varepsilon}(\boldsymbol{x}; \boldsymbol{\theta}_{\boldsymbol{u}}) \, dV, \\
			W_{ext} &= \int_\Omega \boldsymbol{f} \cdot \boldsymbol{u}(\boldsymbol{x}; \boldsymbol{\theta}_{\boldsymbol{u}}) \, dV + \int_{\Gamma^{\boldsymbol{t}}} \bar{\boldsymbol{t}} \cdot \boldsymbol{u}(\boldsymbol{x}; \boldsymbol{\theta}_{\boldsymbol{u}}) \, dS, \\
			\text{s.t. } & u_i(\boldsymbol{x}; \boldsymbol{\theta}_{\boldsymbol{u}}) = \bar{u}_i(\boldsymbol{x}, t^{n+1}), \ \boldsymbol{x} \in \Gamma^{\boldsymbol{u}}; \quad u_i^+ \not\equiv u_i^-, \ \boldsymbol{x} \in \Gamma^c.
		\end{aligned}
		\label{eq:dis_DEM}
	\end{equation}
	Here, $\boldsymbol{\theta}_{\boldsymbol{u}}$ represents the trainable parameters of the displacement neural network $NN(\boldsymbol{x}; \boldsymbol{\theta}_{\boldsymbol{u}})$. 
	XDEM incorporates a crack function to represent the discontinuous field across cracks and an extended function to embed the asymptotic solution near the crack tip into the displacement neural network, detailed in \cite{wang2025towards}
	
	For continuous damage, XDEM employs the classical phase-field fracture model \cite{bourdin2000numerical}, based on the variational principle of Francfort and Marigo \citet{francfort1998revisiting}. The energy functional is extended to include fracture energy:
	\begin{equation}
		\begin{aligned}
			\boldsymbol{u},\phi &= \arg\min_{\boldsymbol{u},\phi}\Pi, \\
			\Pi &= U_{e}+U_{c}-W_{ext}, \\
			U_{e}(\boldsymbol{u},\phi) &= \int_{\Omega} w(\phi)\,\varPsi^{+}(\boldsymbol{\varepsilon}) + \varPsi^{-}(\boldsymbol{\varepsilon})\,dV, \\
			U_{c}(\boldsymbol{u},\phi) &= \frac{G_{c}}{c_{w}}\int_{\Omega}\frac{g(\phi)}{l_{0}}+l_{0}\,(\nabla\phi)\cdot(\nabla\phi)\,dV, \\
			W_{ext} &= \int_{\Omega}\boldsymbol{f}\cdot\boldsymbol{u}\,dV+\int_{\Gamma^{\boldsymbol{t}}}\bar{\boldsymbol{t}}\cdot\boldsymbol{u}\,dS, \\
			\text{s.t. } & u_{i}=\bar{u}_{i}, \ \boldsymbol{x}\in\Gamma^{\boldsymbol{u}};\quad \phi^{n+1}\geq\phi^{n}.
		\end{aligned}
		\label{eq:con_PFM}
	\end{equation}
	where $w(\phi)$ is the degradation function that represents the reduction of material stiffness. It must satisfy the following conditions: $w(0)=1$, $w(1)=0$, $w^{\prime}(1)=0$, and $w^{\prime}(\phi)<0$. A common choice is $w(\phi)=(1-\phi)^{2}$. $\varPsi^{+}(\boldsymbol{\varepsilon})$ and $\varPsi^{-}(\boldsymbol{\varepsilon})$ denote the tensile and compressive contributions of the strain energy, respectively. Typical decompositions include the formulations of Miehe \citet{miehe2010phase} and Amor \citet{amor2009regularized}:  
	\begin{equation}
		\begin{aligned}
			\text{Miehe:} \quad & \varPsi^{+}(\boldsymbol{\varepsilon})=\tfrac{1}{2}\lambda\langle\varepsilon_{ii}\rangle_{+}^{2}
			+ G \sum_{i=1}^{3}\langle\lambda_{i}\rangle_{+}^{2}, \\
			& \varPsi^{-}(\boldsymbol{\varepsilon})=\tfrac{1}{2}\lambda\langle\varepsilon_{ii}\rangle_{-}^{2}
			+ G \sum_{i=1}^{3}\langle\lambda_{i}\rangle_{-}^{2}, \\
			\text{Amor:} \quad & \varPsi^{+}(\boldsymbol{\varepsilon})=\tfrac{1}{2}K\langle\varepsilon_{ii}\rangle_{+}^{2}
			+ G\,\varepsilon_{ij}^{\prime}\varepsilon_{ij}^{\prime}, \\
			& \varPsi^{-}(\boldsymbol{\varepsilon})=\tfrac{1}{2}K\langle\varepsilon_{ii}\rangle_{-}^{2},
		\end{aligned}
	\end{equation}
	where $\lambda$ and $G$ are the Lamé constants, $K=\lambda+2G/3$ is the bulk modulus, $\lambda_{i}$ are the eigenvalues of the strain tensor $\boldsymbol{\varepsilon}$, and $\varepsilon_{ij}^{\prime}=\varepsilon_{ij}-\varepsilon_{kk}\delta_{ij}/3$ is the deviatoric strain tensor. The positive and negative parts of a scalar are defined as $\langle x\rangle_{+}=(x+|x|)/2$ and $\langle x\rangle_{-}=(x-|x|)/2$.   In the manuscript, we use Miehe as the form of the energy decomposition.
	
	In \Cref{eq:con_PFM}, $G_{c}$ is the critical energy release rate, and $c_{w}=4\int_{0}^{1}\sqrt{g(\phi)}\,d\phi$ is a normalization constant. The function $g(\phi)$ denotes the local dissipation function, commonly chosen as in the AT1 ($g=\phi$, $c_{w}=8/3$) or AT2 ($g=\phi^{2}$, $c_{w}=2$) models.  
	
	In practical phase-field simulations, the choices of $w(\phi)$ and $g(\phi)$, the type of energy decomposition, and the length-scale parameter $l_{0}$ must be specified in advance. By minimizing $\Pi$ in \Cref{eq:con_PFM}, both the displacement field $\boldsymbol{u}$ and the phase-field variable $\phi$ can be obtained. Compared with discrete crack models, the phase-field model has the advantage of allowing spontaneous crack nucleation without prescribing a fracture criterion. However, its major drawback is the significantly higher computational cost. It is also important to note that the irreversibility condition $\phi^{n+1}\geq\phi^{n}$ is typically enforced by either a history field approach \citet{miehe2010phase} or a penalization technique \citet{gerasimov2019penalization}

	For the continuous phase-field fracture model, the XDEM optimization problem is given by:  
	\begin{equation}
		\begin{aligned}\{\boldsymbol{u}^{n+1},\phi^{n+1}\} & =\arg\min_{\boldsymbol{\theta}_{\boldsymbol{u}},\boldsymbol{\theta}_{\phi}}\Pi(\boldsymbol{u}(\boldsymbol{x};\boldsymbol{\theta}_{\boldsymbol{u}}),\phi(\boldsymbol{x};\boldsymbol{\theta}_{\phi}))\\
			\Pi & =U_{e}+U_{c}-W_{ext}\\
			U_{e}(\boldsymbol{u},\phi) & =\int_{\Omega}[w(\phi(\boldsymbol{x};\boldsymbol{\theta}_{\phi}))\varPsi^{+}(\boldsymbol{u}(\boldsymbol{x};\boldsymbol{\theta}_{\boldsymbol{u}}))+\varPsi^{-}(\boldsymbol{u}(\boldsymbol{x};\boldsymbol{\theta}_{\boldsymbol{u}}))]dV\\
			U_{c}(\boldsymbol{u},\phi) & =\frac{G_{c}}{c_{w}}\int_{\Omega}\frac{g(\phi(\boldsymbol{x};\boldsymbol{\theta}_{\phi}))}{l_{0}}+l_{0}(\nabla\phi(\boldsymbol{x};\boldsymbol{\theta}_{\phi}))\cdot(\nabla\phi(\boldsymbol{x};\boldsymbol{\theta}_{\phi}))dV\\
			W_{ext} & =\int_{\Omega}\boldsymbol{f}\cdot\boldsymbol{u}(\boldsymbol{x};\boldsymbol{\theta}_{\boldsymbol{u}})dV+\int_{\Gamma^{\boldsymbol{t}}}\bar{\boldsymbol{t}}\cdot\boldsymbol{u}(\boldsymbol{x};\boldsymbol{\theta}_{\boldsymbol{u}})dS.\\
			\text{s.t. } & u_{i}(\boldsymbol{x};\boldsymbol{\theta}_{\boldsymbol{u}})=\bar{u}_{i}(\boldsymbol{x},t^{n+1}),\boldsymbol{x}\in\Gamma^{\boldsymbol{u}};\phi^{n+1}\geq\phi^{n}
		\end{aligned}
		,\label{eq:variational_principle_DEM}
	\end{equation}
	where $\boldsymbol{\theta}_{\boldsymbol{u}}$ and $\boldsymbol{\theta}_{\phi}$ denote the trainable parameters of the displacement neural network $NN(\boldsymbol{x};\boldsymbol{\theta}_{\boldsymbol{u}})$ and the phase-field neural network $NN(\boldsymbol{x};\boldsymbol{\theta}_{\phi})$, respectively. Unlike discrete models, the phase-field formulation does not require a predefined crack propagation criterion, as cracks can nucleate and evolve naturally. However, the irreversibility condition $\phi^{n+1}\geq \phi^{n}$ must be satisfied, ensuring that cracks cannot heal once formed. 
	The continuous phase-field fracture model of XDEM uses KAN for displacement-field approximation and RBF for phase-field approximation, as detailed in \cite{wang2025towards}.
	
	XDEM has been systematically validated on classical fracture mechanics benchmarks, including stress intensity factor predictions and crack path simulations \cite{wang2025towards}.

	\subsubsection{Summary}
	
	Overall, for addressing forward problems, current PINNs, in terms of accuracy and efficiency, still lag
	behind traditional numerical methods \cite{PINN_solid_mechanics}. 
	Recent studies have shown that operator learning can significantly enhance the computational efficiency
	for forward problems. Although traditional operator learning is purely data-driven, current theoretical
	explanations suggest that after extensive data training, operator learning can learn the mappings of
	PDEs families. Additionally, the integration of operator learning with physical equations holds promising
	academic and industrial prospects. This involves using existing data to propose a good initial solution,
	which is then fine-tuned using physical equations. This approach theoretically offers significant
	benefits in computational efficiency and accuracy, especially for complex nonlinear problems. In mathematics, some nonlinear problems can be transformed into iterations
	of nonlinear equation systems, where a good initial solution is vital for reducing iteration time. For
	instance, in hyperelasticity problems, operator learning can provide a good initial solution, followed
	by iterations based on the nonlinear equation system. Here, using finite element methods instead of PINNs
	to solve physical equations could enhance accuracy and efficiency, suggesting a combination of finite
	element methods and operator learning to greatly improve the computational efficiency of nonlinear problems.
	
	This section has introduced the applications of AI for PDEs in addressing forward problems in solid mechanics.
	Next, we will discuss the applications of AI for PDEs in solving forward problems in fluid mechanics.
	
	\subsection{Fluid mechanics}
	
	In this section, we will focus on the latest research on fluid mechanics with AI for PDEs. The algorithms in \Cref{sec:AI4PDEs_method} are very
	popular in fluid mechanics.
	
	\subsubsection{Hydrodynamics}
	
	Hydrodynamics, a sub-discipline of fluid dynamics, focuses on the motion of liquids and the forces acting
	upon them.  
	In hydrodynamics, the most popular governing equations, the Navier-Stokes (NS) equations, can be written
	as 
	\begin{equation}
		\frac{D \left( \rho\boldsymbol{v}\right)}{D t}+\rho\boldsymbol{v} \left( \nabla \cdot \boldsymbol{v}\right)=\nabla\cdot\boldsymbol{\sigma}+\rho\boldsymbol{f}\label{eq:momentum_conservation}
	\end{equation}
	\begin{equation}
		\frac{\partial\rho}{\partial t}+\nabla\cdot(\rho\boldsymbol{v})=0\label{eq:mass_conservation},
	\end{equation}
	where $\rho$ is the density of the fluid, $\boldsymbol{v}$ is the velocity vector in this section,
	$p$ is the pressure, $\boldsymbol{\sigma}$ is the stress tensor and $\boldsymbol{f}$ is the body force.
	\Cref{eq:momentum_conservation} describes the conservation of momentum, while \Cref{eq:mass_conservation} ensures the conservation
	of mass. For incompressible flow, the material derivative of density is zero, i.e.
	$D\rho/Dt=\partial\rho/\partial t+\rho_{,i}v_{i}=0$. Therefore,  \Cref{eq:mass_conservation} can be simplified as
	\begin{equation}
		\nabla\cdot\boldsymbol{v}=0\label{eq:incompressive}.
	\end{equation}
	
	To impose the incompressibility as a hard constraint in 2D problems, the stream function $\varphi$ can
	be used as the direct output of neural networks. In this manner, the velocity can be computed by
	\begin{equation}
		\begin{aligned}u & =\frac{\partial\varphi}{\partial y}\\
			v & =-\frac{\partial\varphi}{\partial x}
		\end{aligned}.
		\label{eq:imcompressive_fai}
	\end{equation}
	It is easy to verify that \Cref{eq:imcompressive_fai} automatically satisfies the hard constraint of incompressibility in \Cref{eq:incompressive}.
	For the sake of simplicity, herein we only introduce the Newtonian fluid model
	\begin{align}
		\sigma_{ij} & =(-p+\lambda S_{kk})\delta_{ij}+2\mu S_{ij}^{'}\label{eq:rheological_model}\\
		S_{ij}^{'} & =S_{ij}-\frac{1}{3}S_{kk}\delta_{ij}\label{eq:shear_stress}\\
		S_{ij} & =\frac{1}{2}(v_{i,j}+v_{j,i})\label{eq:geo_meteric}
	\end{align}
	where $\mu$ and $\lambda$ are the viscosity coefficients. According to \Cref{eq:incompressive}, we
	substitute \Cref{eq:shear_stress} into \Cref{eq:rheological_model}, and obtain:
	\begin{equation}
		\sigma_{ij}=-p\delta_{ij}+\mu(v_{i,j}+v_{j,i}).\label{eq:constituve_law}
	\end{equation}
	
	According to \Cref{eq:incompressive}, substituting \Cref{eq:constituve_law} into \Cref{eq:momentum_conservation}
	yields the following equation:
	\begin{equation}
		\frac{D(\rho\boldsymbol{v})}{Dt}=-\nabla p+\mu\nabla^{2}\boldsymbol{v}+\rho\boldsymbol{f}.\label{eq:imcompressive_VP}
	\end{equation}
	
\begin{figure}
		\begin{centering}
			\includegraphics[scale=0.55]{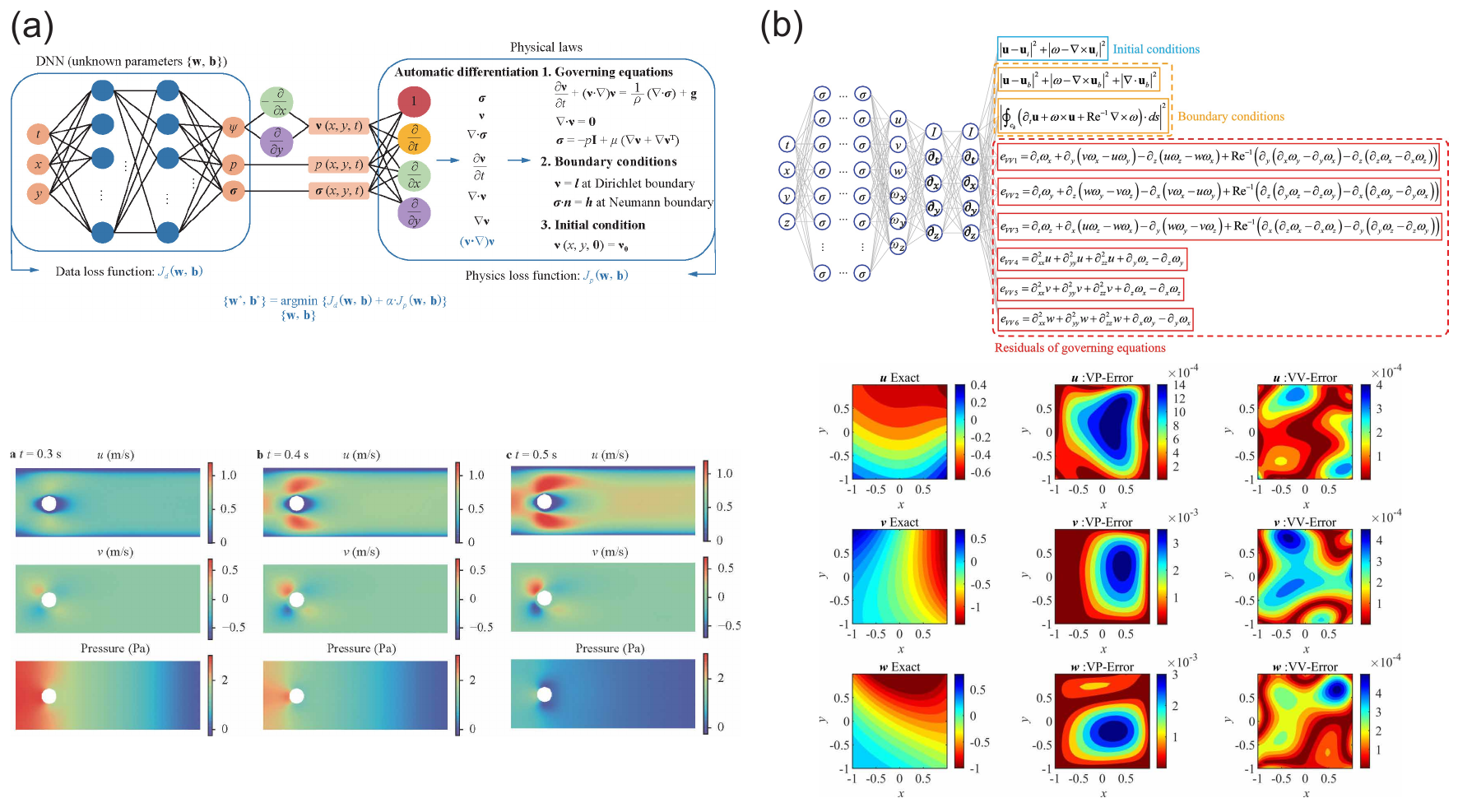}
			\par\end{centering}
		\caption{Different Formats of PINNs for Solving Fluid Dynamics Equations: (a) VP format PINNs for incompressible
			laminar flow \cite{rao2020physics}. (b) VV format PINNs for incompressible Navier-Stokes equations\cite{jin2021nsfnets}.\label{fig:PINNs_fluid_formula}}
	\end{figure}

	Guiding by \Cref{eq:imcompressive_VP}, various formats of PDEs for fluid mechanics have been proposed,
	such as the VP format based on velocity and pressure, and the VV format based on vorticity and velocity.
	Rao et al. \cite{rao2020physics} proposed PINNs framework for incompressible laminar flows based VP
	formulation, as shown in \Cref{fig:PINNs_fluid_formula}a. In the framework, only the conservation of
	momentum is explicitly embedded into the loss function, while the incompressibility is naturally satisfied
	by using \Cref{eq:imcompressive_fai}. Jin et al.\cite{jin2021nsfnets} further provided a PINNs framework
	(NSFnets) informed by vorticity-velocity (VV) formulation for incompressible hydrodynamics and compared
	VV with VP formulation, as shown in \Cref{fig:PINNs_fluid_formula}b. By taking the curl of \Cref{eq:imcompressive_VP},
	VV formulation is obtained:
	\begin{equation}
		\begin{aligned}\nabla\vartimes\frac{D(\boldsymbol{v})}{Dt} & =\nabla\vartimes(-\nabla\frac{p}{\rho}+\mu\nabla^{2}\frac{\boldsymbol{v}}{\rho}+\boldsymbol{f})\\
			\epsilon_{ijk}\dot{v}_{j,i}+\epsilon_{ijk}(v_{s}v_{j,s})_{,i} & =-\frac{1}{\rho}\epsilon_{ijk}p_{,ji}+\nu\epsilon_{ijk}v_{j,ssi}+\epsilon_{ijk}f_{j,i}\\
			\epsilon_{ijk}\dot{v}_{j,i}+\epsilon_{ijk}v_{s,i}v_{j,s}+\epsilon_{ijk}v_{s}v_{j,si} & =\nu\epsilon_{ijk}v_{j,ssi}\\
			\frac{D(\boldsymbol{w})}{Dt} & -\boldsymbol{w}\cdot(\nabla\boldsymbol{u})=\nu\nabla^{2}\boldsymbol{w}
		\end{aligned},
		\label{eq:imcompressive_VV}
	\end{equation}
	where $\nu=\mu/\rho$ is the dynamic viscosity and $\boldsymbol{w}=\nabla\vartimes\boldsymbol{v}$ denotes
	the vorticity. Note that the fluid is still incompressible, so in our derivation, we actually use $\partial\rho/\partial t=0$.
	
	Although the VP and VV formats mathematically address the same physical problem, they are not computationally
	equivalent. For instance, the VV format is more effective at high Reynolds numbers, while the VP format
	can directly describe velocity and pressure. The VV format involves first solving for vorticity
	and then using the resulting vorticity to update the velocity field through the Poisson equation.
	
	Moreover, the adaptive weight technique proposed in \cite{ill_gradient} can be adopted. It is reported
	that PINNs can achieve high accuracy at $Re=100$. Zhang et al. \cite{zhang2022drvn} introduced a PINN
	framework, the so-called dynamics random vorticity network (DRVN), via the random vortex method \cite{majda2002vorticity}.
	The proposed DRVN is capable of effectively addressing fractural and sharp singularity issues, which
	the NSFnet \cite{jin2021nsfnets} struggles with. Rui et al. \cite{rui2024time} implemented the Reynolds-averaged
	Navier-Stokes (RANS) equations. The Reynolds decomposition, which divides flow quantities into time-averaged
	and fluctuating terms, is applied to the NS equations. Thus, the governing equations in RANS can be written as
	\begin{equation}
		\rho\bar{\boldsymbol{v}}\cdot(\nabla\bar{\boldsymbol{v}})=\rho\boldsymbol{f}-\nabla\bar{p}+\mu\nabla\cdot(\nabla\bar{\boldsymbol{v}}+\bar{\boldsymbol{v}}\nabla)-\rho\nabla\cdot\overline{\boldsymbol{v}^{'}\boldsymbol{v}^{'}},\label{eq:RANS}
	\end{equation}
	where $\overline{\boldsymbol{v}}$ and $\overline{\boldsymbol{v}^{'}}$ are the mean (time-averaged)
	velocity and the fluctuating velocity, respectively. Reynolds decomposition refers that we can decompose
	velocity into the time average $\overline{\boldsymbol{v}}$ and the fluctuating velocity $\boldsymbol{v}^{'}$, i.e.
	$\boldsymbol{v}(\boldsymbol{x},t)=\bar{\boldsymbol{v}}(\boldsymbol{x})+\boldsymbol{v}^{'}(\boldsymbol{x},t)$,
	note that although $\overline{\boldsymbol{v}^{'}}=0$, but $\overline{\boldsymbol{v}^{'}\boldsymbol{v}^{'}}\neq0$.
	Thus, $-\rho\overline{\boldsymbol{v}^{'}\boldsymbol{v}^{'}}$ is the Reynolds stress due to the fluctuating
	velocity field. This nonlinear Reynolds stress term requires additional modeling to close the RANS equation 
	and has led to the creation of many different turbulence models. 
	
	Various neural network structures are adopted in the framework of PINN. Wang et al. \cite{wang2023prediction}
	combined the PINNs with long-short term memory (LSTM) \cite{LSTM} model for hydrodynamics. The LSTM has the advantage
	of extracting temporal features in raw data. Thus, the fusion of PINNs and LSTM further enhanced the extrapolation
	ability for hydrodynamics modeling. Han et al. \cite{han2024prediction} used criss-cross convolutional
	neural network \cite{huang2019ccnet} to learn the solution of parametric flow problems with spatial
	heterogeneity, while Cheng and Zhang \cite{cheng2021deep} applied the Resnet \cite{he2016deep} for
	flow modeling.
	
\begin{figure}
		\begin{centering}
			\includegraphics[scale=0.57]{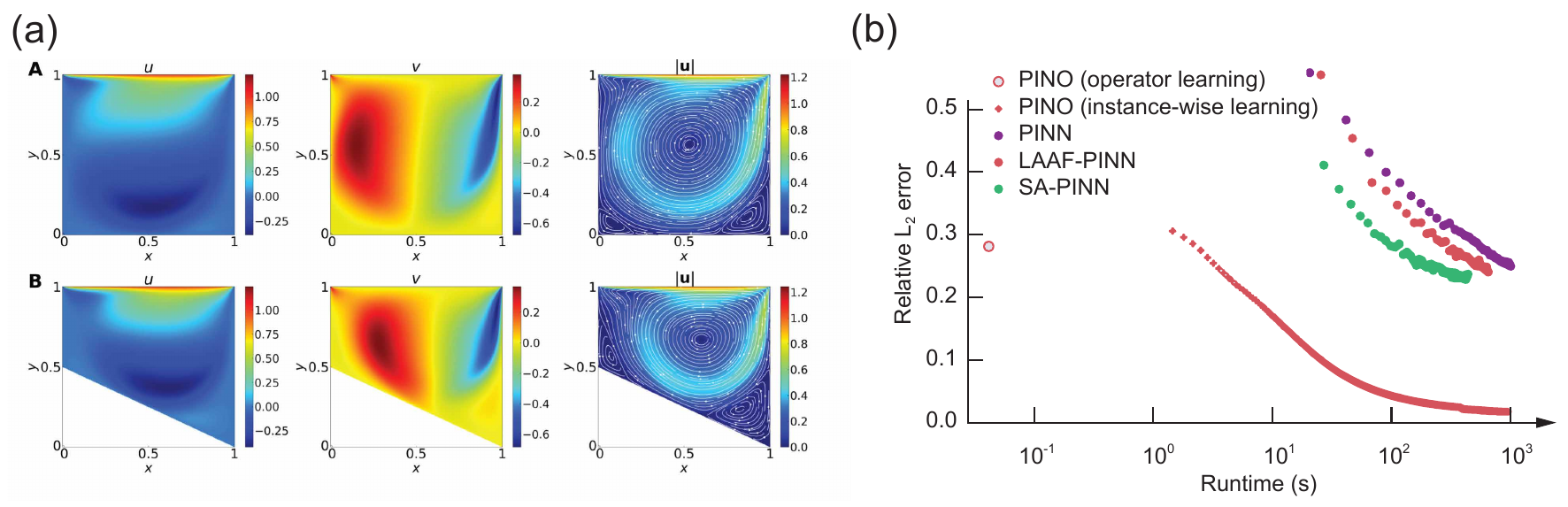}
			\par\end{centering}
		\caption{The results of solving Fluid Dynamics Equations with PINO: (a) Utilizing DeepONet to provide an initial solution, followed
			by constructing PDEs losses using the AD algorithm \cite{zhu2023reliable} (b) Employing FNO to provide
			an initial solution, then constructing PDEs losses using finite differences \cite{li2021physics}.\label{fig:PINOs_fluid}}
	\end{figure}

	In the context of neural operators, hydrodynamics problems have become the most prevailing benchmarks.
	Initially fully data-driven, Li et al. \cite{li2020neural} tested the performances of GNO through the 2D Darcy flow problems. Subsequently, operator learning is combined with physical equations.
	Zhu et al. \cite{zhu2023reliable} utilized the DeepONet framework, initially training it with data
	and subsequently fine-tuning it for specific problems based on physical equations. This process involved
	constructing PDEs loss functions using the automatic differentiation (AD) algorithm. The results of Zhu et al. \cite{zhu2023reliable} are shown in \Cref{fig:PINOs_fluid}a.
	On the other hand, Li et al. \cite{li2021physics} employed the Fourier Neural Operator (FNO) framework,
	initially training it with data and then constructing the PDEs' loss using finite differences for specific
	tasks.
	The results of Li et al. \cite{li2021physics} are shown in \Cref{fig:PINOs_fluid}b.
	In specific fluid dynamics applications, Rosofsky et al.
	\cite{rosofsky2023magnetohydrodynamics} leveraged the PINO to simulate the magnetohydrodynamics. Li
	and Shatarah \cite{li2024operator} proposed a composite neural network, which comprised a series of DeepONet
	and a PINN. The proposed framework could not only accurately predict the velocity field and the particulate
	matter (PM) concentration contours by limited ground truth data, but also seamlessly coped with scenarios
	with different geometries.
	
	\subsubsection{Aerodynamics and shock waves}
	
	The NS equations are also applicable to describe the aerodynamics problems. Unlike hydrodynamics,
	fluids in aerodynamics are generally compressible and inviscidity. Consequently, in aerodynamics,
	the rheological model in \Cref{eq:rheological_model} set the viscosity coefficients $\lambda=\mu=0$.
	It is important to note that since aerodynamic fluids are compressible, $\nabla\cdot\boldsymbol{v}\neq0$.
	Therefore, \Cref{eq:momentum_conservation} is modified according to the compressible and inviscid properties
	of fluids in aerodynamics:
	\begin{equation}
		\begin{aligned}\frac{D(\rho\boldsymbol{v})}{Dt}+\rho\boldsymbol{v}(\nabla\cdot\boldsymbol{v}) & =-\nabla p+\rho\boldsymbol{f}\\
			\frac{\partial\rho}{\partial t}+\nabla\cdot(\rho\boldsymbol{v}) & =0
		\end{aligned}.
		\label{eq:aerodynamic_PDEs}
	\end{equation}
	
\begin{figure}
		\begin{centering}
			\includegraphics[scale=0.55]{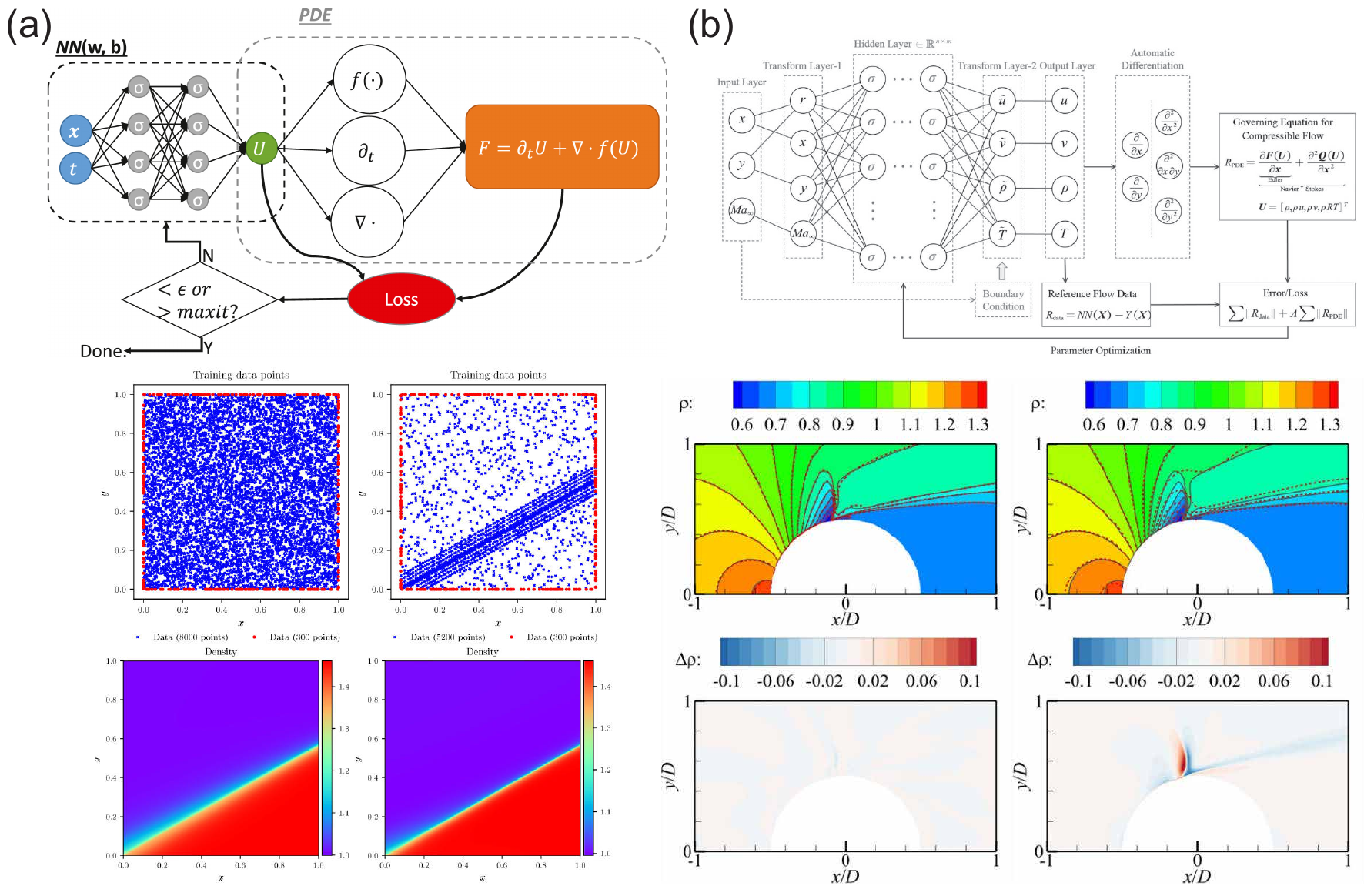}
			\par\end{centering}
		\caption{Applications of PINNs in Aerodynamics: (a) Solving the compressible and inviscid Euler equations \cite{PINNfiuld}.
			(b) Solving the RANS equations \cite{ren2024physics}. \label{fig: PINNs_aerodynamics}}
	\end{figure}

	Mao et al. \cite{PINNfiuld} used \Cref{eq:aerodynamic_PDEs} to model the high-speed compressible inviscid
	aerodynamics problems, for example, the Sod shock-tube problem, as shown \Cref{fig: PINNs_aerodynamics}a.
	The authors also discussed the strategies for selecting the positions of sample points for discontinuous
	problem. Instead of using FNN, Peng et al. \cite{peng2024rapid} applied graph neural network (GNN)
	with the Euler equations to study the transient compressible fields via limited probed data. It has been
	shown that, once the network is well-trained, the predictions can be made within 1 ms and highly reliable
	explosive overpressure can be inferred. Ren et al. \cite{ren2024physics} extend PINNs to learn and forecast
	the steady-state aerodynamics flows around a cylinder with high Reynold numbers. Apart from the limited
	measured data, the RANS equations in \Cref{eq:RANS} and \Cref{eq:aerodynamic_PDEs} served as additional
	knowledge for neural network training \cite{ren2024physics}, as shown in \Cref{fig: PINNs_aerodynamics}b. 
	
	Other works have extended the application of PINNs to aerodynamics. Li et al. \cite{li2022ref} utilized
	PINNs for predicting the gas bearing problems, including the flow field and aerodynamics characteristics.
	Joshi et al. \cite{joshi2023investigation} introduced specific weights (hyperparameters) to balance
	the loss terms from different equations for both low and high speeds. Auddy et al. \cite{auddy2024grinn}
	considered self-gravity of gas flows, which is essential in astrophysics.
	
	\subsubsection{Multiphase and moving boundary problems}
	
	The boundary tracking algorithms are also possible to be modeled by PINNs when facing multiphase and moving
	boundary problems. Wang and Perdikaris \cite{wang2021deep} tested the performances of PINNs with respect
	to simple Stefan problems where moving boundaries or free surfaces exist. In the numerical examples,
	the authors also presented promising results for multi phases scenarios. However, the moving and free
	boundaries considered in that work did not undergo any deformations. The moving
	boundaries have remained the same shape throughout the whole modeling. Jalili et al. \cite{jalili2024physics}
	implemented the well-known Volume-of-Fluid (VOF) method \cite{hirt1981volume}, which is a prevailing boundary
	tracking algorithm in hydrodynamics study, in the PINNs loss function. By doing so, floating bubbles
	under a thermal field were modeled. 
	
\begin{figure}
		\begin{centering}
			\includegraphics[scale=0.54]{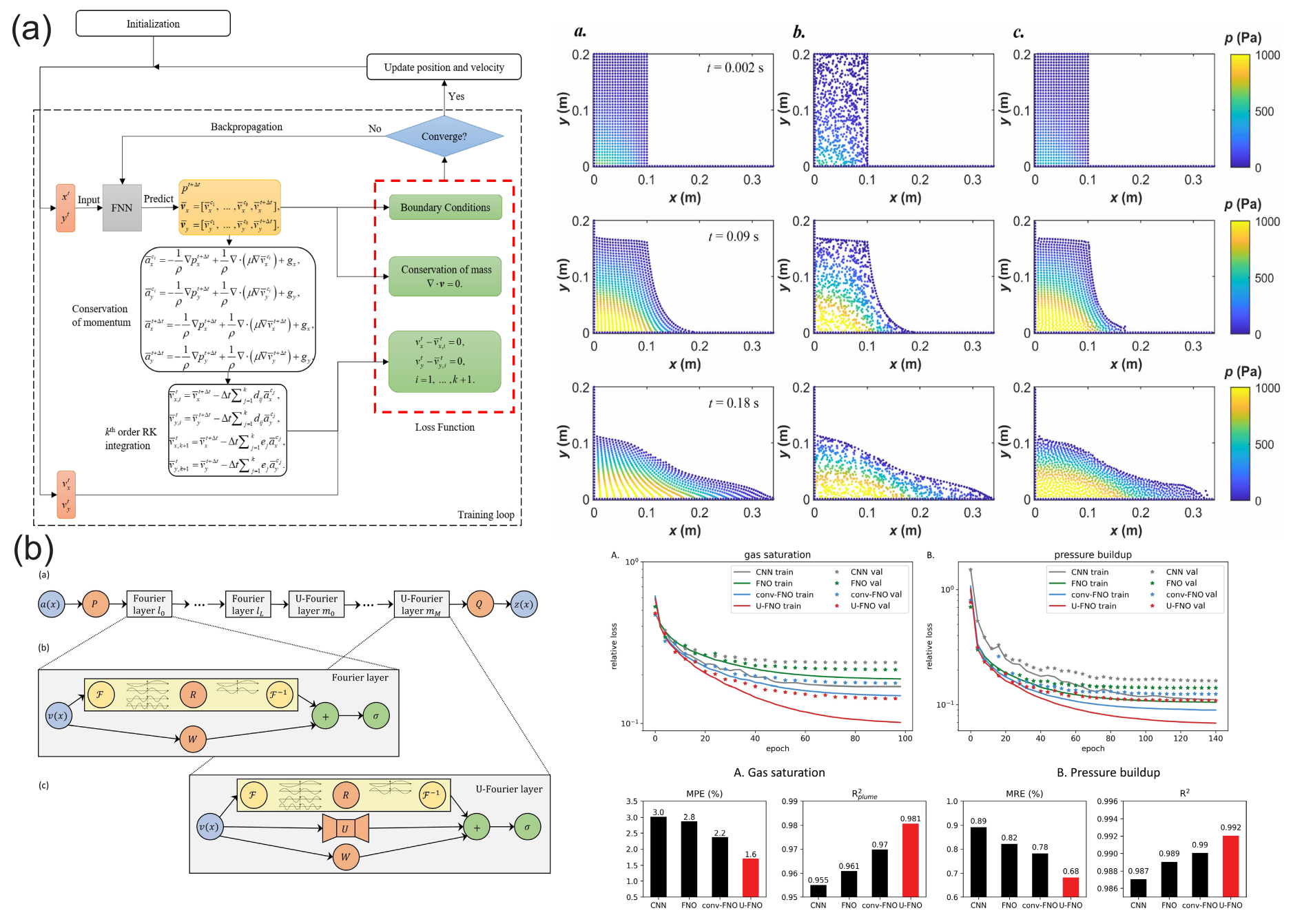}
			\par\end{centering}
		\caption{Applications of PINNs and operator learning in multiphase and moving boundary problems: (a) The flowchart
			of the general neural particle method (gNPM) \cite{bai2022general}: comparisons between the general
			neural particle method (gNPM) and the incompressible smoothed particle hydrodynamics (ISPH) for the 2D
			dam-breaking problem. (b) The schematic of the U-FNO for multiphase flow and error evolution for different
			algorithms \cite{wen2022u}. \label{fig:PINNs_multiphase_moving_boundary}}
	\end{figure}

	Based on the Lagrangian description, some hydrodynamics methods have been also developed. The use of
	the Lagrangian description makes the framework easily track free surface boundaries, which are denoted
	by training sample points, during the modeling. Among them, the neural particle method (NPM) is the
	most typical one for free surface boundaries \cite{wessels2020neural}. As presented, the NPM discretized the temporal domain into
	pieces, while the neural network is utilized to approximate the spatial domain. Later, Bai. et al. \cite{bai2022general}
	extended the framework and proposed the general neural particle method (gNPM). The flowchart of the gNPM
	is presented in \Cref{fig:PINNs_multiphase_moving_boundary}a. In the gNPM, spatial derivatives and outputs
	of neural networks are simplified to speed up the training process. Moreover, a pressure normalization scheme
	is introduced into the framework so that the gNPM can produce a reliable pressure field. Due to the
	use of the Lagrangian particles, Shao et al. \cite{shao2023improved} integrated the alpha-shape algorithm \cite{kirkpatrick1983shape}
	with the NPM to identify the boundary particles, enabling interactions between flow and solid boundaries
	or structures. 
	
	Another way to solve the moving boundaries is to transfer the Eulerian description into the Lagrangian frame
	\cite{huang2023solving}, the transformation is given by
	\begin{equation}
		\frac{\partial}{\partial\boldsymbol{X}^{L}}=\frac{\partial\boldsymbol{x}^{E}}{\partial\boldsymbol{X}^{L}}\frac{\partial}{\partial\boldsymbol{x}^{E}},
	\end{equation}
	where $\boldsymbol{x}^{E}$ and $\boldsymbol{X}^{L}$ are the coordinates under Eulerian and Lagrangian
	descriptions, respectively. Besides, the distance network in \Cref{eq:admissible_distance} is applied to exactly impose the zero-pressure
	condition for the free surface. By exploiting the versatility of neural networks, 
	PINNs-based algorithms are robust and stable for irregular node distributions, as shown in \Cref{fig:PINNs_multiphase_moving_boundary}a.
	
	Operator learning also serves as an alternative way to cope with multiphase and moving boundary
	problems. Wen et al. \cite{wen2022u} integrated the U-Net structure with FNO and proposed the U-FNO
	model, as shown in \Cref{fig:PINNs_multiphase_moving_boundary}b. In their work, the flow of CO$_2$ and
	water in the context of geological storage of CO$_2$ was considered. Compared to the traditional models of deep learning,
	the U-FNO exhibited excellent generalization properties and convergence rates. Furthermore, the U-FNO
	can achieve better accuracy when facing heterogeneous inputs. Diab et al. \cite{diab2024learning}
	proposed physics-informed DeepONet (PI-DeepONet) for multiphase flow in porous media. Using operator learning, the trained PI-DeepONet was capable of predicting solutions under any
	given flux functions (boundary conditions), which greatly alleviated the computational expense of multiphase
	modeling in porous media.
	
	\subsubsection{Multiscale and multiphysics}
	
\begin{figure}
		\begin{centering}
			\includegraphics[scale=0.55]{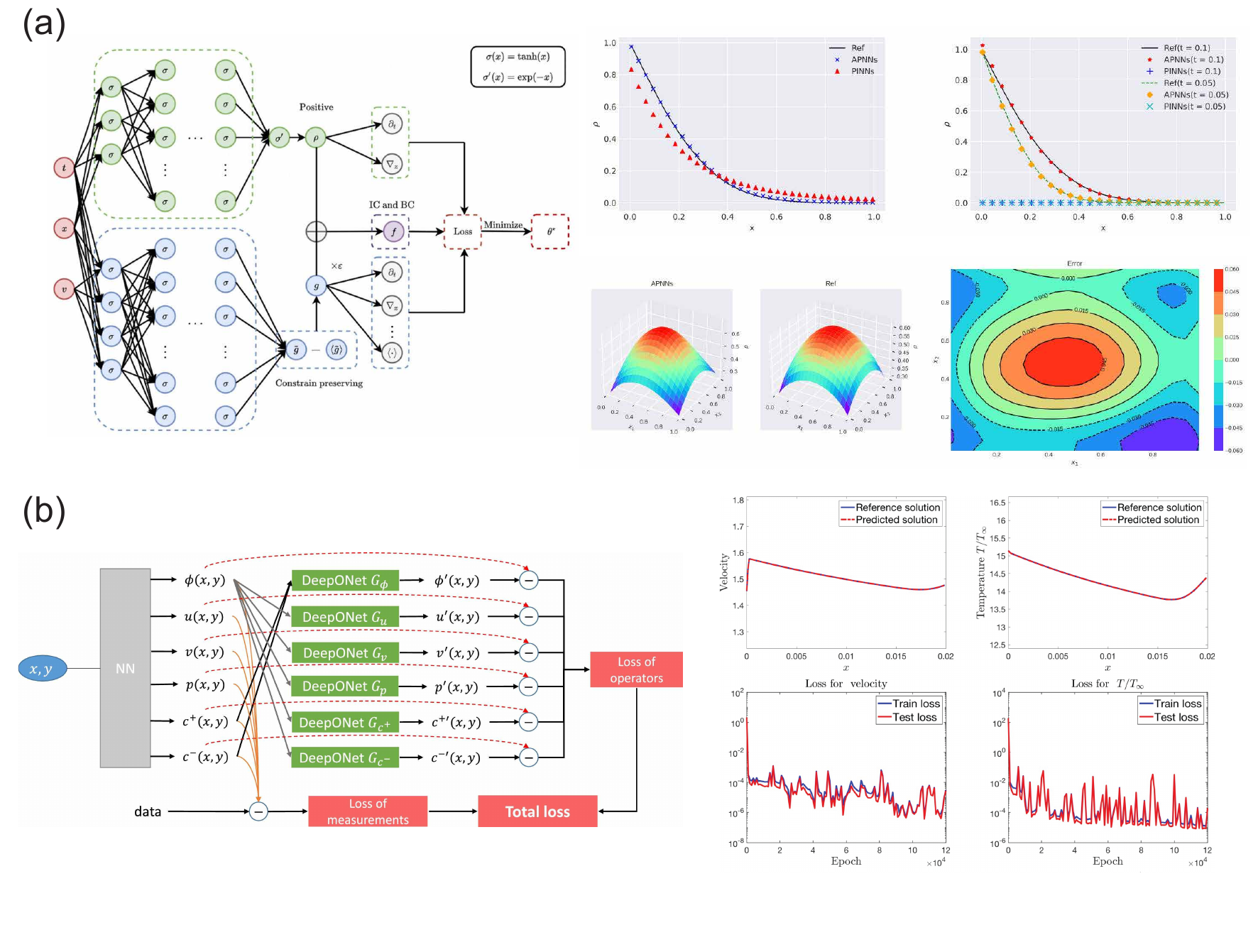}
			\par\end{centering}
		\caption{Applications of PINNs and PINOs in multiscale and multiphysics: (a) Multiscale PINNs for linear transportation equations.
			Two neural networks are designed for micro- and macro-variables, respectively \cite{jin2023asymptotic}. (b) The multiscale physics-informed
			DeepONets for multiscale fluid modeling problems \cite{cai2021deepm,mao2021deepm}. \label{fig:PINOs_multi}}
	\end{figure}

	The multiscale formulation also provides a different way to solve fluid mechanics. In general, the solution
	fields can be decomposed into the fine-scale term and the coarse-scale term. By applying the variational
	multiscale formulation, Hsieh and Huang \cite{hsieh2024multiscale} derived a novel physics-informed
	loss function based on the least squares multiscale functional form. In the case studies, the 2D and
	3D advection-dominated flow problems were conducted to demonstrate the effectiveness of the proposed loss
	function. Apart from using multiscale losses, Jin et al. \cite{jin2023asymptotic} focused on neural
	network structures and developed the asymptotic-preserving neural network (APNN) for linear transportation
	equations, which decomposed the micro- and macro-variables via two independent networks, as shown in
	\Cref{fig:PINOs_multi}a.
	
	Alongside PINNs, operator learning also sheds light on multiscale fluid modeling. Lin et al. \cite{lin2021operator}
	implemented the DeepONet to infer the bubble dynamics across different scales. In general, different
	equations and methods are required to deal with the bubble growth at different scales, for example, using
	the Rayleigh\textendash Plesset (R-P) equation for the macroscale and the dissipative particle dynamics
	(DPD) for microscale. In the paper, the R-P and DPD modeling results were used to train different DeepONets.
	Through numerical tests, the authors demonstrate the possibility of using DeepONets to speed up the solving
	process of the bubble dynamics at different scales. Later, the framework was extended by seamlessly combining
	both the macroscale and microscale knowledge in a single DeepONet to predict the dynamic behaviors of
	bubbles at different scales \cite{lin2021seamless}. It has been demonstrated that DeepONet successfully
	learned the mapping operators for bubble growth dynamics.
	
	More importantly, the DeepONet offered a novel way to form a unified modeling method across scales instead
	of using traditional multiscale modeling approaches. Cai et al. \cite{cai2021deepm} and Mao et al.
	\cite{mao2021deepm} proposed the DeepM\&Mnet, where the \textquotedblleft M\&M\textquotedblright{}
	denotes the words \textquotedblleft multiphysics\textquotedblright{} and \textquotedblleft multiscale\textquotedblright{}
	based on the operator learning technique. The proposed DeepM\&Mnet comprised numbers of pre-trained DeepONets,
	as shown in \Cref{fig:PINOs_multi}b. In Mao\textquoteright s framework, the DeepM\&Mnet was developed
	for hypersonic flow problems with shocks, which may involve sharp density gradients spanning across various
	magnitudes. By preparing data in terms of flow velocity and temperature under scenarios whose densities
	span eight orders of magnitude downstream of the shock, the DeepM\&Mnet accurately and efficiently predicted
	velocity and density fields. To ensure the accuracy of the wide range
	of velocity fields, the mean absolute percentage error (MAPE) was implemented in the loss function. 
	
	Wu et al. \cite{wu2024capturing} proposed two convolutional DeepONets to deal with the linear transportation
	problem. Ahmed and Stinis \cite{ahmed2023multifidelity} developed the multi-fidelity operator network
	(MFON), where an auxiliary term that comes from the Galerkin proper orthogonal decomposition was added
	to a DeepONet. The effectiveness of the proposed MFON was then demonstrated via a viscous Burgers equation
	problem and a 2D vortex merger problem.
	
	\subsubsection{Summary}
	
	The same as aforementioned in solid mechanics, PINNs is currently still in its infancy as a forward
	problem solver in computational fluid mechanics. The computational efficiency still impedes the use of PINNs on many problems of interest. Besides, the variational loss
	is not commonly seen in the context of hydrodynamics problems, while the strong form loss function still
	plays a vital role. As it is well known, with an increasing number of loss terms, more challenges and difficulties
	will be encountered during the training process. Therefore, training robustness also remains a huge issue
	that hinders the applications of PINNs in the context of fluid modeling. However, the great potential of
	PINNs has been found as an effective tool for inverse problems in fluid mechanics, such as field reconstruction and essential
	parameter estimation. The above PINNs frameworks for forward problems can be applied to inverse studies
	with small modifications in fluid mechanics. More details regarding the PINNs for inverse problems can be found in \Cref{subsec:Fluid-mechanics_inverse}.
	
	On the contrary, operator learning serves as a surrogate modeling scheme for fluid flows. Despite its
	data-driven nature, physics knowledge in terms of governing equations can also guide the learning process,
	such as PINO. It has shown excellent generalization properties for predicting fluid flow problems even
	with sharp gradients and complex geometries. Moreover, once the neural operator is trained, the neural operator can
	be reused for different boundary conditions. This is an extraordinary advantage compared to PINNs, because
	a PINNs is generally applicable for one set of boundary conditions. Re-training is necessary for a PINN
	if the boundary conditions of the target fluid problem are changed. Another advantage of operator learning
	is its discretization-invariant properties. In fact, great efforts have been witnessed in fluid modeling
	enhanced by operator learning, trying to reveal details of fluid fields via low-resolution results. By
	using operator learning, it is straightforward and effective to obtain super-resolution results.
	
	In the future, uncertainties, which are crucial in practical applications, should be effectively incorporated into or considered within AI for PDEs (AI4PDEs).

	\subsection{Biomechanics}

	\subsubsection{Soft tissue deformation}
	
	Computational models of soft tissue mechanics have the potential to provide valuable patient-specific
	diagnostic insights. However, their clinical deployment has been limited due to the high computational
	costs associated with traditional numerical solvers for biomechanical simulations. PINNs can be used to simulate the deformation of soft tissues, such as skin or organs,
	under various loads or boundary conditions \cite{liu2020generic}. This capability is critical for understanding
	and predicting tissue behavior in different scenarios.
	
\begin{figure}
		\begin{centering}
			\includegraphics[scale=0.57]{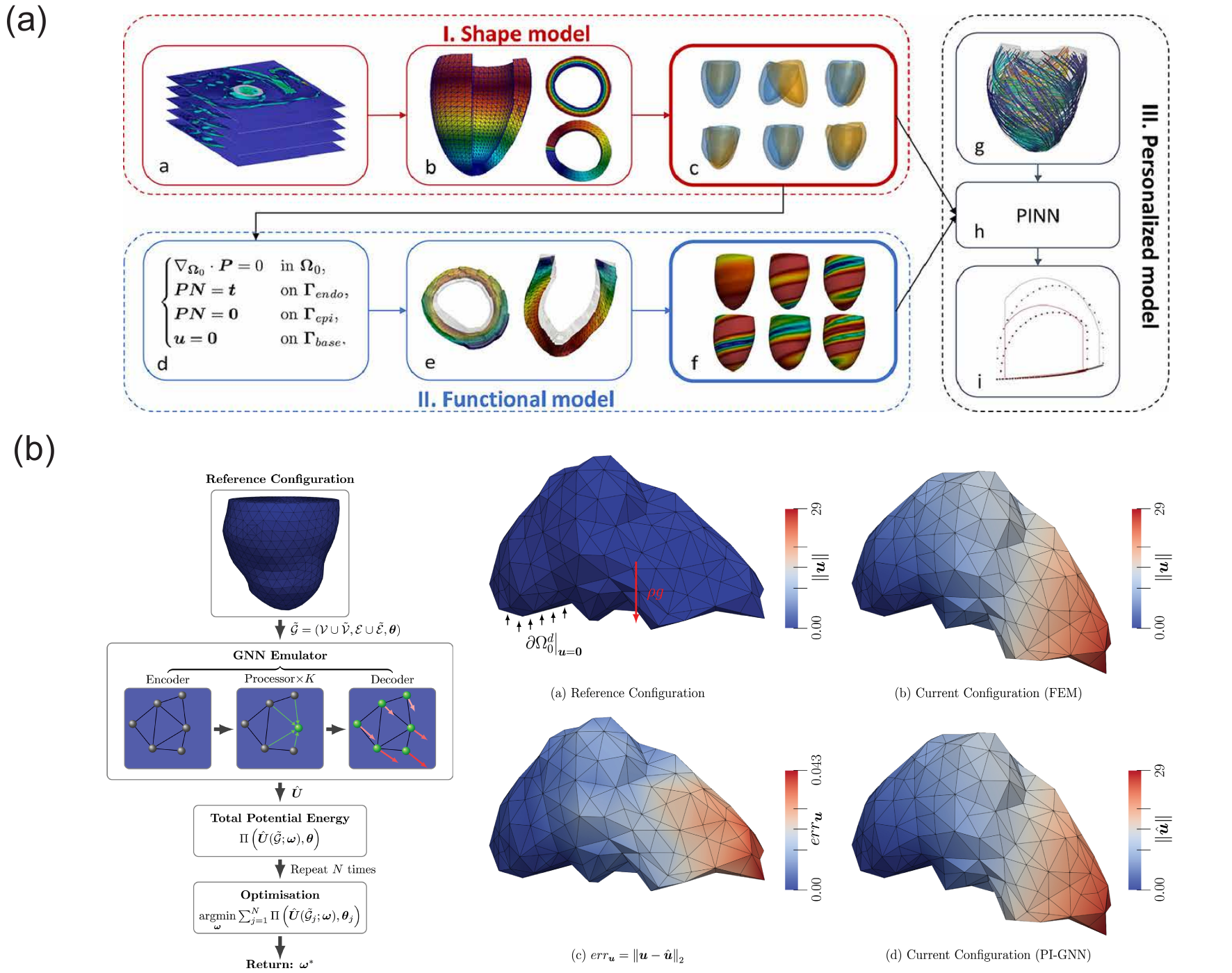}
			\par\end{centering}
		\caption{Applications of PINNs in soft tissue: (a) three main blocks of the framework: generation of the shape
			model (I), generation of the functional model (II), and definition of the personalized biophysical left
			ventricular model \cite{buoso2021personalising}. (b) The schematic of a physics-informed GNN emulator and
			emulation results for Liver and Left Ventricle model \cite{dalton2023physics}. \label{fig:PINNs_soft_issue}}
	\end{figure}

	Buoso et al. \cite{buoso2021personalising} introduce a novel approach to generate realistic numerical phantoms of the left ventricle
	by combining statistical shape learning, biophysical simulations, and tissue texture generation,
	as shown in \Cref{fig:PINNs_soft_issue}a. This study addresses the limitations of previous in-silico
	phantoms by increasing variability and realism. The generated data can be used for standardized performance
	assessment of cardiovascular magnetic resonance acquisition, reconstruction, and processing methods. 
	
	PINNs based on graph convolutional networks (GCNs) and the variational principle has also been
	proposed, which can unify the solving of PDE-governed forward and inverse problems \cite{gao2022physics}.
	Then, Dalton et al. developed an efficient emulation of soft tissue mechanics using the PINNs based on
	graph neural networks (GNNs) \cite{dalton2023physics}, as shown in \Cref{fig:PINNs_soft_issue}b. GNNs
	can handle the unique geometry of a patient's soft tissue without requiring low-order approximations.
	The physics-informed training approach simulates the soft tissue mechanical behavior. With PDEs, the approach based on PINNs accurately predicts the deformation field of the liver,
	left ventricle, and brain models under various loading conditions. PINNs based on GNNs are also used
	to solve brain mechanics, aiding in understanding the microstructure-mechanics relationship in human brain tissue \cite{linka2021unraveling}. 
	
	Lin et al. \citet{lin2026physics} developed the energy-based PINNs for viscoelastic buckling, biological tissue growth and morphological evolution. This study found that the inherent oscillations of the neural network optimizer during training could serve as a natural perturbation, allowing the model to capture creep buckling and structural instabilities without the manual introduction of artificial imperfections. PINNs could be useful for studying the specific folding and wrinkling patterns seen in developing tissues.We found that the approach to using PINNs for simulating soft tissue problems is similar to that for solid mechanical problems, where the minimum potential energy function is used as the loss function.
	
	\subsubsection{Blood flow of biomechanics}
	
	Computational modeling of blood flow and cardiovascular dynamics presents significant challenges due
	to the complexity of the underlying physics and patient-specific geometries. 
	PINNs integrate PDEs and data to provide
	a novel approach for modeling blood flow in biomechanics. This capability has significant implications
	for cardiovascular disease diagnosis, treatment planning, and personalized medicine. 
	
\begin{figure}
		\begin{centering}
			\includegraphics[scale=0.5]{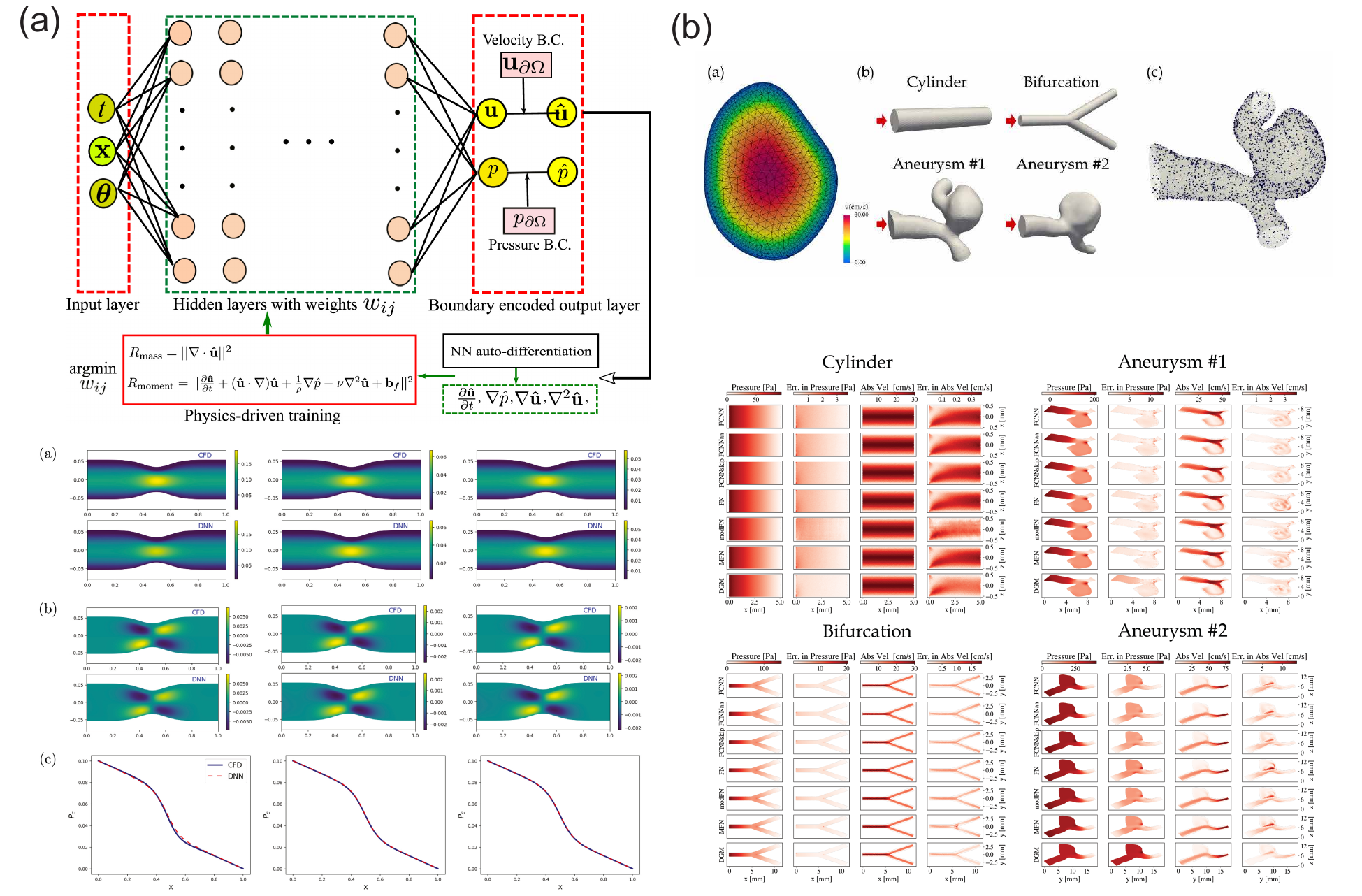}
			\par\end{centering}
		\caption{Applications of PINNs in blood flow of biomechanics: (a) The PINNs framework for surrogate modeling of
			blood flows, and results of PINNs and CFD solutions of idealized stenotic flows at different viscosity
			parameters \cite{sun2020surrogate}. (b) Physics-constrained prediction of idealized stenotic flows
			and aneurysmal flows of different geometries, the simulation geometries included two idealized structures
			(cylinder, bifurcation), and two aneurysms that were segmented from brain digital subtraction angiography
			(DSA) images \cite{moser2023modeling}. \label{fig:PINNs_blood_flow}}
	\end{figure}

	Sun et al. \cite{sun2020surrogate} developed a physics-constrained deep learning approach shown in
	\Cref{fig:PINNs_blood_flow}a for surrogate modeling of fluid flows without relying on any simulation
	data, making it suitable for parametric fluid dynamics problems where data is often scarce. Their numerical
	experiments on various internal flow problems relevant to hemodynamics demonstrate the potential of this
	approach. Additionally, PINNs has been used to model blood flow in patient-specific geometries by incorporating
	clinical data, providing insights into complex cardiovascular dynamics \cite{qiu2022physics}. 
	
	The applications of AI for PDEs in various cardiovascular imaging modalities, including echocardiography, cardiac
	computed tomography (CT), cardiovascular magnetic resonance imaging (CMR), and imaging in the catheterization
	laboratory, have also been explored, demonstrating the versatility of AI for PDEs \cite{lim2020artificial}.
	Furthermore, various neural network architectures, such as fully connected networks, Fourier networks,
	and multiplicative filter networks have been employed to model 3D blood flows using PINNs, showcasing
	the potential for efficient and accurate modeling of complex blood
	flow shown in \Cref{fig:PINNs_blood_flow}b \cite{moser2023modeling}. 
	
	Moreover, optimization techniques, such as variable-separated physics-informed neural networks based
	on adaptive weighted loss functions (AW-vsPINN), have been developed to apply to blood flow models in
	arteries. These methods decompose the blood flow problem into simpler sub-problems, reducing complexity
	and improving the efficiency of the neural network training process \cite{liu2024variable}.
	
	We have found that the simulation of blood flow in biomechanics closely aligns with the governing equations
	used in fluid mechanics. Therefore, the study of blood flow in biomechanics closely resembles the application
	of fluid mechanics to biological vascular flows.
	
	\subsubsection{Summary}

	AI for PDEs is a powerful tool for addressing the forward problem in biomechanics. It has been applied to various biomechanical issues such as tissue deformation, blood flow simulation, and morphodynamics. By enhancing our understanding of these phenomena, AI for PDEs can aid in developing new treatments for diseases like cardiopathy and other cardiovascular conditions. By incorporating the governing equations of biomechanics into neural networks, AI for PDEs can provide more accurate predictions of biological tissue behavior. With proper training, it can solve complex biomechanical problems more efficiently than traditional numerical methods, which are often computationally expensive. Moreover, AI for PDEs can model a wide range of biomechanical phenomena, from simple to complex systems, thereby broadening their applicability in the field.
	
	Biomechanics employs methodologies from both solid and fluid mechanics, such as the use of hyperelastic constitutive equations from solid mechanics. Therefore, the application of AI for PDEs in biomechanics shares similarities with both solid and fluid mechanics. However, solving biomechanical problems requires extra consideration of biochemical factors such as nutrient diffusion, growth factor concentration, and gene expression.

	Traditional modeling approaches often rely on simplified assumptions and conditions that struggle to capture the complexity of morphogenetic processes. However, recent advances in AI for PDEs offer promising avenues for integrating physical principles with data-driven approaches. In the future, AI for PDEs has the potential to combine multiscale data and bridge the gap between molecular, cellular, and tissue-level processes. AI for PDEs could be instrumental in uncovering missing or unknown physical relationships and mechanisms. By integrating PDEs with experimental data, we can potentially gain new insights into the intricate interplay of multiscale processes in biomechanics.
	
	\section{AI for PDEs in the application of inverse problems of computational mechanics} \label{sec:AI4PDEs_inverse}
	
	\subsection{Solid mechanics}
	
	\subsubsection{Identification of Elastic Modulus and Poisson's Ratio}
	
\begin{figure}
		\begin{centering}
			\includegraphics[scale=0.7]{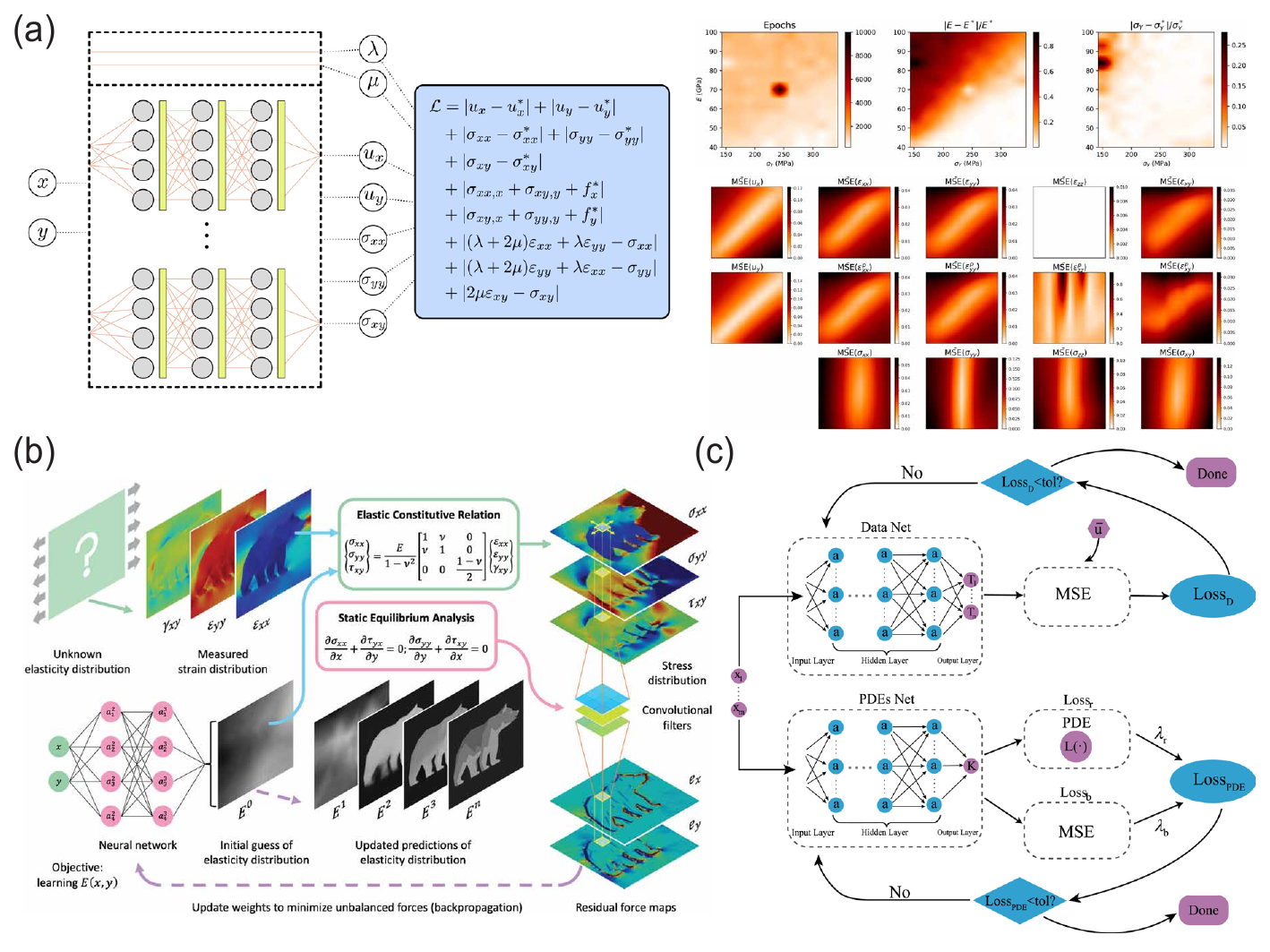}
			\par\end{centering}
		\caption{Applications of PINNs in identifying elastic modulus and poisson\textquoteright s ratio: (a) Identification
			of elastic modulus, Poisson\textquoteright s ratio, and yield stress in homogeneous and isotropic elastoplastic
			materials \cite{PINN_solid_mechanics}, (b) Identification of the elastic modulus field in non-uniform
			isotropic materials \cite{chen2021learning}, (c) Identification of thermal conductivity in non-uniform
			isotropic materials \cite{liu2024multi}.\label{fig:application_inverse_parameter}}
	\end{figure}

	Haghighat et al. (2021) \cite{PINN_solid_mechanics} were the first to propose the use of PINNs
	to solve inverse problems in solid mechanics, addressing both elastic and elastoplastic issues, as illustrated
	in \Cref{fig:application_inverse_parameter}a. In elasticity, the approach starts by gathering data
	through analytical solutions or high-precision numerical solutions. Multiple
	neural networks are used to fit multiple variables (displacement and stress), and a loss function is
	constructed using data loss, equilibrium equations, and constitutive equations. We consider the problem
	of two-dimensional plane stress to explain the idea: 
	\begin{equation}
		\begin{aligned}\mathcal{L} & =|\hat{u}_{x}-u_{x}^{*}|+|\hat{u}_{y}-u_{y}^{*}|+|\hat{\sigma}_{xx}-\sigma_{xx}^{*}|+|\hat{\sigma}_{yy}-\sigma_{yy}^{*}|+|\hat{\sigma}_{xy}-\sigma_{xy}^{*}|\\
			& +|\sigma_{xx,x}+\sigma_{xy,y}-f_{x}^{*}|+|\sigma_{yx,x}+\sigma_{yy,y}-f_{y}^{*}|\\
			& +|(\lambda+2G)\varepsilon_{xx}+\lambda\varepsilon_{yy}-\sigma_{xx}|+|(\lambda+2G)\varepsilon_{yy}+\lambda\varepsilon_{xx}-\sigma_{yy}|+|2G\varepsilon_{xy}-\sigma_{xy}|
		\end{aligned}
		\label{eq:the_loss_inverse_solid_mechanics}
	\end{equation}
	where the superscript $\ensuremath{*}$ indicates the given data, and $\lambda$ and $G$ are Lame
	constants (set as the inverse problem variables to be optimized). The least squares $|\cdot|$ is used
	for fitting, and $\lambda$ and $G$ are obtained by optimization using the given displacement, stress,
	and body force fields. Note that this formulation assumes simultaneous availability of both displacement and stress field data. In practical experimental settings however, stress fields are rarely measured directly and are typically obtained from displacements via constitutive assumptions.

	For elastoplastic mechanics, the approach is similar, although the equations change to J2
	flow theory for plane strain conditions. According to the J2 flow theory, we add plasticity multiplier $\gamma$ and the
	KKT conditions for convex optimization: 
	\begin{equation}
		\begin{aligned}\mathcal{L} & =|\hat{u}_{x}-u_{x}^{*}|+|\hat{u}_{y}-u_{y}^{*}|+|\hat{\sigma}_{xx}-\sigma_{xx}^{*}|+|\hat{\sigma}_{yy}-\sigma_{yy}^{*}|+|\hat{\sigma}_{xy}-\sigma_{xy}^{*}|+|\hat{\sigma}_{zz}-\sigma_{zz}^{*}|\\
			& +|\sigma_{xx,x}+\sigma_{xy,y}-f_{x}^{*}|+|\sigma_{yx,x}+\sigma_{yy,y}-f_{y}^{*}|\\
			& +|(\lambda+\frac{2}{3}G)\varepsilon_{kk}+2G(\varepsilon_{xx}-\varepsilon_{xx}^{*p})-\sigma_{xx}|+|(\lambda+\frac{2}{3}G)\varepsilon_{kk}+2G(\varepsilon_{yy}-\varepsilon_{yy}^{*p})-\sigma_{yy}|\\
			& +|(\lambda+\frac{2}{3}G)\varepsilon_{kk}+2G(\varepsilon_{zz}-\varepsilon_{zz}^{*p})-\sigma_{zz}|+|2G(\varepsilon_{xy}-\varepsilon_{xy}^{*p})-\sigma_{xy}|\\
			& +|(\bar{\varepsilon}-\frac{\sigma_{s}}{3G})-\bar{\varepsilon}^{*p}|+|[1-\text{sign}(\bar{\varepsilon}^{*p})]|\bar{\varepsilon}^{*p}||+|(1+\text{sign}(\mathcal{F}))|\mathcal{F}||+|\bar{\varepsilon}^{*p}\mathcal{F}|
		\end{aligned},
		\label{eq:the_loss_inverse_solid_mechanics_elaplasticity}
	\end{equation}
	where the first line represents data loss (all variables marked with $*$ are given beforehand), the
	second line represents the equilibrium equations, the third and fourth lines are the constitutive equations
	with the superscript $p$ indicating the plastic part, and the last line includes the KKT conditions.
	The first item in the KKT conditions is the J2 theory plasticity multiplier $\gamma=\bar{\varepsilon}^{*p}=\bar{\varepsilon}-\sigma_{s}/(3G)$, $\bar{\varepsilon}=\sqrt{2\varepsilon_{ij}\varepsilon_{ij}/3}$, and $\bar{\varepsilon}^{p}=\sqrt{2\varepsilon_{ij}^{p}\varepsilon_{ij}^{p}/3}$.
	$\mathcal{F}$ is the yield surface function, defined by J2 theory as $\mathcal{F}=\sqrt{3s_{ij}s_{ij}/2}-\sigma_{s}$.
	\text{sign} is as follows:
	
	\begin{equation}
		\text{sign}(x)=\begin{cases}
			1 & x>0\\
			0 & x=0\\
			-1 & x<0
		\end{cases}
	\end{equation}
	
	One issue of the above formulation is the reliance on plastic strain data, which are probably difficult to measure experimentally. Furthermore, the inclusion of KKT conditions via sign functions introduces non-differentiabilities that will complicate gradient-based optimization.
	It is worth noting that the units of stress and displacement in \Cref{eq:the_loss_inverse_solid_mechanics} and \Cref{eq:the_loss_inverse_solid_mechanics_elaplasticity} must be consistent. 
	Clearly, \Cref{eq:the_loss_inverse_solid_mechanics_elaplasticity} encompasses all the PDEs described
	by J2 theory for elastoplastic mechanics. Hence, optimizing \Cref{eq:the_loss_inverse_solid_mechanics_elaplasticity}
	involves fitting displacement and stress using multiple neural networks (similar to \Cref{eq:the_loss_inverse_solid_mechanics}),
	then setting $\lambda$, $G$, and $\sigma_{s}$ as optimization variables for solving the inverse problem.
	If a theory other than J2 is used, the method remains the same, though the specific PDEs would differ.
	Moreover, the plastic strain in \Cref{eq:the_loss_inverse_solid_mechanics_elaplasticity} is difficult to measure experimentally in practice, which precludes the use of this framework for material identification.

	For isotropic but heterogeneous materials, Chen et al. (2021) \cite{chen2021learning}
	suggest identifying the elastic field in isotropic and inhomogeneous materials as shown in \Cref{fig:application_inverse_parameter}b.
	However, instead of mapping coordinates to displacement fields, it maps coordinates to elastic modulus fields. The strain field is calculated from the given displacement field using kinematic equations, and the elastic field is then computed
	at the coordinates of the strain fields using neural networks. Stresses are subsequently calculated using
	isotropic elasticity constitutive relations, and these stresses are used to derive partial differential
	fields through convolutional kernel differentiation. The final loss function is established based on
	equilibrium equations, and it's optimized to tune the neural network parameters in the elastic modulus
	field. It's critical to note that merely using equilibrium equations is insufficient. Theoretically,
	we must also incorporate force boundary conditions or an average modulus of elasticity. This average
	modulus is used to calibrate the learned modulus back to realistic conditions because the modulus field
	learned solely from equilibrium equations can just represent a pattern, whose amplitude is undetermined.
	Thus, an average modulus is required to realign it with reality, assuming prior knowledge of the material's
	properties. If there is no prior knowledge, applying force boundary conditions is necessary. It is important
	to emphasize that these force boundary conditions are not only applied at the boundaries but can also
	be implemented in internal domains if the loading conditions are special. Details can be referred to \cite{chen2021learning}.
	Unlike learning a fixed modulus,
	this neural network mapping establishes a link from coordinates to the elastic modulus field, not merely
	as optimization variables. Additionally, learning physical equations through convolutional neural networks' kernel functions is
	feasible by setting the elastic modulus and strain fields in advance. 
	There are also approaches that learn PDEs. Brunton et al. \cite{brunton2016discovering} proposed SINDy, which discovers governing equations from data for nonlinear dynamical systems.

	Furthermore, Liu et al. (2024) \cite{liu2024multi} presented a similar idea
	for solving the inverse problem of thermal conductivity for isotropic and heterogeneous materials. They
	fit temperature data using Data Net and then used PDEs Net, based on PINNs, to solve for non-uniform
	thermal conductivity, as depicted in \Cref{fig:application_inverse_parameter}c. This method combines
	direct data fitting with physical laws to refine the prediction of material properties. 

	\subsubsection{Identification of Constitutive Equations}
	
\begin{figure}
		\begin{centering}
			\includegraphics[scale=0.70]{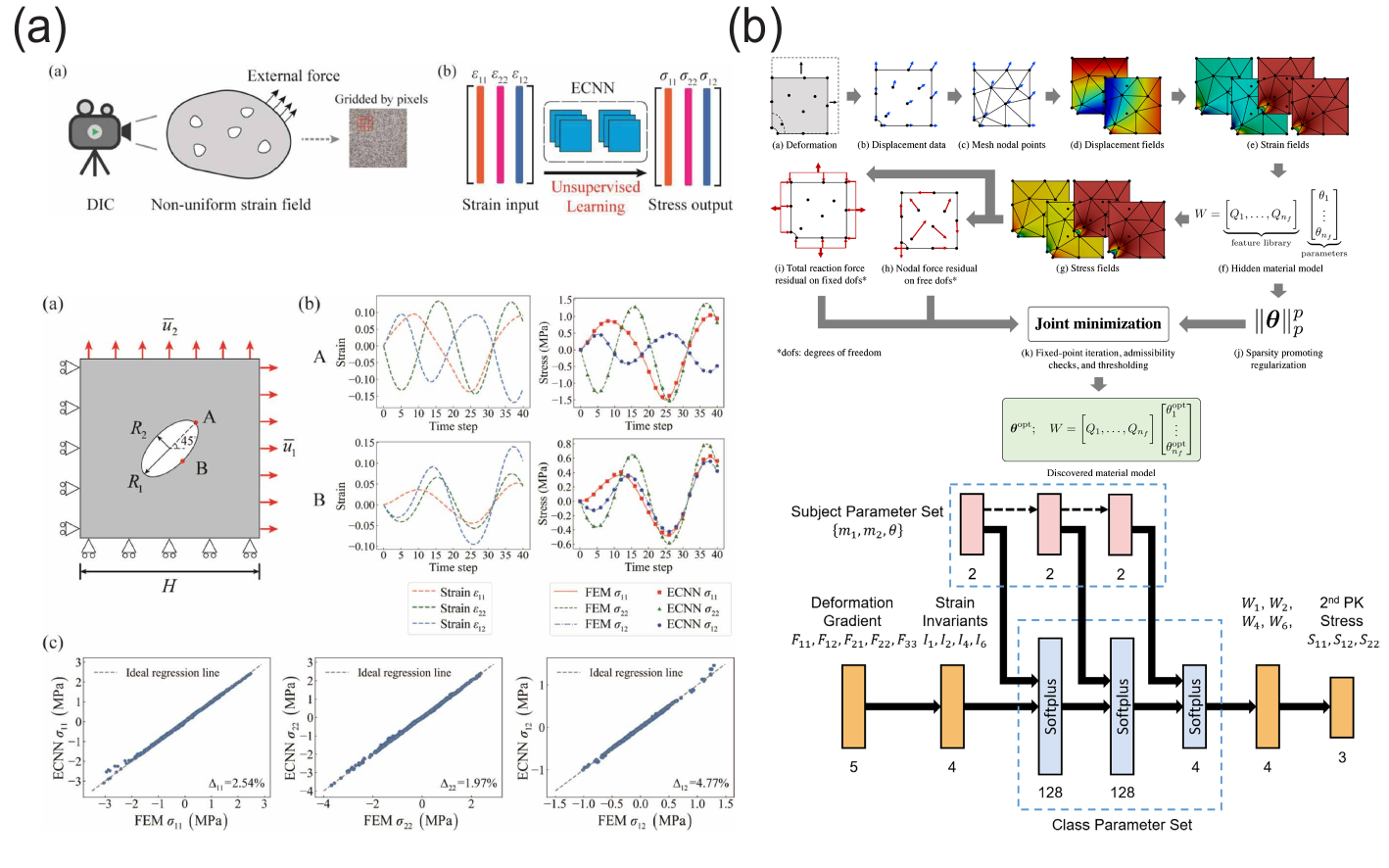}
			\par\end{centering}
		\caption{Applications of AI for PDEs in Constitutive Equation Identification: (a) Non-parametric model: neural network fitting of strain
			and stress relationships in hyperelastic materials \cite{li2022equilibrium}, (b) Parametric model:  predefined hyperelastic
			constitutive forms with fitting of preceding parameters \cite{liu2020generic,hyper_constitutive_data_driven}.\label{fig:application_inverse_constitutive}}
	\end{figure}

	The identification of constitutive equations remains a core issue in solid mechanics, primarily focusing
	on the relationship between stress and strain. Traditional methods usually involve defining a specific
	form of the constitutive equation and then determining the unknown parameters of the model based on experimental
	results (displacement fields and external load forces). Li et al. (2022) \cite{li2022equilibrium} proposed
	using a neural network to replace the constitutive equation, where the relationship from strain to stress
	is determined by the parameters of the neural network, as shown in \Cref{fig:application_inverse_constitutive}a.
	The loss function is defined as:
	
	\begin{equation}
		\begin{aligned}\mathcal{L} & =\int_{\Omega}|\sigma_{ij,j}+f_{i}|^{2}d\Omega+\sum_{s=1}\beta^{s}|\int_{\Gamma^{t_{s}}}(\sigma_{ij}n_{j})d\Gamma-\boldsymbol{T}^{s}|^{2}\\
			\sigma_{ij} & =NN(\varepsilon_{ij};\boldsymbol{\theta})
		\end{aligned},
		\label{eq:inverse_constitutive_law}
	\end{equation}
	where $\boldsymbol{T}^{s}$ is the total force for the s-th force boundary $\Gamma^{t_{s}}$, and $\boldsymbol{T}^{s}$ can
	be easily obtained from universal testing machines for the corresponding load. Additionally, the displacement
	field under $\boldsymbol{T}^{s}$ can be captured using Digital Image Correlation (DIC), and then the strain is calculated
	using kinematic equations and inputted into the neural network $NN(\varepsilon_{ij};\boldsymbol{\theta})$
	to obtain the corresponding stress. The stress field is used to establish the loss of internal balance equation and the force boundary conditions. The loss about force boundary conditions integrates over the force boundary $\Gamma^{t_{s}}$ using experimental load data. Note that non-homogeneous
	force boundary conditions are essential. If only displacement data are available without force boundary conditions, it is impossible to
	determine the constitutive equation theoretically. \Cref{fig:application_inverse_constitutive}a outlines a non-parametric hyperelastic constitutive
	model using a neural network, but a parametric model as shown in \Cref{fig:application_inverse_constitutive}b can also be established. For example, the form of
	the constitutive equation can be pre-determined (e.g., the HGO model for biomaterials by \cite{liu2020generic}),
	considering all possible base functions in the constitutive equation \cite{hyper_constitutive_data_driven},
	and then using optimization techniques to determine the parameters ahead of these base functions to define
	the form of the constitutive equation.
	
	It is important to note that both parametric and non-parametric models require non-homogeneous force
	boundary conditions. 
	It is easy to imagine two different materials subjected to a displacement-controlled tensile test that exhibit exactly the same displacement field. If non-homogeneous force boundary conditions are not provided, it is impossible to distinguish between these two materials solely from the displacement field, without any information about the applied forces. Therefore, non-homogeneous force boundary conditions constitute indispensable information for inverse problems.
	The essential difference between these implementation methods lies in the form of
	$\sigma_{ij}=NN(\varepsilon_{ij};\boldsymbol{\theta})$ in \Cref{eq:inverse_constitutive_law}, where
	the non-parametric model uses a neural network and the parametric model uses a predefined function. Currently,
	AI for PDEs applications in identifying constitutive equations are primarily focused on hyperelastic
	materials, with fewer studies on history- and path-dependent elastoplastic materials, and rate-dependent
	viscoelastic materials. Future research using neural networks to replace constitutive equations will
	gradually address such complex materials requiring solid constitutive theory foundations to establish
	physically meaningful neural network constitutive models. We believe this research area holds significant
	potential and prospects because constitutive equations primarily involve fitting, and neural networks
	inherently possess robust fitting capabilities.
	
	It is important to distinguish between two classes of inverse problems in constitutive modeling: parameter identification, where the functional form of the constitutive law is known a priori and only material constants are unknown; and constitutive law discovery, where the functional form itself is learned from data. The former is well-suited for PINNs with trainable scalar parameters, while the latter often requires more expressive representations such as neural networks with built-in physical constraints such as polyconvexity or thermodynamic consistency \cite{thakolkaran2022nn}. The latter remains an active area of research with significant open challenges, including ensuring physical admissibility of learned constitutive relations.
	
	An alternative to neural-network-based constitutive modeling is sparse regression in the weak form, often referred to as Variational System Identification (VSI). This approach constructs a library of candidate functions such as polynomial, exponential or physics-inspired terms and solves a sparse linear system to identify which terms govern the material response. Unlike PINNs-based methods, VSI does not require neural network training, yields interpretable models and is naturally robust to noise due to integration by parts. A very interesting approach has been recently presented in the context of phase field models for fracture by Livingston et al. \cite{livingston2025inference}, who employed VSI to infer the degradation function, geometric crack function, and crack driving history term from synthetically generated data, demonstrating successful model selection among competing fracture formulations.
	
	\subsubsection{Topology Optimization}
	
	Topology optimization is a classic inverse problem. It modifies the material distribution based on a
	predefined objective function, such as minimizing compliance. Typically, changing the material distribution
	requires sensitivity analysis, which mathematically involves deriving the objective function with respect
	to the density field (the physical representation of material distribution). Notably, the objective function
	alone is insufficient without the stress-deformation state under the current density field, which can
	be mathematically expressed as:
	
	\begin{equation}
		\begin{aligned}\mathcal{\text{Objective Function (Minimize Compliance): }} & \min_{\rho}\mathcal{F}=\frac{1}{2}\int_{\Omega}\boldsymbol{\sigma}^{*}:\boldsymbol{\varepsilon}d\Omega\\
			\text{Material Constraint: } & \int_{\Omega}\rho(\boldsymbol{x})d\Omega=\bar{V}
		\end{aligned},
		\label{eq:inverse_topology_density}
	\end{equation}
	where $\rho$ is the density field (material distribution), a scalar field function ranging from 0 to
	1. $\sigma^{*}$ is the stress field adjusted according to the density field $\rho$, dictating the material
	degradation, typically set as $\boldsymbol{\sigma}^{*}(\boldsymbol{x})=g(\rho)\ast\boldsymbol{\sigma}(\boldsymbol{x})$.
	A widely used choice is the SIMP \cite{bendsoe1999material} interpolation, for which $g(\rho)=\rho^{p}$ with $p>1$ being a penalization parameter.
	$\boldsymbol{\sigma}(\boldsymbol{x})$ is determined by
	substituting the displacement field $\boldsymbol{u}$ into the kinematic equation and then through the
	constitutive relation.  For linear elastic problems,
	the original stress field is modified to $\boldsymbol{\sigma}^{*}(\boldsymbol{x})$ as expressed mathematically:
	
	\begin{equation}
		\begin{aligned}\text{Minimum Potential Energy Principle: } & \min_{\boldsymbol{u}}\mathcal{L}=\int_{\Omega}\frac{1}{2}\boldsymbol{\sigma}^{*}:\boldsymbol{\varepsilon}d\Omega-\int_{\Omega}\boldsymbol{f}\cdot\boldsymbol{u}d\Omega-\int_{\Gamma^{t}}\bar{\boldsymbol{t}}\cdot\boldsymbol{u}d\Gamma\\
			\text{Displacement Constraint: } & \boldsymbol{u}(\boldsymbol{x})=\boldsymbol{\bar{u}}(\boldsymbol{x}), \boldsymbol{x}\in\Gamma^{u}
		\end{aligned},
		\label{eq:inverse_topology_dis}
	\end{equation}
	where $\boldsymbol{\sigma}^{*}(\boldsymbol{x})$ is derived from the displacement field and the current
	density field, with the constraint that the displacement field satisfies essential boundary ($\Gamma^{u}$) conditions.

	Notably, the compliance minimization in \Cref{eq:inverse_topology_density} is not independent of the equilibrium condition, but must be carried out subject to the equilibrium constraint represented by the minimum potential energy principle in \Cref{eq:inverse_topology_dis}.
	These two optimization
	problems differ in that each update to the density field in \Cref{eq:inverse_topology_density} through
	gradient descent, $\rho^{*}=\rho-\alpha\partial\mathcal{F}/\partial\rho$, necessitates a complete optimization
	of the potential energy $\mathcal{L}$. Thus, mathematically, it is a coupled optimization problem under
	material and displacement constraints. While compliance optimization can rely on traditional optimization
	algorithms, the displacement field can be obtained using solvers like the traditional finite element
	method.
	
\begin{figure}
		\begin{centering}
			\includegraphics[scale=0.70]{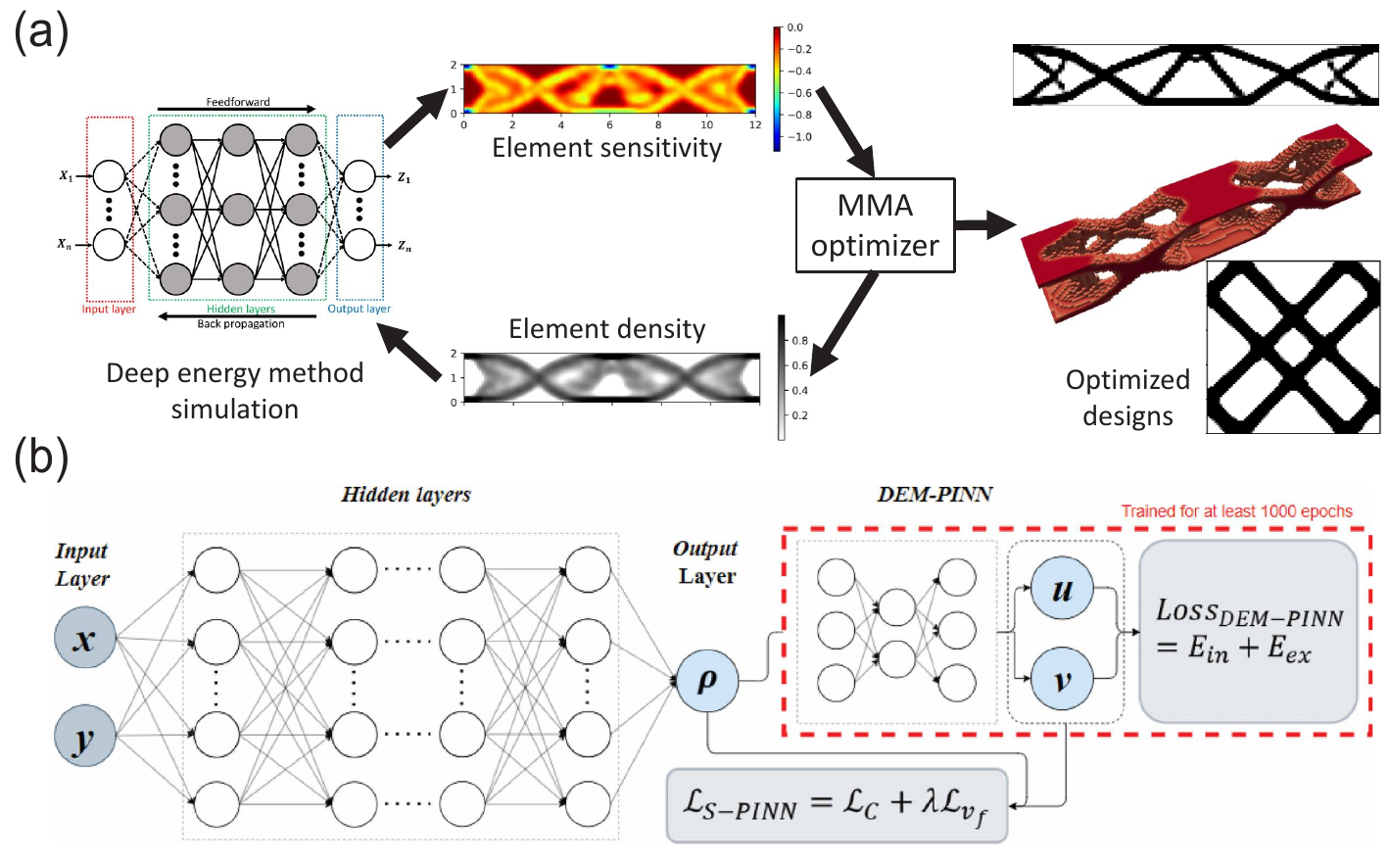}
			\par\end{centering}
		\caption{Applications of DEM in topology optimization: (a) Traditional methods such as the moving asymptote method is used to iterate the density	field for compliance minimization, and DEM is for the displacement field \cite{he2023deep_o}. (b) DEM is used for both the density and displacement field \cite{jeong2023complete} \label{fig:application_topology}}
	\end{figure}

	Recent studies have proposed using DEM to replace compliance and potential energy optimization. He
	et al. (2023) \cite{he2023deep_o} adopted the idea from Zehnder et al. (2021) \cite{zehnder2021ntopo}
	to replace the finite element calculation with the DEM method for minimum potential energy, but the target
	compliance minimization function is still calculated using traditional methods like the moving asymptote
	method, as illustrated in \Cref{fig:application_topology}a. Subsequently, Jeong et al. (2023) \cite{jeong2023complete}
	also replaced the traditional algorithms with an idea similar to DEM for target compliance minimization function,
	establishing a fully connected neural network function mapping coordinates to the density field. Compliance  obtained from the displacement field  and the density
	field  is then optimized once, updating the neural network
	parameters of the density field and this optimized density field is re-entered into the DEM to compute
	the displacement field, repeating until convergence, as shown in \Cref{fig:application_topology}b.
	It is noteworthy that the combination of DEM and topology optimization centralizes some
	optimization problems using neural network optimization algorithms instead of traditional methods, such
	as using PINNs or DEM to replace finite element calculations for PDEs in solid mechanics. With target optimization functions solved using PINNs methods \cite{jeong2023complete}, traditional
	SIMP \cite{jeong2023physics}, or traditional MMA \cite{he2023deep_o}, different algorithms for solving
	target optimization functions and different methods for solving
	mechanical displacement fields form various works combining deep learning and topology optimization.

	Since the bulk of the computational effort comes from continually recalculating the displacement field
	for updated geometric topologies, using operator learning to accelerate the computation of the forward
	problem in topology optimization has significant potential to enhance the total speed of topology optimization.

	\subsubsection{Defect Identification }
	
\begin{figure}
		\begin{centering}
			\includegraphics[scale=0.65]{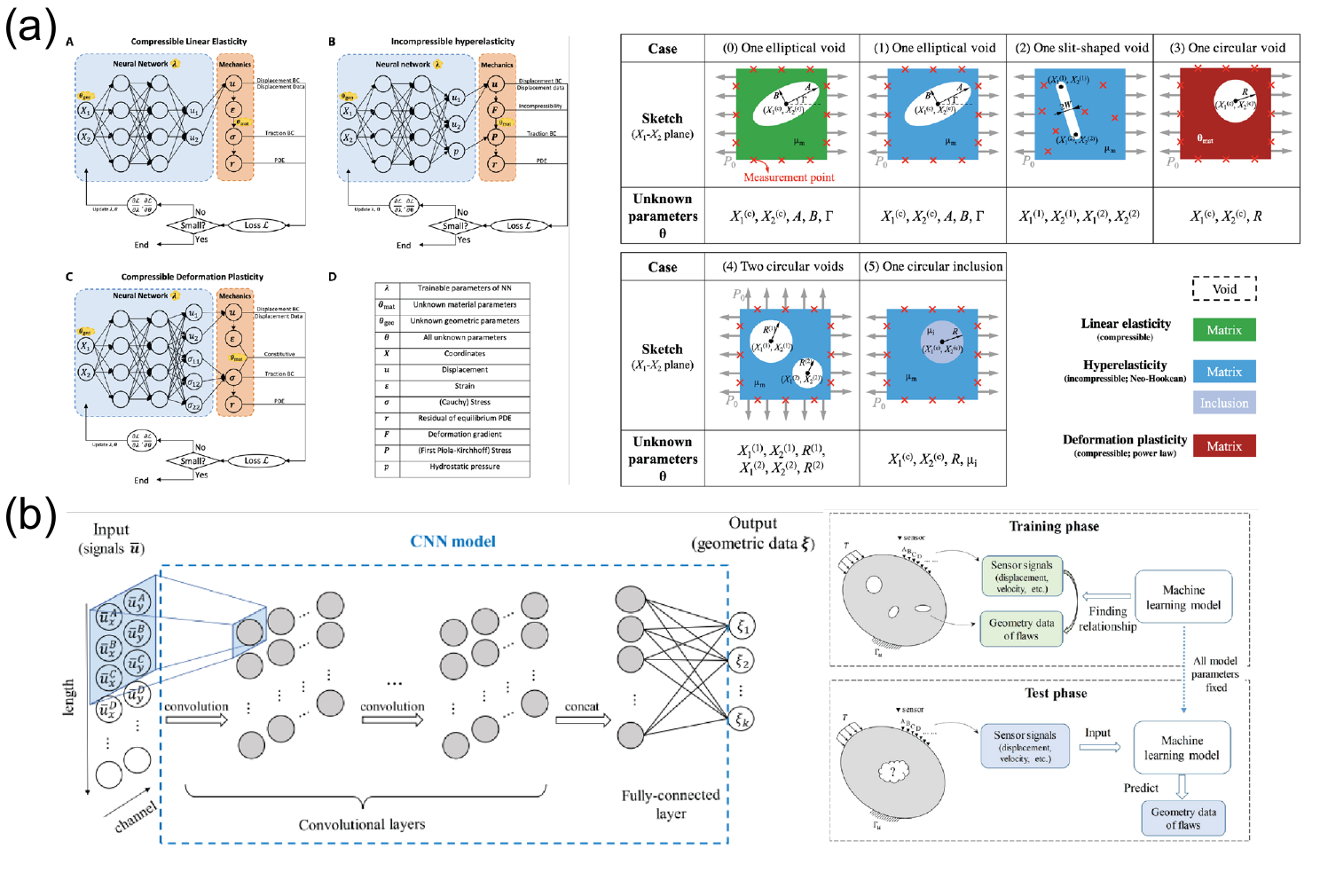}
			\par\end{centering}
		\caption{Applications of AI for PDEs in defect identification: (a) PINNs for defect detection: Utilizing PINNs
			to identify internal defects by analyzing known loads and displacement fields at boundaries, applied
			to linear, hyperelastic, and elastoplastic issues by \cite{zhang2022analyses}. (b) Data-driven defect
			detection: An approach by \cite{sun2023data} employing boundary element methods and convolutional neural
			networks to map displacement fields to detect multiple defects. \label{fig:application_inverse_flaw_detection}}
	\end{figure}

	Defect identification in solid mechanics has garnered significant attention and presents a challenging
	task because it involves solving inverse problems with unknown geometries, topologies (such as the locations of the internal defects), and material properties of inclusions. Therefore, characterizing internal inclusions
	and defects is a classical and complex inverse problem. Zhang et al. (2022) \cite{zhang2022analyses}
	applied PINNs to defect identification. Their approach involves placing displacement sensors along the
	boundaries, using known loads and boundary displacement fields to infer the locations of internal defects
	in problems without body forces. The mathematical expression, similar to classical PINNs that include
	data and PDEs loss, is as follows: 
	\begin{equation}
		\mathcal{L}=\beta_{1}\mathcal{L}_{pdes}+\beta_{2}\mathcal{L}_{u}+\beta_{3}\mathcal{L}_{t}+\beta_{4}\mathcal{L}_{data},
	\end{equation}
	where $\mathcal{L}_{pdes}$, $\mathcal{L}_{u}$, $\mathcal{L}_{t}$, and $\mathcal{L}_{data}$ respectively
	represent the loss from physical equations, displacement boundary conditions, force boundary conditions,
	and boundary data loss functions as follows:
	\begin{equation}
		\begin{aligned}\mathcal{L}_{pdes}=\frac{1}{N_{pdes}}\sum_{n=1}^{N_{pde}}|\sigma(\boldsymbol{x}^{(n)})_{ij,j}|^{2} & ,\boldsymbol{x}^{(n)}\in\Omega(\boldsymbol{\theta}_{geo})\\
			\mathcal{L}_{u}=\frac{1}{N_{u}}\sum_{n=1}^{N_{u}}|u_{i}^{(n)}(\boldsymbol{x}^{(n)})-\bar{u}_{i}^{(n)}(\boldsymbol{x}^{(n)})|^{2} & ,\boldsymbol{x}^{(n)}\in\Gamma^{u}(\boldsymbol{\theta}_{geo})\\
			\mathcal{L}_{t}=\frac{1}{N_{t}}\sum_{n=1}^{N_{t}}|\sigma_{ij}(\boldsymbol{x}^{(n)})n_{j}-\bar{t}_{i}(\boldsymbol{x}^{(n)})|^{2} & ,\boldsymbol{x}^{(n)}\in\Gamma^{t}(\boldsymbol{\theta}_{geo})\\
			\mathcal{L}_{data}=\frac{1}{N_{d}}\sum_{n=1}^{N_{d}}|u_{i}(\boldsymbol{x}^{(n)})-\bar{u}_{i}^{*}(\boldsymbol{x}^{(n)})|^{2} & ,\boldsymbol{x}^{(n)}\in\Gamma^{data}(\boldsymbol{\theta}_{geo})
		\end{aligned}
		\label{eq:inverse_flaw_detection}
	\end{equation}
	where the domain $\Omega(\boldsymbol{\theta}_{geo})$ and the boundaries $\Gamma^{u}(\boldsymbol{\theta}_{geo})$, $\Gamma^{t}(\boldsymbol{\theta}_{geo})$, $\Gamma^{data}(\boldsymbol{\theta}_{geo})$ are functions of the geometric parameters $\boldsymbol{\theta}_{geo}$,
	which denote the internal defect locations. For instance, a circular defect could be described
	by the circle\textquoteright s center and radius, $\boldsymbol{\theta}_{geo}=\{x_{c},y_{c},r\}$. This
	approach's core integrates $\boldsymbol{\theta}_{geo}$ into the sampling coordinates, allowing optimization
	of both the parameters of the neural network for  coordinates-to-displacement field and $\boldsymbol{\theta}_{geo}$
	during backpropagation. Ultimately, extracting the $\boldsymbol{\theta}_{geo}$ parameters serves to
	pinpoint the defect location. Notably, Zhang et al. (2022) \cite{zhang2022analyses} applied PINNs not
	only to linear elasticity but also to hyperelasticity and elastoplasticity problems, detecting internal
	defects' material parameters by treating them as optimization variables. The overall framework of their
	method is shown in \Cref{fig:application_inverse_flaw_detection}a. Sun et al. (2023) \cite{sun2023data}
	proposed a purely data-driven algorithm for identifying multiple defects, using boundary element methods
	to calculate defects and displacement fields and employing convolutional neural networks to map
	boundary displacement fields to defects as shown in \Cref{fig:application_inverse_flaw_detection}b.
	
	However, current defect detection has not yet incorporated operator learning, which could potentially
	accelerate the computation of the forward problem under various defects, thereby enhancing the overall
	speed of defect detection.
	
	\subsubsection{Summary }
	
	We have identified three main approaches for applying AI for PDEs to inverse problems in solid mechanics.  
	The first approach involves setting the variables to be identified in the inverse problem as optimization variables, such as elastic moduli and Poisson's ratios, and using PINNs for tasks like defect detection.  
	The second approach is applied when the inverse problem is more complex, such as when dealing with heterogeneous fields, where a new neural network is constructed to approximate the heterogeneous field and incorporated into the loss function for optimization, e.g., in density function optimization for topology.  
	The third approach leverages neural networks to learn the mapping between variables, such as constitutive equation identification (mapping strain to stress via a neural network).  
	
	For many inverse problems, the first approach offers a simpler framework because it avoids the traditional method of repeatedly solving forward problems to determine the parameters. Instead, the parameters are directly set as optimization variables within the existing deep learning framework, making the program and code structure relatively straightforward. This, however, does not imply that the underlying computations are trivial; they still involve derivatives of the loss function with respect to the optimization variables, which remain computationally intensive and complex. Nevertheless, current deep learning optimization frameworks effectively encapsulate these processes, making them appear simpler.

	In the future, incorporating operator learning into inverse problems
	holds significant potential, as the core of most current inverse problems remains using PINNs to substitute finite element computations. Integrating operator learning could significantly accelerate the computation of each forward problem to enhance
	the computational efficiency of inverse problems substantially.
	
	\subsection{Fluid mechanics\label{subsec:Fluid-mechanics_inverse}}
	
	\subsubsection{Field reconstruction}
	
\begin{figure}
		\begin{centering}
			\includegraphics[scale=0.55]{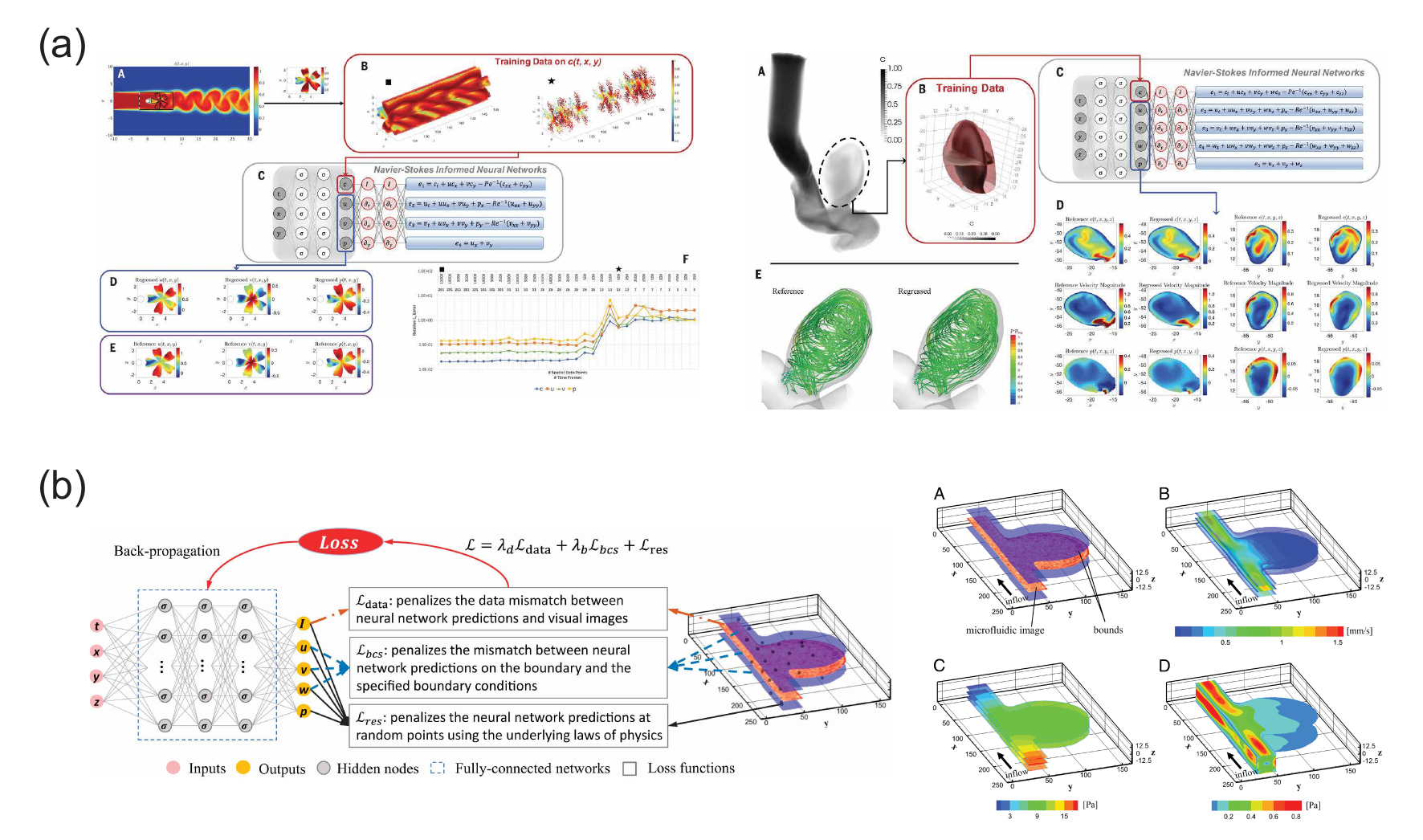}
			\par\end{centering}
		\caption{Applications of PINNs in field reconstruction: (a) The typical example of the hidden fluid mechanics
			(HFM) \cite{raissi2020hidden}. (b) The artificial intelligence velocimetry (AIV) framework \cite{cai2021artificial}.
			\label{fig:PINNs_field_reconstruction}}
	\end{figure}

	Raissi et al. \cite{raissi2020hidden} introduced the hidden fluid mechanics (HFM) that combined the
	knowledge from observation snapshots and the NS equations to evacuate and reconstruct the velocity,
	pressure fields, and even diffusions of passive scalars. A typical example of the HFM is shown in \Cref{fig:PINNs_field_reconstruction}a.
	It has been demonstrated to be effective for studying the 3D hemodynamics of intracranial aneurysms without
	boundary conditions and inlet flow waveform, where only the concentration of a given passive scalar was
	known. Cai et al. \cite{cai2021flow} harnessed PINNs to infer 3D fields on an espresso cup. In that
	work, the tomographic background-oriented Schlieren (Tomo-BOS) imaging technique measures the density and temperature fields above a cup of espresso. Then, the authors embedded the measurements as well as the NS equations
	into the training process, enabling PINNs to reconstruct the 3D field in terms of velocity and pressure. 
	
	Later, a framework based on PINNs named artificial intelligence velocimetry (AIV), as shown in \Cref{fig:PINNs_field_reconstruction}b,
	was proposed to recover the blood flow in physiology problems \cite{cai2021artificial}. Only 2D velocity
	data, which was measured by snapshots of the platelet-tracking technique, was applied without knowledge
	of the inlet and outlet conditions. The AIV successfully reconstructs various 3D fields of quantities, including
	the velocity, pressure, and even the shear stress at vessel boundaries, as shown in \Cref{fig:PINNs_field_reconstruction}b.
	The same framework also was extended to study the murine perivascular flows \cite{boster2023artificial} and non-Newtonian fluids \cite{mahmoudabadbozchelou2022nn}. By combining
	the NS equations, non-linear rheological model, and sparse observation data, the PINNs successfully reconstructed
	both the velocity and stress fields at not only the in-domain area but also boundaries. Molnar et al.
	\cite{schierholz2023engineering} also integrated background-oriented Schlieren techniques with PINN
	to study the global velocity and pressure fields in supersonic flows. In their framework, the Euler equations
	as well as the irrotational equation are treated as additional knowledge for learning.
	
	Turbulent flow estimation is even more challenging due to its complexity. Integrating the NS equation
	and adjoint-variational data assimilation, PINNs has been demonstrated to be effective for evaluating
	the spatial-temporal evolution of turbulent flow with low dimension \cite{du2023state}.  However, one of the disadvantages of the PINNs for turbulence is that it is hard to assess the neural network structure. The accuracy of PINNs is highly related to networks\textquoteright{}
	structure, while the relation remains to be explored \cite{du2023state}. Zhang et al. \cite{zhang2023physics}
	harnessed PINNs to recover the vorticity fields of tornados with limited measurements. The predicted
	velocity can achieve great accuracies at different stages of tornados, including the single-cell stage,
	vortex breakdown stage, and multi-cell stage. 
	
\begin{figure}
		\begin{centering}
			\includegraphics[scale=0.67]{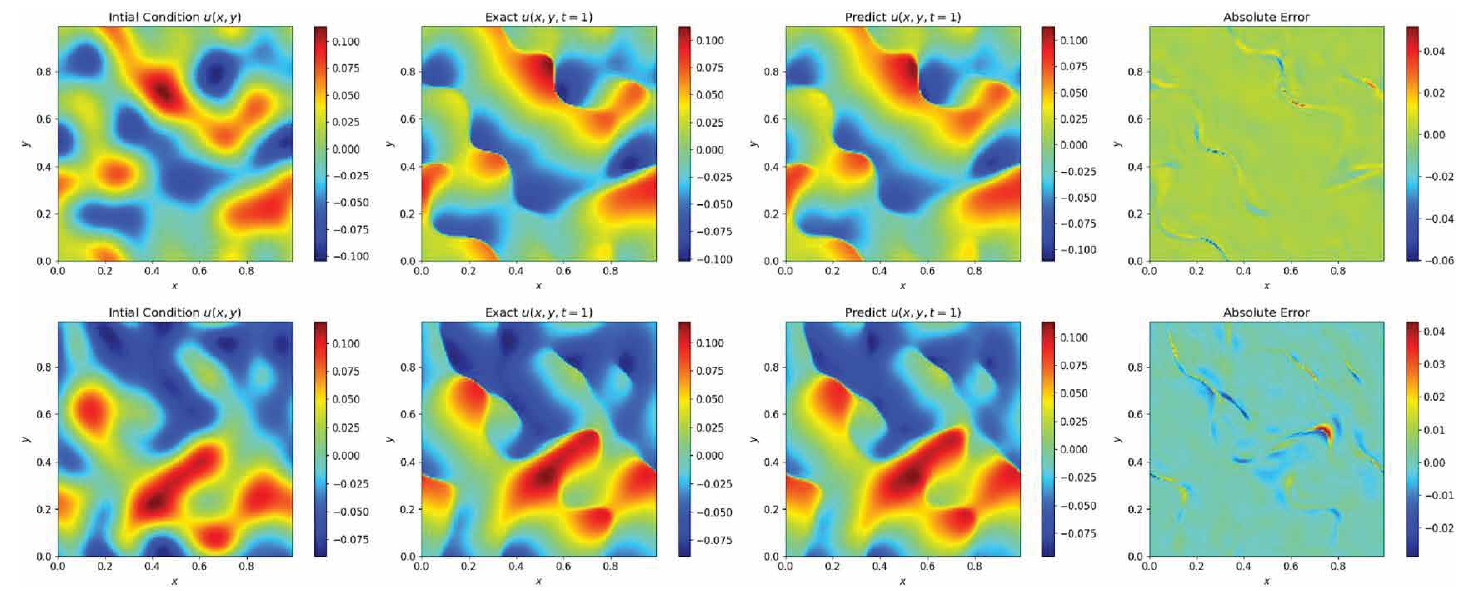}
			\par\end{centering}
		\caption{Applications of FNO and PINO in field reconstruction for 2D inviscid
			Burgers equation \cite{rosofsky2023applications}. \label{fig:PINO_fluid_reconstruction}}
	\end{figure}

	The neural operator is another way to recover the field quantities of fluid flows. Renn et al. \cite{renn2023forecasting}
	leveraged FNO to infer the vortex flow in subcritical cylinder wakes. It has been shown that the FNO
	was able to achieve high accuracy even at a Re number of 3060.
	Meanwhile, the FNO exhibited favorable computational efficiency, potentially enabling faster than 
	real-time modeling and reconstruction. Rosofsky et al. \cite{rosofsky2023applications} offered a wide
	range of inverse fluid benchmarks by using PINO, including wave equations, linear and nonlinear shallow
	water equations, and Burgers equations, as shown in \Cref{fig:PINO_fluid_reconstruction}. Lu et al.
	\cite{lu2022multifidelity} proposed the multifidality DeepONet for the Boltzmann transportation equation.
	It is worth highlighting that, with the fast reconstructed global field quantities and traditional topology
	optimization algorithms, it is straightforward to further guide the design of structures under fluid
	flow conditions for multiple objectives. Li et al. \cite{li2022fourier_eddy} applied FNO to cope with 3D
	turbulent flow and large eddy simulation (LES). In terms of velocity spectrum,
	probability density functions of vorticity, and velocity increments, the results showed that the FNO performed better
	and more efficiently than traditional LES methods. More importantly, FNO can
	be extended to LES and turbulent modeling with high Re numbers.
	
	\subsubsection{Parameter estimation and identification in fluid}
	
	Estimating and identifying parameters of fluids in terms of viscosity, density and other passive scalars
	are also of great importance for fluid mechanics. In general, the unknown parameters can be simultaneously
	tuned along with the field reconstruction. Jagtap et al. \cite{CPINN} elucidated a general form of
	extracting essential parameters from fluid flows. In their work, the governing equations of fluid flows
	were considered to consist of various differential terms and a nonlinear operator
	\begin{equation}
		\begin{aligned}
			\frac{\partial v}{\partial t}=&\mathcal{H}(x,t,v,v_{x},v_{xx},\cdots,v^{2},v^{3},\cdots,v_{x}v^{2},\cdots)\\
			\mathcal{H}=&a_{0}+a_{1}x+a_{2}x^{2}+\cdots+b_{1}v+b_{2}v^{2}+\cdots+c_{1}v_{x}+c_{2}v_{xx}+\cdots+d_{1}v_{x}v^{2}+\cdots,
		\end{aligned}
	\end{equation}
	where $a$, $b$, $c$ and $d$ are unknown parameters that remain to be determined. 
	
\begin{figure}
		\begin{centering}
			\includegraphics[scale=0.7]{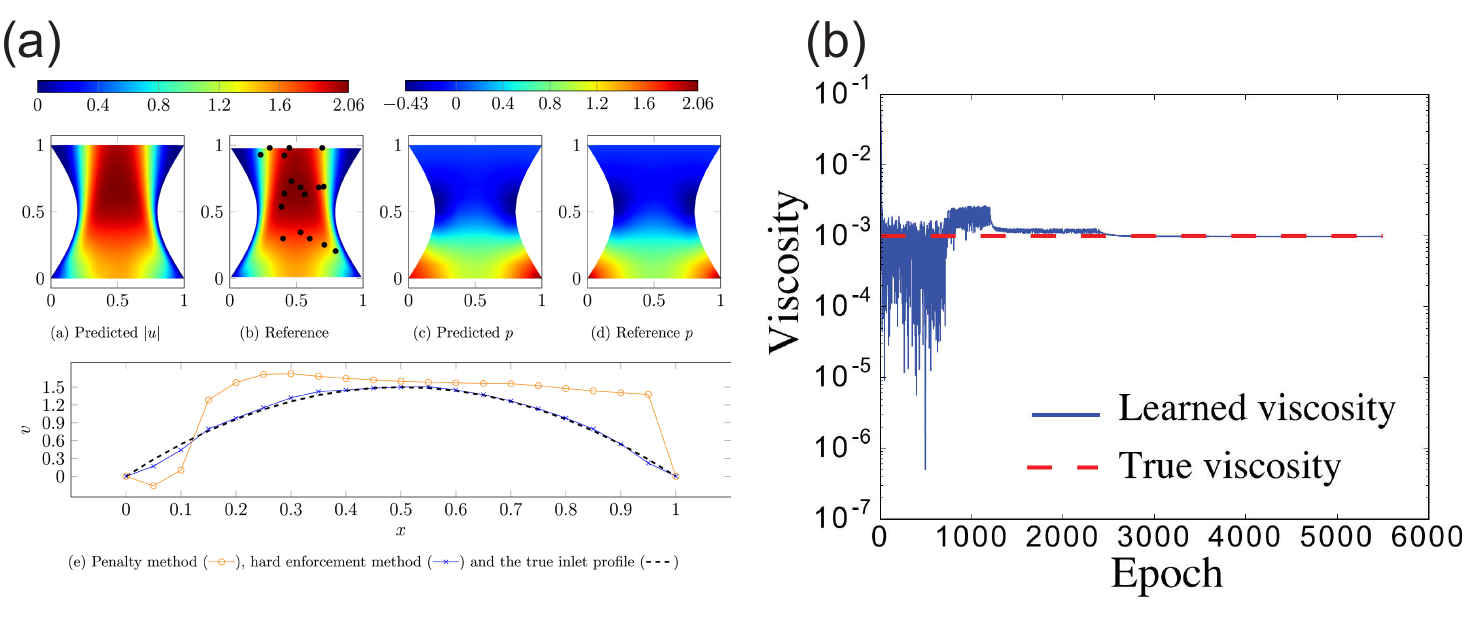}
			\par\end{centering}
		\caption{Applications of PINNs in fluid parameter estimation and identification: (a) Inferring the boundary velocity
			condition of the lid-driven stenosis problem \cite{gao2022physics}. (b) Unveil the viscosity using
			PINNs \cite{arzani2021uncovering}. \label{fig:PINNs_identification_fluid}}
	\end{figure}

	The effectiveness of that PINNs framework was demonstrated by a 1D viscous Burgers equation and a 2D
	inviscid Burgers equation. Yu et al. \cite{yu2022gradient} employed PINNs with a gradient-enhanced
	technique to infer the effective viscosity and permeability in porous media. By numerical examples, the
	proposed PINNs framework provided reliable viscosity as well as permeability with only 5 to 10 sets of
	measured data at different sensor locations. Lou. et al. \cite{lou2021physics} extended this framework
	by the Bhatnagar-Gross-Krook (BGK) collision model. The numerical examples have shown that it is very
	efficient in solving inverse problems, even if the problem is ill-posed. Kou et al. \cite{kou2024physics} utilized PINNs
	to estimate the turbulent viscosity and mass diffusivity for turbulent mass transfer problems.
	Jin et al. \cite{jin2021nsfnets} applied PINNs to infer unknown Re numbers using velocity data and NS equations. More surprisingly,
	the PINNs can be also effective even with missing or noisy boundary conditions. Gao et al. \cite{gao2022physics}
	utilized the graph neural networks to predict the velocity boundary condition of a lid-driven stenosis,
	as shown in \Cref{fig:PINNs_identification_fluid}a. Arzani et al. \cite{arzani2021uncovering} developed
	a PINNs framework that incorporates the NS equations and boundary conditions as soft constraints during
	the training process, enabling the accurate reconstruction of near-wall blood flow from sparse data.
	Moreover, the framework was able to infer the viscosity of fluids, as shown in \Cref{fig:PINNs_identification_fluid}b.
	The proposed framework has potential applications in cardiovascular disease diagnosis and treatment planning.
	
	\subsubsection{Summary}
	
	Field reconstruction and the estimation of the parameters are the major inverse problems in fluid mechanics.
	It has been shown that AI for PDEs is very effective not only for laminar flows, but also for turbulent flows with vorticities.
	
	In summary, the loss functions, which are informed by physics, regulate the reconstructed fields from
	neural networks along with the limited observations. Thus, PINNs offers a novel and effective way to unveil
	the knowledge deeply hidden inside fluid flow, which is considered challenging to measure by conventional
	experimental equipment. As for operator learning, it focuses on learning the mapping between functions
	by available data. Its extraordinary generalization and discretization-invariant characteristics have
	made the operator learning distinguish from other CFD and deep learning-based fluid algorithms. In other
	words, a well-trained operator learning framework for fluid modeling can be generalized to a wide range
	of scenarios (with different boundary conditions, essential parameters, and even problem geometries) as
	well as maintaining accuracy and achieving super-resolutions. Therefore, it can be concluded
	that both PINNs and neural operators have appended the arsenal of computational fluid dynamics (CFD) and serve as additional tools for
	studying fluid phenomena inversely.
	
	\subsection{Biomechanics}
	
	\subsubsection{Modeling Blood Flow}
	
	Accurate modeling of blood flow is essential for understanding and predicting cardiovascular diseases
	such as atherosclerosis and heart failure. Traditional CFD methods require
	high-resolution data, which is often challenging to obtain, especially near vessel walls. AI for PDEs
	can bridge this gap by incorporating governing equations into the learning framework,
	thus enabling accurate predictions from limited data. 
	
\begin{figure}
		\begin{centering}
			\includegraphics[scale=0.5]{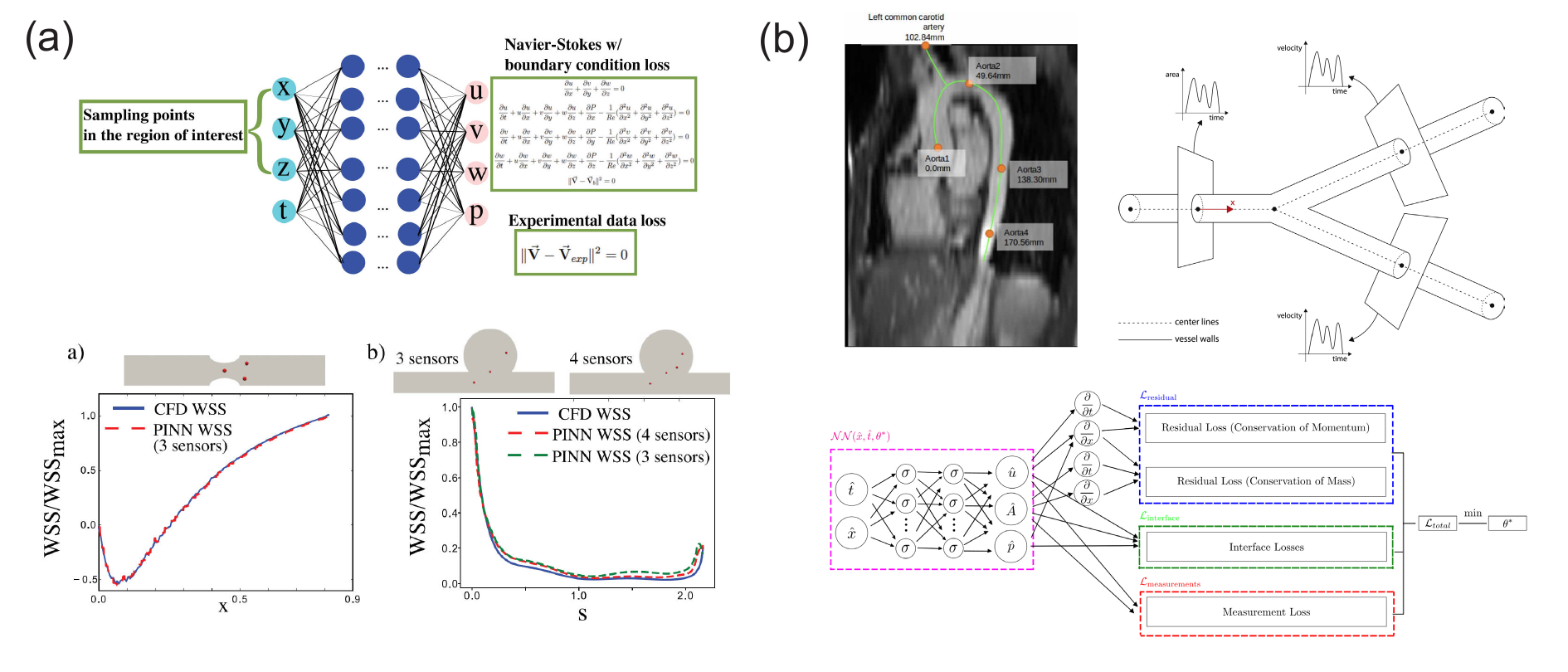}
			\par\end{centering}
		\caption{Applications of PINNs in modeling blood flow: (a) PINNs could be used for solving Navier\textendash Stokes
			equations. A few sensors (red dots) are used in the 2D stenosis model and aneurysm model, and the results
			predicted by PINNs are compared with the reference CFD data \cite{arzani2021uncovering}. (b) Flow through
			the aorta/carotid bifurcation of a healthy human subject \cite{kissas2020machine}. \label{fig:PINNs_blood_flow_inverse}}
	\end{figure}

	Near-wall blood flow is particularly crucial for understanding cardiovascular diseases and designing
	effective treatments. Due to the limitations of medical imaging techniques, obtaining high-resolution
	data near vessel walls is challenging. Arzani et al. \cite{arzani2021uncovering} developed a PINNs
	framework that incorporates the Navier-Stokes equations and boundary conditions as soft constraints during
	the training process, enabling the accurate reconstruction of near-wall blood flow from sparse data,
	as shown in \Cref{fig:PINNs_blood_flow_inverse}a. This PINNs framework accounts for uncertainties in
	the data and enables accurate reconstruction of near-wall blood flow from sparse data. Besides, the coordinate
	transformation technique that maps the irregular physical domain to a rectangular computational domain
	is applied, which can handle complex vessel geometries. This represents a significant advancement over
	traditional CFD methods, which require high-resolution data for accurate simulations. The proposed method
	has potential applications in cardiovascular disease diagnosis and treatment planning.
	
	Accurate prediction of arterial blood pressure is vital for cardiovascular disease diagnosis and treatment
	planning. Traditional methods, such as invasive catheterization or simplified computational models, often
	lack patient specificity. Kissas et al. \cite{kissas2020machine} addressed these limitations by leveraging
	non-invasive 4D flow Magnetic Resonance Imaging (MRI) data and physics-informed machine learning techniques,
	as shown in \Cref{fig:PINNs_blood_flow_inverse}b.. Their PINNs framework incorporates the governing
	equations of fluid dynamics (Navier-Stokes equations) and the arterial Windkessel model into the neural
	network architecture. In addition to predicting blood pressure, PINNs has been used for the super-resolution
	and denoising of 4D flow MRI data, enhancing the quality and utility of MRI data in clinical applications.
	These advanced techniques ensure more accurate and reliable measurements, further supporting cardiovascular
	disease diagnosis and treatment planning \cite{fathi2020super}.
	
	\subsubsection{Material parameter identification in soft tissue}
	
\begin{figure}
		\begin{centering}
			\includegraphics[scale=0.5]{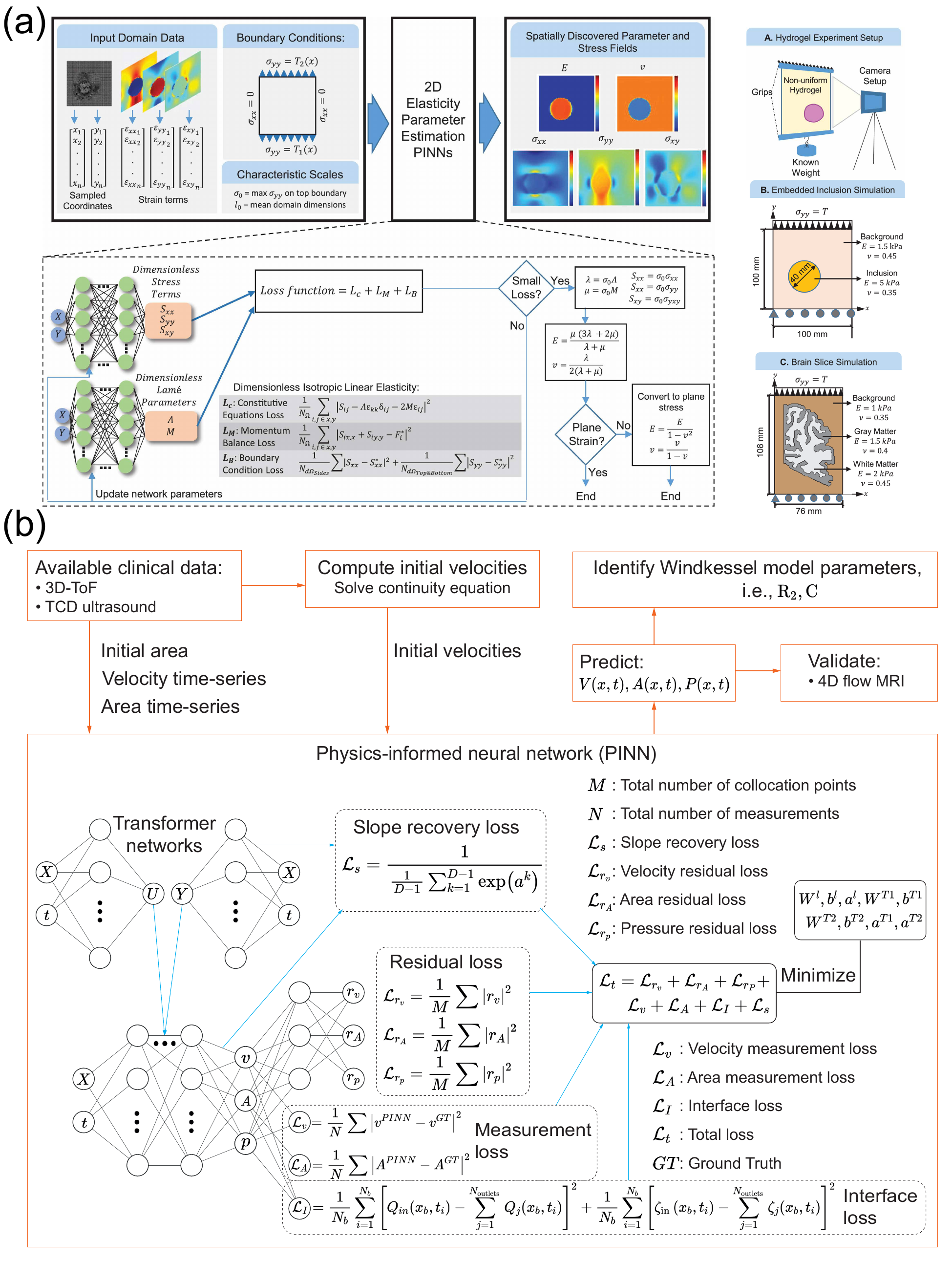}
			\par\end{centering}
		\caption{Applications of PINNs in material parameters identification in soft issue \label{fig:PINN_material_parameter_identification_in_soft_tissue}. (a) Parameter
			estimation PINNs for 2D elasticity and results for the discovery of material parameter and stress distributions
			for the brain \cite{kamali2023elasticity}. (b) Overview of the procedure of brain hemodynamics
			\cite{sarabian2022physics}. }
	\end{figure}

	Elasticity imaging aims to uncover the spatial distribution of mechanical properties such as the elastic
	modulus and Poisson's ratio within biological tissues. This information is critical for various applications,
	including tumor detection and characterization. Traditional methods often rely on simplifying assumptions,
	such as homogeneous distributions for one material parameter, which can result in inaccuracies. Kamali
	et al. \cite{kamali2023elasticity} leverage PINNs to solve linear elasticity problems and discover
	space-dependent distributions of both Young's modulus and Poisson's ratio using strain data, normal stress boundary conditions,
	and the governing physics equations, as shown in \Cref{fig:PINN_material_parameter_identification_in_soft_tissue}a.
	Unlike conventional methods that estimate one material parameter while assuming the other is homogeneous,
	the framework based on PINNs simultaneously estimates the spatial distributions of both elastic modulus
	and Poisson's ratio. This approach was successfully demonstrated on a simulated hydrogel sample containing
	a human brain slice with distinct gray matter and white matter regions, accurately capturing the spatial
	distribution of mechanical properties and tissue interfaces \cite{kamali2023elasticity}. The application
	of PINNs in nonhomogeneous soft tissue identification allows for the precise determination of mechanical
	properties based on full-field displacement measurements under quasi-static loading conditions \cite{zhang2020physics}.
	The PINNs also provides a tool for cardiac electrophysiology characterization and parameter estimation
	from sparse data \cite{herrero2022ep}.
	
	Brain hemodynamics, which involves quantifying blood flow velocity, vessel cross-sectional area, and
	pressure, is crucial for diagnosing and treating cerebrovascular diseases. Transcranial Doppler (TCD)
	ultrasound is a common clinical technique for non-invasively measuring blood flow velocity within cerebral
	arteries. However, TCD is spatially limited due to constrained accessibility through the skull's acoustic
	windows, providing measurements only at select locations across the cerebrovasculature. To address these
	limitations, Sarabian et al. \cite{sarabian2022physics} propose a novel physics-informed deep learning
	framework that integrates real-time TCD velocity measurements with baseline vessel cross-sectional areas
	obtained from 3D angiography images, as shown in \Cref{fig:PINN_material_parameter_identification_in_soft_tissue}b.
	This framework utilizes a physics-informed deep learning model to generate high-resolution maps of velocity,
	area, and pressure throughout the entire brain vasculature. By augmenting sparse clinical measurements
	with one-dimensional (1D) Reduced-Order Model (ROM) simulations, the model ensures that the predicted
	hemodynamic parameters are physically consistent and adhere to the governing equations of fluid dynamics.
	This approach enables the estimation and quantification of subject-specific cerebral hemodynamic variables
	with high accuracy, even without detailed knowledge of inlet and outlet boundary conditions \cite{wu2024machine}.
	
	\subsubsection{Summary}
	
	While AI for PDEs represents significant advancements in the field of biomechanics, offering the potential
	for more accurate and efficient modeling of complex biological systems, several challenges remain. Addressing
	these challenges\textemdash data acquisition, interpretability, validation, model complexity, and generalizability\textemdash will
	be essential for the broader adoption and successful application of AI for PDEs in biomechanics and clinical
	practice. Future research should focus on developing strategies to overcome these barriers, such as improved
	data collection methods, enhanced interpretability techniques, robust validation protocols, simplified
	model implementation processes, and approaches to increase model generalizability. Through these efforts,
	the full potential of AI for PDEs in advancing our understanding and treatment of biomechanical phenomena
	can be realized.

	\section{Conclusion and outlook\label{sec:Conclusion}}
	
	This paper mainly reviews the algorithms of AI for PDEs (including Physics-Informed Neural Networks,
	Operator Learning, and Physics-Informed Neural Operators), the related theoretical research, and applications
	in the forward and inverse problems of computational mechanics, including solid mechanics, fluid mechanics, and biomechanics. Based on the current state of research,
	possible future directions for AI for PDEs in computational mechanics might include:
	
	(1) \textbf{Nonlinear Problems}: The core component of AI for PDEs is neural networks, which have strong
	nonlinear capabilities. Therefore, future research on nonlinear problems theoretically has good prospects.
	For linear elasticity problems in solid mechanics, finite element methods are already quite perfect due
	to the positive definiteness and sparsity of the stiffness matrix, which can be quickly and accurately
	solved by direct matrix inversion. However, for some nonlinear problems, traditional finite element methods
	either solve the nonlinear equation system directly using Newton's iterative method or transform and
	solve it explicitly using incremental steps. Neural networks, due to their inherent nonlinear capabilities,
	theoretically have an advantage in solving nonlinear problems in computational mechanics. Also, by training
	operator neural networks with existing numerical simulations or experimental results, and then using
	operator learning to provide a good initial solution for the initial iteration vector of the nonlinear
	equation system, the computational efficiency could be greatly improved theoretically. Hyperelastic problems,
	for example, are a good entry point because they are path-independent and can be directly formulated
	as a nonlinear equation system.
	
	(2) \textbf{Complex Phenomena with Inadequate Understanding}: For such problems, due to their complexity,
	the mathematical PDEs descriptions of these issues can only be approximations, meaning the simulation
	results still differ from actual experimental results. In this case, we can rely on data to fine-tune
	the results. That is, an approximate solution is first provided by the boundary conditions and an approximate
	physical equation, and then the simulation results are fine-tuned according to the experimental data,
	blending a small amount of data with an approximate physical equation, especially for simulating complex
	phenomena. As humanity's understanding of the phenomenon becomes clearer, only the physical equations
	need to be corrected, and this framework remains unchanged.
	
	(3) \textbf{Constitutive Equations}: Constitutive equations have always been a core issue in mechanics,
	where most of the work involves fitting. Typically, experts first construct a specific form of the constitutive
	model based on some basic physical principles and then fit the parameters in the model according to experimental
	stress-strain points. However, because of the fitting characteristics of neural networks, theoretically,
	they can replace constitutive equations.
	Most current research uses neural networks to directly fit the relationship between strain and stress. However, future studies could integrate fundamental principles of constitutive behavior, such as Noll's three theorems and thermodynamic postulates like Drucker's postulate in plasticity, into the neural network framework for constitutive equations.
	This would improve the convergence speed of the constitutive model and reduce the experimental data needed
	for stress and strain. Currently, most constitutive work in solid mechanics focuses on elasticity and
	hyperelasticity, but future explorations could include rate-dependent viscoelasticity and history-dependent
	elastoplasticity.
	
	(4) \textbf{Foundation Models in Computational Mechanics}: Based on current research on PINNs and operator
	learning, it is very likely that the foundation model in computational mechanics will emerge in the future.
	Although current PINNs and traditional finite element methods still have significant gaps in accuracy
	and efficiency, PINNs may ultimately teach us how to incorporate physical equations into neural networks.
	Recent studies have found that operator learning can significantly speed up the simulation of forward
	problems, and there is theoretical evidence that operator learning can learn the families behind PDEs.
	Therefore, combining operator learning with algorithms for solving PDEs (algorithms for solving PDEs are not limited to PINNs, as finite element methods are also applicable) has a very promising
	academic and industrial future. Not only could it significantly increase the simulation speed for forward
	problems (accelerating by tens of thousands of times), but it could also use experimental data to simulate
	complex phenomena and integrate approximate constitutive equations to form a whole new set of foundation
	models in computational mechanics. These models would not only act as ultra-fast solvers for forward problems, but more importantly, they would serve as powerful engines for scientific discovery. By enabling real-time inverse analysis and learning from the interplay of data and physics, they could unveil new constitutive laws, identify hidden parameters in complex systems and accelerate the design of novel materials and structures. Such a foundation model would be the ultimate computational engine for a digital twin, a living, virtual replica of a physical system, capable of integrating real-time sensor data to provide instantaneous predictions and insights, thereby revolutionizing predictive maintenance, design optimization, and operational control.
	In the past, expert systems were developed, but their performance was often lacking. However, with the advent of neural networks, there is renewed potential to revitalize expert systems. AI for PDEs soon play a crucial role in determining which models, physical laws, discretizations, and other factors are best suited for a given problem.
	Moreover, AI possesses various capabilities that can enhance and improve traditional techniques, solvers, and mesh generators. The recent achievement of AI earning a silver medal in the International Mathematical Olympiad (IMO) demonstrates that AI can perform complex mathematical reasoning \cite{AI4math}. These developments suggest that AI for PDEs holds significant promise in computational mechanics, potentially leading to the emergence of foundation models in this field.

	At the same time, comparisons between AI for PDE and classical numerical methods such as finite element methods (FEM) should be interpreted with caution. Reported advantages in inference speed are often meaningful only when substantial offline training cost is amortized over many repeated-query scenarios, which is particularly relevant for operator-learning-based approaches. For single-instance engineering simulations, classical FEM often remains highly competitive in robustness, interpretability, and reliability.
	Similarly, accuracy comparisons are only fair when discretization level, computational budget, and geometric generalization range are matched consistently. Recent studies have shown that weak numerical baselines or incomplete reporting of training cost may lead to overly optimistic conclusions regarding machine-learning superiority \cite{mcgreivy2024weak}. Therefore, future development of AI for PDEs should place increasing emphasis on fair benchmarking and out-of-distribution generalization.

	Scientific Artificial Intelligence (AI4Science) has garnered significant attention,
	similar to the work on neural networks in computer vision and language over the past decade. Its growth could mirror the rapid
	development seen in the Cambrian explosion of species. Initially, deep learning showed potential in these
	traditional scientific fields, but over time, its limitations became apparent. To overcome these shortcomings,
	traditional domain knowledge has been integrated to enhance AI's effectiveness in these scientific areas.
	AI for PDEs is a prime example of such integration.
	
	PDEs reflect a profound and quantified expression of human understanding of the physical world. Thus,
	PDEs constitute a form of knowledge that can be integrated into neural network training. The core idea
	behind AI for PDEs is the integration of data with physical equations, aimed at scenarios with infinite
	data, where the physical laws derived from human understanding of nature would become unnecessary. However,
	in the current era where data is not infinite, integrating physical equations into data-driven approaches
	provides a useful transition and balancing point. This integration involves using previously calculated
	results for operator learning to pre-train the system, then providing a good initial solution for specific
	problems, and finally fine-tuning with physical equations. Most importantly, the combination of data
	and physical laws enables a self-learning algorithm that becomes faster with more accurate data. All past
	computed results are used to train operator learning, improving the accuracy of operator learning and
	thus reducing the iterations needed for fine-tuning by physical equations. Therefore, AI for PDEs possesses a self-updating
	iterative capability, promising a significant and forward-looking direction in computational mechanics.
	
	AI for PDEs represents the future of computational mechanics. In the past, the emergence of computers brought about computational mechanics. At present, artificial intelligence is actually the previous computer, so artificial intelligence will have a new manifestation in computational mechanics, that is, AI for PDEs.
	
	\section*{Declaration of competing interest}
	The authors declare that they have no known competing financial interests or personal relationships that could
	have appeared to influence the work reported in this paper.
	
	\section*{Acknowledgement}
	The study was supported by the Key Project of the National Natural Science Foundation of China (12332005) and scholarship from Bauhaus University in Weimar. Thank Zhongkai Hao for the helpful discussion.
	
	\section*{Author Contributions Statement}
	Yizheng Wang wrote the methodology, theoretical research, and the application of AI4PDEs to solid mechanics. 
	Jinshuai Bai wrote the application of AI4PDEs to fluid mechanics. 
	Zhongya Lin wrote the application of AI4PDEs to biomechanics.
	Qimin Wang applied the permissions of pictures and drew some pictures.
	Cosmin Anitescu reviewed the paper.  
	Mohammad Sadegh Eshaghi reviewed the paper. 
	Yuantong Gu supervised the project.
	Xiqiao Feng supervised the project.
	Xiaoying Zhuang supervised the project.
	Timon Rabczuk reviewed the paper and supervised the project. Yinghua Liu supervised the project.
	
	\bibliographystyle{elsarticle-num}
	\addcontentsline{toc}{section}{\refname}\bibliography{AI4PDEs_CM.bib}

@article{loss_is_minimum_potential_energy,
  title={An energy approach to the solution of partial differential equations in computational mechanics via machine learning: Concepts, implementation and applications},
  author={Samaniego, Esteban and Anitescu, Cosmin and Goswami, Somdatta and Nguyen-Thanh, Vien Minh and Guo, Hongwei and Hamdia, Khader and Zhuang, X and Rabczuk, T},
  journal={Computer Methods in Applied Mechanics and Engineering},
  volume={362},
  pages={112790},
  year={2020},
  publisher={Elsevier}
}

@article{deep_ritz,
  title={The deep Ritz method: a deep learning-based numerical algorithm for solving variational problems},
  author={Yu, Bing and others},
  journal={Communications in Mathematics and Statistics},
  volume={6},
  number={1},
  pages={1--12},
  year={2018},
  publisher={Springer}
}

@article{PINN_hyperelasticity,
  title={A deep energy method for finite deformation hyperelasticity},
  author={Nguyen-Thanh, Vien Minh and Zhuang, Xiaoying and Rabczuk, Timon},
  journal={European Journal of Mechanics-A/Solids},
  volume={80},
  pages={103874},
  year={2020},
  publisher={Elsevier}
}

@article{PINN_original_paper,
  title={Physics-informed neural networks: A deep learning framework for solving forward and inverse problems involving nonlinear partial differential equations},
  author={Raissi, Maziar and Perdikaris, Paris and Karniadakis, George E},
  journal={Journal of Computational Physics},
  volume={378},
  pages={686--707},
  year={2019},
  publisher={Elsevier}
}

@article{ill_gradient,
   author = {Wang, Sifan and Teng, Yujun and Perdikaris, Paris %J SIAM Journal on Scientific Computing},
	title = {Understanding and mitigating gradient flow pathologies in physics-informed neural networks},
	journal = {SIAM Journal on Scientific Computing},
	volume = {43},
	number = {5},
	pages = {A3055-A3081},
	ISSN = {1064-8275},
	year = {2021},
	type = {Journal Article}
}

@article{hp-VPINN,
  title={hp-VPINNs: Variational physics-informed neural networks with domain decomposition},
  author={Kharazmi, Ehsan and Zhang, Zhongqiang and Karniadakis, George Em},
  journal={Computer Methods in Applied Mechanics and Engineering},
  volume={374},
  pages={113547},
  year={2021},
  publisher={Elsevier}
}

@article{kharazmi2019variational,
  title={Variational physics-informed neural networks for solving partial differential equations},
  author={Kharazmi, Ehsan and Zhang, Zhongqiang and Karniadakis, George Em},
  journal={arXiv preprint arXiv:1912.00873},
  year={2019}
}

@article{CPINN,
  title={Conservative physics-informed neural networks on discrete domains for conservation laws: Applications to forward and inverse problems},
  author={Jagtap, Ameya D and Kharazmi, Ehsan and Karniadakis, George Em},
  journal={Computer Methods in Applied Mechanics and Engineering},
  volume={365},
  pages={113028},
  year={2020},
  publisher={Elsevier}
}

@article{complex_PINN_a_method_to_construct_admissible_function,
  title={A unified deep artificial neural network approach to partial differential equations in complex geometries},
  author={Berg, Jens and Nystr{\"o}m, Kaj},
  journal={Neurocomputing},
  volume={317},
  pages={28--41},
  year={2018},
  publisher={Elsevier}
}

@article{high_dimension,
  title={Solving high-dimensional partial differential equations using deep learning},
  author={Han, Jiequn and Jentzen, Arnulf and Weinan, E},
  journal={Proceedings of the National Academy of Sciences},
  volume={115},
  number={34},
  pages={8505--8510},
  year={2018},
  publisher={National Acad Sciences}
}

@article{PINN_solid_mechanics,
   author = {Haghighat, Ehsan and Raissi, Maziar and Moure, Adrian and Gomez, Hector and Juanes, Ruben},
   title = {A physics-informed deep learning framework for inversion and surrogate modeling in solid mechanics},
   journal = {Computer Methods in Applied Mechanics and Engineering},
   volume = {379},
   pages = {113741},
   ISSN = {0045-7825},
   DOI = {10.1016/j.cma.2021.113741},
   year = {2021},
   type = {Journal Article}
}

@article{the_comparision_of_strong_and_energy_form,
   author = {Li, Wei and Bazant, Martin Z. and Zhu, Juner},
   title = {A physics-guided neural network framework for elastic plates: Comparison of governing equations-based and energy-based approaches},
   journal = {Computer Methods in Applied Mechanics and Engineering},
   volume = {383},
   pages = {113933},
   ISSN = {0045-7825},
   DOI = {10.1016/j.cma.2021.113933},
   year = {2021},
   type = {Journal Article}
}

@article{use_neural_network_to_solve_PDE,
  title={Neural algorithm for solving differential equations},
author={Lee, Hyuk and Kang, In Seok},
journal={Journal of Computational Physics},
volume={91},
number={1},
pages={110--131},
year={1990},
publisher={Elsevier}
}

@article{PINN_review,
   author = {Karniadakis, George Em and Kevrekidis, Ioannis G. and Lu, Lu and Perdikaris, Paris and Wang, Sifan and Yang, Liu},
   title = {Physics-informed machine learning},
   journal = {Nature Reviews Physics},
   volume = {3},
   number = {6},
   pages = {422-440},
   ISSN = {2522-5820},
   DOI = {10.1038/s42254-021-00314-5},
   year = {2021},
   type = {Journal Article}
}

@article{admissible_in_PINN_energy_form,
   author = {Sheng, Hailong and Yang, Chao},
   title = {PFNN: A penalty-free neural network method for solving a class of second-order boundary-value problems on complex geometries},
   journal = {Journal of Computational Physics},
   volume = {428},
   pages = {110085},
   ISSN = {0021-9991},
   year = {2021},
   type = {Journal Article}
}

@article{PINN_energy_form_to_solve_C0_without_subdomains, 
   author = {Wang, Zhongjian and Zhang, Zhiwen},
   title = {A mesh-free method for interface problems using the deep learning approach},
   journal = {Journal of Computational Physics},
   volume = {400},
   pages = {108963},
   ISSN = {0021-9991},
   DOI = {10.1016/j.jcp.2019.108963},
   year = {2020},
   type = {Journal Article}
}

@article{NTK_PINN,
   author = {Wang, Sifan and Wang, Hanwen and Perdikaris, Paris},
   title = {On the eigenvector bias of Fourier feature networks: From regression to solving multi-scale PDEs with physics-informed neural networks},
   journal = {Computer Methods in Applied Mechanics and Engineering},
   volume = {384},
   pages = {113938},
   ISSN = {0045-7825},
   DOI = {10.1016/j.cma.2021.113938},
   year = {2021},
   type = {Journal Article}
}

@article{super_approximation,
  title={Approximation by superpositions of a sigmoidal function},
  author={Cybenko, George},
  journal={Mathematics of control, signals and systems},
  volume={2},
  number={4},
  pages={303--314},
  year={1989},
  publisher={Springer}
}

@article{NTK_to_get_hyperparameter_of_PINN,
  title={When and why PINNs fail to train: A neural tangent kernel perspective},
author={Wang, Sifan and Yu, Xinling and Perdikaris, Paris},
journal={Journal of Computational Physics},
volume={449},
pages={110768},
year={2022},
publisher={Elsevier}
}

@book{finite_element_book,
   author = {Zienkiewicz, Olek C and Taylor, Robert L and Zhu, Jian Z},
   title = {The finite element method: its basis and fundamentals},
   publisher = {Elsevier},
   ISBN = {008047277X},
   year = {2005},
   type = {Book}
}

@book{goodfellow2016deep,
  title={Deep learning},
  author={Goodfellow, Ian and Bengio, Yoshua and Courville, Aaron},
  year={2016},
  publisher={MIT press}
}

@article{Adaptive_activation_functions,
   author = {Jagtap, Ameya D. and Kawaguchi, Kenji and Karniadakis, George Em},
   title = {Adaptive activation functions accelerate convergence in deep and physics-informed neural networks},
   journal = {Journal of Computational Physics},
   volume = {404},
   pages = {109136},
   ISSN = {0021-9991},
   DOI = {10.1016/j.jcp.2019.109136},
   year = {2020},
   type = {Journal Article}
}

@article{hyper_constitutive_data_driven,
  title={Unsupervised discovery of interpretable hyperelastic constitutive laws},
  author={Flaschel, Moritz and Kumar, Siddhant and De Lorenzis, Laura},
  journal={Computer Methods in Applied Mechanics and Engineering},
  volume={381},
  pages={113852},
  year={2021},
  publisher={Elsevier}
}

@article{PINNfiuld,
  title={Physics-informed neural networks for high-speed flows},
  author={Mao, Zhiping and Jagtap, Ameya D and Karniadakis, George Em},
  journal={Computer Methods in Applied Mechanics and Engineering},
  volume={360},
  pages={112789},
  year={2020},
  publisher={Elsevier}
}

@article{paradeepenergy,
   author = {Nguyen-Thanh, Vien Minh and Anitescu, Cosmin and Alajlan, Naif and Rabczuk, Timon and Zhuang, Xiaoying},
   title = {Parametric deep energy approach for elasticity accounting for strain gradient effects},
   journal = {Computer Methods in Applied Mechanics and Engineering},
   volume = {386},
   pages = {114096},
   ISSN = {0045-7825},
   DOI = {10.1016/j.cma.2021.114096},
   year = {2021},
   type = {Journal Article}
}

@article{PINNlibrary,
   author = {Lu, Lu and Meng, Xuhui and Mao, Zhiping and Karniadakis, George Em},
   title = {DeepXDE: A Deep Learning Library for Solving Differential Equations},
   journal = {SIAM Review},
   volume = {63},
   number = {1},
   pages = {208-228},
   ISSN = {0036-1445},
   DOI = {10.1137/19m1274067},
   year = {2021},
   type = {Journal Article}
}

@article{DeepOnet,
   author = {Lu, Lu and Jin, Pengzhan and Pang, Guofei and Zhang, Zhongqiang and Karniadakis, George Em},
   title = {Learning nonlinear operators via DeepONet based on the universal approximation theorem of operators},
   journal = {Nature Machine Intelligence},
   volume = {3},
   number = {3},
   pages = {218-229},
   ISSN = {2522-5839},
   DOI = {10.1038/s42256-021-00302-5},
   year = {2021},
   type = {Journal Article}
}

@article{PINNstrong_form_in_elastodynamics,
	title={Physics-informed deep learning for computational elastodynamics without labeled data},
	author={Rao, Chengping and Sun, Hao and Liu, Yang},
	journal={Journal of Engineering Mechanics},
	volume={147},
	number={8},
	pages={04021043},
	year={2021},
	publisher={American Society of Civil Engineers}
}

@article{admissible_earliest_paper,
	title={Artificial neural networks for solving ordinary and partial differential equations},
	author={Lagaris, Isaac E and Likas, Aristidis and Fotiadis, Dimitrios I},
	journal={IEEE transactions on neural networks},
	volume={9},
	number={5},
	pages={987--1000},
	year={1998},
	publisher={IEEE}
}

@article{boundary_conditions_distance_functions,
	author = {Sukumar, N. and Srivastava, Ankit},
	title = {Exact imposition of boundary conditions with distance functions in physics-informed deep neural networks},
	journal = {Computer Methods in Applied Mechanics and Engineering},
	volume = {389},
	pages = {114333},
	year = {2022},
	type = {Journal Article}
}

@article{LAAF,
	author = {Jagtap, Ameya D. and Kawaguchi, Kenji and Em Karniadakis, George},
	title = {Locally adaptive activation functions with slope recovery for deep and physics-informed neural networks},
	journal = {Proceedings of the Royal Society A: Mathematical, Physical and Engineering Sciences},
	volume = {476},
	number = {2239},
	pages = {20200334},
	ISSN = {1364-5021},
	DOI = {10.1098/rspa.2020.0334},
	year = {2020},
	type = {Journal Article}
}

@article{XPINN,
	title={Extended physics-informed neural networks (xpinns): A generalized space-time domain decomposition based deep learning framework for nonlinear partial differential equations},
	author={Jagtap, Ameya D and Karniadakis, George Em},
	journal={Communications in Computational Physics},
	volume={28},
	number={5},
	pages={2002--2041},
	year={2020}
}

@article{XPINN_parallel,
	author = {Shukla, Khemraj and Jagtap, Ameya D. and Karniadakis, George Em},
	title = {Parallel physics-informed neural networks via domain decomposition},
	journal = {Journal of Computational Physics},
	volume = {447},
	pages = {110683},
	ISSN = {0021-9991},
	DOI = {10.1016/j.jcp.2021.110683},
	year = {2021},
	type = {Journal Article}
}

@article{wang2022cenn,
  title={CENN: Conservative energy method based on neural networks with subdomains for solving variational problems involving heterogeneous and complex geometries},
  author={Wang, Yizheng and Sun, Jia and Li, Wei and Lu, Zaiyuan and Liu, Yinghua},
  journal={Computer Methods in Applied Mechanics and Engineering},
  volume={400},
  pages={115491},
  year={2022},
  publisher={Elsevier}
}

@article{li2019predicting,
  title={Predicting the effective mechanical property of heterogeneous materials by image based modeling and deep learning},
  author={Li, Xiang and Liu, Zhanli and Cui, Shaoqing and Luo, Chengcheng and Li, Chenfeng and Zhuang, Zhuo},
  journal={Computer Methods in Applied Mechanics and Engineering},
  volume={347},
  pages={735--753},
  year={2019},
  publisher={Elsevier}
}

@article{li2022ref,
  title={ReF-nets: Physics-informed neural network for Reynolds equation of gas bearing},
  author={Li, Liangliang and Li, Yunzhu and Du, Qiuwan and Liu, Tianyuan and Xie, Yonghui},
  journal={Computer Methods in Applied Mechanics and Engineering},
  volume={391},
  pages={114524},
  year={2022},
  publisher={Elsevier}
}

@article{bc-pinn,
  title={A novel sequential method to train physics informed neural networks for Allen Cahn and Cahn Hilliard equations},
  author={Mattey, Revanth and Ghosh, Susanta},
  journal={Computer Methods in Applied Mechanics and Engineering},
  volume={390},
  pages={114474},
  year={2022},
  publisher={Elsevier}
}

@article{lu2021physics,
  title={Physics-informed neural networks with hard constraints for inverse design},
  author={Lu, Lu and Pestourie, Raphael and Yao, Wenjie and Wang, Zhicheng and Verdugo, Francesc and Johnson, Steven G},
  journal={SIAM Journal on Scientific Computing},
  volume={43},
  number={6},
  pages={B1105--B1132},
  year={2021},
  publisher={SIAM}
}

@book{darwish2016finitevolumemethod,
  title={The finite volume method in computational fluid dynamics: an advanced introduction with OpenFOAM{\textregistered} and Matlab{\textregistered}},
  author={Darwish, Marwan and Moukalled, Fadl},
  year={2016},
  publisher={Springer}
}

@book{leveque2007finitedifferentialmethod,
  title={Finite difference methods for ordinary and partial differential equations: steady-state and time-dependent problems},
  author={LeVeque, Randall J},
  year={2007},
  publisher={SIAM}
}

@book{zhang2016material,
  title={The material point method: a continuum-based particle method for extreme loading cases},
  author={Zhang, Xiong and Chen, Zhen and Liu, Yan},
  year={2016},
  publisher={Academic Press}
}

@article{LSTM,
  title={Long short-term memory},
  author={Hochreiter, Sepp and Schmidhuber, J{\"u}rgen},
  journal={Neural computation},
  volume={9},
  number={8},
  pages={1735--1780},
  year={1997},
  publisher={MIT Press}
}

@article{lu2022comprehensive,
  title={A comprehensive and fair comparison of two neural operators (with practical extensions) based on fair data},
  author={Lu, Lu and Meng, Xuhui and Cai, Shengze and Mao, Zhiping and Goswami, Somdatta and Zhang, Zhongqiang and Karniadakis, George Em},
  journal={Computer Methods in Applied Mechanics and Engineering},
  volume={393},
  pages={114778},
  year={2022},
  publisher={Elsevier}
}

@article{chen1995universal,
  title={Universal approximation to nonlinear operators by neural networks with arbitrary activation functions and its application to dynamical systems},
  author={Chen, Tianping and Chen, Hong},
  journal={IEEE Transactions on Neural Networks},
  volume={6},
  number={4},
  pages={911--917},
  year={1995},
  publisher={IEEE}
}

@article{wen2022u,
  title={U-FNO—An enhanced Fourier neural operator-based deep-learning model for multiphase flow},
  author={Wen, Gege and Li, Zongyi and Azizzadenesheli, Kamyar and Anandkumar, Anima and Benson, Sally M},
  journal={Advances in Water Resources},
  volume={163},
  pages={104180},
  year={2022},
  publisher={Elsevier}
}

@article{li2020fourier,
  title={Fourier neural operator for parametric partial differential equations},
  author={Li, Zongyi and Kovachki, Nikola and Azizzadenesheli, Kamyar and Liu, Burigede and Bhattacharya, Kaushik and Stuart, Andrew and Anandkumar, Anima},
  journal={arXiv preprint arXiv:2010.08895},
  year={2020}
}

@article{wang2021learning,
  title={Learning the solution operator of parametric partial differential equations with physics-informed DeepONets},
  author={Wang, Sifan and Wang, Hanwen and Perdikaris, Paris},
  journal={Science advances},
  volume={7},
  number={40},
  pages={eabi8605},
  year={2021},
  publisher={American Association for the Advancement of Science}
}

@article{goswami2022physics,
  title={A physics-informed variational DeepONet for predicting crack path in quasi-brittle materials},
  author={Goswami, Somdatta and Yin, Minglang and Yu, Yue and Karniadakis, George Em},
  journal={Computer Methods in Applied Mechanics and Engineering},
  volume={391},
  pages={114587},
  year={2022},
  publisher={Elsevier}
}

@article{hornik1989multilayer,
  title={Multilayer feedforward networks are universal approximators},
  author={Hornik, Kurt and Stinchcombe, Maxwell and White, Halbert},
  journal={Neural networks},
  volume={2},
  number={5},
  pages={359--366},
  year={1989},
  publisher={Elsevier}
}

@article{li2022equilibrium,
  title={Equilibrium-based convolution neural networks for constitutive modeling of hyperelastic materials},
  author={Li, LF and Chen, CQ},
  journal={Journal of the Mechanics and Physics of Solids},
  volume={164},
  pages={104931},
  year={2022},
  publisher={Elsevier}
}

@article{cai2021physics,
  title={Physics-informed neural networks (PINNs) for fluid mechanics: A review},
  author={Cai, Shengze and Mao, Zhiping and Wang, Zhicheng and Yin, Minglang and Karniadakis, George Em},
  journal={Acta Mechanica Sinica},
  volume={37},
  number={12},
  pages={1727--1738},
  year={2021},
  publisher={Springer}
}

@article{raissi2020hidden,
  title={Hidden fluid mechanics: Learning velocity and pressure fields from flow visualizations},
  author={Raissi, Maziar and Yazdani, Alireza and Karniadakis, George Em},
  journal={Science},
  volume={367},
  number={6481},
  pages={1026--1030},
  year={2020},
  publisher={American Association for the Advancement of Science}
}

@article{goswami2020transfer,
  title={Transfer learning enhanced physics informed neural network for phase-field modeling of fracture},
  author={Goswami, Somdatta and Anitescu, Cosmin and Chakraborty, Souvik and Rabczuk, Timon},
  journal={Theoretical and Applied Fracture Mechanics},
  volume={106},
  pages={102447},
  year={2020},
  publisher={Elsevier}
}

@article{abueidda2022deep,
  title={A deep learning energy method for hyperelasticity and viscoelasticity},
  author={Abueidda, Diab W and Koric, Seid and Al-Rub, Rashid Abu and Parrott, Corey M and James, Kai A and Sobh, Nahil A},
  journal={European Journal of Mechanics-A/Solids},
  volume={95},
  pages={104639},
  year={2022},
  publisher={Elsevier}
}

@article{hornik1991approximation,
  title={Approximation capabilities of multilayer feedforward networks},
  author={Hornik, Kurt},
  journal={Neural networks},
  volume={4},
  number={2},
  pages={251--257},
  year={1991},
  publisher={Elsevier}
}

@article{cai2021flow,
  title={Flow over an espresso cup: inferring 3-D velocity and pressure fields from tomographic background oriented Schlieren via physics-informed neural networks},
  author={Cai, Shengze and Wang, Zhicheng and Fuest, Frederik and Jeon, Young Jin and Gray, Callum and Karniadakis, George Em},
  journal={Journal of Fluid Mechanics},
  volume={915},
  pages={A102},
  year={2021},
  publisher={Cambridge University Press}
}

@article{he2023deep,
  title={A deep learning energy-based method for classical elastoplasticity},
  author={He, Junyan and Abueidda, Diab and Al-Rub, Rashid Abu and Koric, Seid and Jasiuk, Iwona},
  journal={International Journal of Plasticity},
  pages={103531},
  year={2023},
  publisher={Elsevier}
}

@article{brunton2016discovering,
  title={Discovering governing equations from data by sparse identification of nonlinear dynamical systems},
  author={Brunton, Steven L and Proctor, Joshua L and Kutz, J Nathan},
  journal={Proceedings of the national academy of sciences},
  volume={113},
  number={15},
  pages={3932--3937},
  year={2016},
  publisher={National Acad Sciences}
}

@article{li2021physics,
  title={Physics-informed neural operator for learning partial differential equations},
  author={Li, Zongyi and Zheng, Hongkai and Kovachki, Nikola and Jin, David and Chen, Haoxuan and Liu, Burigede and Azizzadenesheli, Kamyar and Anandkumar, Anima},
  journal={ACM/JMS Journal of Data Science},
  volume={1},
  number={3},
  pages={1--27},
  year={2024},
  publisher={ACM New York, NY}
}

@book{hughes2012finite,
  title={The finite element method: linear static and dynamic finite element analysis},
  author={Hughes, Thomas JR},
  year={2012},
  publisher={Courier Corporation}
}

@book{bathe2006finite,
  title={Finite element procedures},
  author={Bathe, Klaus-J{\"u}rgen},
  year={2006},
  publisher={Klaus-Jurgen Bathe}
}

@book{reddy2019introduction,
  title={Introduction to the finite element method},
  author={Reddy, Junuthula Narasimha},
  year={2019},
  publisher={McGraw-Hill Education}
}

@article{liu2003mesh,
  title={Mesh free methods: moving beyond the finite element method},
  author={Liu, Gui-Rong and Karamanlidis, D},
  journal={Appl. Mech. Rev.},
  volume={56},
  number={2},
  pages={B17--B18},
  year={2003}
}

@article{rabczuk2004cracking,
  title={Cracking particles: a simplified meshfree method for arbitrary evolving cracks},
  author={Rabczuk, Timon and Belytschko, Ted},
  journal={International journal for numerical methods in engineering},
  volume={61},
  number={13},
  pages={2316--2343},
  year={2004},
  publisher={Wiley Online Library}
}

@article{nguyen2008meshless,
  title={Meshless methods: a review and computer implementation aspects},
  author={Nguyen, Vinh Phu and Rabczuk, Timon and Bordas, St{\'e}phane and Duflot, Marc},
  journal={Mathematics and computers in simulation},
  volume={79},
  number={3},
  pages={763--813},
  year={2008},
  publisher={Elsevier}
}

@article{rabczuk2007three,
  title={A three-dimensional large deformation meshfree method for arbitrary evolving cracks},
  author={Rabczuk, Timon and Belytschko, T23253911128},
  journal={Computer methods in applied mechanics and engineering},
  volume={196},
  number={29-30},
  pages={2777--2799},
  year={2007},
  publisher={Elsevier}
}

@book{rabczuk2019extended,
  title={Extended finite element and meshfree methods},
  author={Rabczuk, Timon and Song, Jeong-Hoon and Zhuang, Xiaoying and Anitescu, Cosmin},
  year={2019},
  publisher={Academic Press}
}

@inproceedings{jasak2007openfoam,
  title={OpenFOAM: A C++ library for complex physics simulations},
  author={Jasak, Hrvoje and Jemcov, Aleksandar and Tukovic, Zeljko and others},
  booktitle={International workshop on coupled methods in numerical dynamics},
  volume={1000},
  pages={1--20},
  year={2007}
}

@article{bi2023accurate,
  title={Accurate medium-range global weather forecasting with 3D neural networks},
  author={Bi, Kaifeng and Xie, Lingxi and Zhang, Hengheng and Chen, Xin and Gu, Xiaotao and Tian, Qi},
  journal={Nature},
  pages={1--6},
  year={2023},
  publisher={Nature Publishing Group UK London}
}

@article{abueidda2021meshless,
  title={Meshless physics-informed deep learning method for three-dimensional solid mechanics},
  author={Abueidda, Diab W and Lu, Qiyue and Koric, Seid},
  journal={International Journal for Numerical Methods in Engineering},
  volume={122},
  number={23},
  pages={7182--7201},
  year={2021},
  publisher={Wiley Online Library}
}

@book{brebbia2012boundary,
  title={Boundary element techniques: theory and applications in engineering},
  author={Brebbia, Carlos Alberto and Telles, Jos{\'e} Claudio Faria and Wrobel, Luiz C},
  year={2012},
  publisher={Springer Science \& Business Media}
}

@article{gao2022physics,
  title={Physics-informed graph neural Galerkin networks: A unified framework for solving PDE-governed forward and inverse problems},
  author={Gao, Han and Zahr, Matthew J and Wang, Jian-Xun},
  journal={Computer Methods in Applied Mechanics and Engineering},
  volume={390},
  pages={114502},
  year={2022},
  publisher={Elsevier}
}

@article{gao2021phygeonet,
  title={PhyGeoNet: Physics-informed geometry-adaptive convolutional neural networks for solving parameterized steady-state PDEs on irregular domain},
  author={Gao, Han and Sun, Luning and Wang, Jian-Xun},
  journal={Journal of Computational Physics},
  volume={428},
  pages={110079},
  year={2021},
  publisher={Elsevier}
}

@article{hestenes1969multiplier,
  title={Multiplier and gradient methods},
  author={Hestenes, Magnus R},
  journal={Journal of optimization theory and applications},
  volume={4},
  number={5},
  pages={303--320},
  year={1969},
  publisher={Springer}
}

@inproceedings{he2023learning,
  title={Learning physics-informed neural networks without stacked back-propagation},
  author={He, Di and Li, Shanda and Shi, Wenlei and Gao, Xiaotian and Zhang, Jia and Bian, Jiang and Wang, Liwei and Liu, Tie-Yan},
  booktitle={International Conference on Artificial Intelligence and Statistics},
  pages={3034--3047},
  year={2023},
  organization={PMLR}
}

@article{sirignano2018dgm,
  title={DGM: A deep learning algorithm for solving partial differential equations},
  author={Sirignano, Justin and Spiliopoulos, Konstantinos},
  journal={Journal of computational physics},
  volume={375},
  pages={1339--1364},
  year={2018},
  publisher={Elsevier}
}

@article{yang2020physics,
  title={Physics-informed generative adversarial networks for stochastic differential equations},
  author={Yang, Liu and Zhang, Dongkun and Karniadakis, George Em},
  journal={SIAM Journal on Scientific Computing},
  volume={42},
  number={1},
  pages={A292--A317},
  year={2020},
  publisher={SIAM}
}

@article{ramabathiran2021spinn,
  title={SPINN: sparse, physics-based, and partially interpretable neural networks for PDEs},
  author={Ramabathiran, Amuthan A and Ramachandran, Prabhu},
  journal={Journal of Computational Physics},
  volume={445},
  pages={110600},
  year={2021},
  publisher={Elsevier}
}

@article{sun2023binn,
  title={BINN: A deep learning approach for computational mechanics problems based on boundary integral equations},
  author={Sun, Jia and Liu, Yinghua and Wang, Yizheng and Yao, Zhenhan and Zheng, Xiaoping},
  journal={Computer Methods in Applied Mechanics and Engineering},
  volume={410},
  pages={116012},
  year={2023},
  publisher={Elsevier}
}

@article{meng2020ppinn,
  title={PPINN: Parareal physics-informed neural network for time-dependent PDEs},
  author={Meng, Xuhui and Li, Zhen and Zhang, Dongkun and Karniadakis, George Em},
  journal={Computer Methods in Applied Mechanics and Engineering},
  volume={370},
  pages={113250},
  year={2020},
  publisher={Elsevier}
}

@article{meng2020composite,
  title={A composite neural network that learns from multi-fidelity data: Application to function approximation and inverse PDE problems},
  author={Meng, Xuhui and Karniadakis, George Em},
  journal={Journal of Computational Physics},
  volume={401},
  pages={109020},
  year={2020},
  publisher={Elsevier}
}

@article{haghighat2021sciann,
  title={SciANN: A Keras/TensorFlow wrapper for scientific computations and physics-informed deep learning using artificial neural networks},
  author={Haghighat, Ehsan and Juanes, Ruben},
  journal={Computer Methods in Applied Mechanics and Engineering},
  volume={373},
  pages={113552},
  year={2021},
  publisher={Elsevier}
}

@article{zubov2021neuralpde,
  title={Neuralpde: Automating physics-informed neural networks (pinns) with error approximations},
  author={Zubov, Kirill and McCarthy, Zoe and Ma, Yingbo and Calisto, Francesco and Pagliarino, Valerio and Azeglio, Simone and Bottero, Luca and Luj{\'a}n, Emmanuel and Sulzer, Valentin and Bharambe, Ashutosh and others},
  journal={arXiv preprint arXiv:2107.09443},
  year={2021}
}

@inproceedings{hennigh2021nvidia,
  title={NVIDIA SimNet: An AI-accelerated multi-physics simulation framework},
  author={Hennigh, Oliver and Narasimhan, Susheela and Nabian, Mohammad Amin and Subramaniam, Akshay and Tangsali, Kaustubh and Fang, Zhiwei and Rietmann, Max and Byeon, Wonmin and Choudhry, Sanjay},
  booktitle={International conference on computational science},
  pages={447--461},
  year={2021},
  organization={Springer}
}

@article{fuhg2022mixed,
  title={The mixed deep energy method for resolving concentration features in finite strain hyperelasticity},
  author={Fuhg, Jan N and Bouklas, Nikolaos},
  journal={Journal of Computational Physics},
  volume={451},
  pages={110839},
  year={2022},
  publisher={Elsevier}
}

@article{wang2023dcm,
  title={DCEM: A deep complementary energy method for solid mechanics},
  author={Wang, Yizheng and Sun, Jia and Rabczuk, Timon and Liu, Yinghua},
  journal={International Journal for Numerical Methods in Engineering},
  year={2024},
  DOI = {10.1002/nme.7585}
}

@article{he2023use,
  title={On the use of graph neural networks and shape-function-based gradient computation in the deep energy method},
  author={He, Junyan and Abueidda, Diab and Koric, Seid and Jasiuk, Iwona},
  journal={International Journal for Numerical Methods in Engineering},
  volume={124},
  number={4},
  pages={864--879},
  year={2023},
  publisher={Wiley Online Library}
}

@article{kovachki2023neural,
  title={Neural Operator: Learning Maps Between Function Spaces With Applications to PDEs.},
  author={Kovachki, Nikola B and Li, Zongyi and Liu, Burigede and Azizzadenesheli, Kamyar and Bhattacharya, Kaushik and Stuart, Andrew M and Anandkumar, Anima},
  journal={J. Mach. Learn. Res.},
  volume={24},
  number={89},
  pages={1--97},
  year={2023}
}

@inproceedings{he2016deep,
  title={Deep residual learning for image recognition},
  author={He, Kaiming and Zhang, Xiangyu and Ren, Shaoqing and Sun, Jian},
  booktitle={Proceedings of the IEEE conference on computer vision and pattern recognition},
  pages={770--778},
  year={2016}
}

@article{hao2020ai,
  title={AI has cracked a key mathematical puzzle for understanding our world},
  author={Hao, Karen},
  journal={MIT Technology Review},
  volume={15},
  pages={2021},
  year={2020}
}

@article{de2022generic,
  title={Generic bounds on the approximation error for physics-informed (and) operator learning},
  author={De Ryck, Tim and Mishra, Siddhartha},
  journal={Advances in Neural Information Processing Systems},
  volume={35},
  pages={10945--10958},
  year={2022}
}

@article{kovachki2021universal,
  title={On universal approximation and error bounds for Fourier neural operators},
  author={Kovachki, Nikola and Lanthaler, Samuel and Mishra, Siddhartha},
  journal={The Journal of Machine Learning Research},
  volume={22},
  number={1},
  pages={13237--13312},
  year={2021},
  publisher={JMLRORG}
}

@article{cuomo2022scientific,
  title={Scientific machine learning through physics--informed neural networks: Where we are and what next},
  author={Cuomo, Salvatore and Di Cola, Vincenzo Schiano and Giampaolo, Fabio and Rozza, Gianluigi and Raissi, Maziar and Piccialli, Francesco},
  journal={Journal of Scientific Computing},
  volume={92},
  number={3},
  pages={88},
  year={2022},
  publisher={Springer}
}

@article{dissanayake1994neural,
  title={Neural-network-based approximations for solving partial differential equations},
  author={Dissanayake, MWMG and Phan-Thien, Nhan},
  journal={communications in Numerical Methods in Engineering},
  volume={10},
  number={3},
  pages={195--201},
  year={1994},
  publisher={Wiley Online Library}
}

@article{yuan2022pinn,
  title={A-PINN: Auxiliary physics informed neural networks for forward and inverse problems of nonlinear integro-differential equations},
  author={Yuan, Lei and Ni, Yi-Qing and Deng, Xiang-Yun and Hao, Shuo},
  journal={Journal of Computational Physics},
  volume={462},
  pages={111260},
  year={2022},
  publisher={Elsevier}
}

@article{zhang2019quantifying,
  title={Quantifying total uncertainty in physics-informed neural networks for solving forward and inverse stochastic problems},
  author={Zhang, Dongkun and Lu, Lu and Guo, Ling and Karniadakis, George Em},
  journal={Journal of Computational Physics},
  volume={397},
  pages={108850},
  year={2019},
  publisher={Elsevier}
}

@article{mcclenny2023self,
  title={Self-adaptive physics-informed neural networks},
  author={McClenny, Levi D and Braga-Neto, Ulisses M},
  journal={Journal of Computational Physics},
  volume={474},
  pages={111722},
  year={2023},
  publisher={Elsevier}
}

@article{liu2021dual,
  title={A dual-dimer method for training physics-constrained neural networks with minimax architecture},
  author={Liu, Dehao and Wang, Yan},
  journal={Neural Networks},
  volume={136},
  pages={112--125},
  year={2021},
  publisher={Elsevier}
}

@article{yang2022diffusion,
  title={Diffusion models: A comprehensive survey of methods and applications},
  author={Yang, Ling and Zhang, Zhilong and Song, Yang and Hong, Shenda and Xu, Runsheng and Zhao, Yue and Zhang, Wentao and Cui, Bin and Yang, Ming-Hsuan},
  journal={ACM Computing Surveys},
  year={2022},
  publisher={ACM New York, NY}
}

@article{dwivedi2020physics,
  title={Physics informed extreme learning machine (pielm)--a rapid method for the numerical solution of partial differential equations},
  author={Dwivedi, Vikas and Srinivasan, Balaji},
  journal={Neurocomputing},
  volume={391},
  pages={96--118},
  year={2020},
  publisher={Elsevier}
}

@article{huang2006extreme,
  title={Extreme learning machine: theory and applications},
  author={Huang, Guang-Bin and Zhu, Qin-Yu and Siew, Chee-Kheong},
  journal={Neurocomputing},
  volume={70},
  number={1-3},
  pages={489--501},
  year={2006},
  publisher={Elsevier}
}

@article{fang2021high,
  title={A high-efficient hybrid physics-informed neural networks based on convolutional neural network},
  author={Fang, Zhiwei},
  journal={IEEE Transactions on Neural Networks and Learning Systems},
  volume={33},
  number={10},
  pages={5514--5526},
  year={2021},
  publisher={IEEE}
}

@article{baydin2018automatic,
  title={Automatic differentiation in machine learning: a survey},
  author={Baydin, Atilim Gunes and Pearlmutter, Barak A and Radul, Alexey Andreyevich and Siskind, Jeffrey Mark},
  journal={Journal of Marchine Learning Research},
  volume={18},
  pages={1--43},
  year={2018},
  publisher={Microtome Publishing}
}

@article{lu2019deeponet,
  title={Deeponet: Learning nonlinear operators for identifying differential equations based on the universal approximation theorem of operators},
  author={Lu, Lu and Jin, Pengzhan and Karniadakis, George Em},
  journal={arXiv preprint arXiv:1910.03193},
  year={2019}
}

@article{jacot2018neural,
  title={Neural tangent kernel: Convergence and generalization in neural networks},
  author={Jacot, Arthur and Gabriel, Franck and Hongler, Cl{\'e}ment},
  journal={Advances in neural information processing systems},
  volume={31},
  year={2018}
}

@article{lauer2008incorporating,
  title={Incorporating prior knowledge in support vector machines for classification: A review},
  author={Lauer, Fabien and Bloch, G{\'e}rard},
  journal={Neurocomputing},
  volume={71},
  number={7-9},
  pages={1578--1594},
  year={2008},
  publisher={Elsevier}
}

@article{lagaris2000neural,
  title={Neural-network methods for boundary value problems with irregular boundaries},
  author={Lagaris, Isaac E and Likas, Aristidis C and Papageorgiou, Dimitris G},
  journal={IEEE Transactions on Neural Networks},
  volume={11},
  number={5},
  pages={1041--1049},
  year={2000},
  publisher={IEEE}
}

@book{mcfall2006artificial,
  title={An artificial neural network method for solving boundary value problems with arbitrary irregular boundaries},
  author={McFall, Kevin S},
  year={2006},
  publisher={Georgia Institute of Technology}
}

@article{mcfall2009artificial,
  title={Artificial neural network method for solution of boundary value problems with exact satisfaction of arbitrary boundary conditions},
  author={McFall, Kevin Stanley and Mahan, James Robert},
  journal={IEEE Transactions on Neural Networks},
  volume={20},
  number={8},
  pages={1221--1233},
  year={2009},
  publisher={IEEE}
}

@article{xu2023transfer,
  title={Transfer learning based physics-informed neural networks for solving inverse problems in engineering structures under different loading scenarios},
  author={Xu, Chen and Cao, Ba Trung and Yuan, Yong and Meschke, G{\"u}nther},
  journal={Computer Methods in Applied Mechanics and Engineering},
  volume={405},
  pages={115852},
  year={2023},
  publisher={Elsevier}
}

@inproceedings{kendall2018multi,
  title={Multi-task learning using uncertainty to weigh losses for scene geometry and semantics},
  author={Kendall, Alex and Gal, Yarin and Cipolla, Roberto},
  booktitle={Proceedings of the IEEE conference on computer vision and pattern recognition},
  pages={7482--7491},
  year={2018}
}

@article{chakraborty2021transfer,
  title={Transfer learning based multi-fidelity physics informed deep neural network},
  author={Chakraborty, Souvik},
  journal={Journal of Computational Physics},
  volume={426},
  pages={109942},
  year={2021},
  publisher={Elsevier}
}

@article{guo2021deep,
  title={A deep collocation method for the bending analysis of Kirchhoff plate},
  author={Guo, Hongwei and Zhuang, Xiaoying and Rabczuk, Timon},
  journal={arXiv preprint arXiv:2102.02617},
  year={2021}
}

@article{zhuang2021deep,
  title={Deep autoencoder based energy method for the bending, vibration, and buckling analysis of Kirchhoff plates with transfer learning},
  author={Zhuang, Xiaoying and Guo, Hongwei and Alajlan, Naif and Zhu, Hehua and Rabczuk, Timon},
  journal={European Journal of Mechanics-A/Solids},
  volume={87},
  pages={104225},
  year={2021},
  publisher={Elsevier}
}

@article{li2022fourier,
  title={Fourier neural operator with learned deformations for pdes on general geometries},
  author={Li, Zongyi and Huang, Daniel Zhengyu and Liu, Burigede and Anandkumar, Anima},
  journal={Journal of Machine Learning Research},
  volume={24},
  number={388},
  pages={1--26},
  year={2023}
}

@book{simo2006computational,
  title={Computational inelasticity},
  author={Simo, Juan C and Hughes, Thomas JR},
  volume={7},
  year={2006},
  publisher={Springer Science \& Business Media}
}

@article{xu2024worth,
  title={Worth of prior knowledge for enhancing deep learning},
  author={Xu, Hao and Chen, Yuntian and Zhang, Dongxiao},
  journal={Nexus},
  year={2024},
  publisher={Elsevier}
}

@article{bourdin2000numerical,
  title={Numerical experiments in revisited brittle fracture},
  author={Bourdin, Blaise and Francfort, Gilles A and Marigo, Jean-Jacques},
  journal={Journal of the Mechanics and Physics of Solids},
  volume={48},
  number={4},
  pages={797--826},
  year={2000},
  publisher={Elsevier}
}

@article{francfort1998revisiting,
  title={Revisiting brittle fracture as an energy minimization problem},
  author={Francfort, Gilles A and Marigo, J-J},
  journal={Journal of the Mechanics and Physics of Solids},
  volume={46},
  number={8},
  pages={1319--1342},
  year={1998},
  publisher={Elsevier}
}

@article{chen2021learning,
  title={Learning hidden elasticity with deep neural networks},
  author={Chen, Chun-Teh and Gu, Grace X},
  journal={Proceedings of the National Academy of Sciences},
  volume={118},
  number={31},
  pages={e2102721118},
  year={2021},
  publisher={National Acad Sciences}
}

@article{zhang2023artificial,
  title={Artificial intelligence for science in quantum, atomistic, and continuum systems},
  author={Zhang, Xuan and Wang, Limei and Helwig, Jacob and Luo, Youzhi and Fu, Cong and Xie, Yaochen and Liu, Meng and Lin, Yuchao and Xu, Zhao and Yan, Keqiang and others},
  journal={arXiv preprint arXiv:2307.08423},
  year={2023}
}

@article{liu2020generic,
  title={A generic physics-informed neural network-based constitutive model for soft biological tissues},
  author={Liu, Minliang and Liang, Liang and Sun, Wei},
  journal={Computer methods in applied mechanics and engineering},
  volume={372},
  pages={113402},
  year={2020},
  publisher={Elsevier}
}

@article{wang2024homogenius,
  title={A Pretraining-Finetuning Computational Framework for Material Homogenization},
  author={Wang, Yizheng and Li, Xiang and Yan, Ziming and Ma, Shuaifeng and Bai, Jinshuai and Liu, Bokai and Rabczuk, Timon and Liu, Yinghua},
  journal={International Journal of Mechanical Sciences},
  volume={314},
  pages={111388},
  year={2026},
  publisher={Elsevier}
}

@article{liu2024multi,
  title={Multi-scale modeling in thermal conductivity of Polyurethane incorporated with Phase Change Materials using Physics-Informed Neural Networks},
  author={Liu, Bokai and Wang, Yizheng and Rabczuk, Timon and Olofsson, Thomas and Lu, Weizhuo},
  journal={Renewable Energy},
  volume={220},
  pages={119565},
  year={2024},
  publisher={Elsevier}
}

@article{he2023deep_o,
  title={Deep energy method in topology optimization applications},
  author={He, Junyan and Chadha, Charul and Kushwaha, Shashank and Koric, Seid and Abueidda, Diab and Jasiuk, Iwona},
  journal={Acta Mechanica},
  volume={234},
  number={4},
  pages={1365--1379},
  year={2023},
  publisher={Springer}
}

@article{jeong2023complete,
  title={A complete Physics-Informed Neural Network-based framework for structural topology optimization},
  author={Jeong, Hyogu and Batuwatta-Gamage, Chanaka and Bai, Jinshuai and Xie, Yi Min and Rathnayaka, Charith and Zhou, Ying and Gu, YuanTong},
  journal={Computer Methods in Applied Mechanics and Engineering},
  volume={417},
  pages={116401},
  year={2023},
  publisher={Elsevier}
}

@article{zehnder2021ntopo,
  title={Ntopo: Mesh-free topology optimization using implicit neural representations},
  author={Zehnder, Jonas and Li, Yue and Coros, Stelian and Thomaszewski, Bernhard},
  journal={Advances in Neural Information Processing Systems},
  volume={34},
  pages={10368--10381},
  year={2021}
}

@article{jeong2023physics,
  title={A physics-informed neural network-based topology optimization (PINNTO) framework for structural optimization},
  author={Jeong, Hyogu and Bai, Jinshuai and Batuwatta-Gamage, Chanaka Prabuddha and Rathnayaka, Charith and Zhou, Ying and Gu, YuanTong},
  journal={Engineering Structures},
  volume={278},
  pages={115484},
  year={2023},
  publisher={Elsevier}
}

@article{zhang2022analyses,
  title={Analyses of internal structures and defects in materials using physics-informed neural networks},
  author={Zhang, Enrui and Dao, Ming and Karniadakis, George Em and Suresh, Subra},
  journal={Science advances},
  volume={8},
  number={7},
  pages={eabk0644},
  year={2022},
  publisher={American Association for the Advancement of Science}
}

@article{sun2023data,
  title={A data-driven multi-flaw detection strategy based on deep learning and boundary element method},
  author={Sun, Jia and Liu, Yinghua and Yao, Zhenhan and Zheng, Xiaoping},
  journal={Computational Mechanics},
  volume={71},
  number={3},
  pages={517--542},
  year={2023},
  publisher={Springer}
}

@article{pinkus1999approximation,
  title={Approximation theory of the MLP model in neural networks},
  author={Pinkus, Allan},
  journal={Acta numerica},
  volume={8},
  pages={143--195},
  year={1999},
  publisher={Cambridge University Press}
}

@inproceedings{cohen2016expressive,
  title={On the expressive power of deep learning: A tensor analysis},
  author={Cohen, Nadav and Sharir, Or and Shashua, Amnon},
  booktitle={Conference on learning theory},
  pages={698--728},
  year={2016},
  organization={PMLR}
}

@article{shin2020convergence,
  title={On the convergence of physics informed neural networks for linear second-order elliptic and parabolic type PDEs},
  author={Shin, Yeonjong and Darbon, Jerome and Karniadakis, George Em},
  journal={arXiv preprint arXiv:2004.01806},
  year={2020}
}

@article{mishra2022estimates,
  title={Estimates on the generalization error of physics-informed neural networks for approximating a class of inverse problems for PDEs},
  author={Mishra, Siddhartha and Molinaro, Roberto},
  journal={IMA Journal of Numerical Analysis},
  volume={42},
  number={2},
  pages={981--1022},
  year={2022},
  publisher={Oxford University Press}
}

@article{psaros2023uncertainty,
  title={Uncertainty quantification in scientific machine learning: Methods, metrics, and comparisons},
  author={Psaros, Apostolos F and Meng, Xuhui and Zou, Zongren and Guo, Ling and Karniadakis, George Em},
  journal={Journal of Computational Physics},
  volume={477},
  pages={111902},
  year={2023},
  publisher={Elsevier}
}

@article{lanthaler2022error,
  title={Error estimates for deeponets: A deep learning framework in infinite dimensions},
  author={Lanthaler, Samuel and Mishra, Siddhartha and Karniadakis, George E},
  journal={Transactions of Mathematics and Its Applications},
  volume={6},
  number={1},
  pages={tnac001},
  year={2022},
  publisher={Oxford University Press}
}

@article{li2020neural,
  title={Neural operator: Graph kernel network for partial differential equations},
  author={Li, Zongyi and Kovachki, Nikola and Azizzadenesheli, Kamyar and Liu, Burigede and Bhattacharya, Kaushik and Stuart, Andrew and Anandkumar, Anima},
  journal={arXiv preprint arXiv:2003.03485},
  year={2020}
}

@article{liu2024kan,
  title={KAN: Kolmogorov-Arnold Networks},
  author={Liu, Ziming and Wang, Yixuan and Vaidya, Sachin and Ruehle, Fabian and Halverson, James and Solja{\v{c}}i{\'c}, Marin and Hou, Thomas Y and Tegmark, Max},
  journal={arXiv preprint arXiv:2404.19756},
  year={2024}
}

@article{rao2020physics,
  title={Physics-informed deep learning for incompressible laminar flows},
  author={Rao, Chengping and Sun, Hao and Liu, Yang},
  journal={Theoretical and Applied Mechanics Letters},
  volume={10},
  number={3},
  pages={207--212},
  year={2020},
  publisher={Elsevier}
}

@article{jin2021nsfnets,
  title={NSFnets (Navier-Stokes flow nets): Physics-informed neural networks for the incompressible Navier-Stokes equations},
  author={Jin, Xiaowei and Cai, Shengze and Li, Hui and Karniadakis, George Em},
  journal={Journal of Computational Physics},
  volume={426},
  pages={109951},
  year={2021},
  publisher={Elsevier}
}

@article{zhang2022drvn,
  title={DRVN (deep random vortex network): A new physics-informed machine learning method for simulating and inferring incompressible fluid flows},
  author={Zhang, Rui and Hu, Peiyan and Meng, Qi and Wang, Yue and Zhu, Rongchan and Chen, Bingguang and Ma, Zhi-Ming and Liu, Tie-Yan},
  journal={Physics of Fluids},
  volume={34},
  number={10},
  year={2022},
  publisher={AIP Publishing}
}

@article{majda2002vorticity,
  title={Vorticity and incompressible flow. Cambridge texts in applied mathematics},
  author={Majda, Andrew J and Bertozzi, Andrea L and Ogawa, A},
  journal={Appl. Mech. Rev.},
  volume={55},
  number={4},
  pages={B77--B78},
  year={2002}
}

@article{rui2024time,
  title={Time-averaged flow field reconstruction based on a multifidelity model using physics-informed neural network (PINN) and nonlinear information fusion},
  author={Rui, En-Ze and Zeng, Guang-Zhi and Ni, Yi-Qing and Chen, Zheng-Wei and Hao, Shuo},
  journal={International Journal of Numerical Methods for Heat \& Fluid Flow},
  volume={34},
  number={1},
  pages={131--149},
  year={2024},
  publisher={Emerald Publishing Limited}
}

@article{wang2023prediction,
  title={The prediction of external flow field and hydrodynamic force with limited data using deep neural network},
  author={Wang, Tong-sheng and Xi, Guang and Sun, Zhong-guo and Huang, Zhu},
  journal={Journal of Hydrodynamics},
  volume={35},
  number={3},
  pages={549--570},
  year={2023},
  publisher={Springer}
}

@article{han2024prediction,
  title={Prediction of Porous Media Fluid Flow with Spatial Heterogeneity Using Criss-Cross Physics-Informed Convolutional Neural Networks},
  author={Han, Jiangxia and Xue, Liang and Jia, Ying and Mwasamwasa, Mpoki Sam and Nanguka, Felix and Sangweni, Charles and Liu, Hailong and Li, Qian},
  journal={CMES-Computer Modeling in Engineering \& Sciences},
  volume={138},
  number={2},
  pages={1323--1340},
  year={2024},
  publisher={TECH SCIENCE PRESS 871 CORONADO CENTER DR, SUTE 200, HENDERSON, NV 89052 USA}
}

@article{cheng2021deep,
  title={Deep learning method based on physics informed neural network with resnet block for solving fluid flow problems},
  author={Cheng, Chen and Zhang, Guang-Tao},
  journal={Water},
  volume={13},
  number={4},
  pages={423},
  year={2021},
  publisher={MDPI}
}

@article{rosofsky2023magnetohydrodynamics,
  title={Magnetohydrodynamics with physics informed neural operators},
  author={Rosofsky, Shawn G and Huerta, EA},
  journal={Machine Learning: Science and Technology},
  volume={4},
  number={3},
  pages={035002},
  year={2023},
  publisher={IOP Publishing}
}

@article{zhu2023reliable,
  title={Reliable extrapolation of deep neural operators informed by physics or sparse observations},
  author={Zhu, Min and Zhang, Handi and Jiao, Anran and Karniadakis, George Em and Lu, Lu},
  journal={Computer Methods in Applied Mechanics and Engineering},
  volume={412},
  pages={116064},
  year={2023},
  publisher={Elsevier}
}

@article{li2024operator,
  title={Operator learning for urban water clarification hydrodynamics and particulate matter transport with physics-informed neural networks},
  author={Li, Haochen and Shatarah, Mohamed},
  journal={Water Research},
  volume={251},
  pages={121123},
  year={2024},
  publisher={Elsevier}
}

@article{peng2024rapid,
  title={Rapid and sparse reconstruction of high-speed steady-state and transient compressible flow fields using physics-informed graph neural networks},
  author={Peng, Jiang-Zhou and Wang, Zhi-Qiao and Rong, Xiaoli and Mei, Mei and Wang, Mingyang and He, Yong and Wu, Wei-Tao},
  journal={Physics of Fluids},
  volume={36},
  number={4},
  year={2024},
  publisher={AIP Publishing}
}

@article{ren2024physics,
  title={Physics-informed neural networks for transonic flow around a cylinder with high Reynolds number},
  author={Ren, Xiang and Hu, Peng and Su, Hua and Zhang, Feizhou and Yu, Huahua},
  journal={Physics of Fluids},
  volume={36},
  number={3},
  year={2024},
  publisher={AIP Publishing}
}

@article{joshi2023investigation,
  title={Investigation of Low and High-Speed Fluid Dynamics Problems Using Physics-Informed Neural Network},
  author={Joshi, Anubhav and Papados, Alexandros and Kumar, Rakesh},
  journal={International Journal of Computational Fluid Dynamics},
  volume={37},
  number={2},
  pages={149--166},
  year={2023},
  publisher={Taylor \& Francis}
}

@article{auddy2024grinn,
  title={GRINN: a physics-informed neural network for solving hydrodynamic systems in the presence of self-gravity},
  author={Auddy, Sayantan and Dey, Ramit and Turner, Neal J and Basu, Shantanu},
  journal={Machine Learning: Science and Technology},
  volume={5},
  number={2},
  pages={025014},
  year={2024},
  publisher={IOP Publishing}
}

@article{wang2021deep,
  title={Deep learning of free boundary and Stefan problems},
  author={Wang, Sifan and Perdikaris, Paris},
  journal={Journal of Computational Physics},
  volume={428},
  pages={109914},
  year={2021},
  publisher={Elsevier}
}

@article{jalili2024physics,
  title={Physics-informed neural networks for heat transfer prediction in two-phase flows},
  author={Jalili, Darioush and Jang, Seohee and Jadidi, Mohammad and Giustini, Giovanni and Keshmiri, Amir and Mahmoudi, Yasser},
  journal={International Journal of Heat and Mass Transfer},
  volume={221},
  pages={125089},
  year={2024},
  publisher={Elsevier}
}

@article{hirt1981volume,
  title={Volume of fluid (VOF) method for the dynamics of free boundaries},
  author={Hirt, Cyril W and Nichols, Billy D},
  journal={Journal of computational physics},
  volume={39},
  number={1},
  pages={201--225},
  year={1981},
  publisher={Elsevier}
}

@article{wessels2020neural,
  title={The neural particle method--an updated Lagrangian physics informed neural network for computational fluid dynamics},
  author={Wessels, Henning and Wei{\ss}enfels, Christian and Wriggers, Peter},
  journal={Computer Methods in Applied Mechanics and Engineering},
  volume={368},
  pages={113127},
  year={2020},
  publisher={Elsevier}
}

@article{bai2022general,
  title={A general Neural Particle Method for hydrodynamics modeling},
  author={Bai, Jinshuai and Zhou, Ying and Ma, Yuwei and Jeong, Hyogu and Zhan, Haifei and Rathnayaka, Charith and Sauret, Emilie and Gu, Yuantong},
  journal={Computer Methods in Applied Mechanics and Engineering},
  volume={393},
  pages={114740},
  year={2022},
  publisher={Elsevier}
}

@article{shao2023improved,
  title={An Improved Neural Particle Method for Complex Free Surface Flow Simulation Using Physics-Informed Neural Networks},
  author={Shao, Kaixuan and Wu, Yinghan and Jia, Suizi},
  journal={Mathematics},
  volume={11},
  number={8},
  pages={1805},
  year={2023},
  publisher={MDPI}
}

@article{huang2023solving,
  title={Solving free-surface problems for non-shallow water using boundary and initial conditions-free physics-informed neural network (bif-PINN)},
  author={Huang, Ying H and Xu, Zheng and Qian, Cheng and Liu, Li},
  journal={Journal of Computational Physics},
  volume={479},
  pages={112003},
  year={2023},
  publisher={Elsevier}
}

@article{diab2024learning,
  title={Learning generic solutions for multiphase transport in porous media via the flux functions operator},
  author={Diab, Waleed and Chaabi, Omar and Alkobaisi, Shayma and Awotunde, Abeeb and Al Kobaisi, Mohammed},
  journal={Advances in Water Resources},
  volume={183},
  pages={104609},
  year={2024},
  publisher={Elsevier}
}

@article{hsieh2024multiscale,
  title={A multiscale stabilized physics informed neural networks with weakly imposed boundary conditions transfer learning method for modeling advection dominated flow},
  author={Hsieh, Tsung-Yeh and Huang, Tsung-Hui},
  journal={Engineering with Computers},
  pages={1--35},
  year={2024},
  publisher={Springer}
}

@article{jin2023asymptotic,
  title={Asymptotic-preserving neural networks for multiscale time-dependent linear transport equations},
  author={Jin, Shi and Ma, Zheng and Wu, Keke},
  journal={Journal of Scientific Computing},
  volume={94},
  number={3},
  pages={57},
  year={2023},
  publisher={Springer}
}

@article{lin2021operator,
  title={Operator learning for predicting multiscale bubble growth dynamics},
  author={Lin, Chensen and Li, Zhen and Lu, Lu and Cai, Shengze and Maxey, Martin and Karniadakis, George Em},
  journal={The Journal of Chemical Physics},
  volume={154},
  number={10},
  year={2021},
  publisher={AIP Publishing}
}

@article{lin2021seamless,
  title={A seamless multiscale operator neural network for inferring bubble dynamics},
  author={Lin, Chensen and Maxey, Martin and Li, Zhen and Karniadakis, George Em},
  journal={Journal of Fluid Mechanics},
  volume={929},
  pages={A18},
  year={2021},
  publisher={Cambridge University Press}
}

@article{mao2021deepm,
  title={DeepM\&Mnet for hypersonics: Predicting the coupled flow and finite-rate chemistry behind a normal shock using neural-network approximation of operators},
  author={Mao, Zhiping and Lu, Lu and Marxen, Olaf and Zaki, Tamer A and Karniadakis, George Em},
  journal={Journal of computational physics},
  volume={447},
  pages={110698},
  year={2021},
  publisher={Elsevier}
}

@article{cai2021deepm,
  title={DeepM\&Mnet: Inferring the electroconvection multiphysics fields based on operator approximation by neural networks},
  author={Cai, Shengze and Wang, Zhicheng and Lu, Lu and Zaki, Tamer A and Karniadakis, George Em},
  journal={Journal of Computational Physics},
  volume={436},
  pages={110296},
  year={2021},
  publisher={Elsevier}
}

@article{wu2024capturing,
  title={Capturing the diffusive behavior of the multiscale linear transport equations by Asymptotic-Preserving Convolutional DeepONets},
  author={Wu, Keke and Yan, Xiong-Bin and Jin, Shi and Ma, Zheng},
  journal={Computer Methods in Applied Mechanics and Engineering},
  volume={418},
  pages={116531},
  year={2024},
  publisher={Elsevier}
}

@article{ahmed2023multifidelity,
  title={A multifidelity deep operator network approach to closure for multiscale systems},
  author={Ahmed, Shady E and Stinis, Panos},
  journal={Computer Methods in Applied Mechanics and Engineering},
  volume={414},
  pages={116161},
  year={2023},
  publisher={Elsevier}
}

@article{cai2021artificial,
  title={Artificial intelligence velocimetry and microaneurysm-on-a-chip for three-dimensional analysis of blood flow in physiology and disease},
  author={Cai, Shengze and Li, He and Zheng, Fuyin and Kong, Fang and Dao, Ming and Karniadakis, George Em and Suresh, Subra},
  journal={Proceedings of the National Academy of Sciences},
  volume={118},
  number={13},
  pages={e2100697118},
  year={2021},
  publisher={National Acad Sciences}
}

@article{boster2023artificial,
  title={Artificial intelligence velocimetry reveals in vivo flow rates, pressure gradients, and shear stresses in murine perivascular flows},
  author={Boster, Kimberly AS and Cai, Shengze and Ladr{\'o}n-de-Guevara, Antonio and Sun, Jiatong and Zheng, Xiaoning and Du, Ting and Thomas, John H and Nedergaard, Maiken and Karniadakis, George Em and Kelley, Douglas H},
  journal={Proceedings of the National Academy of Sciences},
  volume={120},
  number={14},
  pages={e2217744120},
  year={2023},
  publisher={National Acad Sciences}
}

@article{mahmoudabadbozchelou2022nn,
  title={nn-PINNs: Non-Newtonian physics-informed neural networks for complex fluid modeling},
  author={Mahmoudabadbozchelou, Mohammadamin and Karniadakis, George Em and Jamali, Safa},
  journal={Soft Matter},
  volume={18},
  number={1},
  pages={172--185},
  year={2022},
  publisher={Royal Society of Chemistry}
}

@article{schierholz2023engineering,
  title={Engineering-informed design space reduction for PCB based power delivery networks},
  author={Schierholz, Morten and Hassab, Youcef and Schuster, Christian},
  journal={IEEE Transactions on Components, Packaging and Manufacturing Technology},
  year={2023},
  publisher={IEEE}
}

@article{du2023state,
  title={State estimation in minimal turbulent channel flow: A comparative study of 4DVar and PINN},
  author={Du, Yifan and Wang, Mengze and Zaki, Tamer A},
  journal={International Journal of Heat and Fluid Flow},
  volume={99},
  pages={109073},
  year={2023},
  publisher={Elsevier}
}

@article{zhang2023physics,
  title={A physics-informed neural network-based approach to reconstruct the tornado vortices from limited observed data},
  author={Zhang, Han and Wang, Hao and Xu, Zidong and Liu, Zhenqing and Khoo, Boo Cheong},
  journal={Journal of Wind Engineering and Industrial Aerodynamics},
  volume={241},
  pages={105534},
  year={2023},
  publisher={Elsevier}
}

@article{renn2023forecasting,
  title={Forecasting subcritical cylinder wakes with Fourier Neural Operators},
  author={Renn, Peter I and Wang, Cong and Lale, Sahin and Li, Zongyi and Anandkumar, Anima and Gharib, Morteza},
  journal={arXiv preprint arXiv:2301.08290},
  year={2023}
}

@article{rosofsky2023applications,
  title={Applications of physics informed neural operators},
  author={Rosofsky, Shawn G and Al Majed, Hani and Huerta, EA},
  journal={Machine Learning: Science and Technology},
  volume={4},
  number={2},
  pages={025022},
  year={2023},
  publisher={IOP Publishing}
}

@article{lu2022multifidelity,
  title={Multifidelity deep neural operators for efficient learning of partial differential equations with application to fast inverse design of nanoscale heat transport},
  author={Lu, Lu and Pestourie, Rapha{\"e}l and Johnson, Steven G and Romano, Giuseppe},
  journal={Physical Review Research},
  volume={4},
  number={2},
  pages={023210},
  year={2022},
  publisher={APS}
}

@article{li2022fourier_eddy,
  title={Fourier neural operator approach to large eddy simulation of three-dimensional turbulence},
  author={Li, Zhijie and Peng, Wenhui and Yuan, Zelong and Wang, Jianchun},
  journal={Theoretical and Applied Mechanics Letters},
  volume={12},
  number={6},
  pages={100389},
  year={2022},
  publisher={Elsevier}
}

@article{yu2022gradient,
  title={Gradient-enhanced physics-informed neural networks for forward and inverse PDE problems},
  author={Yu, Jeremy and Lu, Lu and Meng, Xuhui and Karniadakis, George Em},
  journal={Computer Methods in Applied Mechanics and Engineering},
  volume={393},
  pages={114823},
  year={2022},
  publisher={Elsevier}
}

@article{lou2021physics,
  title={Physics-informed neural networks for solving forward and inverse flow problems via the Boltzmann-BGK formulation},
  author={Lou, Qin and Meng, Xuhui and Karniadakis, George Em},
  journal={Journal of Computational Physics},
  volume={447},
  pages={110676},
  year={2021},
  publisher={Elsevier}
}

@article{kou2024physics,
  title={Physics-informed neural network integrate with unclosed mechanism model for turbulent mass transfer},
  author={Kou, Chenhui and Yin, Yuhui and Zeng, Yang and Jia, Shengkun and Luo, Yiqing and Yuan, Xigang},
  journal={Chemical Engineering Science},
  volume={288},
  pages={119752},
  year={2024},
  publisher={Elsevier}
}

@article{arzani2021uncovering,
  title={Uncovering near-wall blood flow from sparse data with physics-informed neural networks},
  author={Arzani, Amirhossein and Wang, Jian-Xun and D'Souza, Roshan M},
  journal={Physics of Fluids},
  volume={33},
  number={7},
  year={2021},
  publisher={AIP Publishing}
}

@inproceedings{huang2019ccnet,
  title={Ccnet: Criss-cross attention for semantic segmentation},
  author={Huang, Zilong and Wang, Xinggang and Huang, Lichao and Huang, Chang and Wei, Yunchao and Liu, Wenyu},
  booktitle={Proceedings of the IEEE/CVF international conference on computer vision},
  pages={603--612},
  year={2019}
}

@article{wang2024kolmogorov,
  title={Kolmogorov--Arnold-Informed neural network: A physics-informed deep learning framework for solving forward and inverse problems based on Kolmogorov--Arnold Networks},
  author={Wang, Yizheng and Sun, Jia and Bai, Jinshuai and Anitescu, Cosmin and Eshaghi, Mohammad Sadegh and Zhuang, Xiaoying and Rabczuk, Timon and Liu, Yinghua},
  journal={Computer Methods in Applied Mechanics and Engineering},
  volume={433},
  pages={117518},
  year={2025},
  publisher={Elsevier}
}

@article{buoso2021personalising,
  title={Personalising left-ventricular biophysical models of the heart using parametric physics-informed neural networks},
  author={Buoso, Stefano and Joyce, Thomas and Kozerke, Sebastian},
  journal={Medical Image Analysis},
  volume={71},
  pages={102066},
  year={2021},
  publisher={Elsevier}
}

@article{dalton2023physics,
  title={Physics-informed graph neural network emulation of soft-tissue mechanics},
  author={Dalton, David and Husmeier, Dirk and Gao, Hao},
  journal={Computer Methods in Applied Mechanics and Engineering},
  volume={417},
  pages={116351},
  year={2023},
  publisher={Elsevier}
}

@article{linka2021unraveling,
  title={Unraveling the local relation between tissue composition and human brain mechanics through machine learning},
  author={Linka, Kevin and Reiter, Nina and W{\"u}rges, Jasmin and Schicht, Martin and Br{\"a}uer, Lars and Cyron, Christian J and Paulsen, Friedrich and Budday, Silvia},
  journal={Frontiers in bioengineering and biotechnology},
  volume={9},
  pages={704738},
  year={2021},
  publisher={Frontiers Media SA}
}

@article{sun2020surrogate,
  title={Surrogate modeling for fluid flows based on physics-constrained deep learning without simulation data},
  author={Sun, Luning and Gao, Han and Pan, Shaowu and Wang, Jian-Xun},
  journal={Computer Methods in Applied Mechanics and Engineering},
  volume={361},
  pages={112732},
  year={2020},
  publisher={Elsevier}
}

@article{qiu2022physics,
  title={Physics-informed neural networks for phase-field method in two-phase flow},
  author={Qiu, Rundi and Huang, Renfang and Xiao, Yao and Wang, Jingzhu and Zhang, Zhen and Yue, Jieshun and Zeng, Zhong and Wang, Yiwei},
  journal={Physics of Fluids},
  volume={34},
  number={5},
  year={2022},
  publisher={AIP Publishing}
}

@article{lim2020artificial,
  title={Artificial intelligence in cardiovascular imaging},
  author={Lim, Lisa J and Tison, Geoffrey H and Delling, Francesca N},
  journal={Methodist DeBakey Cardiovascular Journal},
  volume={16},
  number={2},
  pages={138},
  year={2020},
  publisher={Methodist DeBakey Heart \& Vascular Center}
}

@article{moser2023modeling,
  title={Modeling of 3D blood flows with physics-informed neural networks: comparison of network architectures},
  author={Moser, Philipp and Fenz, Wolfgang and Thumfart, Stefan and Ganitzer, Isabell and Giretzlehner, Michael},
  journal={Fluids},
  volume={8},
  number={2},
  pages={46},
  year={2023},
  publisher={MDPI}
}

@article{liu2024variable,
  title={Variable separated physics-informed neural networks based on adaptive weighted loss functions for blood flow model},
  author={Liu, Youqiong and Cai, Li and Chen, Yaping and Ma, Pengfei and Zhong, Qian},
  journal={Computers \& Mathematics with Applications},
  volume={153},
  pages={108--122},
  year={2024},
  publisher={Elsevier}
}

@article{kissas2020machine,
  title={Machine learning in cardiovascular flows modeling: Predicting arterial blood pressure from non-invasive 4D flow MRI data using physics-informed neural networks},
  author={Kissas, Georgios and Yang, Yibo and Hwuang, Eileen and Witschey, Walter R and Detre, John A and Perdikaris, Paris},
  journal={Computer Methods in Applied Mechanics and Engineering},
  volume={358},
  pages={112623},
  year={2020},
  publisher={Elsevier}
}

@article{fathi2020super,
  title={Super-resolution and denoising of 4D-flow MRI using physics-informed deep neural nets},
  author={Fathi, Mojtaba F and Perez-Raya, Isaac and Baghaie, Ahmadreza and Berg, Philipp and Janiga, Gabor and Arzani, Amirhossein and D’Souza, Roshan M},
  journal={Computer Methods and Programs in Biomedicine},
  volume={197},
  pages={105729},
  year={2020},
  publisher={Elsevier}
}

@article{kamali2023elasticity,
  title={Elasticity imaging using physics-informed neural networks: Spatial discovery of elastic modulus and Poisson's ratio},
  author={Kamali, Ali and Sarabian, Mohammad and Laksari, Kaveh},
  journal={Acta biomaterialia},
  volume={155},
  pages={400--409},
  year={2023},
  publisher={Elsevier}
}

@article{zhang2020physics,
  title={Physics-informed neural networks for nonhomogeneous material identification in elasticity imaging},
  author={Zhang, Enrui and Yin, Minglang and Karniadakis, George Em},
  journal={arXiv preprint arXiv:2009.04525},
  year={2020}
}

@article{herrero2022ep,
  title={EP-PINNs: Cardiac electrophysiology characterisation using physics-informed neural networks},
  author={Herrero Martin, Clara and Oved, Alon and Chowdhury, Rasheda A and Ullmann, Elisabeth and Peters, Nicholas S and Bharath, Anil A and Varela, Marta},
  journal={Frontiers in Cardiovascular Medicine},
  volume={8},
  pages={768419},
  year={2022},
  publisher={Frontiers Media SA}
}

@article{sarabian2022physics,
  title={Physics-informed neural networks for brain hemodynamic predictions using medical imaging},
  author={Sarabian, Mohammad and Babaee, Hessam and Laksari, Kaveh},
  journal={IEEE transactions on medical imaging},
  volume={41},
  number={9},
  pages={2285--2303},
  year={2022},
  publisher={IEEE}
}

@article{wu2024machine,
  title={Machine Learning in Biomaterials, Biomechanics/Mechanobiology, and Biofabrication: State of the Art and Perspective},
  author={Wu, Chi and Xu, Yanan and Fang, Jianguang and Li, Qing},
  journal={Archives of Computational Methods in Engineering},
  pages={1--67},
  year={2024},
  publisher={Springer}
}

@article{bai2023physicsrbf,
  title={Physics-informed radial basis network (PIRBN): A local approximating neural network for solving nonlinear partial differential equations},
  author={Bai, Jinshuai and Liu, Gui-Rong and Gupta, Ashish and Alzubaidi, Laith and Feng, Xi-Qiao and Gu, YuanTong},
  journal={Computer Methods in Applied Mechanics and Engineering},
  volume={415},
  pages={116290},
  year={2023},
  publisher={Elsevier}
}

@article{bai2024robust,
  title={A robust radial point interpolation method empowered with neural network solvers (RPIM-NNS) for nonlinear solid mechanics},
  author={Bai, Jinshuai and Liu, Gui-Rong and Rabczuk, Timon and Wang, Yizheng and Feng, Xi-Qiao and Gu, YuanTong},
  journal={Computer Methods in Applied Mechanics and Engineering},
  volume={429},
  pages={117159},
  year={2024},
  publisher={Elsevier}
}

@article{bai2023physics,
  title={A physics-informed neural network technique based on a modified loss function for computational 2D and 3D solid mechanics},
  author={Bai, Jinshuai and Rabczuk, Timon and Gupta, Ashish and Alzubaidi, Laith and Gu, Yuantong},
  journal={Computational Mechanics},
  volume={71},
  number={3},
  pages={543--562},
  year={2023},
  publisher={Springer}
}

@article{bai2023introduction,
  title={An Introduction to Programming Physics-Informed Neural Network-Based Computational Solid Mechanics},
  author={Bai, Jinshuai and Jeong, Hyogu and Batuwatta-Gamage, CP and Xiao, Shusheng and Wang, Qingxia and Rathnayaka, CM and Alzubaidi, Laith and Liu, Gui Rong and Gu, Yuantong},
  journal={International Journal of Computational Methods},
  year={2023},
  publisher={World Scientific Publishing}
}

@book{karniadakis2005spectral,
  title={Spectral/hp element methods for computational fluid dynamics},
  author={Karniadakis, George and Sherwin, Spencer J},
  year={2005},
  publisher={Oxford University Press, USA}
}

@article{goodfellow2013empirical,
  title={An empirical investigation of catastrophic forgetting in gradient-based neural networks},
  author={Goodfellow, Ian J and Mirza, Mehdi and Xiao, Da and Courville, Aaron and Bengio, Yoshua},
  journal={arXiv preprint arXiv:1312.6211},
  year={2013}
}

@article{noor1979computerized,
  title={Computerized symbolic manipulation in structural mechanics—progress and potential},
  author={Noor, Ahmed K and Andersen, CM},
  journal={Computers \& Structures},
  volume={10},
  number={1-2},
  pages={95--118},
  year={1979},
  publisher={Elsevier}
}

@article{kirkpatrick1983shape,
  title={On the shape of a set of points in the plane},
  author={Kirkpatrick, D and Seidel, Raimund},
  journal={IEEE Transactions on Information Theory},
  volume={29},
  number={4},
  pages={551--559},
  year={1983}
}

@article{wang2023scientific,
  title={Scientific discovery in the age of artificial intelligence},
  author={Wang, Hanchen and Fu, Tianfan and Du, Yuanqi and Gao, Wenhao and Huang, Kexin and Liu, Ziming and Chandak, Payal and Liu, Shengchao and Van Katwyk, Peter and Deac, Andreea and others},
  journal={Nature},
  volume={620},
  number={7972},
  pages={47--60},
  year={2023},
  publisher={Nature Publishing Group UK London}
}

@article{bartolucci2024representation,
  title={Representation equivalent neural operators: a framework for alias-free operator learning},
  author={Bartolucci, Francesca and de B{\'e}zenac, Emmanuel and Raonic, Bogdan and Molinaro, Roberto and Mishra, Siddhartha and Alaifari, Rima},
  journal={Advances in Neural Information Processing Systems},
  volume={36},
  year={2024}
}

@article{miehe2010phase,
  title={A phase field model for rate-independent crack propagation: Robust algorithmic implementation based on operator splits},
  author={Miehe, Christian and Hofacker, Martina and Welschinger, Fabian},
  journal={Computer Methods in Applied Mechanics and Engineering},
  volume={199},
  number={45-48},
  pages={2765--2778},
  year={2010},
  publisher={Elsevier}
}

@article{li2002meshfree,
  title={Meshfree and particle methods and their applications},
  author={Li, Shaofan and Liu, Wing Kam},
  journal={Appl. Mech. Rev.},
  volume={55},
  number={1},
  pages={1--34},
  year={2002}
}

@article{zhou2022open,
  title={An open-source unconstrained stress updating algorithm for the modified Cam-clay model},
  author={Zhou, Xin and Lu, Dechun and Zhang, Yaning and Du, Xiuli and Rabczuk, Timon},
  journal={Computer Methods in Applied Mechanics and Engineering},
  volume={390},
  pages={114356},
  year={2022},
  publisher={Elsevier}
}

@article{goswami2020adaptive,
  title={Adaptive fourth-order phase field analysis using deep energy minimization},
  author={Goswami, Somdatta and Anitescu, Cosmin and Rabczuk, Timon},
  journal={Theoretical and Applied Fracture Mechanics},
  volume={107},
  pages={102527},
  year={2020},
  publisher={Elsevier}
}

@article{es4846935deepnetbeam,
  title={Applications of scientific machine learning for the analysis of functionally graded porous beams},
  author={Eshaghi, Mohammad Sadegh and Bamdad, Mostafa and Anitescu, Cosmin and Wang, Yizheng and Zhuang, Xiaoying and Rabczuk, Timon},
  journal={Neurocomputing},
  volume={619},
  pages={129119},
  year={2025},
  publisher={Elsevier}
}

@online{AI4math,
  author       = {AlphaProof and AlphaGeometry teams},
  title        = {AI achieves silver-medal standard solving International Mathematical Olympiad problems},
  year         = {2024},
  url          = {https://deepmind.google/discover/blog/ai-solves-imo-problems-at-silver-medal-level/},
}

@online{AlphaGeometry,
  author       = {Trieu Trinh and Thang Luong},
  title        = {AlphaGeometry: An Olympiad-level AI system for geometry},
  year         = {2024},
  url          = {https://deepmind.google/discover/blog/alphageometry-an-olympiad-level-ai-system-for-geometry/},
}

@article{he2023novel,
  title={Novel DeepONet architecture to predict stresses in elastoplastic structures with variable complex geometries and loads},
  author={He, Junyan and Koric, Seid and Kushwaha, Shashank and Park, Jaewan and Abueidda, Diab and Jasiuk, Iwona},
  journal={Computer Methods in Applied Mechanics and Engineering},
  volume={415},
  pages={116277},
  year={2023},
  publisher={Elsevier}
}

@article{sahin2024solving,
  title={Solving forward and inverse problems of contact mechanics using physics-informed neural networks},
  author={Sahin, Tarik and von Danwitz, Max and Popp, Alexander},
  journal={Advanced Modeling and Simulation in Engineering Sciences},
  volume={11},
  number={1},
  pages={11},
  year={2024},
  publisher={Springer}
}

@article{jiang2025deepseek,
  title={DeepSeek vs. ChatGPT vs. Claude: A comparative study for scientific computing and scientific machine learning tasks},
  author={Jiang, Qile and Gao, Zhiwei and Karniadakis, George Em},
  journal={Theoretical and Applied Mechanics Letters},
  volume={15},
  number={3},
  pages={100583},
  year={2025},
  publisher={Elsevier}
}

@article{yu2021reinforcement,
  title={Reinforcement learning versus PDE backstepping and PI control for congested freeway traffic},
  author={Yu, Huan and Park, Saehong and Bayen, Alexandre and Moura, Scott and Krstic, Miroslav},
  journal={IEEE Transactions on Control Systems Technology},
  volume={30},
  number={4},
  pages={1595--1611},
  year={2021},
  publisher={IEEE}
}

@article{bastek2024physics,
  title={Physics-informed diffusion models},
  author={Bastek, Jan-Hendrik and Sun, WaiChing and Kochmann, Dennis M},
  journal={arXiv preprint arXiv:2403.14404},
  year={2024}
}

@inproceedings{pfaff2020learning,
  title={Learning mesh-based simulation with graph networks},
  author={Pfaff, Tobias and Fortunato, Meire and Sanchez-Gonzalez, Alvaro and Battaglia, Peter},
  booktitle={International conference on learning representations},
  year={2020}
}

@article{lu2024ai,
  title={The ai scientist: Towards fully automated open-ended scientific discovery},
  author={Lu, Chris and Lu, Cong and Lange, Robert Tjarko and Foerster, Jakob and Clune, Jeff and Ha, David},
  journal={arXiv preprint arXiv:2408.06292},
  year={2024}
}

@article{wang2026deep,
  title={Deep Energy Method with Large Language Model assistance: an open-source Streamlit-based platform for solving variational PDEs},
  author={Wang, Yizheng and Anitescu, Cosmin and Eshaghi, Mohammad Sadegh and Zhuang, Xiaoying and Rabczuk, Timon and Liu, Yinghua},
  journal={arXiv preprint arXiv:2602.07838},
  year={2026}
}

@article{mcgreivy2024weak,
  title={Weak baselines and reporting biases lead to overoptimism in machine learning for fluid-related partial differential equations},
  author={McGreivy, Nick and Hakim, Ammar},
  journal={Nature machine intelligence},
  volume={6},
  number={10},
  pages={1256--1269},
  year={2024},
  publisher={Nature Publishing Group UK London}
}

@article{wang2025towards,
  title={Towards Unified AI-Driven Fracture Mechanics: The Extended Deep Energy Method (XDEM)},
  author={Wang, Yizheng and Lin, Yuzhou and Goswami, Somdatta and Zhao, Luyang and Zhang, Huadong and Bai, Jinshuai and Anitescu, Cosmin and Eshaghi, Mohammad Sadegh and Zhuang, Xiaoying and Rabczuk, Timon and others},
  journal={arXiv preprint arXiv:2511.05888},
  year={2025}
}

@article{eshaghi2025variational,
  title={Variational physics-informed neural operator (VINO) for solving partial differential equations},
  author={Eshaghi, Mohammad Sadegh and Anitescu, Cosmin and Thombre, Manish and Wang, Yizheng and Zhuang, Xiaoying and Rabczuk, Timon},
  journal={Computer Methods in Applied Mechanics and Engineering},
  volume={437},
  pages={117785},
  year={2025},
  publisher={Elsevier}
}

@article{wang2026pretrain,
  title={Pretrain Finite Element Method: A Pretraining and Warm-start Framework for PDEs via Physics-Informed Neural Operators},
  author={Wang, Yizheng and Hao, Zhongkai and Eshaghi, Mohammad Sadegh and Anitescu, Cosmin and Zhuang, Xiaoying and Rabczuk, Timon and Liu, Yinghua},
  journal={arXiv preprint arXiv:2601.03086},
  year={2026}
}

@article{zhao2025denns,
  title={DENNs: Discontinuity-Embedded Neural Networks for fracture mechanics},
  author={Zhao, Luyang and Shao, Qian},
  journal={Computer Methods in Applied Mechanics and Engineering},
  volume={446},
  pages={118184},
  year={2025},
  publisher={Elsevier}
}

@article{erdogan1963crack,
  title={On the crack extension in plates under plane loading and transverse shear},
  author={Erdogan, Fazil and Sih, GC},
  journal={Journal of basic engineering},
  volume={85},
  number={4},
  pages={519--525},
  year={1963},
  publisher={American Society of Mechanical Engineers Digital Collection}
}

@inproceedings{hussain1974strain,
  title={Strain energy release rate e for a Crack Under Combined Mode I and Mode II},
  author={Hussain, MA and Pu, SL and Underwood$^1$, J},
  booktitle={Fracture analysis: Proceedings of the 1973 national symposium on fracture mechanics, part II},
  volume={560},
  pages={2},
  year={1974},
  organization={ASTM International}
}

@article{sih1974strain,
  title={Strain-energy-density factor applied to mixed mode crack problems},
  author={Sih, George C},
  journal={International Journal of fracture},
  volume={10},
  number={3},
  pages={305--321},
  year={1974},
  publisher={Springer}
}

@article{chen2024crack,
  title={Crack propagation simulation and overload fatigue life prediction via enhanced physics-informed neural networks},
  author={Chen, Zhiying and Dai, Yanwei and Liu, Yinghua},
  journal={International Journal of Fatigue},
  volume={186},
  pages={108382},
  year={2024},
  publisher={Elsevier}
}

@article{kiyani2025predicting,
  title={Predicting crack nucleation and propagation in brittle materials using deep operator networks with diverse trunk architectures},
  author={Kiyani, Elham and Manav, Manav and Kadivar, Nikhil and De Lorenzis, Laura and Karniadakis, George Em},
  journal={Computer Methods in Applied Mechanics and Engineering},
  volume={441},
  pages={117984},
  year={2025},
  publisher={Elsevier}
}

@article{zheng2022physics,
  title={Physics-informed machine learning model for computational fracture of quasi-brittle materials without labelled data},
  author={Zheng, Bin and Li, Tongchun and Qi, Huijun and Gao, Lingang and Liu, Xiaoqing and Yuan, Li},
  journal={International Journal of Mechanical Sciences},
  volume={223},
  pages={107282},
  year={2022},
  publisher={Elsevier}
}

@article{manav2024phase,
  title={Phase-field modeling of fracture with physics-informed deep learning},
  author={Manav, Manav and Molinaro, Roberto and Mishra, Siddhartha and De Lorenzis, Laura},
  journal={Computer Methods in Applied Mechanics and Engineering},
  volume={429},
  pages={117104},
  year={2024},
  publisher={Elsevier}
}

@article{bendsoe1999material,
  title={Material interpolation schemes in topology optimization},
  author={Bends{\o}e, Martin P and Sigmund, Ole},
  journal={Archive of applied mechanics},
  volume={69},
  number={9},
  pages={635--654},
  year={1999},
  publisher={Springer}
}

@article{lin2026physics,
  title={A physics-informed neural network framework for simulating creep buckling in growing viscoelastic biological tissues},
  author={Lin, Zhongya and Bai, Jinshuai and Li, Shuang and Chen, Xindong and Li, Bo and Feng, Xi-Qiao},
  journal={Computer Methods in Applied Mechanics and Engineering},
  volume={452},
  pages={118715},
  year={2026},
  publisher={Elsevier}
}

@article{wang2025transfer,
  title={Transfer learning in physics-informed neurals networks: full fine-tuning, lightweight fine-tuning, and low-rank adaptation},
  author={Wang, Yizheng and Bai, Jinshuai and Eshaghi, Mohammad Sadegh and Anitescu, Cosmin and Zhuang, Xiaoying and Rabczuk, Timon and Liu, Yinghua},
  journal={International Journal of Mechanical System Dynamics},
  volume={5},
  number={2},
  pages={212--235},
  year={2025},
  publisher={Wiley Online Library}
}

@article{amor2009regularized,
  title={Regularized formulation of the variational brittle fracture with unilateral contact: Numerical experiments},
  author={Amor, Hanen and Marigo, Jean-Jacques and Maurini, Corrado},
  journal={Journal of the Mechanics and Physics of Solids},
  volume={57},
  number={8},
  pages={1209--1229},
  year={2009},
  publisher={Elsevier}
}

@article{gerasimov2019penalization,
  title={On penalization in variational phase-field models of brittle fracture},
  author={Gerasimov, Tymofiy and De Lorenzis, Laura},
  journal={Computer Methods in Applied Mechanics and Engineering},
  volume={354},
  pages={990--1026},
  year={2019},
  publisher={Elsevier}
}

@article{hu2021lora,
  title={Lora: Low-rank adaptation of large language models.},
  author={Hu, Edward J and Shen, Yelong and Wallis, Phillip and Allen-Zhu, Zeyuan and Li, Yuanzhi and Wang, Shean and Wang, Liang and Chen, Weizhu and others},
  journal={Iclr},
  volume={1},
  number={2},
  pages={3},
  year={2022}
}

@article{eshaghi2025nows,
  title={Nows: Neural operator warm starts for accelerating iterative solvers},
  author={Eshaghi, Mohammad Sadegh and Anitescu, Cosmin and Valizadeh, Navid and Wang, Yizheng and Zhuang, Xiaoying and Rabczuk, Timon},
  journal={arXiv preprint arXiv:2511.02481},
  year={2025}
}

@article{kahana2023geometry,
  title={On the geometry transferability of the hybrid iterative numerical solver for differential equations},
  author={Kahana, Adar and Zhang, Enrui and Goswami, Somdatta and Karniadakis, George and Ranade, Rishikesh and Pathak, Jay},
  journal={Computational Mechanics},
  volume={72},
  number={3},
  pages={471--484},
  year={2023},
  publisher={Springer}
}

@inproceedings{hao2023gnot,
  title={Gnot: A general neural operator transformer for operator learning},
  author={Hao, Zhongkai and Wang, Zhengyi and Su, Hang and Ying, Chengyang and Dong, Yinpeng and Liu, Songming and Cheng, Ze and Song, Jian and Zhu, Jun},
  booktitle={International conference on machine learning},
  pages={12556--12569},
  year={2023},
  organization={PMLR}
}

@article{wu2024transolver,
  title={Transolver: A fast transformer solver for pdes on general geometries},
  author={Wu, Haixu and Luo, Huakun and Wang, Haowen and Wang, Jianmin and Long, Mingsheng},
  journal={arXiv preprint arXiv:2402.02366},
  year={2024}
}

@article{hughes2005isogeometric,
  title={Isogeometric analysis: CAD, finite elements, NURBS, exact geometry and mesh refinement},
  author={Hughes, Thomas JR and Cottrell, John A and Bazilevs, Yuri},
  journal={Computer methods in applied mechanics and engineering},
  volume={194},
  number={39-41},
  pages={4135--4195},
  year={2005},
  publisher={Elsevier}
}

@inproceedings{deng2009imagenet,
  title={Imagenet: A large-scale hierarchical image database},
  author={Deng, Jia and Dong, Wei and Socher, Richard and Li, Li-Jia and Li, Kai and Fei-Fei, Li},
  booktitle={2009 IEEE conference on computer vision and pattern recognition},
  pages={248--255},
  year={2009},
  organization={Ieee}
}

@article{duprez2025varphi,
  title={$\varphi$-FEM-FNO: a new approach to train a neural operator as a fast PDE solver for variable geometries},
  author={Duprez, Michel and Lleras, Vanessa and Lozinski, Alexei and Vigon, Vincent and Vuillemot, Killian},
  journal={Communications in Nonlinear Science and Numerical Simulation},
  pages={109131},
  year={2025},
  publisher={Elsevier}
}

@article{burman2015cutfem,
  title={CutFEM: discretizing geometry and partial differential equations},
  author={Burman, Erik and Claus, Susanne and Hansbo, Peter and Larson, Mats G and Massing, Andr{\'e}},
  journal={International Journal for Numerical Methods in Engineering},
  volume={104},
  number={7},
  pages={472--501},
  year={2015},
  publisher={Wiley Online Library}
}

@article{thakolkaran2022nn,
  title={NN-EUCLID: Deep-learning hyperelasticity without stress data},
  author={Thakolkaran, Prakash and Joshi, Akshay and Zheng, Yiwen and Flaschel, Moritz and De Lorenzis, Laura and Kumar, Siddhant},
  journal={Journal of the Mechanics and Physics of Solids},
  volume={169},
  pages={105076},
  year={2022},
  publisher={Elsevier}
}

@article{livingston2025inference,
  title={Inference of phase field fracture models},
  author={Livingston, Elizabeth and Srivastava, Siddhartha and Holber, Jamie and Mourad, Hashem M and Garikipati, Krishna},
  journal={Journal of the Mechanics and Physics of Solids},
  pages={106495},
  year={2025},
  publisher={Elsevier}
}

\end{document}